\documentclass[twocolumn]{aastex631}

\usepackage{booktabs}
\usepackage{amsmath}

\begin{document}

\title{SMMAN: quasi-Simultaneous Multi-wavelength Monitoring of gamma-ray-loud AGNs with the Nanshan 26-m radio telescope}

\correspondingauthor{Lang Cui}
\email{cuilang@xao.ac.cn}
\shortauthors{Cui et al.}

\author[0000-0003-0721-5509]{Lang Cui}
\affiliation{State Key Laboratory of Radio Astronomy and Technology, Xinjiang Astronomical Observatory, CAS, 150 Science 1-Street, Urumqi 830011, China}
\affiliation{School of Astronomy and Space Science, University of Chinese Academy of Sciences, No.1 Yanqihu East Road, Beijing 101408, China}
\affiliation{Xinjiang Key Laboratory of Radio Astrophysics, 150 Science 1-Street, Urumqi 830011, China}

\author[0000-0002-2665-0680]{Krishna Mohana A}
\affiliation{State Key Laboratory of Radio Astronomy and Technology, Xinjiang Astronomical Observatory, CAS, 150 Science 1-Street, Urumqi 830011, China}

\author[0000-0001-8221-9601]{Xin Wang}
\affiliation{State Key Laboratory of Radio Astronomy and Technology, Xinjiang Astronomical Observatory, CAS, 150 Science 1-Street, Urumqi 830011, China}
\affiliation{School of Astronomy and Space Science, University of Chinese Academy of Sciences, No.1 Yanqihu East Road, Beijing 101408, China}

\author[0000-0002-8684-7303]{Ning Chang}
\affiliation{State Key Laboratory of Radio Astronomy and Technology, Xinjiang Astronomical Observatory, CAS, 150 Science 1-Street, Urumqi 830011, China}

\author{Guiping Tan}
\affiliation{State Key Laboratory of Radio Astronomy and Technology, Xinjiang Astronomical Observatory, CAS, 150 Science 1-Street, Urumqi 830011, China}

\author[0000-0001-9815-2579]{Xiang Liu}
\affiliation{State Key Laboratory of Radio Astronomy and Technology, Xinjiang Astronomical Observatory, CAS, 150 Science 1-Street, Urumqi 830011, China}

%% Note that the \and command from previous versions of AASTeX is now
%% depreciated in this version as it is no longer necessary. AASTeX
%% automatically takes care of all commas and "and"s between authors names.

%% AASTeX 6.31 has the new \collaboration and \nocollaboration commands to
%% provide the collaboration status of a group of authors. These commands
%% can be used either before or after the list of corresponding authors. The
%% argument for \collaboration is the collaboration identifier. Authors are
%% encouraged to surround collaboration identifiers with ()s. The
%% \nocollaboration command takes no argument and exists to indicate that
%% the nearby authors are not part of surrounding collaborations.

%% Mark off the abstract in the ``abstract'' environment.

\begin{abstract}
Active Galactic Nuclei (AGNs) are characterized by strong temporal flux density variability across the electromagnetic spectrum, offering insights into the complex physical processes governing accretion and plasma outflows. To systematically investigate AGNs flux density variability in radio bands, a long-term program was initiated in late 2016: quasi-Simultaneous Multiwavelength Monitoring of gamma-ray-loud AGNs with the Nanshan 26-m radio telescope (SMMAN). This work presents the first data release of the SMMAN program, spanning over eight years from 2016 to 2024 with observations at 4.8 and 23.6 GHz. The SMMAN sample includes 131 northern ($\delta >\sim0^{\circ}$) sources selected from the Fermi Large Area Telescope third source catalog. The characteristics of variability, spectral index, luminosity, and $\gamma$-ray loudness factor are examined for different AGN classes within the sample. Target sources exhibit stronger variability at 23.6 GHz compared to 4.8 GHz, with BL Lac objects being more variable than flat-spectrum radio quasars (FSRQs). BL Lacs generally have flatter radio spectra, while FSRQs, blazar candidates of uncertain type (BCUs), and radio galaxies (RDGs) span a wider range from flat to steep. FSRQs are more radio-luminous than BL Lacs and other classes, with BCUs intermediate and RDGs generally fainter. FSRQs and BL Lacs have higher $\gamma$-ray loudness factors than RDGs, while BCUs have intermediate values. The SMMAN dataset, incorporated with other historical and ongoing monitoring programs, will provide a unique opportunity to investigate the evolution of spectral energy distributions, search for quasi-periodic oscillations, and analyze supermassive black hole binary systems.
\end{abstract}

%% Keywords should appear after the \end{abstract} command.
%% The AAS Journals now uses Unified Astronomy Thesaurus concepts:
%% https://astrothesaurus.org
%% You will be asked to selected these concepts during the submission process
%% but this old "keyword" functionality is maintained in case authors want
%% to include these concepts in their preprints.

\keywords{\href{https://astrothesaurus.org/uat/83}{Astronomy databases (83)};
          \href{https://astrothesaurus.org/uat/16}{Active galactic nuclei (16)};
          \href{https://astrothesaurus.org/uat/1390}{Relativistic jets (1390)};
          \href{https://astrothesaurus.org/uat/1340}{Radio continuum emission (1340)};
          \href{https://astrothesaurus.org/uat/637}{Gamma-rays (637)}
}

%% From the front matter, we move on to the body of the paper.
%% Sections are demarcated by \section and \subsection, respectively.
%% Observe the use of the LaTeX \label
%% command after the \subsection to give a symbolic KEY to the
%% subsection for cross-referencing in a \ref command.
%% You can use LaTeX's \ref and \label commands to keep track of
%% cross-references to sections, equations, tables, and figures.
%% That way, if you change the order of any elements, LaTeX will
%% automatically renumber them.
%%
%% We recommend that authors also use the natbib \citep
%% and \citet commands to identify citations.  The citations are
%% tied to the reference list via symbolic KEYs. The KEY corresponds
%% to the KEY in the \bibitem in the reference list below.

\section{Introduction} \label{sec:intro}
Galaxies that contain a highly bright, compact region at their center, called active galactic nuclei (AGNs), are known as active galaxies \citep{urry1995}. It is found that the high luminosity at the central compact region is powered by supermassive black holes (SMBH; mass $\sim10^{8}-10^{9}M_{\odot}$) via processes known as accretion \citep{frank1992book}. Active galaxies are also found to possess relativistic jet structures emanating from the central region \citep{rees1966Natur}.
Active galaxy's multiwavelength emission is dominated by non-thermal radiation
extending from radio to $\gamma$-ray bands exhibiting intense and rapid flux density variations. In particular, a class of AGNs called blazars with a relativistic jet pointing towards Earth show extreme variability \citep{blandford1978,blandford1979ApJ}.
Due to the small jet viewing angle, the blazar's observed emission is highly Doppler boosted and outshines all other components, thus making them ideal candidates to study the physical conditions and emission processes in relativistic jets.
From an observational standpoint, blazars are categorized into two sub-classes: flat-spectrum radio quasars (FSRQs) and BL Lacertae (BL Lacs) objects. FSRQs exhibit emission line equivalent widths (EW) \textgreater $5$ \AA, while BL Lacs show EW \textless $5$ \AA ~\citep{stocke1991,stickel1991}. As per the classification suggested by \citet{ghisellini2011}, which relies on the luminosity of the broad-line region (BLR), FSRQs exhibit a higher BLR luminosity ($L_{\mbox{BLR}}/L_{\mbox{Edd}} >5\times10^{-4}$) compared to BL Lacs, with $L_{\mbox{Edd}}$ representing the Eddington luminosity of the supermassive black hole located at the core of the AGN. The flux variability timescales in blazars range from minutes to years \citep[e.g.,][and references therein]{boettcher2004,foschini2013,orienti2014,boettchergalaxy2019} and were classified into three types caused by different mechanisms \citep{fan2005ChJAS}. These variations occur as flares or the quiescent/low-activity states. Another key feature of active galaxies is that their broadband spectral energy distribution (SED) from radio to $\gamma$-rays exhibits two-hump structure \citep[e.g.][]{fossati1998}. In addition, it is observed that AGNs/blazars exhibit high optical and radio polarization during flaring periods along with their polarization, showing intensive variability not only in terms of fraction but also in terms of the orientation of the polarization plane \citep[e.g.][]{jorstad2007AJ,marscher2010ApJ,blinov2021MNRAS}.\\

Studying the physical properties of AGNs through multi-frequency data has been found to be one of the most effective ways in the past, which has been employed in several studies (\citealt{marscher2016Galax,boettchergalaxy2019}, and the references therein).
A key feature of AGNs is that the flux density observed from them shows variations across different wavebands over varying time intervals. Further, it is observed that the behavior of the broadband
flares of blazars is quite complex. The literature study points out that such flaring
activities are not always simultaneous in different wavebands. Also, it is important to study the long-term activity of blazars. From the multiband observations of the AGNs, it is noticed that the variability patterns in different wavebands are sometimes correlated and sometimes not (e.g. \citealt{liodakis2018}, and references therein). Comprehending the physical basis for the observed variability remains one of the key unresolved issues in the study of AGNs. Temporal studies can be an efficient method to extract the time evolution trends of observed variability and to identify plausible connections across various blazar emission bands (e.g. \citealt{liodakis2018}, and references therein).
Therefore, a highly effective approach to comprehending jet emission and its dynamics involves examining the alterations in the physical characteristics of AGNs/blazars that are necessary to generate variations in multi-frequency light curves. Hence, simultaneous monitoring of the AGNs across different wavelengths is important to discern the physical mechanisms in them.\\

The ``{\sl Fermi Gamma-ray Space Telescope}'' (hereafter {\sl Fermi}) was launched in
June of 2008 and the ``Large Area Telescope'' (LAT;~\citealt{atwood2009ApJ}) on-board provides a unique opportunity for the systematic study of AGNs/blazar variability \citep{boettchergalaxy2019,principe2022,casaburo2025}. {\sl Fermi}-LAT with its initial four years of observation, the source catalog (3FGL;~\citealt{acero3fgl}) includes blazars that account for more than $\sim60\%$ of the identified sources. The most recent update of the fourth source catalog (4FGL-DR4) spans more than 14 years and includes a total of 7,195 sources within the energy range of 50 MeV to 1 TeV \citep{abdollahi2020ApJS,ballet2023}. Approximately 56\% of these sources are associated with AGNs, including blazars. Conducting a comprehensive monitoring campaign of AGNs across multiple bands (radio to $\gamma$-ray) and carrying out temporal and spectral analysis is essential to understand the causes of the underlying flux density variations in them. Under this motivation, prior to the launch of the {\sl Fermi} and after it, there are a few ongoing and historical AGNs multiband monitoring programs\footnote{\url{https://www.cv.nrao.edu/MOJAVE/blazarlist.html}}.
A few of the major radio band surveys are briefly introduced below, with emphasis on the importance of multi-frequency radio monitoring projects. The University of Michigan Radio Astronomy Observatory's (UMRAO;~\citealt{aller1985ApJS}) 26-m paraboloid telescope was devoted to monitoring the flux density and polarization of blazar jets from the mid-1960s until June 2012 \citep{aller2017Galax}. This facility provided centimeter-band variability over decades at 14.5, 8.0, and 4.8 GHz. The UMRAO data have been used in numerous extensive studies concerning flaring blazars and to acquire information regarding various jet and shock parameters that are not directly observable \citep{aller2014ApJ,hughes2015ApJ,aller2016Galax} and also in several multiband temporal and spectral studies \citep[e.g.][]{krishna2024MNRAS,mohana2025ApJ}. UMRAO observations were also used for the calibration and interpretation of the less frequently sampled Very Long Baseline Array (VLBA) imaging data (\citealt{aller2017Galax}, and references therein). It is to be noted that the long-term monitoring data set could help to search for a supermassive black hole binary (SMBHB)
candidates by identifying the periodic nature in the light curves \citep[e.g.][]{readhead2025,molina2025}.\\

The radio telescope RATAN-600 \citep{parijskij1993IAPM} in Zelenchukskaya, Russia, which is an ongoing survey, has regularly monitored AGNs, including blazars, since 2005.
The observations were carried out with the six frequency bands at 1.25, 2.25, 4.7, 8.2, 11.2, and 22.3 GHz, simultaneously (within 1--2 min) when a source moves along the focal line where the receivers are located. These data formed the basis for the first version of the BLcat \citep{mingaliev2014}, and the latest updated version of this catalog is ``RATAN-600 multi-frequency catalog of blazars'' \citep{sotnikova2022AstBu}, which includes about 1830 source. Using these catalogs with information of flux densities, spectral indices, variability, and radio luminosity for the large blazar list, several studies to understand the flare activity and synchrotron radio spectra of AGNs were reported \citep[e.g.][]{mufakharov2015MNRAS,sotnikova2019AstBu,larinov2020MNRAS,sotnikova2021MNRAS,krishna2024MNRAS,kunert2025ApJS}. The Fermi-GST AGN Multi-frequency Monitoring Alliance (F-GAMMA) programme \citep{fuhrmann2007AIPC,angelakis2010,fuhrmann2014MNRAS}, conducted at the Effelsberg telescope and IRAM 30 m telescope in Germany, performed nearly monthly broadband radio monitoring of selected blazars from January 2007 to January 2015 \citep{angelakis2019}. The monitoring was carried out at multi-frequency from 2.64 to 43 GHz, which includes the frequencies of 2.64, 4.85, 8.35, 10.45, 14.6, 23.05, 32, 43, 86.2, 142.3, and 228.9 GHz. Using initial 3.5~yr data, a cross-correlation analysis was carried out by \citet {fuhrmann2014MNRAS} between the radio and $\gamma$-ray light curves of 54 {\sl Fermi}-bright blazars, with the aim of determining the exact location of the $\gamma$-ray emission region. \citet{fuhrmann2016A&A} reported a multi-frequency study of AGNs in {\sl Fermi} era to examine whether the $\gamma$-ray loudness is related to the radio variability. The 2007-2015 multi-frequency radio data published in \citet{angelakis2019} is used in several temporal and spectral studies of blazars \citep[e.g.][]{paliya2021ApJS,kim2022MNRAS,krishna2024MNRAS}.\\

The Owens Valley Radio Observatory (OVRO) blazar monitoring  program \citep{richards2011ApJS} offers an extensive data collection to examine AGNs variability at 15 GHz. It stands as the most comprehensive and sensitive radio monitoring survey of blazars, which has been in progress since 2008. Currently, the complete sample encompasses approximately 1850 sources, each of which is observed at a frequency of roughly twice a week. Observations from OVRO helped make progress in blazar jet physics through cross-correlations in the time domain between {\sl Fermi}-LAT $\gamma$-ray and 15 GHz light curves. It was found that there is a close connection between $\gamma$-ray and 15 GHz blazar emission \citep{maxmoerbeck2014MNRAS,hovatta2015MNRAS,pushkarev2019MNRAS}.
In 2004, a program was initiated to monitor blazars at the 32 m radio antennas of the INAF-Institute of Radioastronomy (IRA), located in Medicina and Noto, Italy \citep{bach2007A&A}.
Beginning in 2004, 47 targets were observed with a monthly cadence at 5, 8, 24, and 43 GHz.
The Radio Observations of Blazars with INaf telescopes (ROBIN) program has continued to thrive over the years and remains active at frequencies of 8 and 24 GHz, having reached 20 years of operation \citep{marchili2025}. The data accumulated through the program have been used in numerous studies and publications, primarily in relation to Whole Earth Blazar Telescope (WEBT) multi-frequency campaigns \citep[e.g.][]{raiteri2007A&A,abdo2010Natur,carnerero2015MNRAS,raiteri2017Natur,raiteri2024A&A}.
The monitoring program has facilitated the examination of how the variability characteristics and SED of blazars evolve over time (\citealt{marchili2025}, and references therein).\\

At the Aalto University Mets\"ahovi Radio Observatory (MRO, Finland), several hundred AGNs and a few galactic microquasars have been monitored at 22, 37, and 87 GHz since 1980 using the 13.7 m radio telescope \citep{teraesranta1998}. At present, there are 271 sources in the MRO monitoring program with data at 22 and 37 GHz, where the latter includes observations spanning up to 42 years. Data from this ongoing program have contributed to research on the long-term behavior, statistical properties, source populations, and radio variability models of AGNs. This monitoring has enabled studies of the radio to $\gamma$-ray connection in blazars, as well as radio versus optical behavior and multi-frequency variability models \citep[e.g.][]{hovatta2007A&A, kankkunen2025A&A}. The Submillimeter Array \citep[SMA;][]{gurwell2007} in Hawaii, United States, has been monitoring the quasars to determine the flux densities and variability at 230 and 345 GHz since the beginning of 2003 with 437 sources at present. The observations from this program were used in different long-term timing and spectral studies of AGNs \citep[e.g.][]{villata2009A&A,fuhrmann2014MNRAS,jorstad2016Galax}.\\

The above-mentioned radio monitoring programs were successful in providing data, complementing other multiband observations. The studies above show that a comprehensive investigation into radio and radio/$\gamma$-ray populations helps to address the enduring questions about AGNs physical attributes. These include the location, structure, and radiative characteristics of the $\gamma$-ray emission region. Additionally, such research addresses the collimation, composition, particle acceleration, and emission mechanisms in AGNs/blazar jets. Also, to detect the presence of shorter and longer time scales of quasi-periodic oscillations (QPOs) in the AGNs light curves, it is essential to have long-term adequate continuous data \citep[e.g.][]{king2013MNRAS,li2023ApJ}.
Blazars have also been identified as candidate sources of PeV-energy neutrinos \citep{icecube2018Sci,buson2022ApJ}. To thoroughly explore blazars and the link between their radio emissions and neutrinos, multiband, high-cadence, long-term observations are essential \citep{kovalev2020AdSpR,plavin2020ApJ,hovatta2021A&A}.
Motivated by this, the quasi-Simultaneous Multiwavelength Monitoring of gamma-ray-loud AGNs with the Nanshan 26-m radio telescope (SMMAN) program operated by Xinjiang Astronomical Observatory (XAO) in China started in late 2016 and remains active. The information obtained from the SMMAN program, along with {\sl Fermi}-LAT observations, will enable to systematically determine the radio and its connection with other multiband observational characteristics of the gamma-ray-loud AGNs population. In the following, we discuss the SMMAN dataset in 4.8 and 23.6 GHz. The details of the source sample and the radio data analysis are described in Section~\ref{sec:observation}. In Section~\ref{sec:results_and_discussion}, we present and discuss the results. Finally, we summarize and conclude the findings of our work in Section~\ref{sec:summary}.\\

\section{Observation and data reduction}\label{sec:observation}
\subsection{Facility}
The SMMAN program is carried out with the NanShan 26-m Radio Telescope (NSRT), which is located on the Nanshan Mountain, with the geographic coordinates of 43°28'17"N, 87°10'41"E, and an altitude of 2080 meters, $\sim70$ km southwest of Urumqi city, China. Apart from single-dish scientific observations, NSRT has been involved in a large number of Very Long Baseline Interferometry (VLBI) observations in the last $\sim30$ years, as a formal member of the European VLBI Network (EVN). The NSRT has an on-axis, single-beam optical configuration and four receivers operating at L-band (1-2 GHz), S/X-band (2.3/8.6 GHz), C-band (4-8 GHz), and K-band (22-24 GHz). The receivers used in the SMMAN program operate in the C and K bands, centered at 4800 MHz and 23600 MHz, with terminal recording bandwidths of 500 MHz for both. The main technical specifications of the receivers used in the SMMAN program are listed in Table~\ref{tab:recerver_params}.

\begin{table}[h!]
\caption{Main technical specifications of the receivers used in the SMMAN program.}
\centering
\begin{tabular}{lrcc}
\hline \hline
Parameters       &               & C-band           & K-band         \\
\hline
Center frequency &[MHz]      &  4800   &  23600  \\
Terminal bandwidth &[MHz]            &  500   &  500   \\
System temperature &[K]   &  26  &  45   \\
Reference signal &[K]     &  1.7   &  2.5   \\
FWHP/Beam             &[arcsec]           &  546  &  105   \\
Aperture efficiency &[\%]  &  65  &  62   \\
Conversion Factor &[K/Jy]    &  0.174  &  0.208  \\
\hline
\end{tabular}
\label{tab:recerver_params}
\end{table}

\subsection{Source selection and monitoring schedule}
\begin{figure}
\centering
\includegraphics[width=\columnwidth]{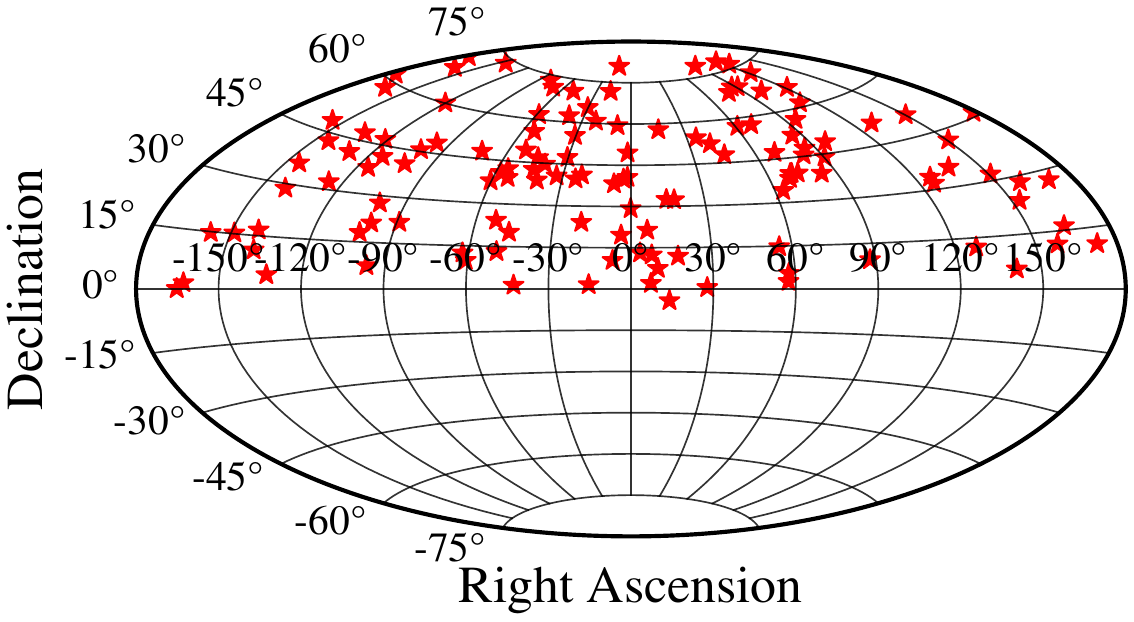}\\
\includegraphics[width=\columnwidth]{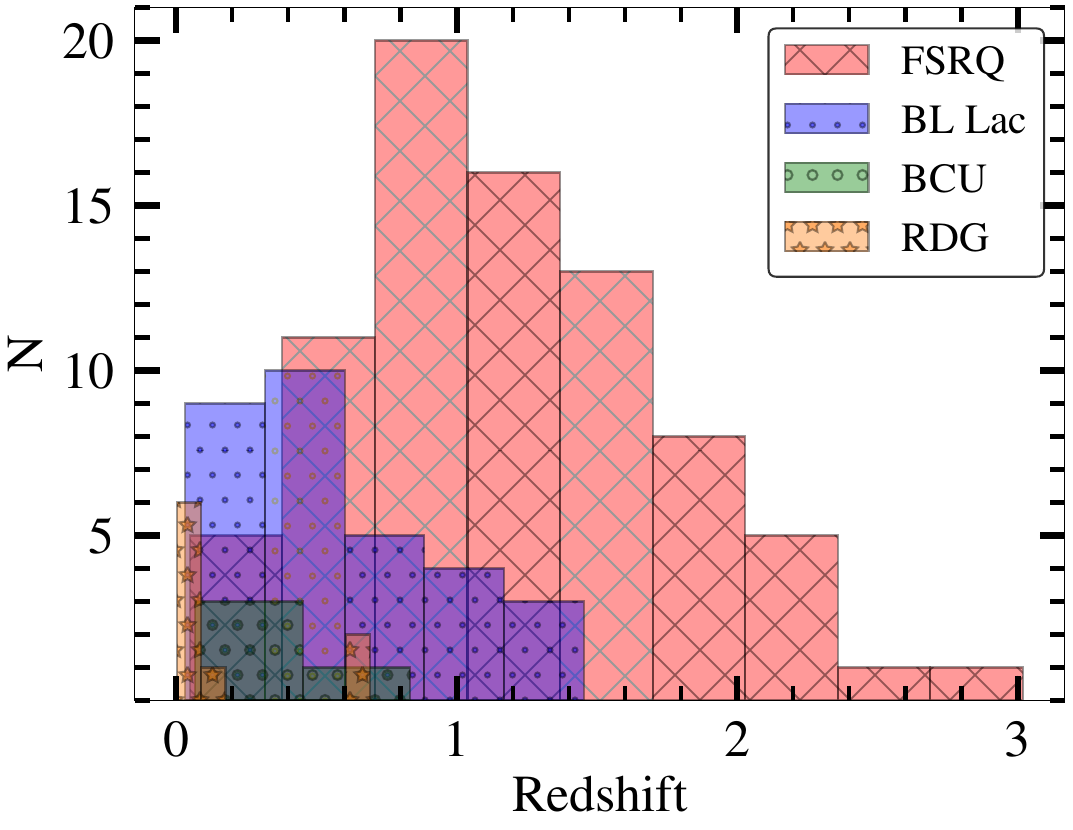}
\caption{Top: the Hammer-Aitoff projection of the SMMAN monitored sources in sky distribution (in equatorial coordinates). Bottom: the redshift distribution of the sources in the SMMAN program. The redshift values were obtained from the NASA/IPAC Extragalactic Database (NED) and the SIMBAD astronomical database \citep{wenger2000}. Redshift measurements are available for 124 of the 131 sources in our monitored sample, of which 121 are spectroscopic and three are photometric.}
\label{fig:sky_dist_and_redshift}
\end{figure}

Our goal is to monitor those bright variable blazars, the majority of which have been captured by the {\sl Fermi}-LAT. Therefore, the initial source sample was established by filtering the {\sl Fermi}-LAT 3FGL catalog \citep{acero3fgl} and cross-matching with the NRAO VLA Sky Survey (NVSS)\footnote{\url{https://www.cv.nrao.edu/nvss/}} by putting a limit on flux density at 1.4 GHz ($S_{1.4\text{ GHz}}) > 300\text{ mJy}$, point-like structure, and source declination (J2000) $>\sim0^\circ$.
Finally, our SMMAN sample contains 131 gamma-ray-loud AGNs, which include 113 blazars (FSRQs = 81 and BL Lacs = 32), nine active galaxies of uncertain type (bcu), seven radio galaxies (rdg), one compact steep spectrum quasar (css), and one steep spectrum radio quasar (ssrq), respectively.
The distribution of SMMAN sources in the sky and their redshift distribution is illustrated in Figure~\ref{fig:sky_dist_and_redshift}.
The redshift values were obtained from the NASA/IPAC Extragalactic Database (NED) and the SIMBAD astronomical database \citep{wenger2000}. Redshift measurements are available for 124 of the 131 sources in our monitored sample, of which 121 are spectroscopic and three are photometric.
These sources were quasi-simultaneously monitored in the C and K bands using the 26-meter NSRT for the last approximately eight years (2016-2024), with a cadence of about one month.\\

\subsection{Data Acquisition and Reduction}
Flux density measurements with single-dish antennas can be performed through multiple modes, such as on-off, mapping, and cross-scan. Considering the monitoring efficiency and measurement accuracy, our SMMAN program has adopted the cross-scan mode. The cross-scan mode is achieved by performing multiple sub-scans of the target source in two orthogonal directions. Our observations were carried out with 4+4 sub-scans in azimuth and elevation over the target’s position, respectively. The data were acquired in this manner at both C-band (4.8 GHz) and K-band (23.6 GHz). For a single sub-scan, the scan width is set to 5 times the beam width, which is $\sim50$ and $\sim10$ arcmin for C-band and K-band, respectively. The total sampling time is set to 20 to 30 seconds for each sub-scan, while the sampling time for each data point is fixed at 64 ms during scanning.\\

\begin{figure*}[ht!]
\centering
\includegraphics[scale=0.65]{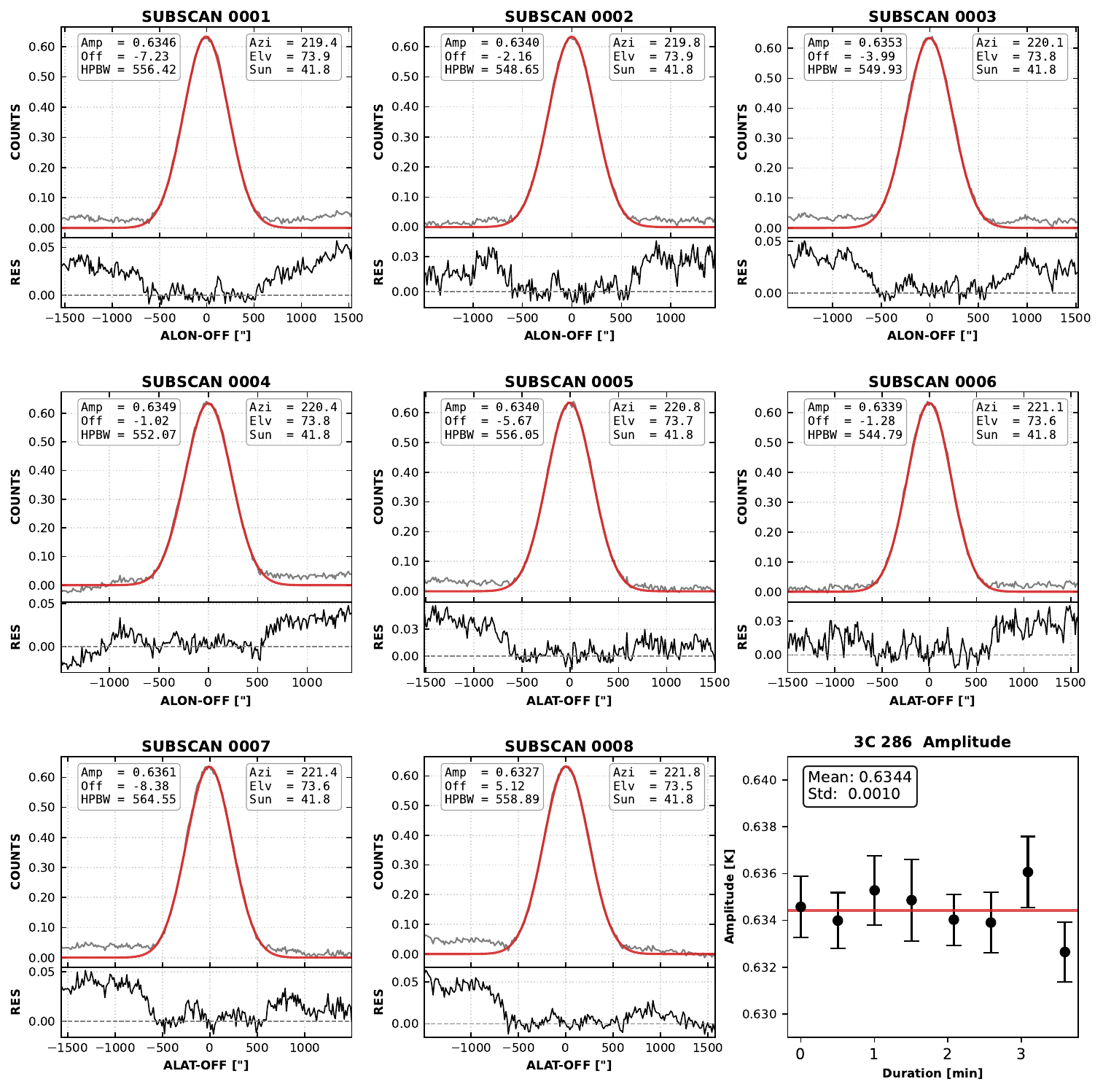}
\caption{Example of a 4+4 cross-scan observation of the calibrator source 3C 286 at 4.8 GHz. The first four subplots display longitude (azimuthal) subscans, followed by the next four showing latitude (elevational) subscans. For each of these eight subplots, the upper panel presents the subscan results, with the red curve representing the Gaussian fit model and the grey line indicating the actual observed data. The lower panel beneath each subscan plot shows the residuals of the model fit. The final subplot summarizes the fitted Gaussian amplitudes from all individual subscans, with a solid red horizontal line indicating the mean amplitude.}
\label{fig:gaussian_fit}
\end{figure*}

The target samples are mainly point sources for the 26-meter NSRT at C and K bands. A sub-scan across the target is simply equal to a beam-wide Gaussian, which is the convolution of the source and the telescope beam. Given the structural perturbations induced by the telescope scanning motions, a Gaussian kernel with a full width at half maximum (FWHM) comparable to the beam size was adopted for scan data model fitting. During cross-scanning, the response attenuation in the orthogonal direction is determined by the Gaussian peak offset along both scan axes, thereby allowing for correcting flux density measurement reduction caused by pointing deviation. As an example, such a cross-scan mode is illustrated in Figure~\ref{fig:gaussian_fit}.\\

The data reduction and calibration process includes Gaussian fitting, pointing correction, gain correction, and absolute flux density calibration. Gaussian fitting is used to subtract the baseline of the system temperature drifting and extract the Gaussian-like signals from the sub-scans. Pointing correction is to calibrate the antenna pointing error with each other direction's offset in the scans. After that, we subsequently conducted atmospheric opacity and antenna gain calibration to further correct the elevation gain effect. Note that the systematic error possibly arising from solar elongation dependence was also accounted for during the pointing correction step. Finally, the absolute flux density was tuned using the calibration factors generated by three primary calibrators, 3C 48, NGC 7027, and 3C 286 \citep{1977A&A....61...99B,1994A&A...284..331O,2008ApJ...681.1296Z,2015A&A...575A..55A}. The observed average flux density for these calibrator sources during the SMMAN program is presented in Table~\ref{tab:calibrators}. \\

\begin{table}[h!]
  \caption{The observed average flux densities (in unit of Jy) of the standard calibrators during the SMMAN program.}
  \label{tab:calibrators}
  \centering
  \begin{tabular}{cccc}
    \hline\hline
    $\nu$ &\multicolumn{3}{c} {Source} \\
    (GHz)       &3C\,48  &3C\,286  &NGC\, 7027\\
    \hline
    4.8  &5.63 (0.11)    &7.51 (0.16)   &5.37 (0.11)   \\
    23.6  &1.26 (0.19)   &2.43 (0.22)     &5.40 (0.39)     \\
    \hline
  \end{tabular}
  \tablecomments{The values quoted inside the parentheses are the standard deviation of the measurements.}
\end{table}

All the data reduction and calibration processes above are conducted with the NSRT radio continuum data processing software \texttt{Data Analysis Application for Radio Continuum (DAARC)}\footnote{\url{https://github.com/junliu/daarc}}, which is developed by our team in Python language.\\

\subsection{$\gamma$-ray data}\label{data:gamma_ray}
We used the monthly binned $\gamma$-ray flux and spectral index data from the {\sl Fermi}-LAT Light Curve Repository\footnote{\url{https://fermi.gsfc.nasa.gov/ssc/data/access/lat/LightCurveRepository/}} (LCR;~\citealt{abdollahi2023ApJS}) in the energy range of $0.1–100$ GeV, covering the same time period as the radio observations (2016 November 1 to 2024 November 30) presented in this study. For a detailed description of the unbinned maximum-likelihood data analysis, we refer to \citet{abdollahi2020ApJS}. In this work, we used the light curve obtained by allowing the spectral slope to vary freely during the fitting process and the fit results of only the convergent likelihood analysis.
The significance of a source detection is evaluated using the test statistic ${\rm TS} = 2\times {\rm log} (L_1/L_0)$, where $L_0$ is the likelihood of the null model (no source) and $L_1$ is the likelihood of the model that includes the source. In the monthly binned light curve, we only consider those bins with detection significance of $\geq 2\sigma$ (${\rm TS}~\geq~4$), and the upper limit values (for non-detection bins) were not included. It is important to note that performing a full {\sl Fermi}-LAT data analysis for 131 SMMAN sources is computationally demanding and time-consuming. Therefore, we used the available light curves from the LCR. These light curves of various $\gamma$-ray sources from this repository have been used in previous research studies \citep[e.g.][]{digesu2023NatAs,wang2023RAA}.\\

\section{Results and Discussion}\label{sec:results_and_discussion}
\begin{figure}[ht!]
\centering
\includegraphics[width=\columnwidth]{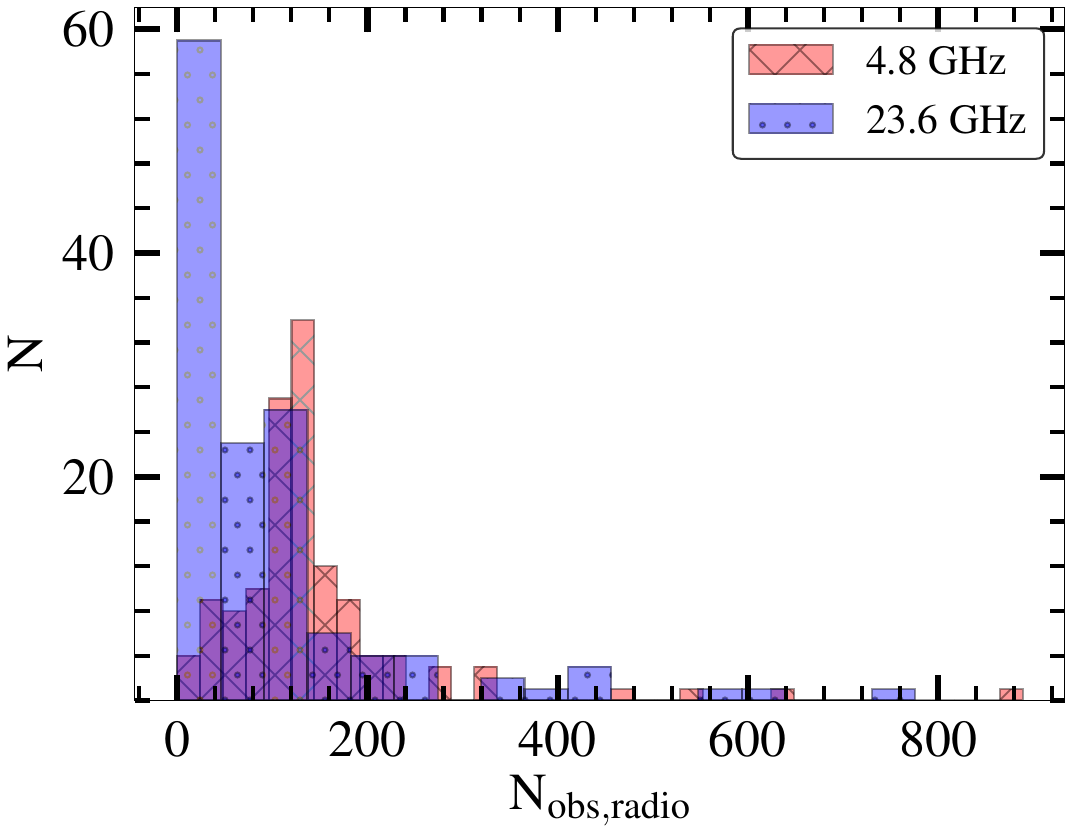}
\caption{The number of radio measurements for the SMMAN sources (131 objects).}
\label{fig:obs_dist}
\end{figure}

\begin{figure*}[ht!]
	\centering
	\includegraphics[scale=0.45]{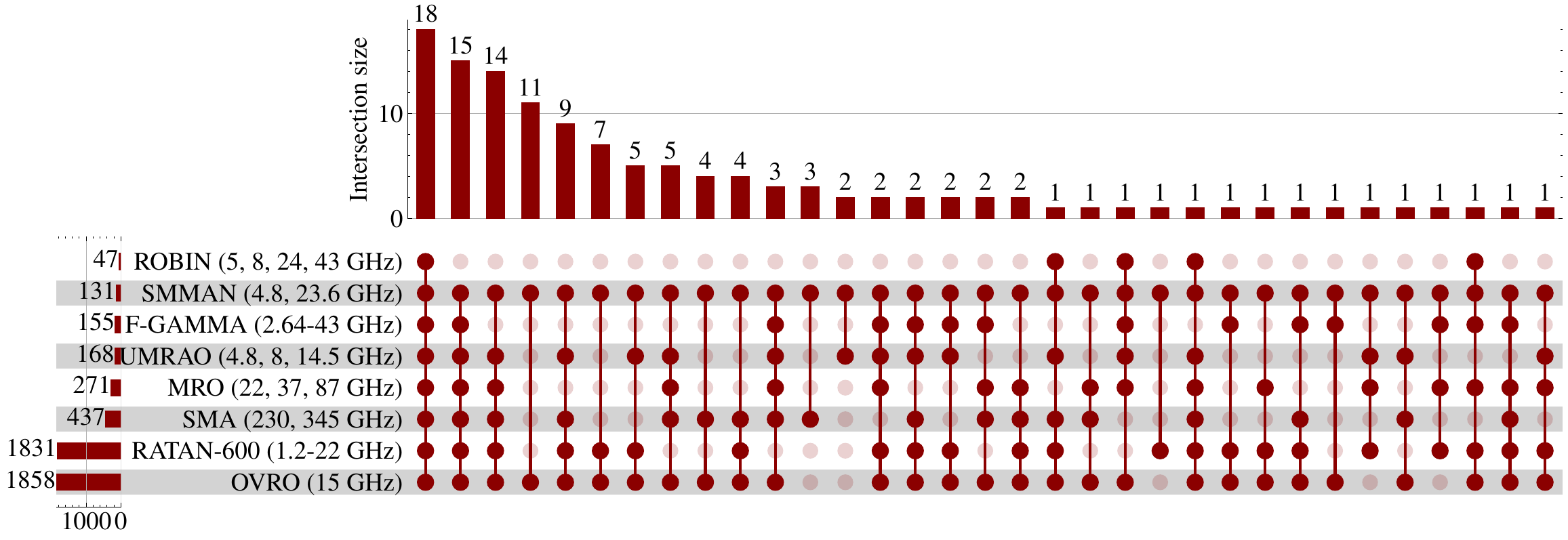}
	\caption{UpSet plot describing the overlapping of the SMMAN sources with other monitoring programs. Reading chart: the horizontal set size bar chart displays the total number of sources within each distinct monitoring program. Vertical intersection size bar chart, positioned at the top, illustrates the number of sources of each particular intersection or combination of monitoring programs. The solid dots and lines (linked by a vertical line) signify the sources that belong to a particular intersection.}
    \label{fig:upset_plot_SMMAN_srcs}
\end{figure*}

The basic information about the sources that were monitored in this work is listed in Table~\ref{tab:list_of_monitored_srcs} of Appendix~\ref{Appendix A:SMMAN_srcs}.
Note that in our analysis and discussion part, we have considered the different AGN classes from Table~\ref{tab:list_of_monitored_srcs} of the Appendix~\ref{Appendix A:SMMAN_srcs} as FSRQ (FSRQ+fsrq), BL Lac (BLL+bll), BCU (bcu), and RDG (RDG+rdg+css+ssrq). Here, the AGNs class is defined according to the {\sl Fermi} Large Area Telescope Fourth Source Catalog Data Release 4 (4FGL-DR4;~\citealt{ballet2023}), with designations in capital letters representing confirmed identifications, and those in lower-case letters denoting associations.
The number of radio measurements for the SMMAN sources between 1 November 2016 and 30 November 2024, as presented in this work, is shown in Figure~\ref{fig:obs_dist}.
Most sources have more observations at 23.6 GHz than at 4.8 GHz, and some sources have exceptionally large datasets with more than 400 measurements.
All histogram figures presented in this work use binning based on Knuth's optimal, data-driven method \citep{knuth2006}. The overlap of the SMMAN sources with other historical surveys or ongoing monitoring programs is displayed in Figure~\ref{fig:upset_plot_SMMAN_srcs}.
From this plot, one can easily identify the number of sources having counter monitoring data for the SMMAN sample, which is useful for further studies that could cover more multiband radio observations.
The sources that overlap among the SMMAN targets and UMRAO, F-GAMMA, RATAN-600, OVRO, ROBIN, MRO, and SMA monitoring programs are shown in  Table~\ref{tab:smman_overlap_srcs} of Appendix~\ref{Appendix A:SMMAN_srcs}.\\

\begin{figure*}[ht!]
	\centering
	\includegraphics[scale=0.55]{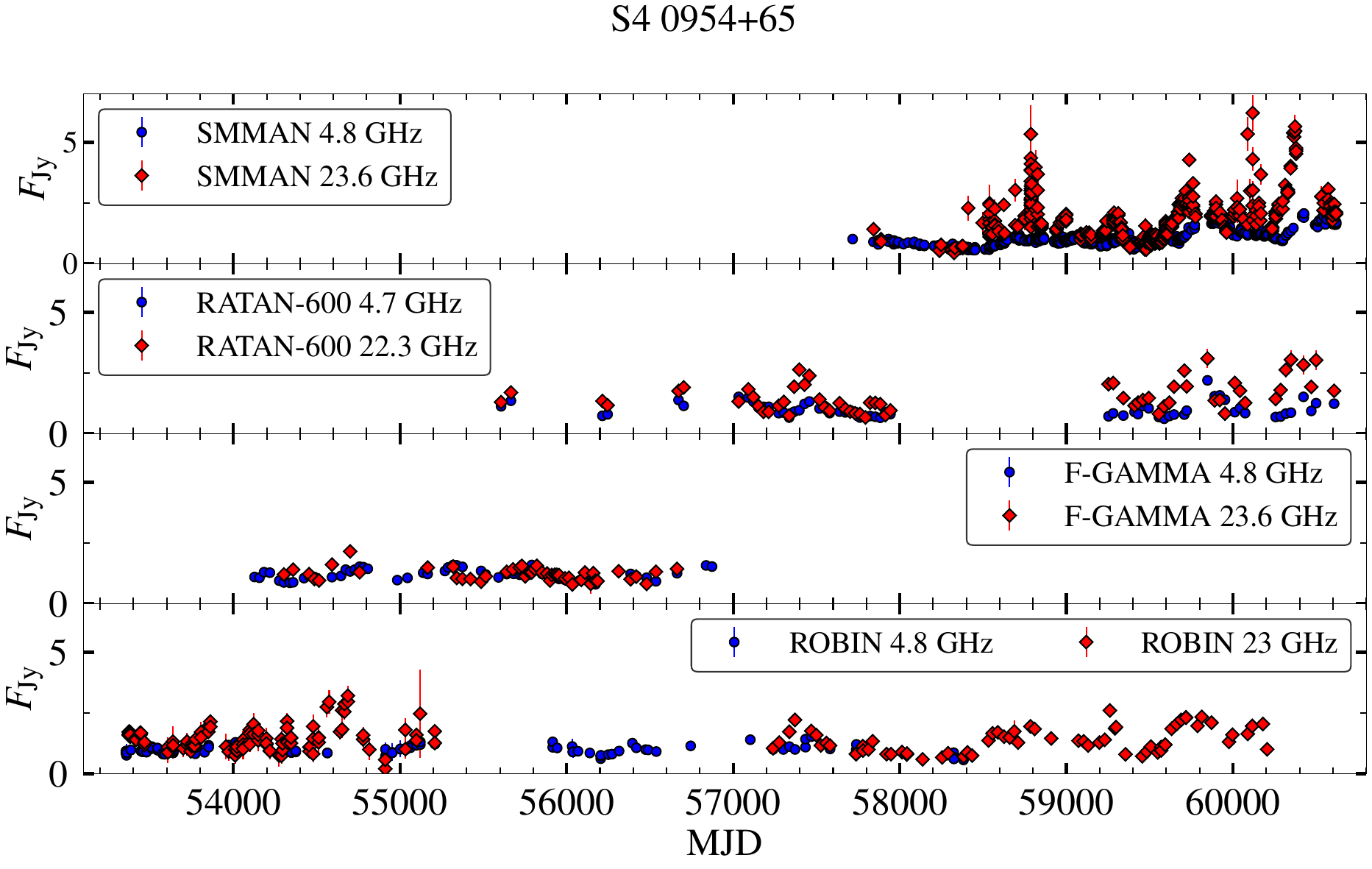}
	\caption{Multiband light curve of S4 0954+65 using SMMAN, RATAN-600, F-GAMMA, and ROBIN monitoring program data at  $\sim4.8$ and $\sim23$ GHz.}
    \label{fig:multiplot_S40954+65}
\end{figure*}

To depict the importance of the SMMAN monitoring program, a long-term multiband light curve of the BL Lac object S4 0954+65 using the data from the SMMAN, RATAN-600\footnote{\url{https://www.sao.ru/blcat/}}\citep{mingaliev2014,sotnikova2022AstBu}, F-GAMMA~\citep{angelakis2019,2019yCat..36260060A} and ROBIN~\citep{marchili2025,2025yCat..37020140M} monitoring program is shown in Figure~\ref{fig:multiplot_S40954+65}. The plot clearly shows that the SMMAN program has covered a greater number of observations after 2016 November compared to the other surveys (publicly available), which is important due to the fact that adequate continuous data are essential for long-term temporal and spectral studies, and the detection of QPOs. In addition, a part of the data from the SMMAN program has been used in systematic studies of a few individual blazars in earlier research works \citep[e.g.][]{kun2022ApJ,krishna2024MNRAS,kun2024ApJ,mohana2025ApJ}, which also showcase the significance of the SMMAN program along with other multiwavelength observations.\\

\subsection{Variability and Other Physical Parameter Estimations}
To quantify the variability of sources, at each frequency band, the variability index ($V_{S}$) is calculated using the formula adopted from \citet{Aller1992ApJ}, defined as
\begin{equation}
\label{variability}
{V}_{S}=\frac{(S_{i}-\sigma_{i})_{\rm max}-(S_{i}+\sigma_{i})_{\rm min}}
{(S_{i}-\sigma_{i})_{\rm max}+(S_{i}+\sigma_{i})_{\rm min}},
\end{equation}
where $S_{\rm max}$ and $S_{\rm min}$ are the maximum and minimum values of the flux density over all epochs of observations, while $\sigma_{\rm max}$ and $\sigma_{\rm min}$ are their root-mean-square errors.
The uncertainty $\Delta\,V_S$ of the variability index was calculated as
\begin{equation}
    \Delta\,V_S = \frac{2\,S_{\rm min}\,(\sigma_{S_{\rm min}}+\sigma_{S_{\rm max}})}{(S_{\rm min}+S_{\rm max})^{2}}
\end{equation}

\begin{figure}[h!]
	\centering
	\includegraphics[width=\columnwidth]{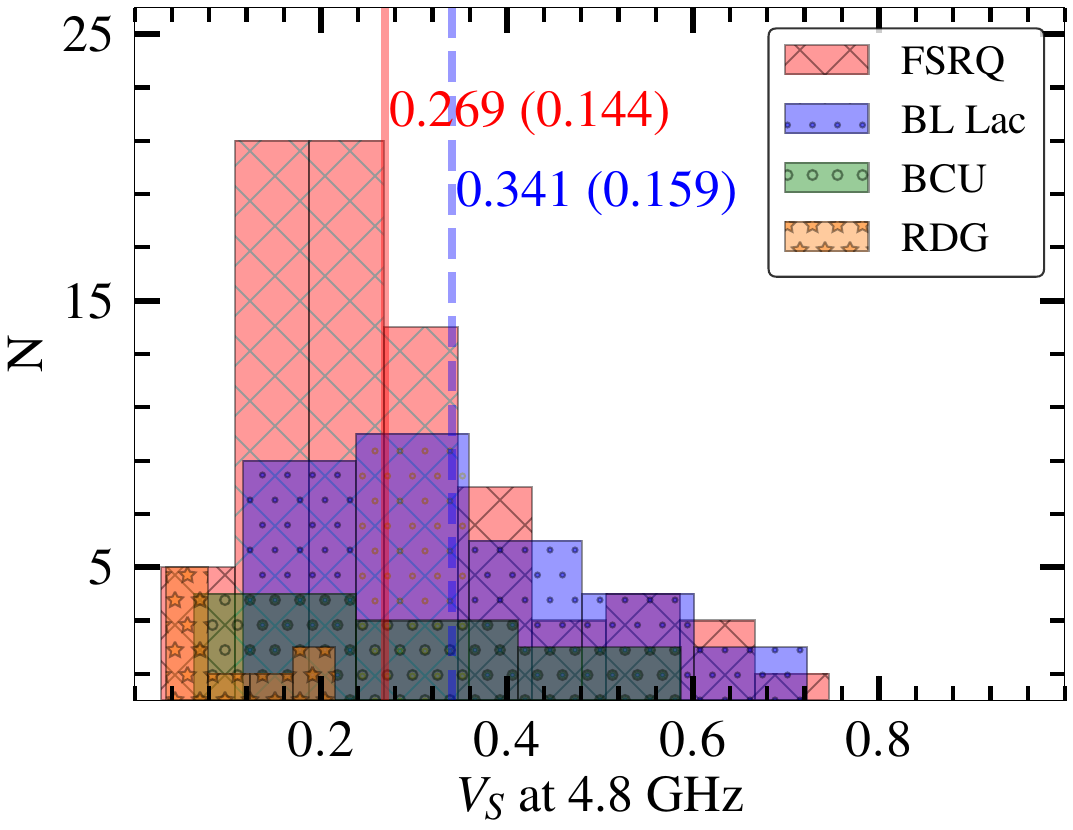}\\
        \includegraphics[width=\columnwidth]{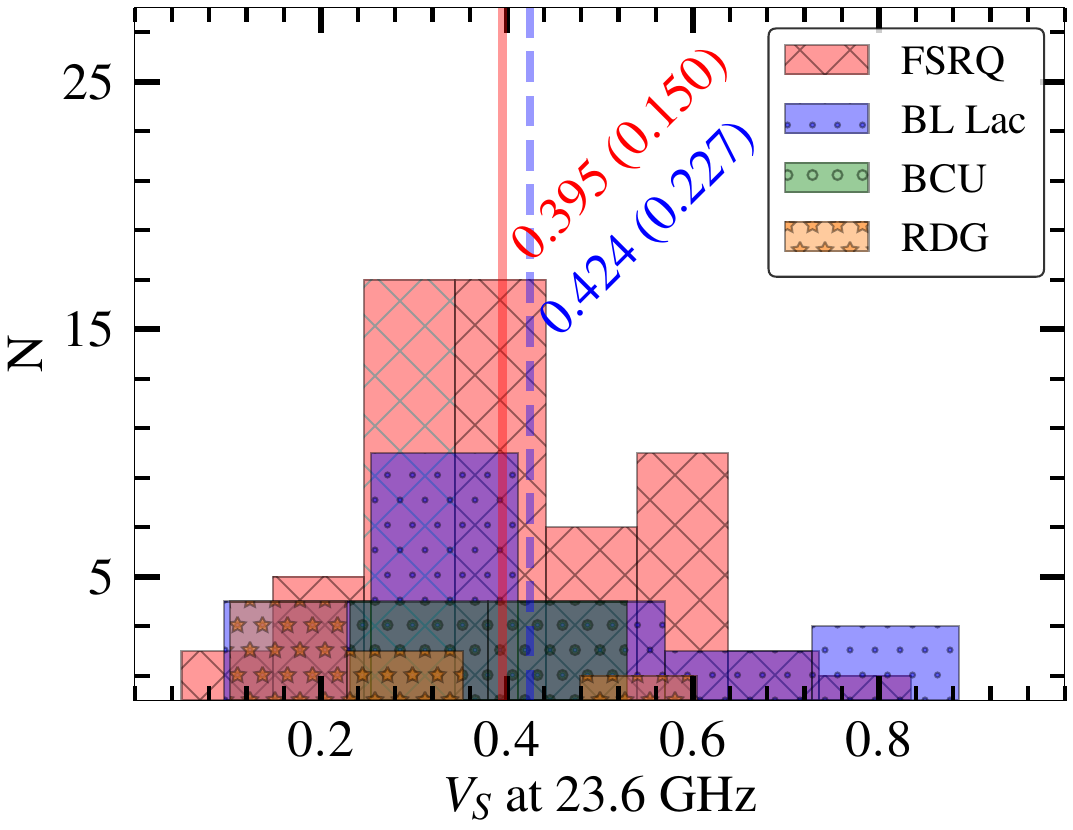}
	\caption{Histogram of the variability index at 4.8 GHz (top panel) and 23.6 GHz (bottom panel). The objects have been categorized according to their type of AGNs classification. The transparent solid red and dashed blue vertical lines represent the average values with standard deviation inside parentheses for FSRQs and BL Lacs, respectively.}
    \label{fig:var_index_hist}
\end{figure}

The variability index is based solely on the two extreme flux density values. Therefore, it is necessary to characterize the variability of the light curve as a whole, incorporating all available flux density measurements. Hence, to identify the presence of true intrinsic variability in the light curve, we have used the fractional root mean square variability parameter ($F_{var}$), since it considers the uncertainties in flux density measurements. Here we use the relation given in \citet{vaughan2003MNRAS}, which is defined as:
\begin{equation} \label{Eq:frac_var}
 F_{var} = \sqrt{\frac{S^2 - \sigma^2}{r^2}} \\
\end{equation}

\begin{equation}
 err(F_{var}) = \sqrt{  \Big(\sqrt{\frac{1}{2N}}. \frac{\sigma^2}{r^2F_{var}} \Big)^2 + \Big( \sqrt{\frac{\sigma^2}{N}}. \frac{1}{r} \Big)^2     } \\
\end{equation}
where, $\sigma^2_{XS}$ = S$^{2}$ -- $\sigma^2$, is called excess variance, S$^{2}$ is the sample variance, $\sigma^2$ is the mean square uncertainties
of each observation and $r$ is the sample mean.

\begin{figure}[h!]
	\centering
	\includegraphics[width=\columnwidth]{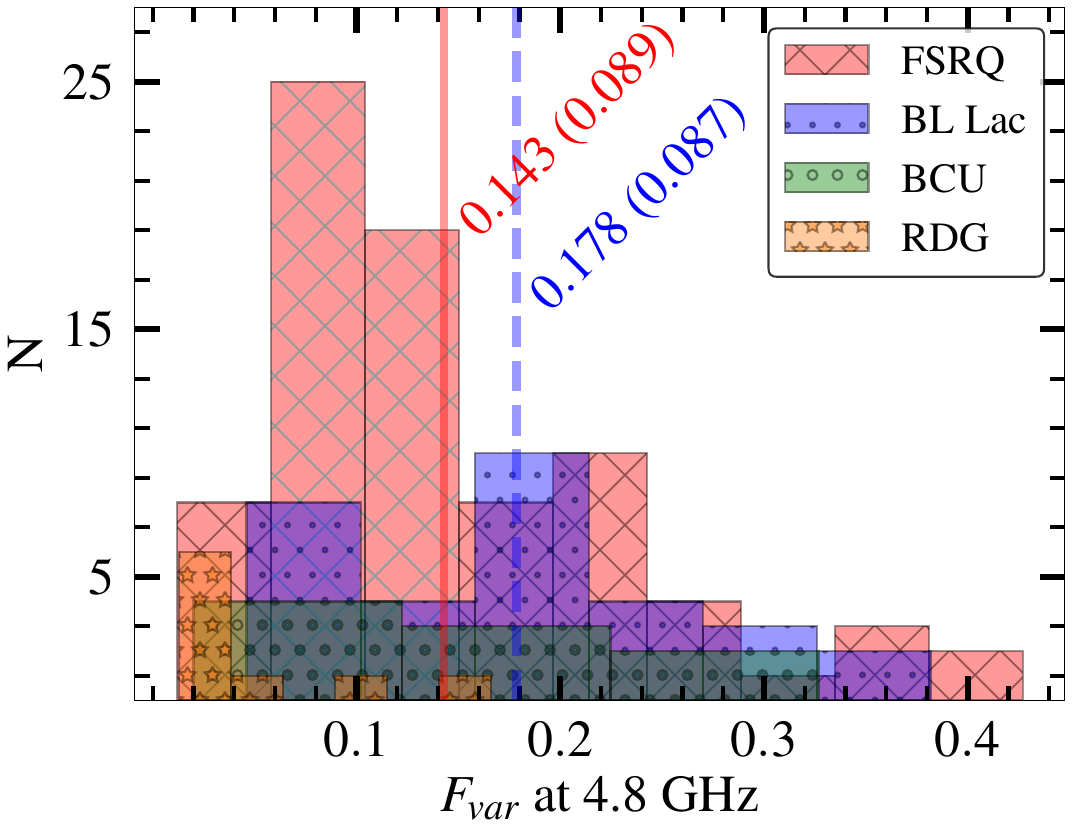}\\
        \includegraphics[width=\columnwidth]{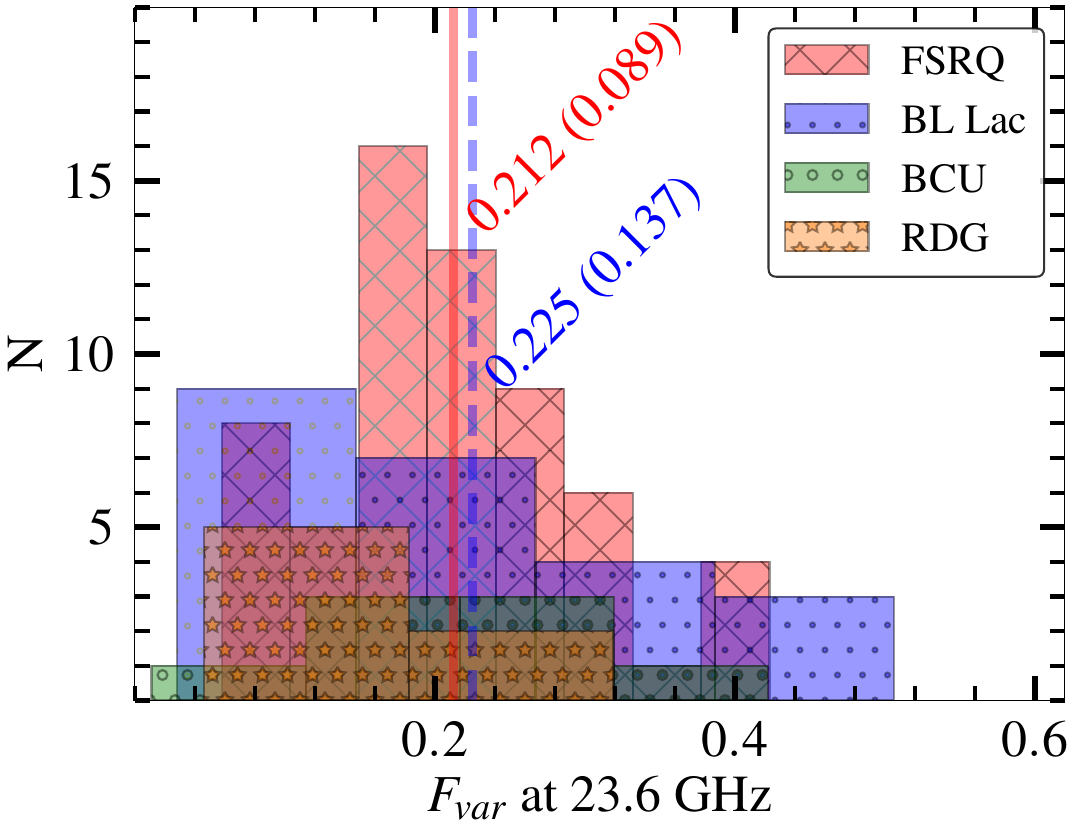}
	\caption{Histogram of the fractional variability at 4.8 GHz (top panel) and 23.6 GHz (bottom panel). The transparent solid red and dashed blue vertical lines represent the average values with standard deviation inside parentheses for FSRQs and BL Lacs, respectively.}
    \label{fig:fvar_hist}
\end{figure}

The radio spectral index is calculated in two frequency intervals (4.8 and 23.6 GHz). The radio spectral index $\alpha$ defined from the power law $S_{\nu} \sim \nu^{\alpha}$, where $S_{\nu}$ is the flux density at frequency $\nu$, and $\alpha$ is a spectral slope. It is calculated using the formula:

\begin{equation}
\label{spindex}
\alpha=\frac{\log S_{2} - \log S_{1}}{\log{\nu}_{2} - \log{\nu}_{1}},
\end{equation}
where $S_{1}$ and $S_{2}$ are the flux densities at frequencies $\nu_{1}$ and $\nu_{2}$, respectively. The uncertainty in the spectral index was determined using Gaussian error propagation as:
\begin{equation}
\Delta\alpha=K\,\sqrt{\left(\frac{\Delta S_{4.8\,\rm{GHz}}}{S_{4.8\,\rm{GHz}}}\right)^2+\left(\frac{\Delta S_{23.6\,\rm{GHz}}}{S_{23.6\,\rm{GHz}}}\right)^2},
\label{spindex_err}
\end{equation}

where the factor of $K\approx|[\log(4.8\,\rm{GHz}/23.6\,\rm{GHz})\cdot \ln(10)]^{-1}|$ arises from the derivation of the spectral index as defined in Equation~\ref{spindex}.\\

The radio luminosity was calculated at 4.8 and 23.6 GHz following the equation:
\begin{equation}\label{Eq:radio_lum}
P = 4 \pi D_{L}^2 S~(1+z)^{-\alpha -1},
\end{equation}

where $S$ is the measured flux density at 4.8 or 23.6 GHz, $z$ is the redshift, $\alpha$ is the average spectral index in the interval 4.8 and 23.6~GHz, and $D_{L}$ is the luminosity distance. For calculation of luminosity distances, we used the $\Lambda$CDM cosmology with $H_0 = 67.74$~km\,s$^{-1}$\,Mpc$^{-1}$, $\Omega_m=0.3089$, and $\Omega_\Lambda=0.6911$ \citep{planck2016A&A}.\\

\subsection{Characteristics of Radio Variability, Spectral index and Luminosity: Relation to AGNs Class}
\begin{figure*}
	\centering
	\includegraphics[width=\columnwidth]{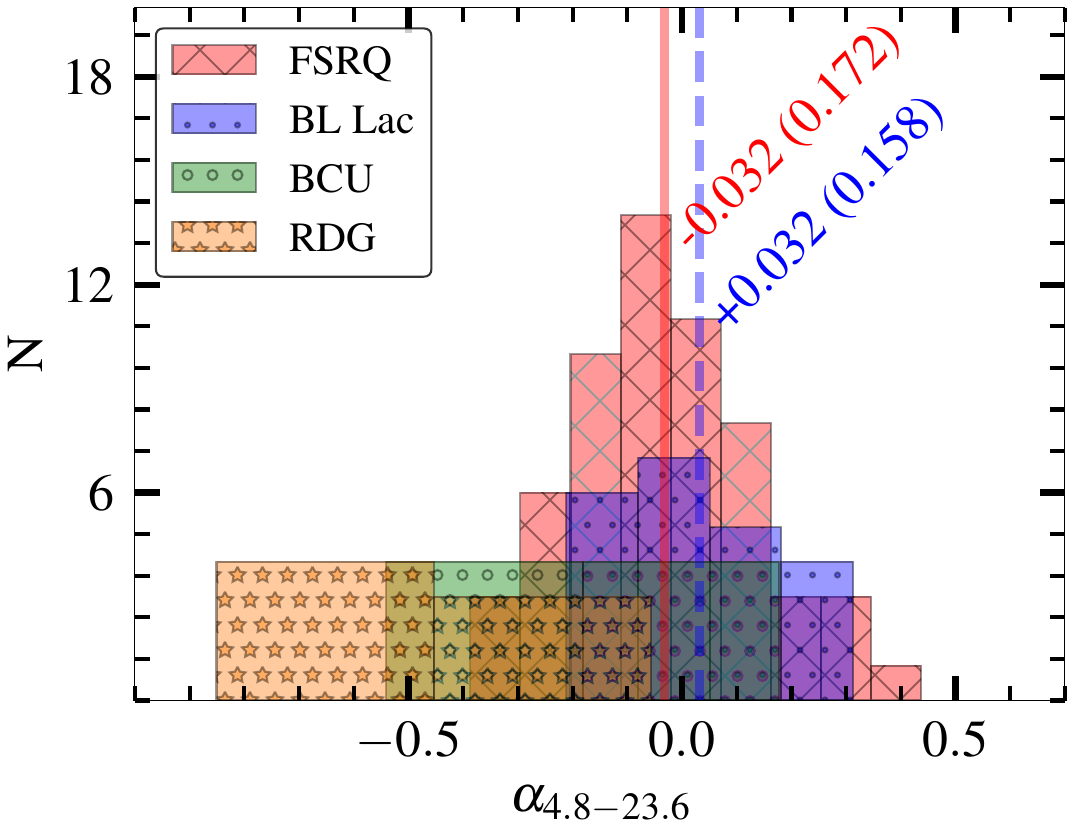}
    \includegraphics[width=\columnwidth]{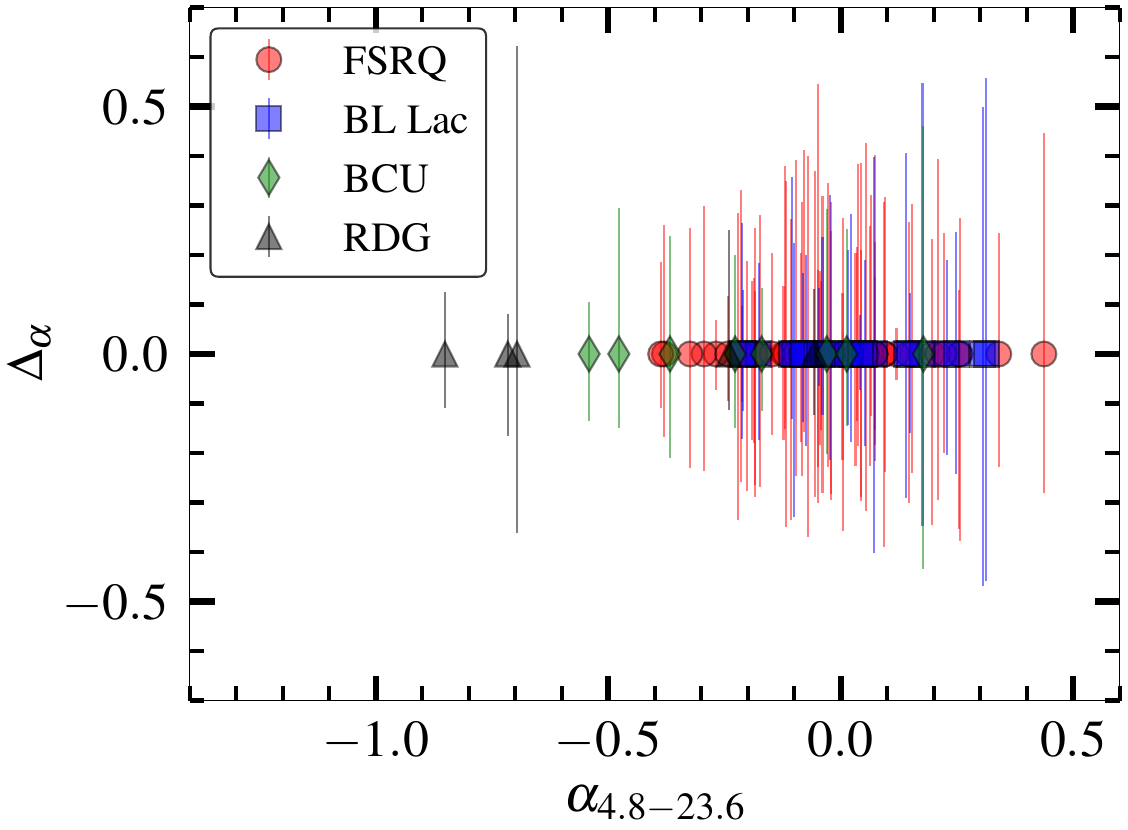}

\caption{Histogram of average radio spectral index in the spectral region 4.8-23.6 GHz (left panel).
The transparent solid red and dashed blue vertical lines represent the average values with standard deviation inside parentheses for FSRQs and BL Lacs, respectively.
The spread of spectral indices, estimated as the difference between maximum and minimum values in the region 4.8-23.6 GHz, as a function of average spectral index (right panel).}
    \label{fig:alpha_dist}
\end{figure*}

\begin{figure*}
	\centering
	\includegraphics[width=\columnwidth]{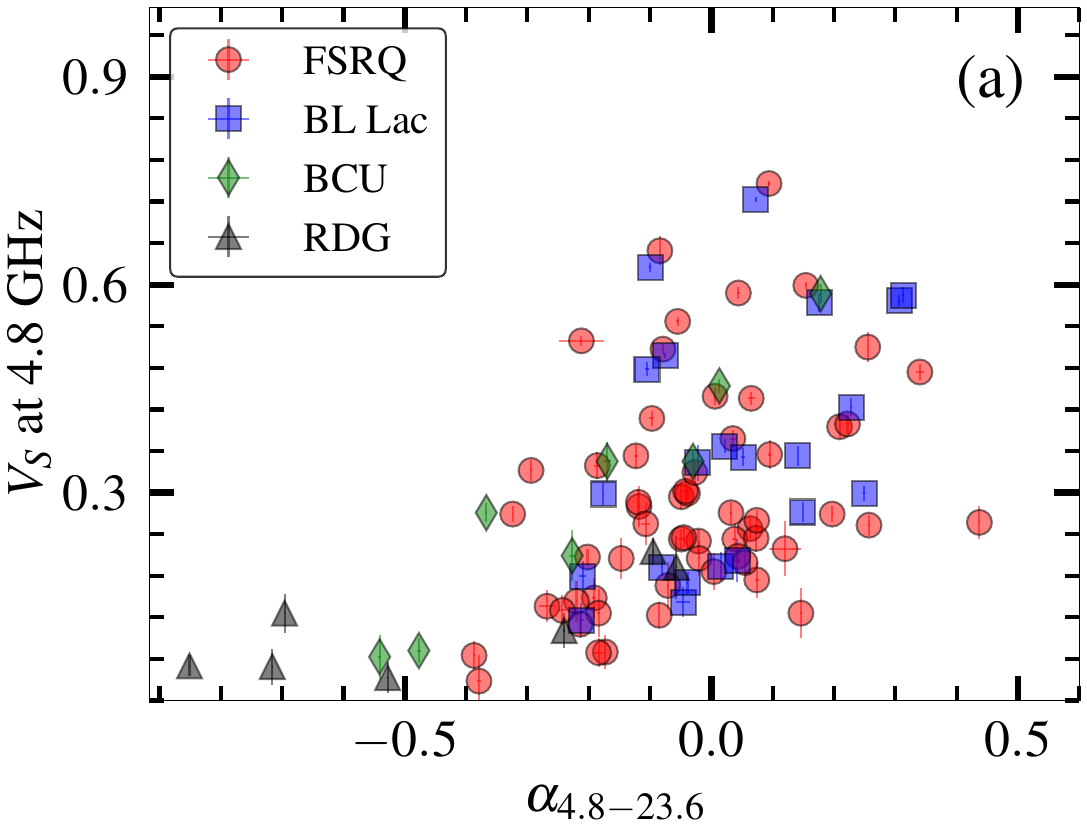}
        \includegraphics[width=\columnwidth]{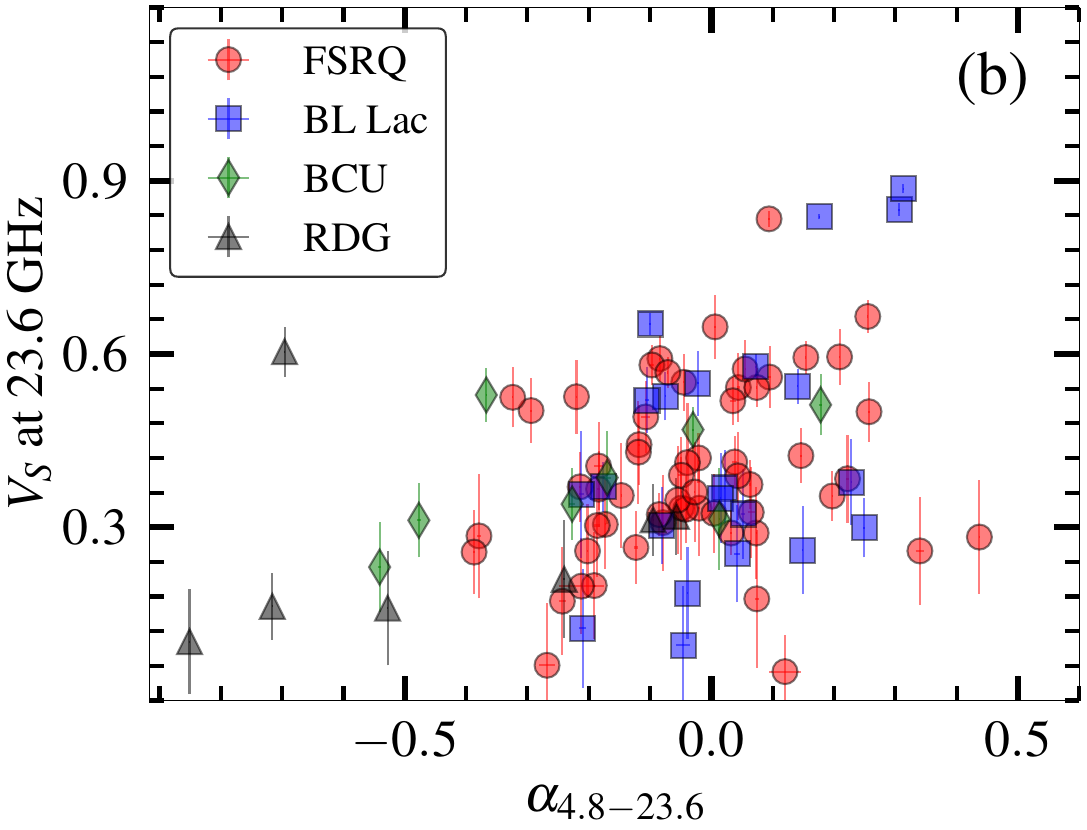}
	\caption{Variability index vs. average spectral index at 4.8 GHz (a) and 23.6 GHz (b). }
    \label{fig:alpha_var_index}
\end{figure*}

The estimated variability index values for different sources at 4.8 and 23.6 GHz are presented in
Table~\ref{tab:properties_of_smman_srcs} of Appendix~\ref{Appendix A:SMMAN_srcs}. The histograms of $V_{S}$
at two frequencies are shown in Figure~\ref{fig:var_index_hist}.
At 23.6 GHz $\sim97\%$ of the sources exhibit variability above the $10\%$ level ($V_{S}>0.1$), whereas at lower frequency 4.8 GHz $\sim90\%$ sources show significant variability. The range in the variability for FSRQs and BL Lacs at 4.8 GHz is  comparable (varying from $<0.1$ up to $\sim0.74$) with the mean value for FSRQs at $0.269~(\sigma=0.144)$ and for BL Lacs at $0.341~(\sigma=0.159)$. The BCU and RDG classes vary from $<0.1$ up to maximum values of $0.21$ and $0.59$, respectively. Similarly, at 23.6 GHz, the range of variability in FSRQs and BL Lacs is comparable (varying from $<0.1$ up to $\sim0.85$) with the mean value for FSRQs at $0.395~(\sigma=0.150)$ and for BL Lacs at $0.424~(\sim0.227)$. The BCU and RDG classes vary from $0.1$ up to maximum values of $0.53$ and $0.60$, respectively.\\

\begin{figure*}
	\centering
	\includegraphics[width=\columnwidth]{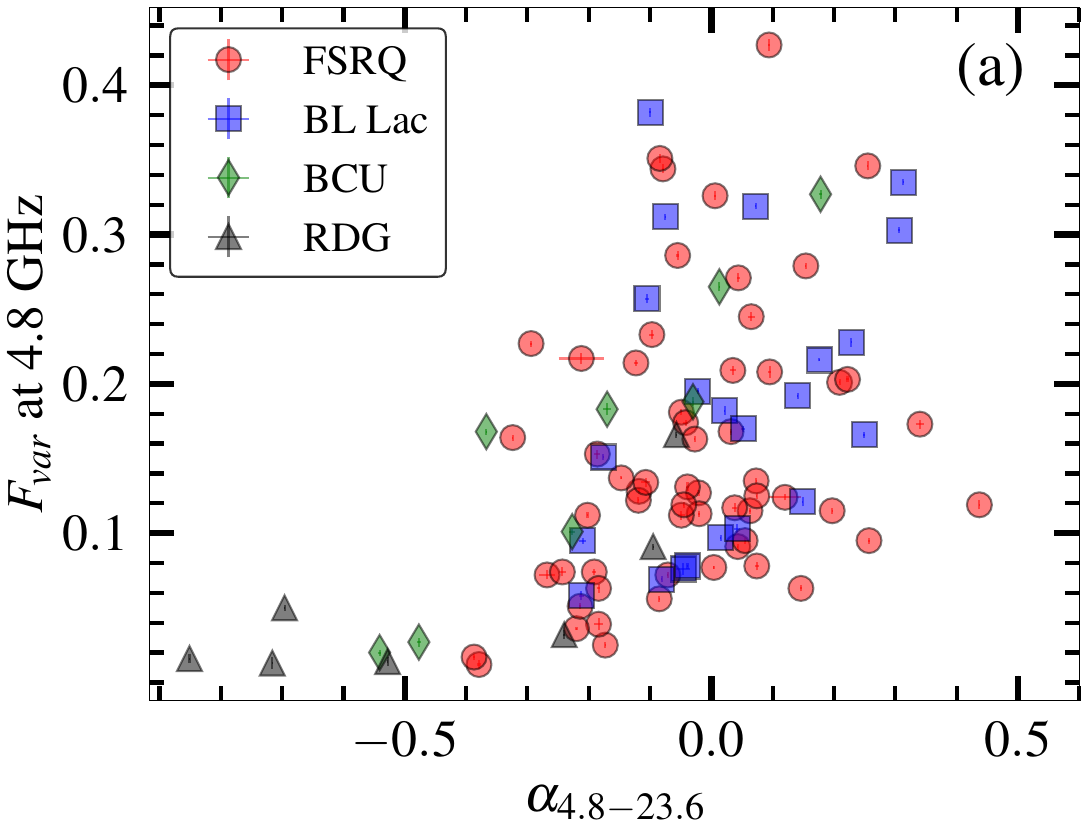}
        \includegraphics[width=\columnwidth]{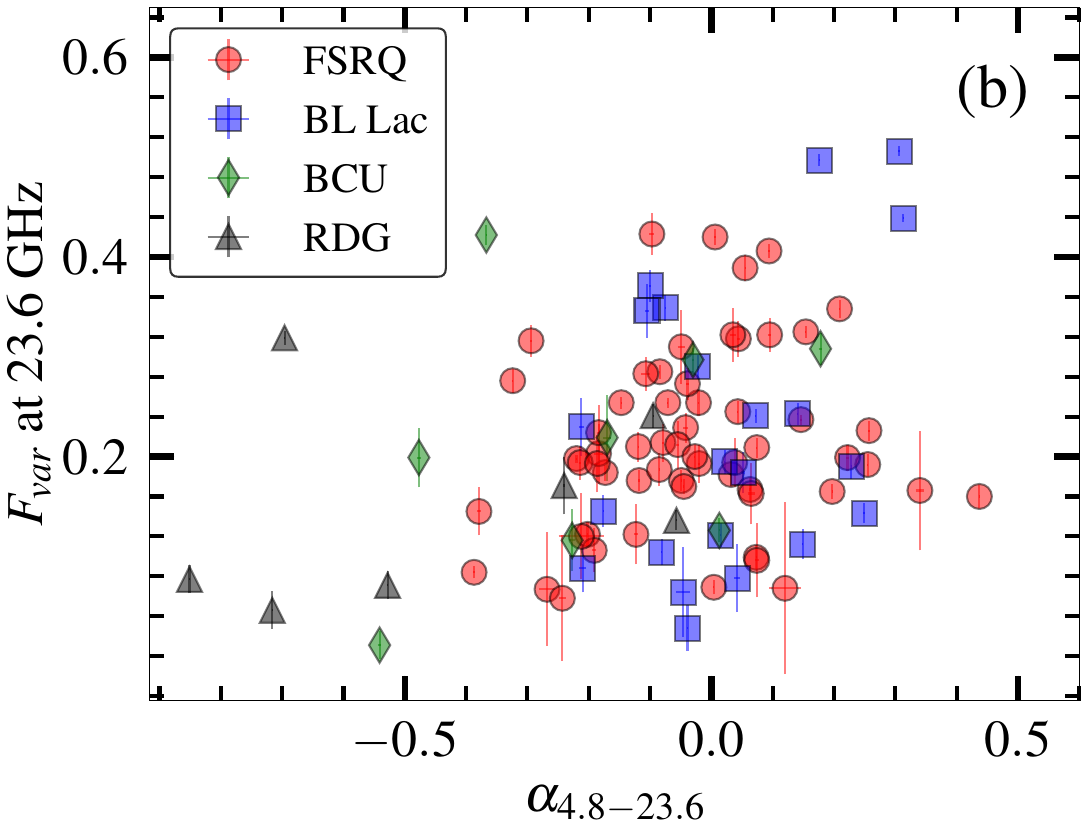}
	\caption{Fractional variability vs. average spectral index at 4.8 GHz (a) and 23.6 GHz (b). }
    \label{fig:alpha_Fvar_index}
\end{figure*}

\begin{figure*}
	\centering
	\includegraphics[width=\columnwidth]{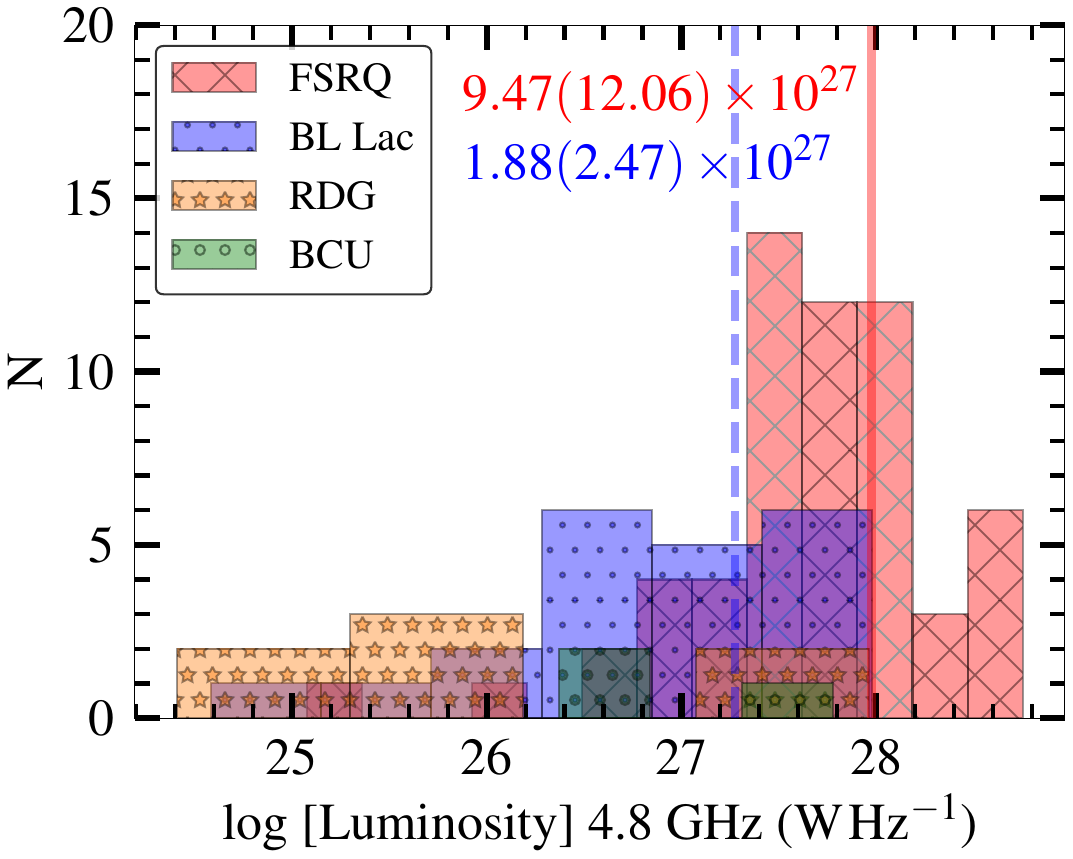}
        \includegraphics[width=\columnwidth]{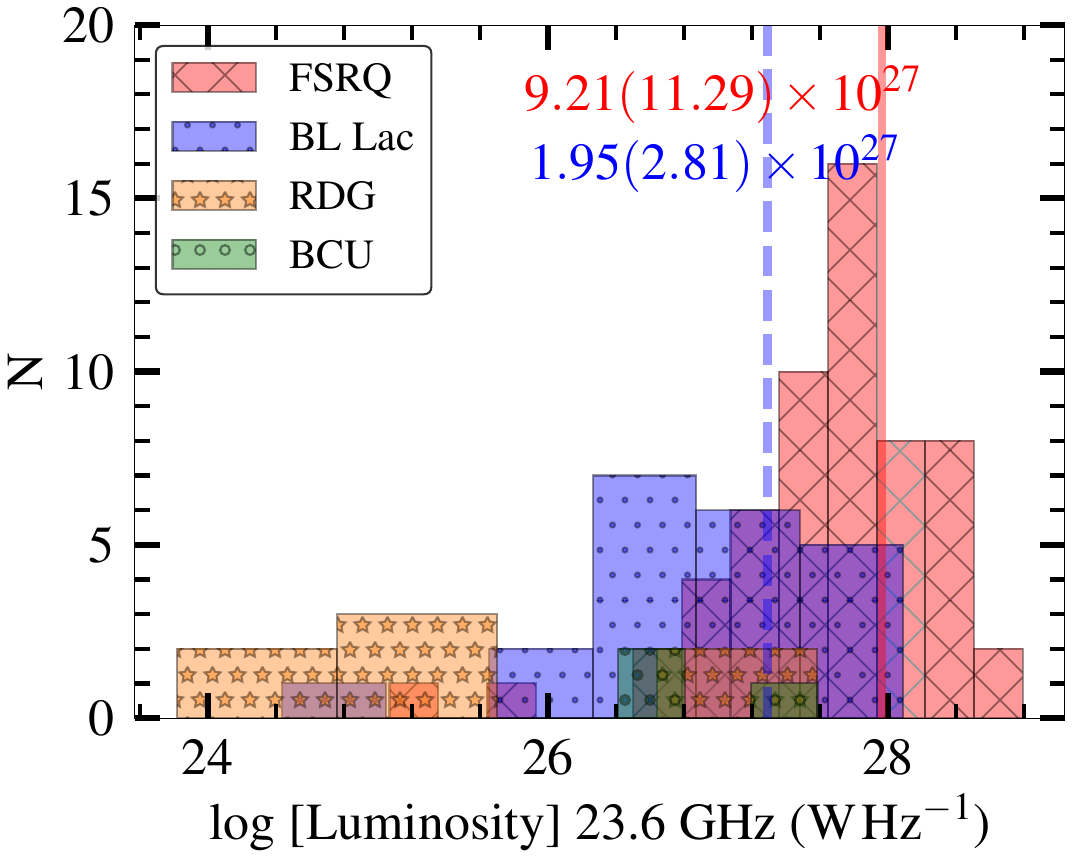}
	\caption{Distribution of the radio luminosity at 4.8 GHz (left panel) and 23.6 GHz (right panel). The transparent solid red and dashed blue vertical lines represent the average values with standard deviation inside parentheses for FSRQs and BL Lacs, respectively.}
    \label{fig:hist_radio_luminosity}
\end{figure*}

The calculated fractional variability values for different
SMMAN sources in the C and K bands are given in Table~\ref{tab:properties_of_smman_srcs} of  the Appendix~\ref{Appendix A:SMMAN_srcs}. In Figure~\ref{fig:fvar_hist}, we show fractional variability histograms at 4.8 and 23.6 GHz for different classes of AGNs in our sample. It is observed that at 4.8 GHz, $\sim60\%$ of the sources show fractional variability
$>10\%$. The $F_{var}$ value of the FSRQs found vary from $<10\%$ to a maximum of $\sim42\%$ with a mean value of $0.143~(\sigma=0.089)$, BL Lacs show a range between $<10\%$ and a maximum value of $\sim30\%$ with a mean value of $0.178~(\sigma=0.087)$.
A similar trend of variation is observed in active galaxies of uncertain type with a maximum value up to $\sim33\%$ and in radio galaxies with maximum variation up to $~17\%$. In contrast to lower frequency behavior, $\sim84\%$ of the sources exhibit fractional variability $>10\%$. Here, the different classes show variation of $F_{var}>10\%$ to maximum values of $\sim42\%$, $\sim50\%$, $\sim42\%$, and $\sim32\%$ in FSRQs, BL Lacs, BCUs, and RDGs, respectively. At 23.6 GHz, the mean value of $F_{var}$ for FSRQs and BL Lacs is found to be $0.212~(\sigma=0.089)$ and $0.225~(\sigma=0.137)$, respectively.
It is noticed that from both the estimation of variability index and fractional variability values at 4.8 and 23.6 GHz, a trend for sources to exhibit a higher degree of variability at 23.6 GHz than at 4.8 GHz, with a mean value for BL Lacs higher than the FSRQs. These results are consistent with the pattern reported by \citet{Aller1992ApJ} and \citet{sotnikova2022AstBu} for a sample comprising 62 extragalactic objects (at 4.8 and 14.5 GHz) and 1700 blazars (in frequency range 1.2-22.3 GHz), respectively.\\

The average spectral index with associated error and the range in spectral index (minimum and maximum values) over the entire observing period are presented in Table~\ref{tab:properties_of_smman_srcs} of the Appendix~\ref{Appendix A:SMMAN_srcs}. It is worth noting that when estimating the spectral index, monthly averaged flux densities were used to ensure temporal alignment between the two frequencies.
In Figure~\ref{fig:alpha_dist}, the left panel shows the distribution of average spectral indices, while the right panel displays the range of spectral indices versus their average values for the various AGNs classes in the SMMAN sample. BL Lac objects tend to exhibit comparatively flatter spectra than the other classes, whereas FSRQs, BCUs, and RDGs show a broader distribution, spanning from flat to steep.
The mean spectral indices for FSRQs is $-0.032~(\sigma=0.172)$ with values ranging between $-0.387$ and $+0.437$. For BL Lacs, the spectral index ranges from $-0.212$ to $+0.313$, with a mean of $+0.032~(\sigma=0.158)$. Radio galaxies generally display steep spectra (ranging between $-0.851$ to $0.178$), and the BCUs are also skewed toward steeper spectral indices (varying from $-0.851$ to $-0.057$). Also, the BL Lacs and FSRQs were found to show a higher degree of variability trend at both frequency bands compared to the other classes.\\

Figure~\ref{fig:alpha_var_index} shows variability indices versus average spectral index in the range of 4.8-23.6 GHz. At both 4.8 and 23.6 GHz, $\sim90$ of the sources with $\alpha\geq-0.4$ show variability greater than $10\%$ level. The majority of these sources belong to the FSRQ and BL Lac classes. Also, sources with $\alpha\leq-0.4$ exhibit a low degree of variability at both frequency bands. Interestingly, the radio galaxy 3C 111 with $\alpha=-0.696$ shows variability above $10\%$ at both 4.8 and 23.6 GHz with a maximum value of $V_{S}=0.604$ at 23.6 GHz. Such large variability could be due to the major outburst, which is usually associated with the release of radio knots from the core~\citep{kadler2008ApJ,grossberger2012AcPol}. Similar features are noticed in the case of fractional variability versus the average spectral index plot (Figure~\ref{fig:alpha_Fvar_index}).\\

The radio luminosity was calculated using the
Equation~\ref{Eq:radio_lum} for different classes of AGNs in our sample, the distribution is presented in Figure~\ref{fig:hist_radio_luminosity}, and the values are given in the
Table~\ref{tab:properties_of_smman_srcs} of the Appendix~\ref{Appendix A:SMMAN_srcs}. It should be noted that, among the redshifts used in our calculations, three out of the 124 measurements ($\sim2.4\%$) are photometric, while the remaining measurements are spectroscopic.
The distribution ranges from $2.58\times10^{24}$ to $5.65\times10^{28}$ W Hz$^{-1}$ at 4.8 GHz, and from $6.60\times10^{23}$ to $3.68\times10^{31}$ W Hz$^{-1}$ at 23.6 GHz. In both frequency bands, FSRQs exhibit higher radio luminosity than BL Lacs and other classes. Specifically, at 4.8 GHz, FSRQs have radio luminosities from $1.19\times10^{25}$ to $5.65\times10^{28}$ W Hz$^{-1}$ (mean $9.47\times10^{27}$, $\,\sigma = 12.06\times10^{27}$ W Hz$^{-1}$), while BL Lacs range from $3.84\times10^{24}$ to $9.55\times10^{27}$ W Hz$^{-1}$ (mean $1.88\times10^{27}$, $\,\sigma = 2.47\times10^{27}$ W Hz$^{-1}$). At 23.6 GHz, the BL Lacs span from $2.74\times10^{24}$ to $1.23\times10^{28}$ W Hz$^{-1}$ (mean $1.95\times10^{27}$, $\,\sigma = 2.81\times10^{27}$ W Hz$^{-1}$) and FSRQs from $1.17\times10^{25}$ to $6.20\times10^{28}$ W Hz$^{-1}$ (mean $9.21\times10^{27}$, $\,\sigma = 11.29\times10^{27}$ W Hz$^{-1}$). In both bands, BCUs have radio luminosities intermediate between BL Lacs and FSRQs. RDGs exhibit lower luminosities, typically below the mean value of FSRQs.\\

\subsection{Gamma-ray and Radio Properties of the SMMAN Sample}~\label{sec:gamma_radio_properties}
\begin{figure}[h!]
	\centering
	\includegraphics[width=\columnwidth]{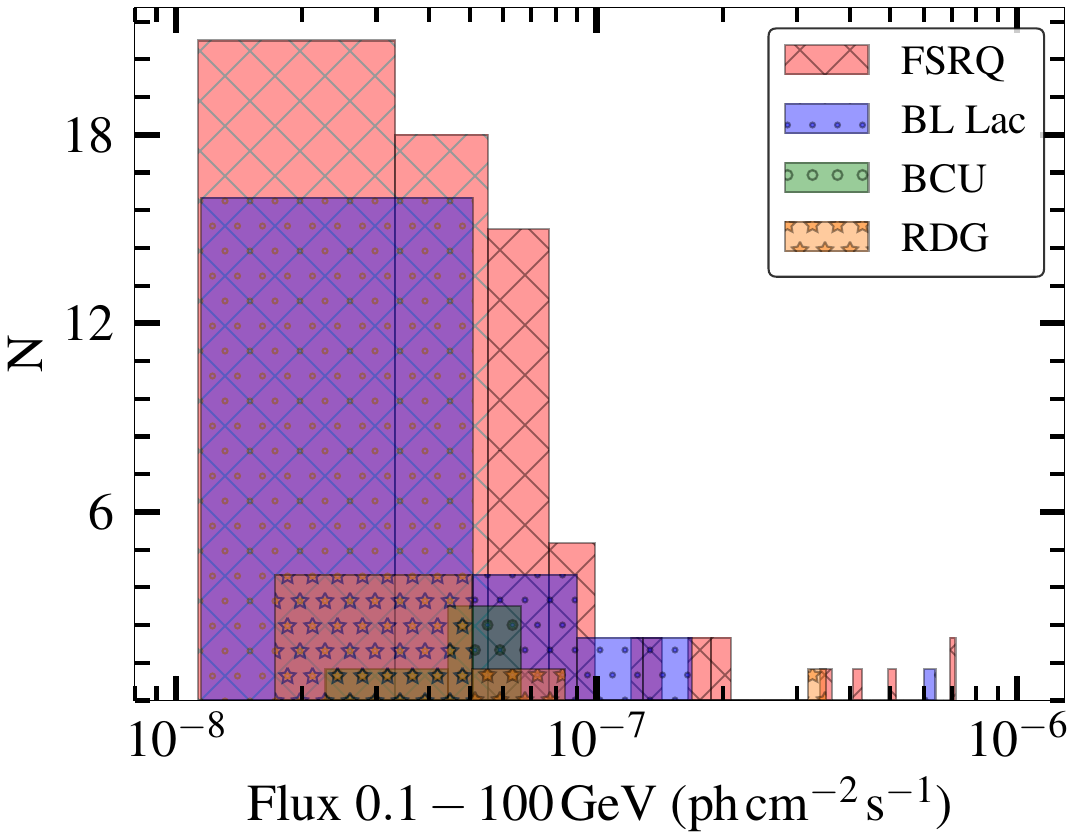}\\
    \includegraphics[width=\columnwidth]{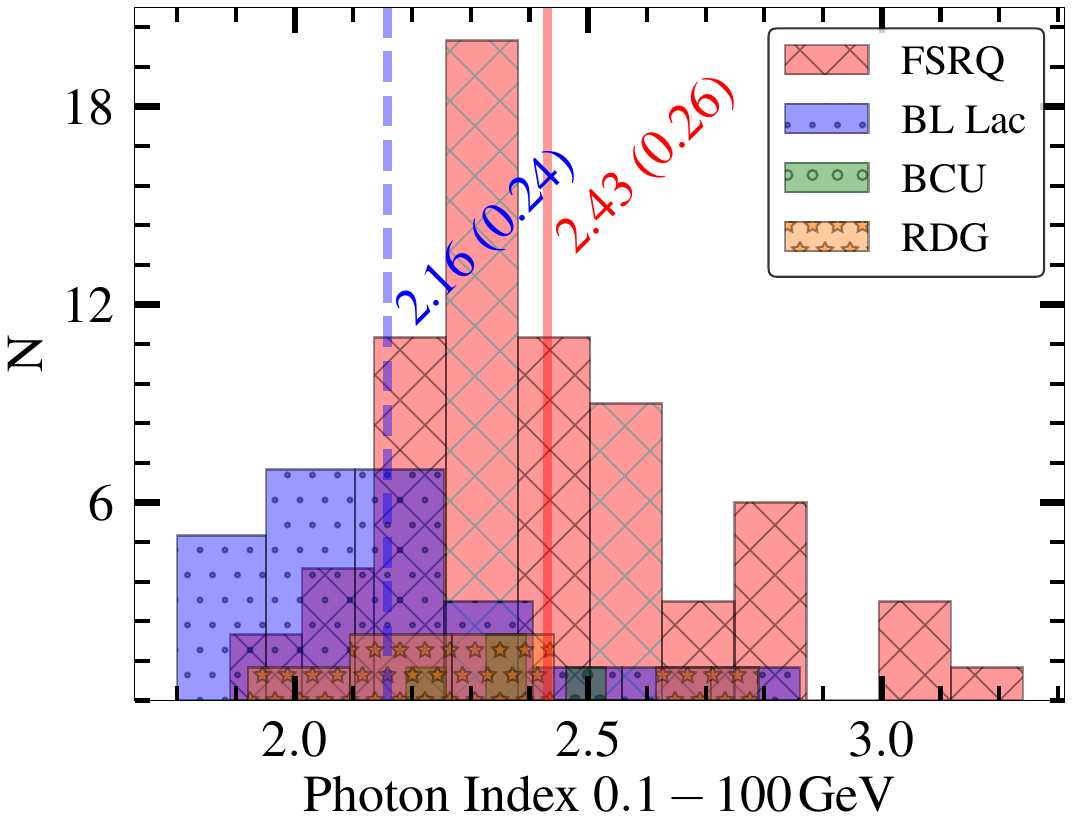}
	\caption{Top: {\sl Fermi}-LAT flux distribution of the sources. Bottom: {\sl Fermi}-LAT $\gamma$-ray photon index distribution of the sources. The transparent solid red and dashed blue vertical lines represent the average values with standard deviation inside parentheses for FSRQs and BL Lacs, respectively.}
    \label{fig:hist_fermi_flux_and_index}
\end{figure}

To investigate the connection between radio and $\gamma$-ray emissions, we use {\sl Fermi}-LAT data in the
energy range of $0.1-100$ GeV covering the same time period (2016 November 1 to 2024 November 30) as that of the radio observation presented in this work. The details of the $\gamma$-ray data are provided in Section~\ref{sec:observation}.
The LCR provides light curves for all sources in the 4FGL-DR2 catalog with a variability index greater than 21.67~\citep{abdollahi2023ApJS}. This index serves as a proxy for the average fractional variability $\delta F/F$, where $\delta F$ is measured on timescales of one year. Consequently, among the 131 SMMAN sources, we obtained monthly binned light curves and spectral indices for 105.
From the light curves, we calculated the average flux and spectral indices; the values are presented in Table~\ref{tab:properties_of_smman_srcs} of the Appendix~\ref{Appendix A:SMMAN_srcs}.
Further, following the procedures
described in \citet{bock2016A&A} with the assumption that sources have a power-law photon flux spectrum (Equation~\ref{Eq:pl_spec}), the $\gamma$-ray energy
flux and luminosities for the 105 sources were calculated as follows.

\begin{equation}~\label{Eq:pl_spec}
\begin{split}
N_\mathrm{ph}(E) &= S_\mathrm{ph} \frac{1-\Gamma}{E_0} \left[\left(\frac{E_\mathrm{max}}{E_0}\right)^{1-\Gamma} -\left(\frac{E_\mathrm{min}}{E_0}\right)^{1-\Gamma} \right]^{-1} \\\\
&\quad \left( \frac{E}{E_0}\right)^{-\Gamma}
\end{split}
\end{equation}

where $S_\mathrm{ph}$ is the estimated photon flux in the energy band
from $E_\mathrm{min}$ to $E_\mathrm{max}$ ($E_0$ is the reference
energy used to make the expression dimensionless while applying a non-integer exponent),
for photon index $\Gamma\ne 1$. Then the energy flux in that band
is given by

\begin{equation}~\label{Eq:energy_flux}
    S_\mathrm{E} = S_\mathrm{ph} \;E_0 \; \frac{1-\Gamma}{2-\Gamma} \;
    \frac{(E_\mathrm{max}/E_0)^{2-\Gamma}-(E_\mathrm{min}/E_0)^{2-\Gamma}}
    {(E_\mathrm{max}/E_0)^{1-\Gamma}-(E_\mathrm{min}/E_0)^{1-\Gamma}}
\end{equation}
for $\Gamma\ne2$.

Because the sources in the sample are distributed across various redshifts, the measured luminosities were corrected using the K-correction as described by \citet{ghisellini2009MNRAS}, i.e.,
\begin{equation}~\label{Eq:gamma_lum}
    L_E = 4\pi d_\mathrm{L}^2 \frac{S_\mathrm{E}}{\left( 1+z\right)^{2-\Gamma}}
\end{equation}

where $z$ is the redshift of the source, $S_\mathrm{E}$ the energy
flux, and $\Gamma$ the photon spectral index in the $\gamma$-ray band.
The luminosity distance, $d_\mathrm{L}$, was calculated assuming a
flat universe with the cosmological parameters same as those used in the estimation of the radio luminosity.\\

The $\gamma$-ray flux distribution for the SMMAN sample is shown in the top panel of Figure~\ref{fig:hist_fermi_flux_and_index}, while the bottom panel presents the corresponding distribution of spectral indices. On average, BL Lacs tend to exhibit harder spectra than FSRQs, consistent with earlier results based on the 3FGL sources (see Figure 7 of \citealt{ackermann2015ApJ}). We performed the two sample
Kolmogorov–Smirnov (KS) test to check whether the BL Lac objects have the same distribution as those of FSRQs. The test shows a strong statistical result with a $p$-value of $7\times10^{-6}$.
This result is in agreement with earlier studies, which have shown a relationship between the $\gamma$-ray spectral index and the peak frequency of the synchrotron component in the spectral energy distribution.
It is to be noted that blazars were also classified into three subclass based on the  synchrotron peak frequency ($\nu_{\rm p}$): low-synchrotron-peaked (LSP; ${\rm log}~(\nu_{\rm p})<14~{\rm Hz}$), intermediate-synchrotron-peaked (ISP; $14~{\rm Hz}<{\rm log}~(\nu_{\rm p})<15~{\rm Hz}$), and high-synchrotron-peaked (HSP; ${\rm log}~(\nu_{\rm p})>15~{\rm Hz}$) blazars \citep{abdo2010b,fan2016,yang2022}.
As the synchrotron peak frequency drops, their $\gamma$-ray spectra become softer (see, e.g., Figure 17 of \citet{ackermann2011ApJ} and Figure 10 of \citet{ackermann2015ApJ}).\\

\begin{figure}
	\centering
	\includegraphics[width=\columnwidth]{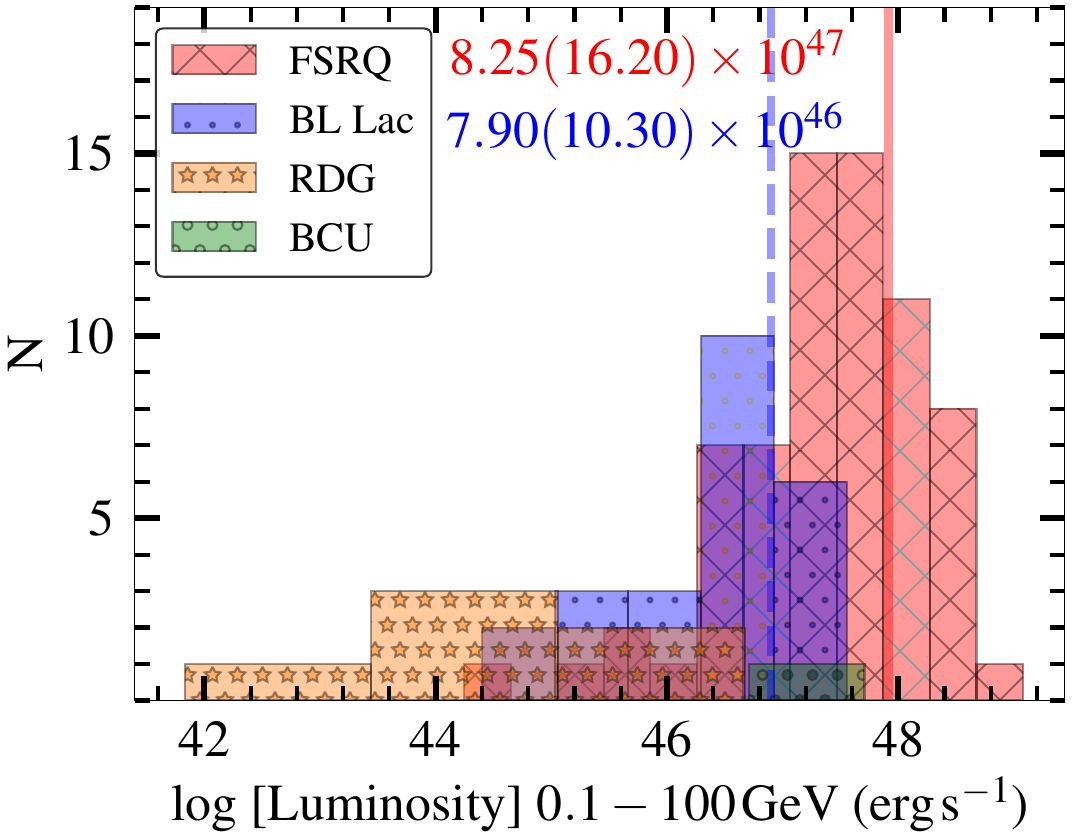}\\
    \includegraphics[width=\columnwidth]{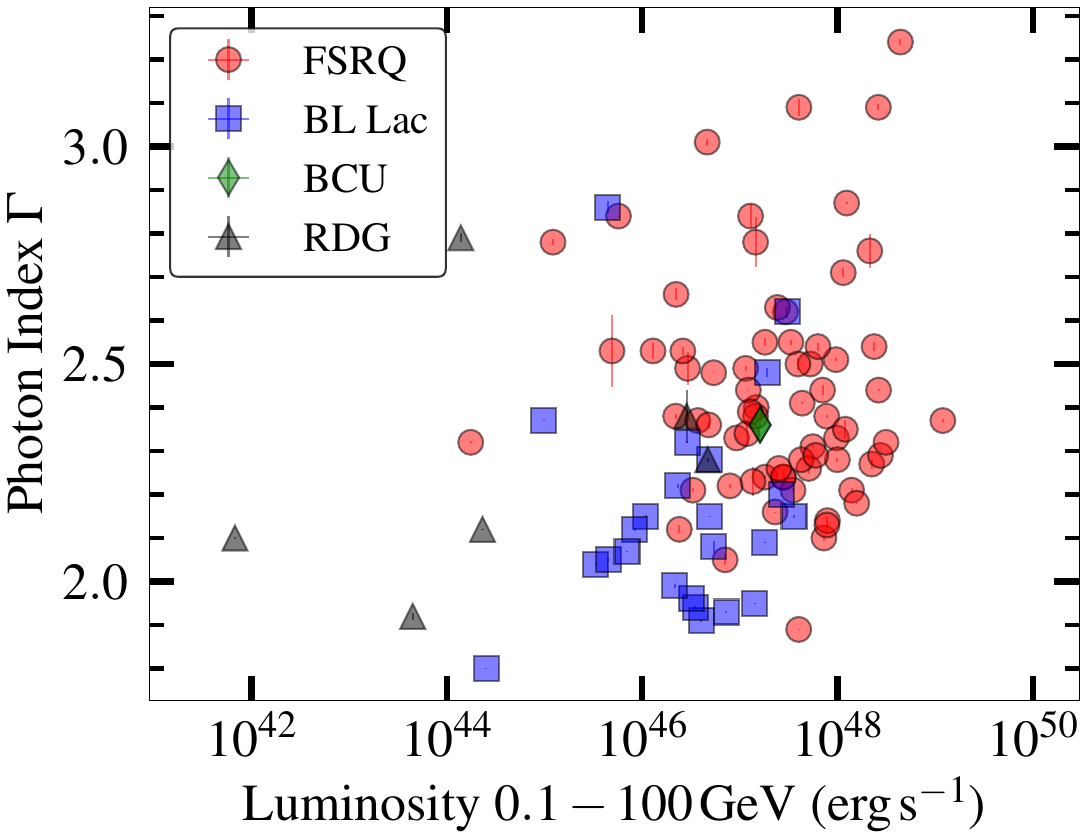}
	\caption{{\sl Fermi}-LAT $\gamma$-ray luminosity distribution of the sources (top panel). The transparent solid red and dashed blue vertical lines represent the average values with standard deviation inside parentheses for FSRQs and BL Lacs, respectively. $\gamma$-ray luminosities Vs. photon indices for different source classes (bottom panel).}
    \label{fig:hist_fermi_lum_and_ind}
\end{figure}

The observed $\gamma$-ray luminosities, as calculated from Equation~\ref{Eq:gamma_lum} and given in Table~\ref{tab:properties_of_smman_srcs} of the Appendix~\ref{Appendix A:SMMAN_srcs}, exhibit a clear dependence on the classification of the source (top panel of Figure~\ref{fig:hist_fermi_lum_and_ind}).
The luminosities of radio galaxies are low, those of BL Lac objects are in between, while FSRQs exhibit high luminosities. The bottom panel of Figure~\ref{fig:hist_fermi_lum_and_ind} shows the relation between the $\gamma$-ray luminosity and the spectral index. A trend of the spectral index becoming harder with decreasing luminosity is noticed. This relation for all {\sl Fermi}-LAT detected
AGNs in 3FGL is presented in Figure~14 of \citet{ackermann2015ApJ} and in the Figure~10 of \citet{ajello2020ApJ} for the fourth {\sl Fermi}-Large Area Telescope Source Catalog (4FGL;~\citealt{abdollahi2020ApJS}).
This pattern has been widely addressed in the perspective of ``blazar sequence'' in the earlier studies (\citealt{ackermann2015ApJ}, and references therein).\\

\begin{figure}[h!]
	\centering
	\includegraphics[width=\columnwidth]{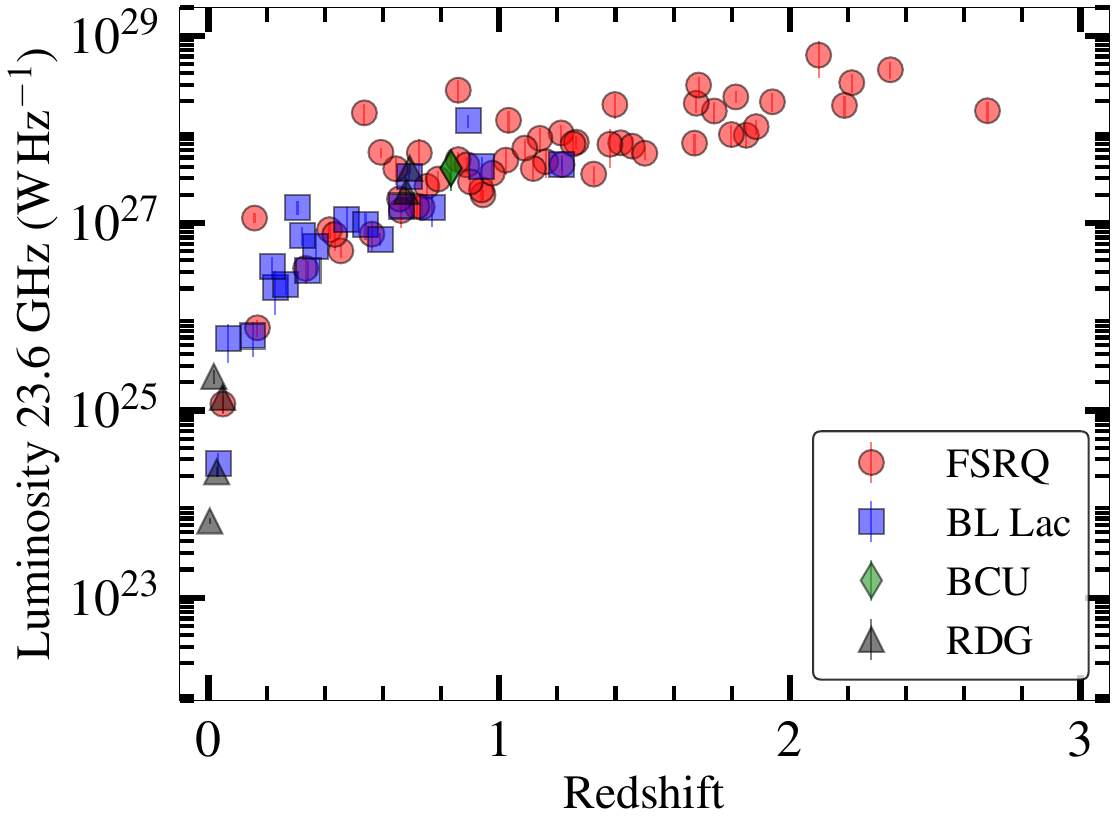}\\
    \includegraphics[width=\columnwidth]{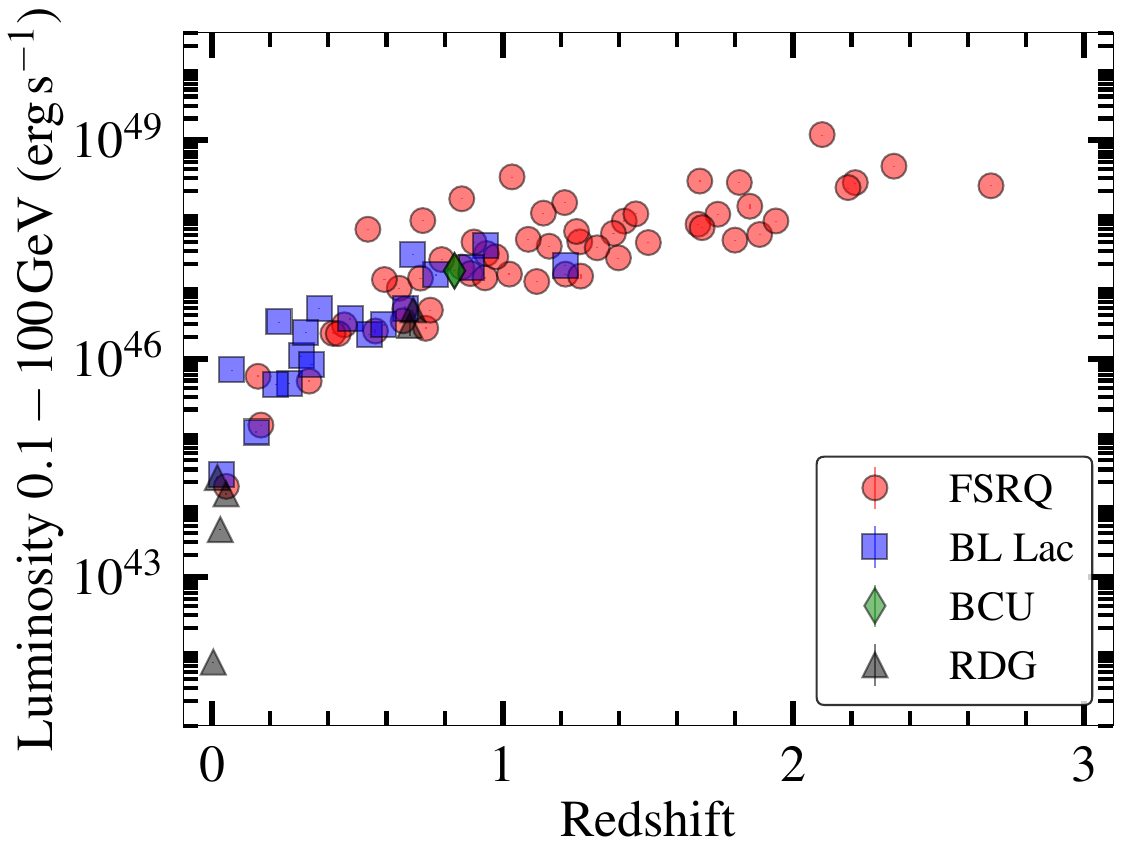}
    \caption{Radio luminosity at 23.6 GHz as a function of redshift (top panel) and $\gamma$-ray luminosity as a function of redshift (bottom panel).}
    \label{fig:lum_vs_redshift}
\end{figure}

\begin{figure}[h!]
	\centering
	\includegraphics[width=\columnwidth]{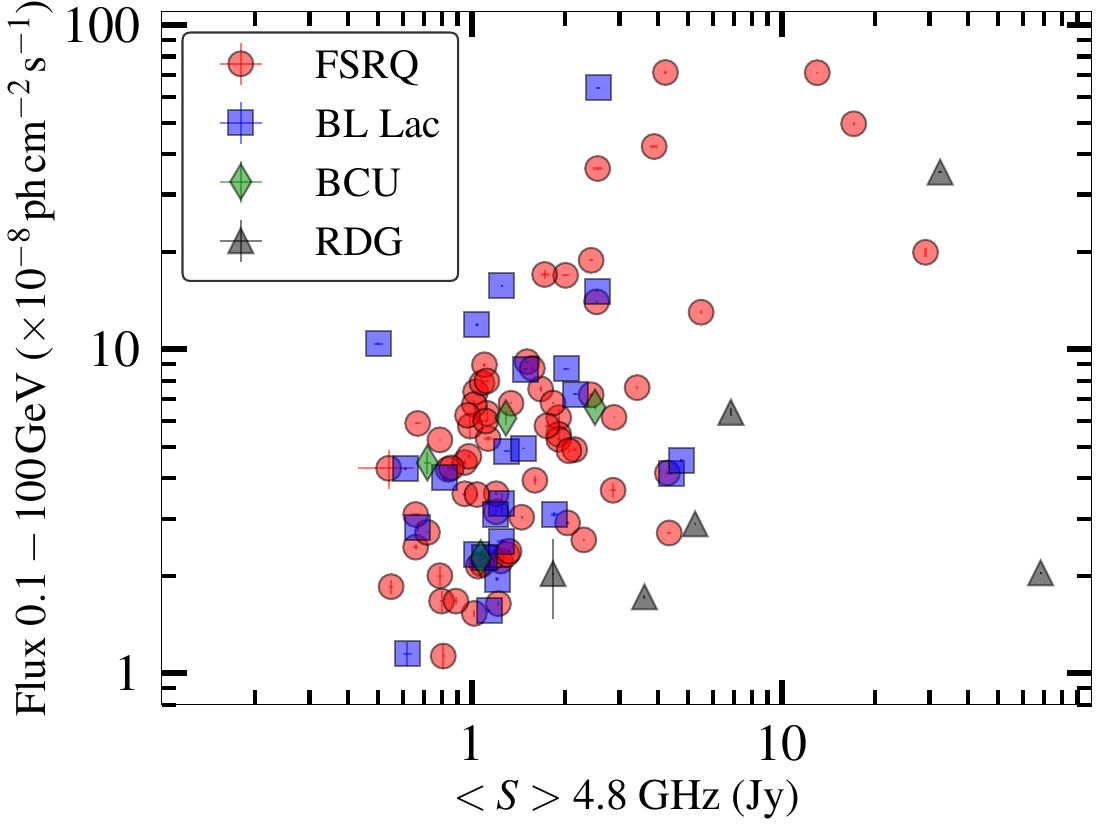}\\
    \includegraphics[width=\columnwidth]{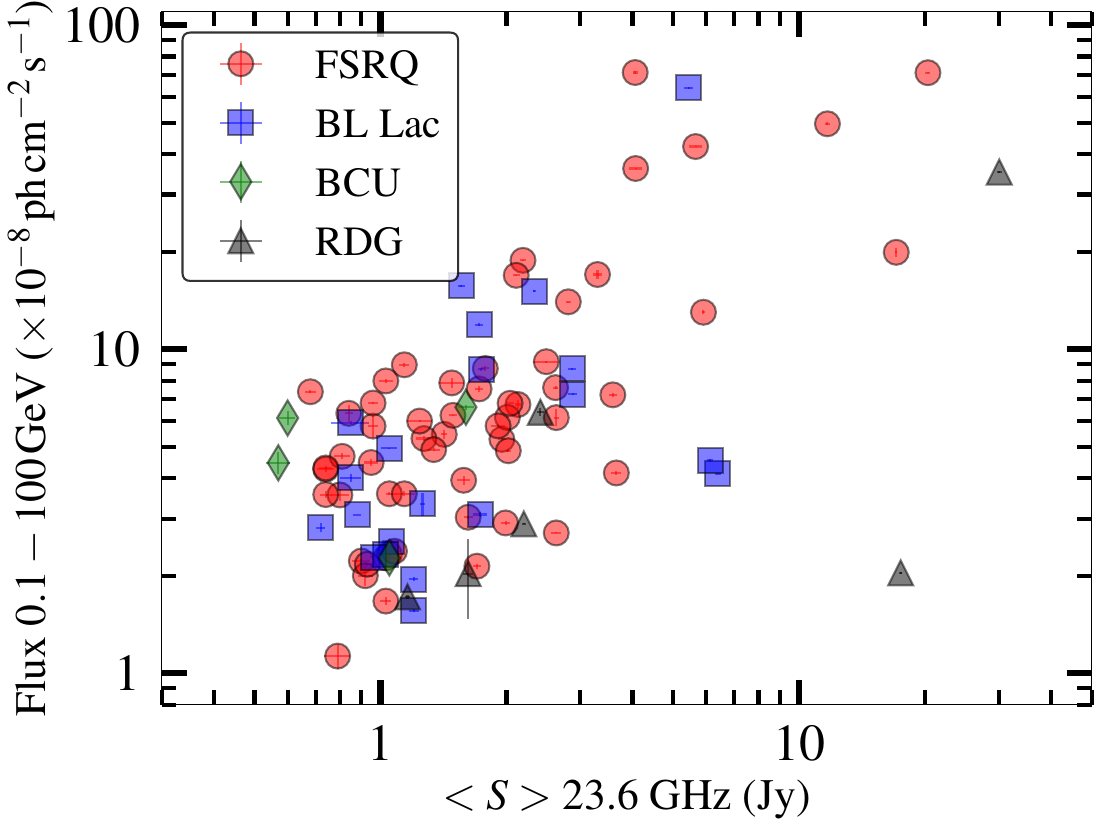}
    \caption{Relation between $\gamma$-ray flux and radio flux density at 4.8 GHz (top panel) and 23.6 GHz (bottom panel).}
    \label{fig:fermi_flux_vs_4p8GHz_and_23p6GHz}
\end{figure}

\begin{figure}[h!]
	\centering
	\includegraphics[width=\columnwidth]{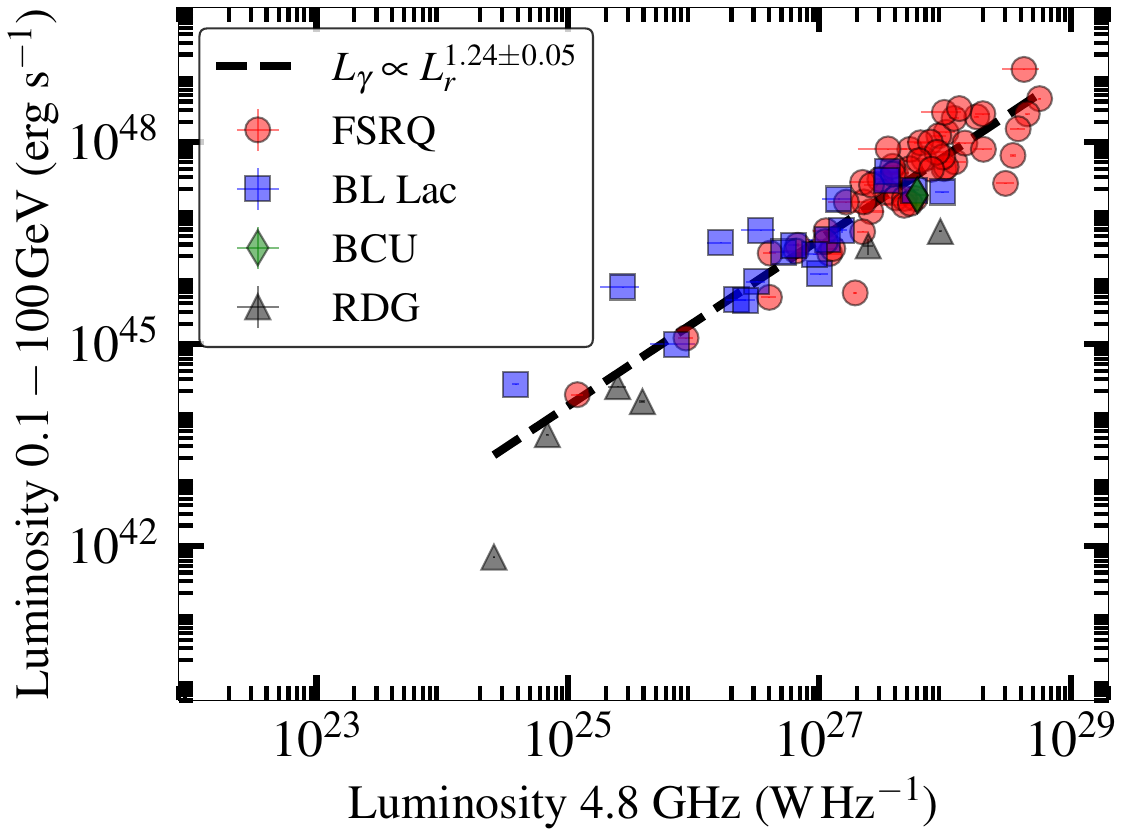}
    \includegraphics[width=\columnwidth]{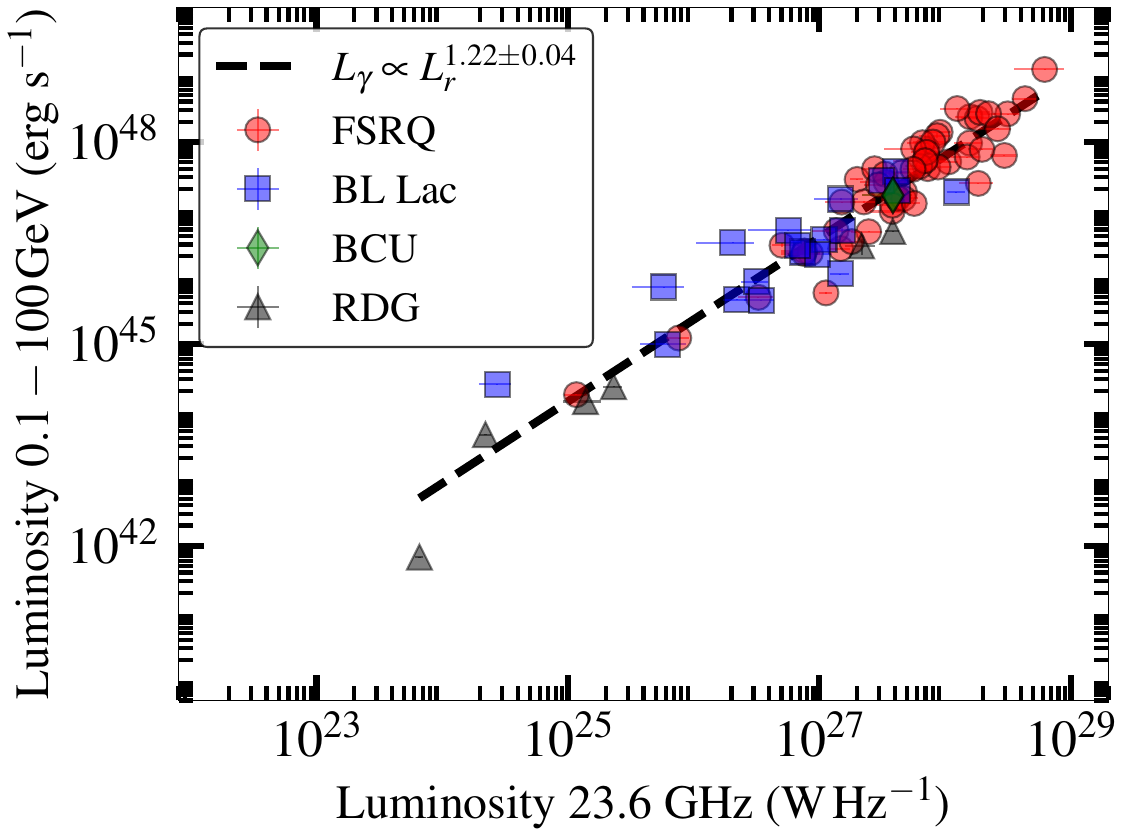}
    \caption{The relationship between $\gamma$-ray luminosity and radio luminosity is illustrated. The dashed line represents the fitted power-law relation between the two luminosities.}
    \label{fig:radio_lum_4p8_and_23p6_vs_gamma_lum}
\end{figure}

Figure~\ref{fig:lum_vs_redshift} shows the radio luminosity at 23.6 GHz as a function of redshift (top panel) and the $\gamma$-ray luminosity as a function of redshift (bottom panel) for the sample of SMMAN sources having radio and $\gamma$-ray luminosities. Clearly, a selection effect is evident: sources with higher luminosities are preferentially detected at higher redshifts in both the radio and $\gamma$-ray bands. FSRQs exhibit higher radio and $\gamma$-ray luminosities and are predominantly found at higher redshifts, whereas BL Lacs tend to have lower radio and $\gamma$-ray luminosities and are generally located at lower redshifts. Radio galaxies span a broader range, extending from the lowest luminosities and redshifts to intermediate values between the FSRQ and BL Lac regimes. Such trend is also noticed for the radio luminosity at 4.8 GHz as a function of redshift.
Figure~\ref{fig:fermi_flux_vs_4p8GHz_and_23p6GHz} shows the $\gamma$-ray flux and radio flux density relation (top panel for 4.8 GHz and bottom panel for 23.6 GHz). The Kendall's $\tau$ rank correlation coefficient is 0.29 with a $p$-value of $1.1\times10^{-5}$ and 0.37 with a $p$-value of $3.6\times10^{-7}$ for the cases of 4.8 and 23.6 GHz, respectively, thus confirming the correlation between these quantities. The $\gamma$-ray luminosities were calculated as explained in Section~\ref{sec:observation}. The
relation between $\gamma$-ray and radio luminosity is presented in Figure~\ref{fig:radio_lum_4p8_and_23p6_vs_gamma_lum}.
A linear fit in log-log space gives $L_{\gamma} \propto L_{r}^{1.24\pm0.05}$ (Pearson correlation coefficient = 0.68) and $L_{\gamma} \propto L_{r}^{1.22\pm0.04}$ (Pearson correlation coefficient = 0.86) for radio luminosities of 4.8 and 23.6 GHz, respectively.  Such a correlation could be induced by observational effects, which depend on the source distance \citep[e.g.,][]{bock2016A&A}. This is further supported by Figure~\ref{fig:lum_vs_redshift}, which indicates that the observed correlation in Figure~\ref{fig:radio_lum_4p8_and_23p6_vs_gamma_lum} is likely largely driven by the fact that our sample is both radio flux-limited and $\gamma$-ray flux-limited. Consequently, a strong correlation is expected as a result of Malmquist bias.\\

\begin{figure}[h!]
	\centering
	\includegraphics[width=\columnwidth]{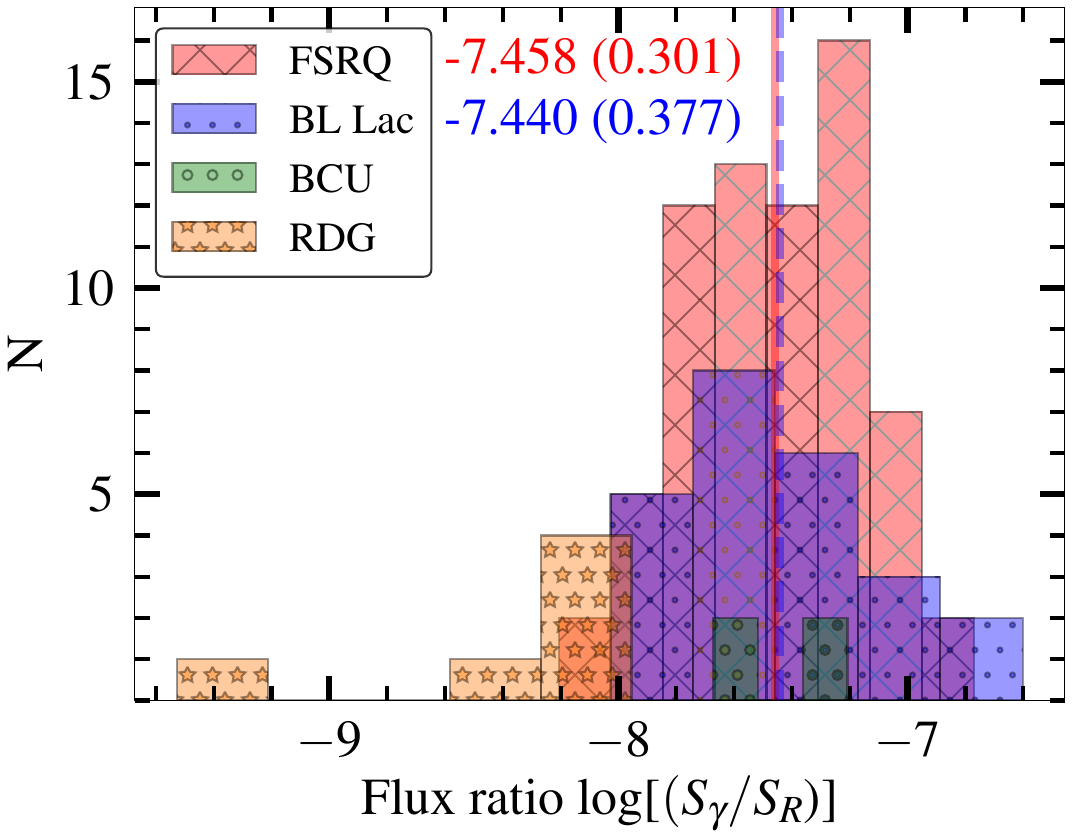}\\
    \includegraphics[width=\columnwidth]{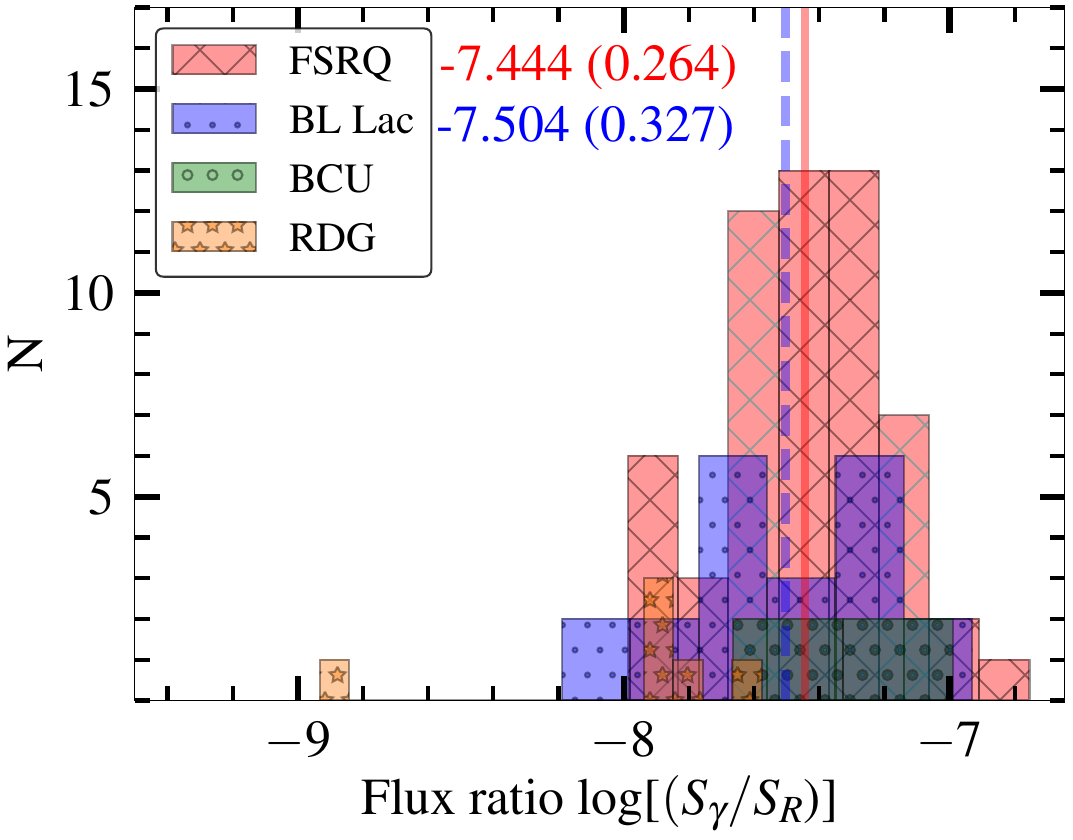}
    \caption{Top: distribution of the ratio between $\gamma$-ray flux and radio flux density (4.8 GHz). Bottom: distribution of the ratio between $\gamma$-ray flux and radio flux density (23.6 GHz). The transparent solid red and dashed blue vertical lines represent the average values with standard deviation inside parentheses for FSRQs and BL Lacs, respectively.}
    \label{fig:gamma_loudness_with_4p8GHz_and_23p6}
\end{figure}

Following the definition of $\gamma$-ray loudness in \citet{bock2016A&A}
as the ratio of the total $\gamma$-ray flux to the radio flux density
($S_{\gamma}/S_{R}$), the distribution of $\gamma$-ray loudness is shown in Figure~\ref{fig:gamma_loudness_with_4p8GHz_and_23p6} for both C and K band data.
Taking into account the case of both frequency bands, the $\gamma$-ray loudness value ranges from $\sim-9.53$ to $\sim-6.60$. It is noted that both FSRQ and BL Lac classes have a higher loudness factor compared to radio galaxies, and the value for BCUs falls between them. The mean value of $\gamma$-ray loudness for FSRQs and BL Lacs is at $-7.46~(\sigma=0.30)$ and $-7.44~(\sigma=0.38)$, respectively, in the C band case. For the K band, the mean values are $-7.44~(\sigma=0.26)$ and $-7.50~(\sigma=0.33)$ for FSRQs and BL Lacs, respectively.\\

\begin{figure}[h!]
	\centering
	\includegraphics[width=\columnwidth]{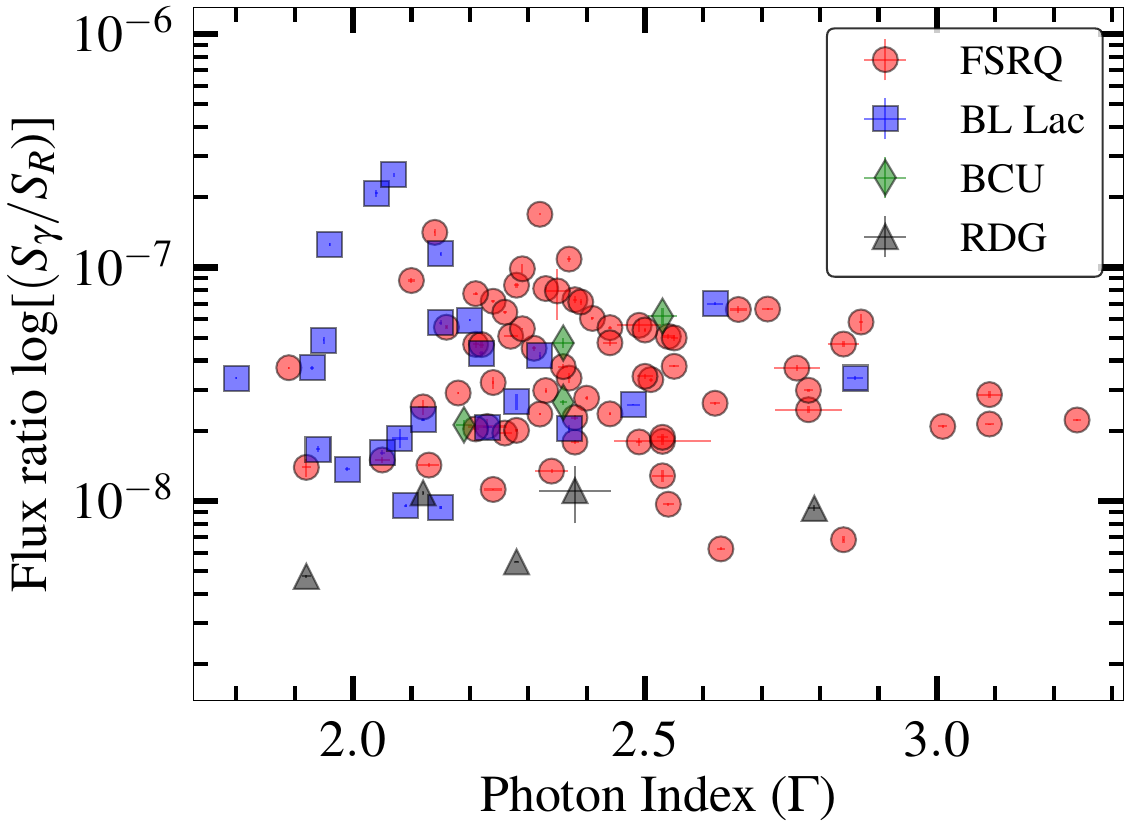}\\
    \includegraphics[width=\columnwidth]{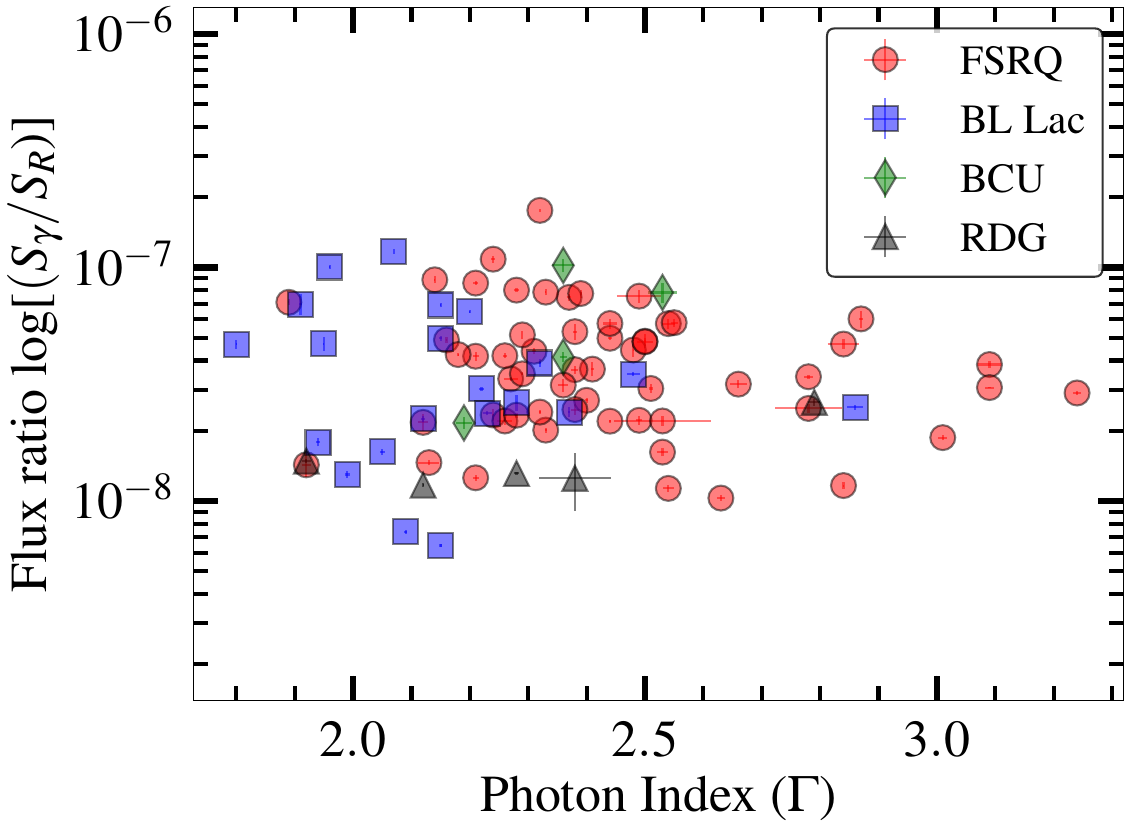}
    \caption{Scatter plot of $\gamma$-ray loudness and $\gamma$-ray spectral index using 4.8 GHz data (top panel) and 23.6 GHz data (bottom panel).}
    \label{fig:gamma_loudness_vs_lat_index}
\end{figure}

The distribution of the $\gamma$-ray loudness shows a similar dependence on the source class as the $\gamma$-ray spectral index distribution.
Both BL Lacs and FSRQs have a wider distribution range; however, BL Lacs appear to have slightly higher $\gamma$-ray loudness factor compared to FSRQs. This relationship is consistent with shifts in the peak frequencies of the SEDs. When the synchrotron peak moves to higher frequencies, the radio-band flux density decreases. In contrast, a shift in the high-energy SED peak to higher frequencies increases the observed $\gamma$-ray flux and results in a more intense spectrum. This interpretation is further supported by the observed weak anticorrelation (Kendall $\tau$ rank correlation = -0.17 at 4.8 GHz and -0.19 at 23 GHz) between the $\gamma$-ray spectral index and the $\gamma$-ray loudness as shown in Figure~\ref{fig:gamma_loudness_vs_lat_index}. The weak anticorrelation observed here may be due to the consideration of the average $\gamma$-ray spectral index value calculated from the $\gamma$-ray monthly light curve taken from the {\sl Fermi} LCR. This pattern is in agreement with the study reported by \citet{bock2016A&A} with a strong anticorrelation (see Figure 9).\\

\section{Summary and Conclusions}\label{sec:summary}
We present a first release of a dataset of 131 gamma-ray-loud AGNs monitored using the Nanshan 26-m Radio Telescope (NSRT) during the November 2016 to November 2024 period in the frequency band of 4.8 and 23.6 GHz. We summarize and conclude our findings as follows.
\begin{enumerate}
           \item The main objective of the SMMAN program was primarily to offer essential multi-frequency radio monitoring at the lower frequency bands that complements the $\gamma$-ray observations conducted by {\sl Fermi}-LAT along with the other multiband monitoring program and to investigate all pertinent aspects of radio physics of gamma-ray-loud AGNs.

           \item The light curves presented here are unbinned, and each data point was subjected to proper post-measurement corrections, including Gaussian fitting, pointing calibration, atmospheric opacity, gain calibration,  and absolute flux density calibration. The complete dataset has undergone a final quality check. The light curves of 131 SMMAN sources and three calibrator sources are presented in Figure~\ref{fig:SMMAN_sources_lc} of Appendix~\ref{Appendix B:SMMAN_lcs}.

           \item From the variability analysis, we find a trend in which sources show a higher degree of variability at 23.6 GHz than at 4.8 GHz, with BL Lacs exhibiting higher mean variability levels than FSRQs. BCUs and radio galaxies show lower variability compared to BL Lacs and FSRQs. In both frequency bands, FSRQs show higher radio luminosities than BL Lacs and other source classes. BCUs display radio luminosities intermediate between those of BL Lacs and FSRQs, while RDGs have lower luminosities, generally below the mean luminosity of FSRQs.

           \item The radio spectral index indicates that BL Lac objects generally have flatter spectra, whereas FSRQs, BCUs, and RDGs span a wider range from flat to steep. At 4.8 and 23.6 GHz, about 90\% of sources with $\alpha\geq-0.4$ display variability above 10\%, predominantly among FSRQs and BL Lacs. Sources with $\alpha\leq-0.4$, however, show little variability at either frequency.

           \item The $\gamma$-ray loudness factor was estimated for different sources, and it was found that FSRQs and BL Lacs are louder than radio galaxies, whereas BCUs lie between these two populations.
           The distribution of $\gamma$-ray loudness exhibits a dependence on source class similar to that of the $\gamma$-ray spectral index distribution. In addition, a weak anticorrelation is observed between the $\gamma$-ray spectral index and the $\gamma$-ray loudness factor.\\

           \item The data release of long-term light curve from this monitoring program in the lower frequency bands provides a unique coverage of the radio part of the SED and is a good addition to the other historical and ongoing programs covering the frequency range from low to high frequency bands.

\end{enumerate}

The users using the SMMAN radio data in their publications are kindly requested to acknowledge the source of the data by citing this paper. To use the data and for details, please contact Lang Cui (cuilang@xao.ac.cn).

\section*{Acknowledgements}
We are thankful to S\'andor Frey, Timur Mufakharov, and Junhui Fan for carefully reading the manuscript and providing insightful comments.
We also thank the anonymous reviewer for valuable comments, which helped us to improve the manuscript.
This work was supported by the Tianshan Talent Training Program (grant No. 2023TSYCCX0099) and the National Key R\&D Program of China (grant Nos. 2024YFA1611500 and 2022SKA0120102). This work was also partly supported by the Urumqi Nanshan Astronomy and Deep Space Exploration Observation and Research Station of Xinjiang (XJYWZ2303) and the Central Guidance for Local Science and Technology Development Fund (grant No. ZYYD2026JD01). The Nanshan 26-m Radio Telescope (NSRT) is operated by the Xinjiang Astronomical Observatory (XAO), CAS. This research has made use of the SIMBAD database, operated at CDS, Strasbourg, France. This research has made use of the NASA/IPAC Extragalactic Database (NED), which is operated by the Jet Propulsion Laboratory, California Institute of Technology, under contract with the National Aeronautics and Space Administration. This research has used the RATAN-600 data, provided by the Special Astrophysical Observatory of the Russian Academy of Sciences (SAO RAS).

%%
%% pdflatex sample631.tex
%% bibtext sample631
%% pdflatex sample631.tex
%% pdflatex sample631.tex

%% Appendix material should be preceded with a single \appendix command.
%% There should be a \section command for each appendix. Mark appendix
%% subsections with the same markup you use in the main body of the paper.

%% Each Appendix (indicated with \section) will be lettered A, B, C, etc.
%% The equation counter will reset when it encounters the \appendix
%% command and will number appendix equations (A1), (A2), etc. The
%% Figure and Table counter will not reset.

\appendix
\section{Source sample}\label{Appendix A:SMMAN_srcs}
Table~\ref{tab:list_of_monitored_srcs} lists all the sources included in the current data release. In Table~\ref{tab:smman_overlap_srcs}, the information on whether the data for SMMAN sources is available in the other historical or the ongoing monitoring programs, such as UMRAO, F-GAMMA, RATAN-600, OVRO, ROBIN, MRO, and SMA, is provided with the corresponding name in that monitoring program.

\begin{table*}[h!]
\centering
\caption{Information regarding the NSRT monitored sources in the SMMAN program.}
\begin{tabular}{lcccccrcrc}\hline\hline
Name 3FGL  &  Assoc. Name      &\multicolumn{2}{c}{Coordinates (J2000)}  &AGN    &z &\multicolumn{2}{c}{4.8 GHz} &\multicolumn{2}{c}{23.6 GHz}\\
 &     &RA   &Dec  &Class   &  &$N$  & $\langle S\rangle$ &$N$  & $\langle S\rangle$\\
 &     &     &   &  &   &  &(Jy) &  &(Jy)\\
(1) &(2)     &(3)     &(4)   &(5)  &(6)   &(7)  &(8) &(9)  &(10)\\
\hline
J0012.4+7040 &TXS 0008+704  &00 11 31.90   &+70 45 31.62   &bcu  & ...  & 163  & 0.75 (0.25) & 68 & 0.88 (0.30)  \\
J0014.6+6119	&4C +60.01     &00 14 48.79   &+61 17 43.54   &bcu  & ... &226  &1.97 (0.07) &52 &0.83 (0.10)   \\
J0059.6+0003	 &PKS 0056-00   &00 59 05.51   &+00 06 51.62   &fsrq  &0.719  &100  &1.39 (0.11) & ... & ...  \\
J0102.8+5825	 &TXS 0059+581  &01 02 45.76   &+58 24 11.14   &fsrq  &0.644  &227 &2.43 (0.68) &227 &3.59 (1.21)  \\
J0108.7+0134	 &4C +01.02     &01 08 38.77   &+01 35 00.32   &fsrq  &2.099  &146 &3.88 (1.27)  &157 &5.66 (2.44) \\
J0110.2+6806	 &4C +67.04     &01 10 12.87   &+68 05 41.22   &bll  &0.290  &173 &0.89 (0.05) & ... & ... \\
J0130.8+1441	&4C +14.06     &01 29 55.35   &+14 46 47.84   &fsrq  &1.630 &39 &0.57 (0.05) & ... & ...  \\
J0137.0+4752	&OC 457        &01 36 58.59   &+47 51 29.10   &fsrq  &0.859  &201 &1.91 (0.23) &171 &2.63 (0.51)  \\
J0205.0+1510	&4C +15.05     &02 04 50.41   &+15 14 11.04   &bcu  &0.834 &157 &2.50 (0.42) &93 &1.60 (0.69)  \\
J0217.5+7349	&S5 0212+73   &02 17 30.81   &+73 49 32.62   &fsrq  &2.346  &211 &3.42 (0.12) &226 &2.62 (0.56) \\
J0221.1+3556	&B0218+357     &02 21 05.47   &+35 56 13.72   &FSRQ  &0.944 &124 &1.03 (0.08) &15 &0.68 (0.08)  \\
J0222.6+4301	&3C 66A        &02 22 39.61   &+43 02 07.80   &BLL  &0.370 & ... & ... &5 &0.85 (0.20)  \\
J0237.9+2848	&4C +28.07     &02 37 52.41   &+28 48 08.99  &FSRQ  &1.213 &175 &2.43 (0.44) &140 &2.19 (0.46)  \\
J0238.6+1636	&AO 0235+164   &02 38 38.93   &+16 36 59.28  &BLL  &0.940 &284 &1.49 (0.48) &213 &1.74 (0.45)   \\
J0242.3+1059	&OD 166        &02 42 29.17   &+11 01 00.73  &fsrq  &2.680 &52 &0.84 (0.04) &29 &0.74 (0.18)  \\
J0303.6+4716	&4C +47.08     &03 03 35.24   &+47 16 16.27  &bll  &0.475 &184 &1.85 (0.58) &105 &1.73 (0.64)  \\
J0308.6+0408	&NGC 1218      &03 08 26.22   &+04 06 39.30  &rdg  &0.028 &144 &3.61 (0.10) &83 &1.16 (0.13)  \\
J0319.8+4130	&NGC 1275      &03 19 48.16   &+41 30 42.11  &RDG  &0.018 &140 &32.46 (5.45) &145 &30.08 (5.15) \\
J0336.5+3210	&NRAO 140      &03 36 30.11   &+32 18 29.34  &fsrq  &1.265 &156 &1.91 (0.22) &117 &1.42 (0.23) \\
J0354.1+4643	&B3 0350+465   &03 54 30.01   &+46 43 18.75  &bcu  & ... &56 &0.72 (0.13) &16 &0.57 (0.15) \\
J0358.8+6002	&TXS 0354+599  &03 59 02.64   &+60 05 22.07  &fsrq	  &0.455 &181 &1.12 (0.09) &59 &0.84 (0.14) \\
J0418.5+3813c	 &3C 111       &04 18 21.28   &+38 01 35.80  &rdg  &0.048  &275 &6.86 (0.39) &271 &2.41 (0.81) \\
J0423.8+4150	&4C +41.11     &04 23 56.01   &+41 50 02.71  &bll  &$0.830^\dag$  &129 &1.58 (0.24) &83 &1.18 (0.22) \\
J0510.0+1802	&PKS 0507+17   &05 10 02.37   &+18 00 41.58  &fsrq  &0.416 &117 &1.02 (0.12) &117 &2.13 (0.42)  \\
J0512.9+4038	&B3 0509+406   &05 12 52.54   &+40 41 43.62  &bcu  & ...  &140 &1.07 (0.21) &80 &1.05 (0.33) \\
J0517.4+4540	&4C +45.08     &05 17 28.90   &+45 37 04.86  &FSRQ  &0.839 &101 &0.79 (0.08) & ... & ... \\
J0519.3+2746	&4C +27.15     &05 19 33.02    &+27 44 04.27   &bcu  &$0.068^\dag$ &106 &0.72 (0.04) & ... & ...  \\
J0521.7+2113	&TXS 0518+211  &05 21 45.96   &+21 12 51.45  &bll  &0.108 &27 &0.50 (0.09) & ... & ...  \\
J0532.7+0732	&OG 050        &05 32 39.00   &+07 32 43.34  &FSRQ  &1.254  &125 &1.67 (0.20) &95 &1.72 (0.34) \\
J0533.2+4822	 &TXS 0529+483  &05 33 15.86   &+48 22 52.81  &fsrq  &1.160 &105 &1.13 (0.31) &68 &1.27 (0.44)  \\
J0603.8+2155	 &4C +22.12     &06 03 51.56   &+21 59 37.70  &bcu  & ... &113 &1.29 (0.05) &32 &0.60 (0.15)  \\
J0638.6+7324	 &S5 0633+73    &06 39 21.96   &+73 24 58.04  &fsrq  &1.850 &161 &0.99 (0.13) &121 &0.96 (0.27) \\
J0710.5+4732	 &S4 0707+47    &07 10 46.10   &+47 32 11.14  &bll  &1.292 &74 &0.61 (0.06) & ... & ...  \\
J0721.9+7120	 &S5 0716+71    &07 21 53.45   &+71 20 36.36  &BLL  &0.230 &648 &1.25 (0.27) &626  &1.56 (0.80) \\
J0750.6+1232	&OI 280         &07 50 52.05   &+12 31 04.83  &fsrq  &0.889 &127 &1.91 (0.32) &95 &1.95 (0.41) \\
J0757.0+0956	 &PKS 0754+100  &07 57 06.64   &+09 56 34.85  &bll  &0.266  &126 &1.21 (0.22) &84 &1.20 (0.26) \\
J0758.7+3747	 &NGC 2484      &07 58 28.11   &+37 47 11.81  &rdg  &0.041  &103 &1.22 (0.04) & ... & ... \\
J0805.4+6144	&TXS 0800+618   &08 05 18.18   &+61 44 23.70   &fsrq  &3.018  &66 &0.66 (0.09) & ... & ... \\
J0809.5+4045	&S4 0805+41     &08 08 56.65   &+40 52 44.89   &fsrq  &1.418  &118 &1.08 (0.14) &3 &1.48 (0.18) \\
J0818.2+4223	&S4 0814+42    &08 18 16.00   &+42 22 45.42   &bll  &0.530  &120 &1.30 (0.28) & ... & ... \\
J0824.9+5551	&OJ 535     &08 24 47.24   &+55 52 42.67   &fsrq  &1.419  &166 &1.20 (0.13) & ... & ... \\
J0824.9+3916	&4C +39.23  &08 24 55.48   &+39 16 41.91   &fsrq  &1.216  &93 &1.24 (0.08) &31 &0.90 (0.24) \\
J0830.7+2408	&OJ 248     &08 30 52.09   &+24 10 59.82   &FSRQ  &0.939  &99 &0.95 (0.11) &35 &0.95 (0.22) \\
\hline
\end{tabular}
\label{tab:list_of_monitored_srcs}
\tablecomments{The 3FGL name (Col. 1); an association (often more common) name (Col. 2); the coordinates in RA-Dec (Cols. 3 and 4); the AGN class (Col. 5) according to {\sl Fermi} Large Area Telescope Fourth Source Catalog Data Release 4 (4FGL-DR4;~\citealt{ballet2023}) with designations in capital letters represent confirmed identifications, while those in lower case letters denote associations; the redshift (Col. 6) taken from NASA/IPAC Extragalactic Database (NED) and the SIMBAD astronomical database \citep{wenger2000}. Here, the $\dag$ symbol denotes a photometric redshift; otherwise, the redshift is spectroscopic}; the number of observations in 4.8 and 23.6 GHz for each source (Col. 7 and 9), and the average flux density calculated over these observations (Col. 8 and 10).
\end{table*}

\addtocounter{table}{-1}
\begin{table*}[h!]
\centering
\caption{Continued.}
\begin{tabular}{lcccccrcrc}\hline\hline
Name 3FGL  &  Assoc. Name      &\multicolumn{2}{c}{Coordinates (J2000)}  &AGN    &z &\multicolumn{2}{c}{4.8 GHz} &\multicolumn{2}{c}{23.6 GHz}\\
 &     &RA   &Dec  &Class   &  &$N$  & $\langle S\rangle$ &$N$  & $\langle S\rangle$\\
 &     &     &   &  &   &  &(Jy) &  &(Jy)\\
(1) &(2)     &(3)     &(4)   &(5)  &(6)   &(7)  &(8) &(9)  &(10)\\
\hline
J0832.6+4914	&OJ 448     &08 32 23.22   &+49 13 21.04   &bll  &0.548  &43 &0.56 (0.10) & ... & ... \\
J0840.8+1315	&3C 207     &08 40 47.59   &+13 12 23.56   &ssrq &0.680  &123 &1.83 (0.17) &89 &1.62 (0.43) \\
J0841.4+7053	&0836+710   &08 41 24.37   &+70 53 42.17   &FSRQ  &2.213 &889 &2.88 (0.14) &776 &2.01 (0.46)  \\
J0854.8+2006	&OJ 287     &08 54 48.87   &+20 06 30.64   &BLL  &0.306 &138 &4.40 (0.74) &126 &6.40 (1.12)  \\
J0903.1+4649	&S4 0859+47 &09 03 03.99   &+46 51 04.14   &fsrq  &1.465 &139 &1.39 (0.08) &77 &0.96 (0.23)  \\
J0904.3+4240	&S4 0900+42    &09 04 15.63   &+42 38 04.76   &fsrq  &1.342 &61 &0.85 (0.08) & ... & ... \\
J0909.1+0121	&PKS 0906+01   &09 09 10.09   &+01 21 35.62   &fsrq  &1.023 &86 &1.60 (0.21) &77 &1.58 (0.48) \\
J0916.3+3857	&S4 0913+39    &09 16 48.90   &+38 54 28.15   &fsrq &1.268  &58 &0.80 (0.06) & ... & ... \\
J0920.9+4442	&S4 0917+44    &09 20 58.46   &+44 41 53.98   &fsrq  &2.188  &123 &1.34 (0.13) &132 &2.04 (0.51)\\
J0921.8+6215	&OK 630        &09 21 36.23   &+62 15 52.18   &fsrq  &1.458  &172 &1.10 (0.13) &77 &1.14 (0.28) \\
J0948.6+4041	&4C +40.24     &09 48 55.34   &+40 39 44.58   &fsrq  &1.250  &112 &1.22 (0.11) & ... & ... \\
J0957.6+5523	&4C +55.17     &09 57 38.18   &+55 22 57.77   &fsrq  &0.901  &159 &1.83 (0.05) &55 &0.96 (0.20) \\
J0957.4+4728	&OK 492        &09 58 19.67   &+47 25 07.84   &fsrq  &1.884 &115 &1.16 (0.25) &38 &1.03 (0.19)  \\
J0958.6+6534	&S4 0954+65    &09 58 47.25   &+65 33 54.82   &BLL  &0.369 &542 &1.04 (0.32) &565 &1.72 (0.90)  \\
J1023.1+3952	&4C +40.25     &10 23 11.57   &+39 48 15.38   &fsrq  &1.254 &86 &0.72 (0.07) & ... & ...  \\
J1033.2+4116	&S4 1030+41    &10 33 03.71   &+41 16 06.23   &fsrq  &1.117 &115 &1.30 (0.45) &57 &1.05 (0.26)  \\
J1033.8+6051	&S4 1030+61    &10 33 51.43   &+60 51 07.33   &FSRQ  &1.408 &73 &0.67 (0.13) & ... & ...  \\
J1043.1+2407	&B2 1040+24A    &10 43 09.04   &+24 08 35.41   &fsrq  &0.562 &84 &0.79 (0.17) &74 &0.92 (0.32) \\
J1044.4+8058	&S5 1039+81    &10 44 23.06   &+80 54 39.44   &fsrq  &0.168 &175 &1.20 (0.16) &141 &1.05 (0.22) \\
J1048.4+7144	&S5 1044+71    &10 48 27.62   &+71 43 35.94   &FSRQ  &1.140 &206 &2.01 (0.71) &235 &2.11 (0.64) \\
J1058.5+0133	&4C +01.28     &10 58 29.61   &+01 33 58.82   &BLL  &0.894  &106 &4.75 (0.59) &93 &6.13 (0.99) \\
J1128.0+5921	&TXS 1125+596  &11 28 13.34   &+59 25 14.80   &fsrq  &1.799 &159 &1.08 (0.13) &109 &0.93 (0.24) \\
J1131.4+3819	&B2 1128+38    &11 30 53.28   &+38 15 18.55   &fsrq  &1.740 &120 &1.75 (0.31) &36 &1.91 (0.47)  \\
J1132.8+1015	&4C +10.33     &11 32 59.49   &+10 23 42.26    &FSRQ  &0.540 &16 &0.45 (0.04) & ... & ...  \\
J1145.1+1935	&3C 264        &11 45 05.01   &+19 36 22.74   &rdg  &0.022 &121 &2.04 (0.06) & ... & ...  \\
J1146.8+3958	&S4 1144+40    &11 46 58.30   &+39 58 34.30   &fsrq  &1.088 &121 &1.51 (0.26) &12 &2.49 (0.59)  \\
J1153.4+4932	&OM 484        &11 53 24.47   &+49 31 08.83   &FSRQ  &0.334 &100 &1.32 (0.18) &71 &1.08 (0.34)  \\
J1153.4+4033	&B3 1151+408   &11 53 54.66   &+40 36 52.62   &fsrq  &0.925 &225 &1.02 (0.10) & ... & ...  \\
J1159.5+2914	&Ton 599       &11 59 31.83   &+29 14 43.83   &fsrq  &0.725 &142 &2.55 (1.09) &139 &4.07 (1.69) \\
J1213.7+1306	&4C +13.46     &12 13 32.16   &+13 07 20.56   &fsrq  &1.140 &42 &0.74 (0.06) & ... & ...  \\
J1224.9+2122	&4C +21.35     &12 24 54.46   &+21 22 46.39   &FSRQ  &0.434  &122 &2.15 (0.49) &81 &1.34 (0.46) \\
J1229.1+0202	&3C 273     &12 29 06.70   &+02 03 08.60   &FSRQ  &0.158 &282 &29.12 (0.85) &421 &17.07 (2.17) \\
J1230.9+1224	 &M 87     &12 30 49.42   &+12 23 28.04  &rdg  &0.004 &122 &68.46 (1.35) &118 &17.49 (1.32)  \\
J1239.4+0727	&PKS 1236+077  &12 39 24.59   &+07 30 17.19   &bll  &0.400 &62 &0.74 (0.11) & ... & ...  \\
J1256.1-0547	&3C 279   &12 56 11.17   &-05 47 21.52   &FSRQ  &0.536 &178 &13.03 (0.89) &335 &20.32 (5.12)  \\
J1258.1+3233	&ON 393        &12 57 57.23   &+32 29 29.33   &fsrq  &0.806 &42 &0.66 (0.15) & ... & ...  \\
J1302.6+5748	&TXS 1300+580  &13 02 52.47   &+57 48 37.61   &bll  &0.950 &43 &0.62 (0.14) & ... & ...  \\
J1309.5+1154	&4C +12.46     &13 09 33.93   &+11 54 24.55   &BLL  &1.453  &41 &0.93 (0.07) &22 &0.87 (0.12)\\
J1310.6+3222	&OP 313        &13 10 28.66   &+32 20 43.78   &fsrq  &1.678 &133 &1.72 (0.60) &64 &3.30 (0.69) \\
J1351.1+0030	&PKS 1348+007  &13 51 04.43    &+00 31 19.40    &fsrq  &2.084 &4 &0.54 (0.22) & ... & ...  \\
J1419.9+5425	&OQ 530        &14 19 46.60   &+54 23 14.79   &bll  &0.152 &124 &1.25 (0.48) &71 &1.06 (0.42)  \\
J1442.6+5156	&3C 303        &14 43 02.78    &+52 01 37.27  &rdg  &0.141  &113 &1.16 (0.05) &36 &0.85 (0.19) \\
J1500.6+4750	&TXS 1459+480  &15 00 48.65   &+47 51 15.54   &bll  &1.059 &33 &0.47 (0.09) & ... & ...  \\
J1540.8+1449	&4C +14.60     &15 40 49.49   &+14 47 45.88   &bll  &0.606 &109 &1.24 (0.23) & ... & ...  \\
J1549.4+0237	&PKS 1546+027  &15 49 29.44   &+02 37 01.16   &fsrq  &0.414  &106 &2.86 (0.61) & ... & ... \\
\hline
\end{tabular}
%\label{tab:list_of_monitored_srcs}
\end{table*}

\addtocounter{table}{-1}
\begin{table*}[h!]
\centering
\caption{Continued.}
\begin{tabular}{lcccccrcrc}\hline\hline
Name 3FGL  &  Assoc. Name      &\multicolumn{2}{c}{Coordinates (J2000)}  &AGN    &z &\multicolumn{2}{c}{4.8 GHz} &\multicolumn{2}{c}{23.6 GHz}\\
 &     &RA   &Dec  &Class   &  &$N$  & $\langle S\rangle$ &$N$  & $\langle S\rangle$\\
 &     &     &   &  &   &  &(Jy) &  &(Jy)\\
(1) &(2)     &(3)     &(4)   &(5)  &(6)   &(7)  &(8) &(9)  &(10)\\
\hline
J1550.5+0526	&4C +05.64     &15 50 35.27   &+05 27 10.45   &fsrq  &1.420 &98 &2.30 (0.08) & ... & ...  \\
J1604.6+5714	&GB6 J1604+5714 &16 04 37.35   &+57 14 36.66   &fsrq  &0.720  & ... & ... &40 &0.80 (0.25) \\
J1613.8+3410	&OS 319       &16 13 41.06   &+34 12 47.91   &fsrq  &1.398  &132 &4.34 (0.72) &118 &2.63 (0.79)\\
J1635.2+3809	&4C +38.41   &16 35 15.49   &+38 08 04.50   &FSRQ  &1.814 &120 &2.53 (0.35) &105 &2.81 (0.40)  \\
J1637.7+4715	&4C +47.44  &16 37 45.13   &+47 17 33.83   &fsrq  &0.735  &91 &0.89 (0.07) &34 &1.03 (0.18) \\
J1637.9+5719	&OS 562  &16 38 13.46   &+57 20 23.98   &fsrq  &0.751 &144 &1.45 (0.14) &127 &1.62 (0.44)  \\
J1640.6+3945	&NRAO 512      &16 40 29.63   &+39 46 46.03   &FSRQ  &1.672 &91 &0.98 (0.15) &30 &0.81 (0.20)  \\
J1642.9+3950	&3C 345      &16 42 58.81   &+39 48 36.99   &FSRQ  &0.593  &313 &5.50 (0.45) &437 &5.91 (0.72) \\
J1653.9+3945	&Mkn 501       &16 53 52.22   &+39 45 36.61   &BLL  &0.033  &113 &1.47 (0.09) &41 &1.05 (0.30) \\
J1715.7+6837	&S4 1716+68    &17 16 13.94   &+68 36 38.74   &fsrq  &0.777 &55 &0.55 (0.05) & ... & ... \\
J1727.1+4531	&S4 1726+45    &17 27 27.65   &+45 30 39.73   &fsrq  &0.717  &132 &1.12 (0.32) &49 &1.03 (0.27) \\
J1734.3+3858  &B2 1732+38A   &17 34 20.58   &+38 57 51.44   &fsrq  &0.976  &215 &0.97 (0.20) &251 &1.49 (0.56) \\
J1740.3+4736	&S4 1738+47    &17 39 57.13   &+47 37 58.36   &fsrq  & ... &95 &0.81 (0.09) &25 &0.79 (0.28)  \\
J1740.3+5211	&4C +51.37     &17 40 36.98   &+52 11 43.41   &fsrq  &1.381  &123 &1.11 (0.26) &42 &1.24 (0.55) \\
J1748.6+7005	&S4 1749+70    &17 48 32.84   &+70 05 50.77   &bll  &0.770 &136 &0.82 (0.21) &44 &0.85 (0.33)  \\
J1751.5+0939	&OT 081        &17 51 32.82   &+09 39 00.73   &bll &0.322  &112 &2.02 (0.46) &121 &2.87 (0.64) \\
J1800.5+7827	&S5 1803+784   &18 00 45.68   &+78 28 04.02   &bll  &0.691  &170 &2.54 (0.21) &183 &2.33 (0.24) \\
J1801.5+4403	&S4 1800+44    &18 01 32.31   &+44 04 21.90   &fsrq  &0.663  &103 &0.95 (0.20) &32 &1.14 (0.40) \\
J1806.7+6949	&3C 371   &18 06 50.68   &+69 49 28.11   &fsrq
&0.049  &184 &2.06 (0.25) &266 &2.02 (0.42) \\
J1824.2+5649	&4C +56.27     &18 24 07.07   &+56 51 01.49   &bll  &0.664  &141 &1.25 (0.13) &131 &1.26 (0.21) \\
J1829.6+4844	&3C 380        &18 29 31.78   &+48 44 46.16   &css  &0.692  &132 &5.26 (0.14) &127 &2.20 (0.28) \\
J1849.2+6705   &S4 1849+67    &18 49 16.07   &+67 05 41.68   &FSRQ  &0.657 &136 &1.05 (0.21) &124 &1.70 (0.41)  \\
J1927.7+6118	&S4 1926+61    &19 27 30.44   &+61 17 32.88   &bll  &$0.540^\dag$ &156 &1.14 (0.23) &131 &1.20 (0.38)  \\
J2005.2+7752	&S5 2007+77    &20 05 30.93   &+77 52 43.14   &bll  &0.342  &146 &1.04 (0.18) &68 &1.03 (0.25) \\
J2007.3+6605	&TXS 2007+659  &20 07 28.77   &+66 07 22.54   &fsrq  &1.325 &133 &0.86 (0.21) &39 &0.74 (0.17)  \\
J2010.3+7228	&4C +72.28     &20 09 52.30   &+72 29 19.35   &bll  & ...  &161 &1.10 (0.08) &117 &0.96 (0.15) \\
J2022.5+7612	&S5 2023+760   &20 22 35.58   &+76 11 26.17   &bll  &0.590 &76 &0.67 (0.07) &40 &0.72 (0.11) \\
J2023.2+3154	&4C +31.56     &20 23 19.02   &+31 53 02.31   &bcu  &0.356 &136 &1.48 (0.15) &36 &1.13 (0.21)  \\
J2025.2+3340	&B2 2023+33    &20 25 10.84   &+33 43 00.21   &bll  &0.219 &138 &2.16 (0.42) &120 &2.88 (0.77)  \\
J2031.8+1223	&PKS 2029+121  &20 31 54.99   &+12 19 41.34   &bll  &1.215  &238 &1.19 (0.12) &50 &0.88 (0.13) \\
J2038.8+5113	&3C 418        &20 38 37.03   &+51 19 12.66   &fsrq  &1.686  &135 &4.28 (0.26) &73 &3.66 (0.84) \\
J2110.3+3540	&B2 2107+35A   &21 09 31.88   &+35 32 57.60   &bcu  &0.202  &136 &2.25 (0.60) &112 &2.49 (0.41) \\
J2123.6+0533	&OX 036  &21 23 44.52   &+05 35 22.09   &fsrq  &1.940  &165 &2.04 (0.34) &96 &1.99 (0.46) \\
J2201.7+5047	&NRAO 676      &22 01 43.54   &+50 48 56.39   &fsrq  &1.899  &102 &0.79 (0.07) & ... & ... \\
J2202.7+4217	&BL Lacertae   &22 02 43.29   &+42 16 39.98   &BLL  &0.067 &457 &2.56 (0.86) &371 &5.45 (2.44) \\
J2203.7+3143	&4C +31.63     &22 03 14.98   &+31 45 38.27   &FSRQ  &0.295  &69 &2.30 (0.34) & ... & ... \\
J2212.0+2355	&PKS 2209+236  &22 12 05.97   &+23 55 40.54   &fsrq  &1.127  &42 &0.80 (0.17) &21 &0.59 (0.11) \\
J2232.5+1143	&CTA 102    &22 32 36.41   &+11 43 50.90   &FSRQ  &1.032  &318 &4.22 (0.33) &362 &4.06 (1.11) \\
J2236.3+2829	&B2 2234+28A   &22 36 22.47   &+28 28 57.41   &fsrq  &0.790  &144 &1.57 (0.20) &109 &1.78 (0.41)\\
J2254.0+1608	&3C 454.3  &22 53 57.75   &+16 08 53.56   &FSRQ  &0.859  &336 &17.08 (2.39) &448 &11.69 (3.27) \\
J2321.9+2732	&4C +27.50  &23 21 59.86   &+27 32 46.44   &fsrq  &1.255  &138 &1.23 (0.12) &103 &1.44 (0.59) \\
J2330.5+1104	&4C +10.73     &23 30 40.85   &+11 00 18.71   &fsrq  &1.501  &103 &1.04 (0.08) &24 &0.74 (0.12) \\
J2354.1+4605	&4C +45.51     &23 54 21.68   &+45 53 04.24   &fsrq  &1.992  &97 &1.11 (0.08) & ... & ... \\
\hline
\end{tabular}
%\label{tab:list_of_monitored_srcs}
\end{table*}

\begin{table*}[h!]
\caption{List of SMMAN sources overlapping with UMRAO, F-GAMMA, RATAN-600, OVRO, ROBIN, MRO and SMA.}
\centering
\begin{tabular}{cccccccc}\hline\hline
SMMAN Source   &UMRAO &F-GAMMA  &RATAN-600  &OVRO  &ROBIN &MRO &SMA\\
(1) &(2)     &(3)     &(4)   &(5)  &(6)   &(7)  &(8) \\
\hline
TXS 0008+704  		&...	&...	&...	&...    &... &... &... \\
4C +60.01     		&...	&...	&...	&...   &... &... &0014+612 \\
PKS 0056-00   		&...	&...	&5BZQJ0059+0006 &...	&... &... &...\\
TXS 0059+581  		&0059+581	&J0102+5824	&5BZUJ0102+5824	 &0059+581 &... &0059+581 &0102+584 \\
4C +01.02     		&0106+013	&J0108+0135	&5BZQJ0108+0135	&J0108+0135  &... &0106+013 &0108+015 \\
4C +67.04     		&...	&...	&5BZUJ0110+6805	&TB0110+6805  &...	&... &...\\
4C +14.06    		&...	&...	&...	&...   &... &... &...\\
OC 457        		&0133+476	&J0136+4751	&5BZQJ0136+4751	&J0136+4751	&...  &0133+476 &...\\
4C +15.05               &0202+149 	&...	&5BZUJ0204+1514	&J0204+1514	&... &... &0204+152 \\
S5 0212+73              &0212+735	&J0217+734	&5BZQJ0217+7349	&J0217+7349  &...  &0212+735 &0217+738\\
B2 0218+357               &0218+357	&J0221+3556	&5BZUJ0221+3556	&J0221+3556  &... &0218+357 &...\\
3C 66A                  &...	&J0222+4302	&5BZBJ0222+4302	&3C 66A	& 0219+428 &3C 66A &...\\
4C +28.07     		&0234+285	&J0237+2848	&5BZQJ0237+2848	&J0237+2848	&...  &0234+285 &0237+288\\
AO 0235+164             &0235+164	&J0238+1636	&5BZBJ0238+1636	&J0238+1636	&0235+164  &0235+164 &0238+166\\
OD 166                  &...	&...	&...	&J0242+1101	&...  &... &0242+110\\
4C +47.08               &0300+470	&J0303+4716	&5BZBJ0303+4716	&0300+471 &...  &... &0303+472\\
NGC 1218      		    &...	&...	&... &NGC1218	&... &... &...	\\
NGC 1275                &0316+413	&J0319+4130	&5BZUJ0319+4130	&J0319+4130	&0316+413  &3C 84 &0319+415\\
NRAO 140                &0333+321	&J0336+3218	&5BZQJ0336+3218	&0333+321	&... &0333+321  &0336+323\\
B3 0350+465             &...	&...	&...	&...  &... &...  &0354+467\\
TXS 0354+599    	    &...	&...	&...	&TXS0354+599 &... &... &...\\
3C 111                  &0415+379	&J0418+3801	&...	&0415+379	&... &0415+379 &0418+380\\
4C +41.11               &...	&...	&5BZBJ0423+4150	&TB0423+4150  &...  &... &0423+418\\
PKS 0507+17             &... &J0510+1800	&5BZQJ0510+1800	&J0510+1800	&...  &0507+179 &0510+180\\
B3 0509+406   		    &...	&...	&...  &B30509+406	&... &...	&... \\
4C +45.08               &...	&...	&...	&4C+45.08	&... &... &...\\
4C +27.15               &...	&...	&...	&...  &... &... &... \\
TXS 0518+211            &...	&J0521+2112	&5BZBJ0521+2112	&...  &...  &RGBJ0521+212  &...\\
OG 050                  &0529+075	&J0532+0732	&5BZQJ0532+0732	&J0532+0732	&... &... &0532+075\\
TXS 0529+483            &...	&...	&5BZQJ0533+4822	&TXS0529+483	&...  &... &0533+483\\
4C +22.12               &...	&...	&...	&... &...  &... &... \\
S5 0633+73              &...	&...	&5BZQJ0639+7324	&J0639+7324	&...  &... &... \\
S4 0707+47              &0707+476	&...	&5BZBJ0710+4732	&B30707+476	&... &0707+476 &0710+475\\
S5 0716+71              &0716+714	&J0721+7120	&5BZBJ0721+7120	&J0721+7120	&0716+714  &0716+714  &0721+713\\
OI 280                  &0748+126	&J0750+1231	&5BZQJ0750+1231	&J0750+1231	&...  &0748+126  &0750+125\\
PKS 0754+100            &0754+100	&...	&5BZBJ0757+0956	&J0757+0956  &...  &0754+100  &0757+099	\\
NGC 2484                &0755+379	&...	&...	&...  &...  &... &... \\
TXS 0800+618            &...	&...	&...	&J0805+6144	&...  &... &... \\
S4 0805+41              &0805+410	&...	&5BZQJ0808+4052	&J0808+4052	&... &... &... \\
S4 0814+42              &0814+425	&J0818+4222	&...	&J0818+4222	&...  &0814+425  &0818+423\\
OJ 535                  &0820+560	&J0824+5552	&5BZQJ0824+5552	&J0824+5552	&...  &0820+560 &0824+558\\
4C +39.23               &0821+394	&...	&5BZQJ0824+3916	&J0824+3916	&... &0821+394 &0824+392\\
OJ 248                  &0827+243	&J0830+2410	&5BZQJ0830+2410	&J0830+2410	&0827+243  &OJ 248 &0830+241\\
OJ 448                  &0828+493	&...	&5BZBJ0832+4913	&...   &...  &0828+493 &... \\
3C 207                  &0838+133	&...	&5BZUJ0840+1312	&3C 207	&0838+133  &... &3C207\\
0836+710                &0836+710	&J0841+7053	&...  &0836+710	&0836+710	&0836+710 &... \\
OJ 287                  &0851+202	&J0854+2006	&5BZBJ0854+2006	&J0854+2006	&0851+202  &OJ 287	&0854+201\\
S4 0859+47              &0859+470	&...	&5BZQJ0903+4651	&S4 0859+47	&... &0859+470 &0903+468\\
\hline
\end{tabular}
\label{tab:smman_overlap_srcs}
\tablecomments{SMMAN program monitored sources name (Col. 1); Col. 2, Col .3, Col. 4, Col. 5, Col. 6, Col. 7, and Col. 8 represent the corresponding name of SMMAN source in UMRAO, F-GAMMA, RATAN-600, OVRO, ROBIN, MRO, and SMA monitoring programs, respectively.}
\end{table*}

\addtocounter{table}{-1}
\begin{table*}[h!]
\caption{Continued.}
\centering
\begin{tabular}{cccccccc}\hline\hline
SMMAN Sources   &UMRAO &F-GAMMA  &RATAN-600  &OVRO &ROBIN  &MRO  &SMA\\
(1) &(2)     &(3)     &(4)   &(5)  &(6)   &(7)  &(8) \\
\hline
S4 0900+42             &...	&...	&...	&S4 0900+42	&... &... &... \\
PKS 0906+01            &0906+015	&...	&...	&J0909+0121	&... &0909+013 &0909+013\\
S4 0913+39             &...	 &...	&5BZQJ0916+3854	&S40913+39	&... &... &...\\
S4 0917+44             &0917+449	&J0920+4441	&5BZQJ0920+4441	&J0920+4441	&... &0917+449 &0920+446\\
OK 630                 &0917+624	&...	&5BZQJ0921+6215	&J0921+6215	&... &0921+622 &0921+622\\
4C +40.24              &0945+408 	&...	&...	&J0948+4039	&... &0945+408 &0948+406\\
4C +55.17              &0954+556	&...	&...	&C0957+5522	&... &S4 0954+556 &0957+553\\
OK 492                 &0955+476	&...	&5BZQJ0958+4725	&J0958+4725	&... &0958+474 &0958+474\\
S4 0954+65             &0954+658	&J0958+6533	&5BZBJ0958+6533	&J0958+6533 &0954+658	&S4 0954+65 &0958+655\\
4C +40.25              &1020+400	&...	&5BZQJ1023+3948	&J1023+3948	&...  &... &...\\
S4 1030+41             &1030+415	&...	&5BZQJ1033+4116	&J1033+4116	&... &1030+415 &...\\
S4 1030+61             &...	&J1033+6051	&5BZQJ1033+6051	&J1033+6051	&... &...  &1033+608\\
B2 1040+24             &...	&...	&5BZQJ1043+2408	&J1043+2408	&... &TEX 1040+244  &1043+241\\
S5 1039+81             &...	&...	&...	&J1044+8054	&... &1039+811  &1044+809\\
S5 1044+71             &...	&...	&5BZUJ1048+7143	&J1048+7143	&... &... &1048+717\\
4C +01.28              &1055+018	&J1058+0133	&5BZUJ1058+0133	&J1058+0133 &... &1055+018  &1058+015 \\
TXS 1125+596           &...	&J1128+5925	&...	&J1128+5925	&... &... &... \\
B2 1128+38             &1128+385	&...	&5BZQJ1130+3815	&J1130+3815	&... &... &...\\
4C +10.33              &...	&...	&...	&...	&... &... &... \\
3C 264                 &...	&...	&...	&3C 264	&... &... &... \\
S4 1144+40             &1144+402	&...	&5BZQJ1146+3958	&J1146+3958	&... &... &1146+399\\
OM 484                 &1150+497 	&...	&5BZQJ1153+4931	&OM484	&... &1150+497  &1153+495 \\
B3 1151+408            &...	&...	&...	&...  &...  &... &... \\
Ton 599                &1156+295	&J1159+2914	&5BZQJ1159+2914	&J1159+2914	&1156+295  &4C 29.45 &1159+292\\
4C +13.46              &...	&...	&5BZQJ1213+1307	&4C +13.46	&... &... &... \\
4C +21.35              &1222+216	&J1224+2122	&5BZQJ1224+2122  &C1224+2122 &1222+21	&PKS 1222+216 &1224+213 \\
3C 273                 &1226+023	&J1229+0203	&5BZQJ1229+0203	&J1229+0203	&1226+023 &3C 273 &1229+020 \\
 M 87                  &...	&J1230+1223	&...	&3C 274 &... &3C 274  &1230+123\\
PKS 1236+077           &1236+077	&...	&5BZUJ1239+0730	&J1239+0730	&... &... &1239+075\\
3C 279                 &1256-057	&J1256-0547	&5BZQJ1256-0547	&J1256-0547	&1253–055 &3C 279 &1256-057 \\
ON 393                 &...	&...	&...	&J1257+3229	 &... &... &... \\
TXS 1300+580           &...	&...	&5BZUJ1302+5748	&J1302+5748	&... &1300+580  &... \\
4C +12.46              &...	&...	&5BZBJ1309+1154	&J1309+1154	&... &MC 21307+12 &1309+119 \\
OP 313                 &1308+326	&J1330+3220	&5BZUJ1310+3220	&J1310+3220	&...  &1308+326 &1310+323 \\
PKS 1348+007           &...	&...	&5BZQJ1351+0031	&PKS 1348+007	&... &... &... \\
OQ 530                 &1418+546	&...	&5BZBJ1419+5423	&J1419+5423	&... &OQ 530  &1419+543\\
3C 303                 &1441+522	&...	&...  &...  &...  &... &... \\
TXS 1459+480           &...	&...	&5BZQJ1500+4751	&J1500+4751	&... &... &... \\
4C +14.60              &1538+149	&...	&5BZBJ1540+1447	&J1540+1447	&... &4C +14.60 &1540+147 \\
PKS 1546+027           &1546+027	&...	&5BZQJ1549+0237	&J1549+0237	&... &1546+027 &1549+026\\
4C +05.64              &1548+056	&J1550+0527	&5BZQJ1550+0527	&J1550+0527	&... &1548+056 &1550+054\\
GB6 J1604+5714         &...	&...	&...	&J1604+5714	&... &... &... \\
OS 319                 &1611+343	&J1613+3412	&5BZQJ1613+3412	&J1613+3412	&1611+343 &DA 406 &1550+054\\
4C +38.41              &1633+382	&J1635+3808	&5BZQJ1635+3808	&J1635+3808	&1633+382 &4C +38.41 &1635+381\\
4C +47.44              &...	&...	&5BZQJ1637+4717	&J1637+4717	&... &... &1637+472\\
OS 562                 &1637+574	&J1638+5720	&...	&J1638+5720	&... &1637+574 &1638+573\\
NRAO 512               &1638+398	&...	&5BZQJ1640+3946	&J1640+3946	&...  &1638+398 &1640+397\\
\hline
\end{tabular}
%\label{tab:smman_overlap_srcs}
\end{table*}

\addtocounter{table}{-1}
\begin{table*}[h!]
\caption{Continued.}
\centering
\begin{tabular}{cccccccc}\hline\hline
SMMAN Sources   &UMRAO &F-GAMMA  &RATAN-600  &OVRO &ROBIN &MRO  &SMA\\
(1) &(2)     &(3)     &(4)   &(5)  &(6)   &(7)  &(8) \\
\hline
3C 345     	           &1641+399	&J1642+3948	&5BZQJ1642+3948	&J1642+3948	&1641+399 &3C 345 &1642+398 \\
Mkn 501                &1652+398 	&J1653+3945	&5BZBJ1653+3945	 &J1653+3945 &1652+398 &MARK 501 &1653+397\\
S4 1716+68             &...	&...	&...	&J1716+6836	&...  &... &1716+686 \\
S4 1726+45             &1726+455	&...	&5BZQJ1727+4530	&J1727+4530	 &...  &... &1727+455 \\
B2 1732+38A     &1732+389	&...	&...	&J1734+3857	&... &... &1734+389 \\
S4 1738+47             &1738+476 	&...	&5BZBJ1739+4737	&J1739+4737  &...  &... &... \\
4C +51.37              &1739+522	&...	&5BZQJ1740+5211	&J1740+5211	&... &S4 1739+52 &1740+521 \\
S4 1749+70             &1749+701	&J1748+7005	&5BZBJ1748+7005	&J1748+7005 &...	&... &... \\
OT 081                 &1749+096	&J1751+0939	&5BZBJ1751+0939	&J1751+0939	&...  &PKS 1749+096 &1751+096 \\
S5 1803+784            &1803+784	&J1800+7828	&5BZBJ1800+7828	 &J1800+7828	&...  &S5 1803+784 &1800+784 \\
S4 1800+44             &1800+440	&...	&5BZQJ1801+4404	&J1801+4404	&...  &... &1801+440\\
3C 371                 &1807+698 	&J1806+6949	&5BZBJ1806+6949	&J1806+6949	 &1807+698  &3C 371 &1806+698 \\
4C +56.27              &1823+568	&J1824+5651	&5BZBJ1824+5651	 &J1824+5651  &...	&4C 56.27 &1824+568\\
3C 380                 &1828+487 	&...	&5BZUJ1829+4844	&3C 380	&1828+487 &1828+487 &1829+487 \\
S4 1849+67             &...	&J1849+6705	&...	&J1849+6705	&... &J1849+6705 &1849+670 \\
S4 1926+61             &1926+611	&...	&5BZBJ1927+6117	&J1927+6117	&... &... &1927+612 \\
S5 2007+77             &2007+777	&...	&...	&J2005+7752	&... &S52007+77 &2005+778 \\
TXS 2007+659           &...	&...	&...	&J2007+6607	&... &... &... \\
4C +72.28              &2010+723	&...	&5BZBJ2009+7229	&J2009+7229	&... &... &2009+724\\
S5 2023+760            &...	&...	&5BZBJ2022+7611	&J2022+7611	&... &... &... \\
4C +31.56              &...	&...	&...	&2021+317 &... &... &2023+318 \\
B2 2023+33             &...	&...	&...	&3EGJ2027+3429	&... &... &2025+337 \\
PKS 2029+121           &2029+121	&J2031+1219	&5BZUJ2031+1219	 &J2031+1219 &... &...  &... \\
3C 418                 &2037+511 	&...	&...	&2037+511  &... &2037+511 &2038+513\\
B2 2107+35A            &...	&...	&...	&... &... &... &2109+355 \\
OX 036                 &2121+053	&...	&5BZQJ2123+0535	&J2123+0535	&... &2121+053 &2123+055 \\
NRAO 676               &...	&...	&...	&NRAO 676	&... &... &... \\
BL Lacertae            &2200+420	&J2202+4216	&5BZBJ2202+4216	&BL Lacertae &2200+420	&BL LAC &2202+422\\
4C +31.63              &2201+315	&J2203+3145	&5BZQJ2203+3145	&J2203+3145	&... &2201+315 &2203+317 \\
PKS 2209+236           &...	&J2212+2355	&5BZUJ2212+2355	&J2212+2355	&...  &... &... \\
CTA 102                &2230+114	&J2232+1143	&5BZQJ2232+1143	&2230+114	&2230+114 &2230+114 &2232+117\\
B2 2234+28A            &2234+282 	&...	&5BZQJ2236+2828	 &J2236+2828 &... &2234+282 &2236+284\\
3C 454.3               &2251+158 	&J2253+1608	&5BZQJ2253+1608	&J2253+1608	&2251+158  &3C 454.3 &2253+161\\
4C +27.50              &2319+272	&...	&5BZQJ2321+2732	&J2321+2732	&... &... &2321+275\\
4C +10.73              &2328+107	&...	&5BZQJ2330+1100	&J2330+1100	&... &... &... \\
4C +45.51              &2351+456	&...	&5BZQJ2354+4553	&2351+456  &...  &... &2354+458\\
\hline
\end{tabular}
%\label{tab:smman_overlap_srcs}
\end{table*}

\begin{longrotatetable}
\begin{table*}[h!]
\centering
\caption{Variability and spectral indices of the SMMAN sources.}
\resizebox{20cm}{!}{
\begin{tabular}{lcccccccccc}\hline\hline
SMMAN source  &\multicolumn{2}{c}{${F}_{var}$}   &\multicolumn{2}{c}{${V}_{S}$} &$\alpha_{4.8-23.6}$  &$P_{4.8}$ &$P_{23.6}$ &Flux (0.1--100\,GeV) &LAT $\Gamma$ &$L_\gamma$ (0.1--100\,GeV)\\
              &4.8 GHz &23.6 GHz                 &4.8 GHz &23.6 GHz &  &W Hz$^{-1}$ &W Hz$^{-1}$ &($10^{-8}\,\mathrm{ph}\,\mathrm{cm}^{-2}\,\mathrm{s}^{-1}$) & (photon index) & ($\mathrm{erg}\,\mathrm{s}^{-1}$)\\
(1) &(2)     &(3)     &(4)   &(5)  &(6)   &(7)  &(8) &(9)  &(10) &(11)\\
\hline
TXS 0008+704  &$0.327\pm0.003$  &$0.308\pm0.016$	 &$0.587\pm0.014$    &$0.512\pm0.053$   &$+0.178\pm0.003~(-0.256,+0.638)$ &...  &... &...	&...  &... \\	
4C +60.01 	  &$0.020\pm0.002$  &$0.011\pm0.014$   &$0.063\pm0.032$   &$0.231\pm0.078$   &$-0.541\pm0.003~(-0.676,-0.435)$  &...  &... &... &...  &... \\	
PKS 0056-00   &$0.073\pm0.003$  &...	 &$0.168\pm0.026$     &...  &...  &...  &...   &... &... &...\\	
TXS 0059+581  &$0.279\pm0.002$  &$0.325\pm0.006$	 &$0.599\pm0.005$     &$0.594\pm0.028$  &$+0.154\pm0.001~(-0.087,+0.457)$  &$2.57~(0.72)\times10^{27}$ &3.80$~(1.28)\times10^{27}$ &$7.23\pm0.09$	&$2.330\pm0.002$  &$(9.272\pm0.118)\times10^{46}$\\	
4C +01.02 	  &$0.326\pm0.003$  &$0.420\pm0.008$   &$0.439\pm0.012$    &$0.647\pm0.056$  &$+0.006\pm0.001~(-0.352,+0.280)$ &$4.25~(1.39)\times10^{28}$  &$6.20~(2.67)\times10^{28}$  &$42.20\pm0.03$ &$2.370\pm0.004$  &$(1.200\pm0.001)\times10^{49}$ \\	
4C +67.04 	  &$0.046\pm0.002$  &...   &$0.117\pm0.031$     &...   &...  &...  &... &... &... &...\\
4C +14.06 	  &$0.084\pm0.005$  &...   &$0.138\pm0.021$   &...   &... &...  &... &... &... &...\\
OC 457 		  &$0.115\pm0.002$  &$0.165\pm0.008$   &$0.269\pm0.011$   &$0.354\pm0.043$  &$+0.197\pm0.001~(-0.149,+0.430)$  &$3.45~(0.42)\times10^{27}$  &$4.76~(0.92)\times10^{27}$ &$6.15\pm0.37$  &$2.240\pm0.002$  &$(1.816\pm0.110)\times10^{47}$\\
4C +15.05 	  &$0.168\pm0.002$  &$0.422\pm0.010$   &$0.271\pm0.014$    &$0.529\pm0.046$  &$-0.367\pm0.001~(-0.578,-0.128)$ &$6.01~(1.01)\times10^{27}$  &$3.85~(1.66)\times10^{27}$  &$6.62\pm0.12$  &$2.360\pm0.006$  &$(1.614\pm0.030)\times10^{47}$\\	
S5 0212+73 	  &$0.025\pm0.002$  &$0.184\pm0.008$   &$0.070\pm0.024$     &$0.305\pm0.077$  &$-0.173\pm0.001~(-0.441,+0.108)$  &$5.65~(0.20)\times10^{28}$  &$4.33~(0.92)\times10^{28}$  &$7.61\pm0.08$  &$3.240\pm0.007$  &$(4.414\pm0.045)\times10^{48}$\\
B2 0218+357 	  &$0.072\pm0.003$  &$0.067\pm0.057$   &$0.136\pm0.023$     &$0.061\pm0.108$  &$-0.268\pm0.013~(-0.341,-0.199)$  &$3.03~(0.24)\times10^{27}$  &$2.00~(0.24)\times10^{27}$ &$7.39\pm0.09$  &$2.240\pm0.003$  &$(2.783\pm0.033)\times10^{47}$\\
3C 66A 		  &...  &$0.209\pm0.055$   &...       &$0.152\pm0.080$  &...  &...  &... &$5.93\pm0.06$  &$1.910\pm0.002$  &$(4.029\pm0.038)\times10^{46}$ \\
4C +28.07 	  &$0.181\pm0.002$  &$0.176\pm0.011$   &$0.294\pm0.017$   &$0.333\pm0.089$  &$-0.049\pm0.001~(-0.351,+0.120)$  &$1.02~(0.18)\times10^{28}$  &$9.15~(1.92)\times10^{27}$  &$18.80\pm0.07$ &$2.210\pm0.004$  &$(1.396\pm0.005)\times10^{48}$\\
AO 0235+164   &$0.319\pm0.002$  &$0.241\pm0.007$ 	 &$0.723\pm0.004$     &$0.578\pm0.022$  &$+0.072\pm0.001~(-0.330,+0.470)$ &$3.47~(1.12)\times10^{27}$  &$4.05~(1.05)\times10^{27}$ &$8.66\pm0.11$  &$2.150\pm0.003$  &$(3.595\pm0.045)\times10^{47}$\\
OD 166 		  &$0.039\pm0.004$  &$0.203\pm0.030$   &$0.069\pm0.021$     &$0.406\pm0.076$  &$-0.184\pm0.007~(-0.448,-0.056)$ &$1.79~(0.09)\times10^{28}$  &$1.58~(0.38)\times10^{28}$ &$4.26\pm0.08$  &$2.540\pm0.011$  &$(2.376\pm0.044)\times10^{48}$\\
4C +47.08 	  &$0.312\pm0.002$  &$0.349\pm0.013$   &$0.498\pm0.004$     &$0.527\pm0.042$  &$-0.075\pm0.001~(-0.260,+0.125)$ &$1.16~(0.37)\times10^{27}$       &$1.09~(0.40)\times10^{27}$ &$3.09\pm0.05$  &$1.940\pm0.003$  &$(3.514\pm0.061)\times10^{46}$ \\
NGC 1218 	  &$0.013\pm0.004$  &$0.046\pm0.019$   &$0.049\pm0.026$     &$0.163\pm0.058$  &$-0.716\pm0.001~(-0.882,-0.636)$ &$6.87~(0.19)\times10^{24}$   &$2.21~(0.25)\times10^{24}$ &$1.72\pm0.03$  &$1.920\pm0.008$  &$(4.480\pm0.071)\times10^{43}$ \\	
NGC 1275 	  &$0.166\pm0.002$  &$0.136\pm0.010$   &$0.192\pm0.045$     &$0.319\pm0.067$  &$-0.057\pm0.001~(-0.180,+0.074)$ &$2.49~(0.42)\times10^{25}$   &$2.31~(0.40)\times10^{25}$  &$35.20\pm0.34$ &$2.120\pm0.001$  &$(2.321\pm0.023)\times10^{44}$ \\
NRAO 140 	  &$0.112\pm0.002$  &$0.122\pm0.012$   &$0.207\pm0.012$     &$0.259\pm0.082$  &$-0.202\pm0.001~(-0.479,-0.014)$ &$9.81~(1.13)\times10^{27}$   &$7.29~(1.18)\times10^{27}$  &$5.46\pm0.17$  &$3.090\pm0.020$  &$(4.031\pm0.125)\times10^{47}$ \\
B3 0350+465   &$0.183\pm0.004$  &$0.219\pm0.043$	 &$0.345\pm0.028$     &$0.386\pm0.081$  &$-0.170\pm0.007~(-0.285,-0.037)$ &...  &... &$4.46\pm0.35$ &$2.530\pm0.025$ &... \\
TXS 0354+599  &$0.074\pm0.002$  &$0.106\pm0.022$ 	 &$0.148\pm0.029$     &$0.199\pm0.092$  &$-0.191\pm0.003~(-0.365,-0.043)$ &$6.73~(0.54)\times10^{26}$   &$5.05~(0.84)\times10^{26}$ &$6.34\pm0.40$  &$2.490\pm0.038$  &$(2.919\pm0.186)\times10^{46}$\\
3C 111 	     &$0.050\pm0.002$  &$0.319\pm0.007$   &$0.126\pm0.028$     &$0.604\pm0.043$  &$-0.696\pm0.001~(-1.058,-0.073)$ &$3.92~(0.22)\times10^{25}$   &$1.38~(0.46)\times10^{25}$  &$6.39\pm0.18$  &$2.790\pm0.010$  &$(1.390\pm0.039)\times10^{44}$\\
4C +41.11 	  &$0.151\pm0.002$  &$0.145\pm0.016$   &$0.298\pm0.016$     &$0.370\pm0.043$  &$-0.176\pm0.001~(-0.350,+0.008)$ &$3.35~(0.51)\times10^{27}$   &$2.50~(0.47)\times10^{27}$ &... &... &... \\
PKS 0507+17   &$0.119\pm0.003$  &$0.160\pm0.012$	 &$0.257\pm0.024$     &$0.283\pm0.098$  &$+0.437\pm0.001~(+0.156,+0.884)$ &$4.07~(0.48)\times10^{26}$   &$8.50~(1.68)\times10^{26}$  &$6.75\pm0.23$  &$2.660\pm0.014$  &$(2.228\pm0.077)\times10^{46}$\\
B3 0509+406   &$0.188\pm0.003$  &$0.297\pm0.012$	 &$0.345\pm0.018$    &$0.469\pm0.039$  &$-0.030\pm0.001~(-0.232,+0.263)$ &...  &... &$2.27\pm0.10$ &$2.190\pm0.014$ &... \\
4C +45.08 	  &$0.092\pm0.003$  &...   &$0.187\pm0.015$    &...  &... &...  &... &... &... &... \\
4C +27.15 	  &$0.050\pm0.003$  &...   &$0.119\pm0.028$     &...  &... &...  &...  &... &... &...\\
TXS 0518+211  &$0.167\pm0.007$  &...	 &$0.254\pm0.022$     &...  &... &...  &... &$10.40\pm0.11$ &$2.040\pm0.003$  &$(3.350\pm0.037)\times10^{45}$ \\
OG 050 		  &$0.115\pm0.002$  &$0.167\pm0.012$   &$0.248\pm0.017$     &$0.374\pm0.043$ &$+0.063\pm0.001~(-0.164,+0.321)$ &$6.79~(0.81)\times10^{27}$   &$7.00~(1.3) \times10^{27}$  &$7.54\pm0.17$  &$2.310\pm0.003$  &$(5.595\pm0.129)\times10^{47}$ \\
TXS 0529+483  &$0.271\pm0.003$  &$0.318\pm0.018$	 &$0.588\pm0.009$     &$0.542\pm0.059$  &$+0.044\pm0.002~(-0.252,+0.254)$ &$4.03~(1.10)\times10^{27}$   &$4.53~(1.5) \times10^{27}$  &$5.30\pm0.09$  &$2.210\pm0.005$  &$(3.507\pm0.057)\times10^{47}$ \\
4C +22.12 	  &$0.027\pm0.003$  &$0.199\pm0.030$   &$0.072\pm0.018$    &$0.312\pm0.064$   &$-0.477\pm0.003~(-0.626,-0.182)$ &...  &... &$6.13\pm0.31$ &$2.36\pm0.019$ &...\\
S5 0633+73 	  &$0.127\pm0.002$  &$0.254\pm0.013$   &$0.230\pm0.018$    &$0.420\pm0.043$   &$-0.021\pm0.001~(-0.304,+0.287)$ &$8.92~(1.17)\times10^{27}$   &$8.65~(2.4) \times10^{27}$ &$5.79\pm0.47$  &$2.870\pm0.003$  &$(1.243\pm0.101)\times10^{48}$ \\
S4 0707+47 	  &$0.085\pm0.004$  &...   &$0.193\pm0.022$    &...   &... &...  &... &$4.28\pm0.04$  &$2.620\pm0.014$  &$(3.070\pm0.026)\times10^{47}$\\	
S5 0716+71 	  &$0.216\pm0.001$  &$0.497\pm0.006$   &$0.574\pm0.013$     &$0.838\pm0.005$ &$+0.176\pm0.001~(-0.171,+0.724)$ &$1.64~(0.35)\times10^{26}$   &$2.05~(1.0) \times10^{26}$ &$15.70\pm0.12$ &$1.960\pm0.001$  &$(3.196\pm0.024)\times10^{46}$ \\
OI 280		  &$0.168\pm0.002$  &$0.182\pm0.012$   &$0.271\pm0.016$     &$0.290\pm0.038$  &$+0.032\pm0.001~(-0.194,+0.237)$  &$4.09~(0.69)\times10^{27}$   &$4.18~(0.88)\times10^{27}$ &$5.27\pm0.08$  &$2.400\pm0.003$  &$(1.476\pm0.023)\times10^{47}$\\
PKS 0754+100  &$0.182\pm0.003$  &$0.195\pm0.012$	 &$0.367\pm0.010$     &$0.368\pm0.065$  &$+0.022\pm0.001~(-0.155,+0.305)$ &$2.21~(0.40)\times10^{26}$   &$2.20~(0.48)\times10^{26}$  &$1.95\pm0.03$  &$2.050\pm0.004$  &$(4.520\pm0.068)\times10^{45}$ \\
NGC 2484 	  &$0.025\pm0.003$  &...   &$0.074\pm0.011$     &...   &... &...   &... &... &... &...\\
TXS 0800+618  &$0.138\pm0.004$  &...	 &$0.309\pm0.020$     &...   &...  &...  &... &$2.45\pm0.05$  &$2.760\pm0.039$  &$(2.143\pm0.048)\times10^{48}$  \\
S4 0805+41 	  &$0.124\pm0.002$  &$0.068\pm0.086$   &$0.219\pm0.04$     &$0.050\pm0.064$  &$+0.120\pm0.026~(+0.067,+0.172)$ &$5.25~(0.68)\times10^{27}$   &$7.19~(0.88)\times10^{27}$  &$7.87\pm0.26$  &$2.380\pm0.003$  &$(7.764\pm0.260)\times10^{47}$\\
S4 0814+42 	  &$0.213\pm0.003$  &...   &$0.362\pm0.023$    &...  &...  &...  &... &$4.84\pm0.02$  &$1.930\pm0.003$  &$(7.301\pm0.026)\times10^{46}$\\
OJ 535 		&$0.101\pm0.002$  &$0.073\pm0.077$     &$0.216\pm0.019$     &...  &...  &...  &... &$3.15\pm0.04$  &$2.620\pm0.008$  &$(2.944\pm0.035)\times10^{47}$\\
4C +39.23 	 &$0.063\pm0.003$  &$0.224\pm0.028$    &$0.126\pm0.035$     &$0.366\pm0.066$  &$-0.184\pm0.003~(-0.473,+0.070)$  &$5.79~(0.37)\times10^{27}$   &$4.21~(1.12)\times10^{27}$ &$2.22\pm0.04$  &$2.380\pm0.007$  &$(1.450\pm0.028)\times10^{47}$\\
OJ 248 		 &$0.117\pm0.003$  &$0.194\pm0.025$    &$0.233\pm0.023$     &$0.413\pm0.045$   &$+0.038\pm0.004~(-0.147,+0.421)$ &$2.26~(0.26)\times10^{27}$   &$2.26~(0.52)\times10^{27}$ &$4.47\pm0.11$  &$2.840\pm0.027$  &$(1.290\pm0.031)\times10^{47}$\\
OJ 448 		 &$0.180\pm0.005$  &...    &$0.274\pm0.023$     &...  &... &...  &... &... &... &... \\
3C 207 		 &$0.091\pm0.002$  &$0.241\pm0.012$    &$0.215\pm0.02$     &$0.313\pm0.062$ &$-0.095\pm0.001~(-0.268,+0.128)$  &$2.45~(0.23)\times10^{27}$   &$2.17~(0.58)\times10^{27}$  &$2.03\pm0.56$  &$2.380\pm0.061$  &$(2.860\pm0.788)\times10^{46}$ \\
0836+710 	 &$0.036\pm0.001$  &$0.198\pm0.004$    &$0.143\pm0.03$     &$0.526\pm0.064$  &$-0.220\pm0.001~(-0.556,+0.064)$  &$4.51~(0.22)\times10^{28}$   &$3.15~(0.72)\times10^{28}$ &$6.16\pm0.02$  &$3.090\pm0.007$  &$(2.626\pm0.009)\times10^{48}$ \\
OJ 287 		&$0.166\pm0.002$  &$0.143\pm0.010$     &$0.299\pm0.011$     &$0.299\pm0.051$  &$+0.249\pm0.001~(+0.007,+0.496)$  &$1.01~(0.17)\times10^{27}$   &$1.47~(0.26)\times10^{27}$ &$4.13\pm0.02$  &$2.150\pm0.003$  &$(1.095\pm0.004)\times10^{46}$ \\
S4 0859+47  &$0.051\pm0.002$  &$0.194\pm0.017$     &$0.110\pm0.017$     &$0.370\pm0.061$   &$-0.214\pm0.002~(-0.472,+0.118)$  &$9.69~(0.56)\times10^{27}$   &$6.69~(1.60)\times10^{27}$ &... &... &... \\
\hline
\end{tabular}}
\label{tab:properties_of_smman_srcs}
\tablecomments{SMMAN source name (same as the association name in Col. 2 of Table~\ref{Appendix A:SMMAN_srcs}) (Col. 1); fractional variability and variability index values for C and K band data sets (Col. 2, 3, 4, and 5); average radio spectral index and the range in spectral index (minimum and maximum values) inside the paranthesis (Col. 6); radio luminosity for each source (with measured redshift) in C and K band (Col. 7 and 8); the estimated {\sl Fermi}-LAT flux, photon index and luminosity for each source (Col. 9, 10, and 11).}
\end{table*}
\end{longrotatetable}

\begin{longrotatetable}
\addtocounter{table}{-1}
\begin{table*}[h!]
\centering
\caption{Continued.}
\resizebox{20cm}{!}{
\begin{tabular}{lcccccccccc}\hline\hline
SMMAN source    &\multicolumn{2}{c}{${F}_{var}$}  &\multicolumn{2}{c}{${V}_{S}$} &$\alpha_{4.8-23.6}$ &$P_{4.8}$ &$P_{23.6}$ &Flux (0.1--100\,GeV) &LAT $\Gamma$ &$L_\gamma$ (0.1--100\,GeV)\\
 &4.8 GHz &23.6 GHz &4.8 GHz &23.6 GHz & &W Hz$^{-1}$ &W Hz$^{-1}$ &($10^{-8}\,\mathrm{ph}\,\mathrm{cm}^{-2}\,\mathrm{s}^{-1}$) & (photon index) & ($\mathrm{erg}\,\mathrm{s}^{-1}$)\\
 (1) &(2)     &(3)     &(4)   &(5)  &(6)   &(7)  &(8) &(9)  &(10) &(11)\\
\hline
S4 0900+42 	 &$0.095\pm0.004$ &...     &$0.180\pm0.026$     &...   &...   &... &... &... &... &... \\
PKS 0906+01  &$0.131\pm0.003$ &$0.273\pm0.016$	 &$0.299\pm0.009$     &$0.412\pm0.103$  &$-0.039\pm0.002~(-0.320,+0.281)$  &$4.75~(0.62)\times10^{27}$   &$4.69~(1.42)\times10^{27}$ &$3.94\pm0.13$  &$2.780\pm0.057$  &$(1.451\pm0.049)\times10^{47}$ \\
S4 0913+39 	 &$0.074\pm0.004$ &...    &$0.130\pm0.016$     &...  &...  &...   &... &$1.67\pm0.13$  &$2.230\pm0.033$  &$(1.364\pm0.109)\times10^{47}$\\
S4 0917+44 	 &$0.095\pm0.002$ &$0.226\pm0.010$    &$0.253\pm0.015$     &$0.500\pm0.052$  &$+0.257\pm0.002~(-0.120,+0.531)$  &$1.18~(0.11)\times10^{28}$   &$1.80~(0.45)\times10^{28}$ &$6.81\pm0.01$  &$2.270\pm0.012$  &$(2.245\pm0.001)\times10^{48}$\\
OK 630 		 &$0.113\pm0.002$ &$0.193\pm0.019$    &$0.205\pm0.025$     &$0.333\pm0.083$  &$-0.020\pm0.002~(-0.315,+0.228)$  &$6.38~(0.75)\times10^{27}$   &$6.61~(1.62)\times10^{27}$ &$8.95\pm0.13$  &$2.330\pm0.001$  &$(9.771\pm0.146)\times10^{47}$ \\
4C +40.24 	 &$0.086\pm0.003$ &...    &$0.160\pm0.032$     &...  &...  &... &... &$1.64\pm0.03$  &$2.340\pm0.028$  &$(1.182\pm0.019)\times10^{47}$ \\
4C +55.17 	 &$0.012\pm0.003$ &$0.145\pm0.024$    &$0.028\pm0.037$     &$0.285\pm0.108$  &$-0.379\pm0.003~(-0.547,-0.119)$  &$5.24~(0.14)\times10^{27}$   &$2.75~(0.57)\times10^{27}$  &$6.82\pm0.06$  &$1.890\pm0.001$  &$(4.012\pm0.036)\times10^{47}$ \\
OK 492 		 &$0.214\pm0.002$ &$0.122\pm0.030$    &$0.353\pm0.017$     &$0.265\pm0.063$  &$-0.123\pm0.003~(-0.296,+0.014)$  &$1.20~(0.26)\times10^{28}$   &$1.07~(0.20)\times10^{28}$ &$2.27\pm0.05$  &$2.260\pm0.013$  &$(5.086\pm0.118)\times10^{47}$ \\
S4 0954+65 	 &$0.303\pm0.002$ &$0.506\pm0.005$    &$0.577\pm0.008$     &$0.850\pm0.011$   &$+0.306\pm0.001~(-0.163,+0.806)$ &$3.43~(1.05)\times10^{26}$   &$5.66~(2.96)\times10^{26}$ &$11.90\pm0.13$ &$2.150\pm0.001$  &$(4.909\pm0.053)\times10^{46}$\\
4C +40.25 	 &$0.086\pm0.003$ &...    &$0.211\pm0.020$     &...  &...  &... &... &$2.72\pm0.00$  &$2.550\pm0.009$  &$(1.812\pm0.003)\times10^{47}$\\
S4 1030+41 	 &$0.344\pm0.003$ &$0.214\pm0.019$    &$0.507\pm0.008$     &$0.308\pm0.083$ &$-0.079\pm0.001~(-0.237,+0.333)$  &$4.73~(1.64)\times10^{27}$   &$3.82~(0.95)\times10^{27}$  &$2.33\pm0.06$  &$2.490\pm0.005$  &$(1.149\pm0.032)\times10^{47}$\\
S4 1030+61 	 &$0.192\pm0.005$ &$0.225\pm0.058$    &$0.352\pm0.052$     &$0.210\pm0.067$  &...  &... &...  &$5.90\pm0.01$  &$2.100\pm0.006$  &$(7.277\pm0.010)\times10^{47}$\\
B2 1040+24 	 &$0.208\pm0.004$ &$0.322\pm0.017$    &$0.355\pm0.021$     &$0.560\pm0.053$  &$+0.095\pm0.002~(-0.143,+0.412)$  &$6.54~(1.41)\times10^{26}$   &$7.62~(2.65)\times10^{26}$ &$2.00\pm0.14$  &$2.120\pm0.009$  &$(2.401\pm0.172)\times10^{46}$\\
S5 1039+81 	 &$0.128\pm0.002$ &$0.176\pm0.011$    &$0.280\pm0.030$     &$0.443\pm0.069$  &$-0.118\pm0.001~(-0.468,+0.232)$ &$8.70~(1.16)\times10^{25}$   &$7.62~(1.60)\times10^{25}$  &$3.58\pm0.04$  &$2.780\pm0.007$  &$(1.223\pm0.014)\times10^{45}$\\
S5 1044+71 	 &$0.351\pm0.003$ &$0.285\pm0.007$    &$0.649\pm0.021$     &$0.592\pm0.039$  &$-0.084\pm0.001~(-0.331,+0.223)$ &$7.64~(2.70)\times10^{27}$   &$8.02~(2.43)\times10^{27}$  &$16.90\pm0.02$ &$2.280\pm0.005$  &$(9.997\pm0.014)\times10^{47}$\\
4C +01.28 	 &$0.121\pm0.003$ &$0.112\pm0.015$    &$0.272\pm0.015$     &$0.260\pm0.076$  &$+0.149\pm0.001~(-0.011,+0.273)$  &$9.55~(1.19)\times10^{27}$   &$1.23~(0.20)\times10^{28}$  &$4.54\pm0.05$  &$2.090\pm0.002$  &$(1.811\pm0.018)\times10^{47}$\\
TXS 1125+596 &$0.122\pm0.002$ &$0.210\pm0.015$	 &$0.286\pm0.019$     &$0.430\pm0.089$  &$-0.119\pm0.001~(-0.271,+0.261)$  &$1.02~(0.12)\times10^{28}$   &$8.81~(2.27)\times10^{27}$ &$2.17\pm0.02$  &$2.280\pm0.003$  &$(4.255\pm0.035)\times10^{47}$\\
B2 1128+38 	 &$0.174\pm0.002$ &$0.229\pm0.015$    &$0.302\pm0.019$     &$0.331\pm0.059$  &$-0.042\pm0.004~(-0.195,+0.106)$ &$1.44~(0.26)\times10^{28}$   &$1.57~(0.39)\times10^{28}$  &$5.80\pm0.07$  &$2.510\pm0.003$  &$(9.651\pm0.123)\times10^{47}$\\
4C +10.33 	 &$0.072\pm0.009$ &...    &$0.109\pm0.029$     &...  &...  &... &... &... &... &...\\
3C 264 		 &$0.016\pm0.003$ &...    &$0.062\pm0.018$     &...  &...  &... &... &... &... &...\\
S4 1144+40 	 &$0.173\pm0.003$ &$0.166\pm0.060$    &$0.474\pm0.012$     &$0.259\pm0.094$ &$+0.340\pm0.007~(+0.112,+0.585)$  &$3.83~(0.66)\times10^{27}$   &$6.31~(1.50)\times10^{27}$  &$9.15\pm0.07$  &$2.410\pm0.002$  &$(4.362\pm0.032)\times10^{47}$ \\
OM 484 		 &$0.134\pm0.003$ &$0.283\pm0.017$    &$0.255\pm0.031$     &$0.491\pm0.062$  &$-0.107\pm0.007~(-0.443,+0.166)$ &$4.01~(0.55)\times10^{26}$   &$3.28~(1.03)\times10^{26}$ &$2.38\pm0.07$  &$2.530\pm0.083$  &$(4.916\pm0.142)\times10^{45}$\\
B3 1151+408  &$0.097\pm0.002$ &...	 &$0.191\pm0.015$     &...   &...  &... &... &$1.53\pm0.04$  &$2.050\pm0.013$  &$(7.083\pm0.187)\times10^{46}$\\
Ton 599 	 &$0.427\pm0.004$ &$0.406\pm0.007$    &$0.746\pm0.004$     &$0.834\pm0.014$  &$+0.094\pm0.001~(-0.296,+0.402)$ &$3.52~(1.50)\times10^{27}$   &$5.62~(2.33)\times10^{27}$  &$36.10\pm0.16$ &$2.140\pm0.001$  &$(7.898\pm0.036)\times10^{47}$\\
4C +13.46 	 &$0.078\pm0.004$ &...    &$0.153\pm0.020$     &...   &... &... &... &... &... &... \\
4C +21.35 	 &$0.227\pm0.002$ &$0.316\pm0.016$    &$0.332\pm0.015$    &$0.502\pm0.056$ &$-0.294\pm0.001~(-0.531,+0.004)$ &$1.21~(0.28)\times10^{27}$   &$7.56~(2.59)\times10^{26}$ &$4.90\pm0.05$  &$2.380\pm0.006$  &$(2.206\pm0.022)\times10^{46}$ \\
3C 273 		 &$0.017\pm0.002$ &$0.084\pm0.006$    &$0.065\pm0.021$     &$0.257\pm0.073$  &$-0.387\pm0.001~(-0.497,-0.202)$ &$1.93~(0.06)\times10^{27}$   &$1.13~(0.14)\times10^{27}$  &$19.90\pm0.67$ &$2.840\pm0.001$  &$(5.725\pm0.194)\times10^{45}$\\
M\,87 		&$0.016\pm0.003$ &$0.077\pm0.014$     &$0.050\pm0.015$     &$0.102\pm0.091$  &$-0.851\pm0.001~(-0.960,-0.725)$ &$2.58~(0.05)\times10^{24}$   &$6.60~(0.50)\times10^{23}$  &$2.04\pm0.02$  &$2.100\pm0.003$  &$(6.776\pm0.054)\times10^{41}$\\
PKS 1236+077 &$0.145\pm0.003$ &...	 &$0.269\pm0.019$     &...  &...  &... &.... &... &... &... \\
3C 279 		&$0.063\pm0.002$ &$0.237\pm0.005$     &$0.126\pm0.036$     &$0.424\pm0.047$   &$+0.146\pm0.001~(-0.156,+0.413)$ &$9.60~(0.66)\times10^{27}$   &$1.50~(0.38)\times10^{28}$ &$71.30\pm0.17$ &$2.290\pm0.001$  &$(5.997\pm0.015)\times10^{47}$\\
ON 393 		&$0.222\pm0.005$ &...     &$0.338\pm0.021$     &...  &... &... &...  &$3.09\pm0.01$  &$2.220\pm0.005$  &$(7.935\pm0.028)\times10^{46}$\\
TXS 1300+580 &$0.222\pm0.004$ &...	 &$0.412\pm0.017$     &...   &... &... &... &$1.15\pm0.10$  &$2.080\pm0.014$  &$(5.424\pm0.486)\times10^{46}$\\
4C +12.46 	&$0.076\pm0.004$ &$0.064\pm0.045$     &$0.142\pm0.022$     &$0.096\pm0.102$  &$-0.046\pm0.011~(-0.110,+0.088)$ &$5.48~(0.41)\times10^{27}$   &$5.13~(0.71)\times10^{27}$ &... &... &... \\
OP 313 		&$0.346\pm0.003$ &$0.192\pm0.011$     &$0.510\pm0.022$     &$0.665\pm0.028$   &$+0.255\pm0.001~(-0.099,+0.384)$ &$9.88~(3.45)\times10^{27}$   &$1.90~(0.40)\times10^{28}$  &$17.00\pm0.53$ &$2.290\pm0.000$  &$(2.758\pm0.086)\times10^{48}$\\
PKS 1348+007 &$0.398\pm0.030$ &...	 &$0.375\pm0.038$     &...   &... &... &... &$4.29\pm0.59$  &$2.350\pm0.022$  &$(1.204\pm0.166)\times10^{48}$\\
OQ 530 		&$0.382\pm0.003$ &$0.371\pm0.016$     &$0.625\pm0.006$     &$0.652\pm0.019$  &$-0.100\pm0.002~(-0.430,+0.125)$ &$7.35~(2.82)\times10^{25}$   &$6.23~(2.47)\times10^{25}$ &$2.55\pm0.05$  &$2.370\pm0.002$  &$(9.819\pm0.206)\times10^{44}$\\
3C 303 		&$0.032\pm0.003$ &$0.171\pm0.029$     &$0.101\pm0.025$     &$0.210\pm0.102$    &$-0.240\pm0.002~(-0.354,+0.011)$ &$5.95~(0.26)\times10^{25}$   &$4.36~(0.97)\times10^{25}$ &... &... &...\\
TXS 1459+480 &$0.183\pm0.006$ &...	 &$0.357\pm0.016$     &...  &... &... &...  &... &... &...\\
4C +14.60 	 &$0.186\pm0.003$ &...    &$0.376\pm0.016$     &...  &... &... &...  &... &... &...\\
PKS 1546+027 &$0.213\pm0.003$ &...	 &$0.339\pm0.029$     &...  &...  &...  &...  &$3.67\pm0.20$  &$2.530\pm0.018$  &$(1.293\pm0.069)\times10^{46}$ \\
4C +05.64 	 &$0.028\pm0.003$ &...    &$0.076\pm0.019$     &...  &... &...  &...  &$2.58\pm0.02$  &$2.240\pm0.015$  &$(2.803\pm0.017)\times10^{47}$ \\
GB6 J1604+5714 &... &$0.272\pm0.027$	 &...     &$0.412\pm0.075$  &... &... &... &$3.55\pm0.15$  &$2.480\pm0.001$  &$(5.402\pm0.230)\times10^{46}$\\
OS 319 	&$0.164\pm0.002$ &$0.276\pm0.012$	     &$0.269\pm0.012$    &$0.525\pm0.052$  &$-0.324\pm0.001~(-0.554,-0.069)$  &$3.03~(0.50)\times10^{28}$   &$1.84~(0.55)\times10^{28}$ &$2.71\pm0.01$  &$2.630\pm0.007$  &$(2.428\pm0.007)\times10^{47}$\\
4C +38.41 	&$0.135\pm0.002$ &$0.099\pm0.012$     &$0.234\pm0.020$     &$0.290\pm0.079$  &$+0.073\pm0.001~(-0.143,+0.299)$ &$2.00~(0.28)\times10^{28}$   &$2.22~(0.32)\times10^{28}$ &$14.00\pm0.02$ &$2.440\pm0.003$  &$(2.638\pm0.004)\times10^{48}$ \\
4C +47.44 	&$0.078\pm0.003$ &$0.096\pm0.037$     &$0.174\pm0.026$    &$0.176\pm0.119$  &$+0.074\pm0.003~(-0.079,+0.230)$ &$1.28~(0.10)\times10^{27}$   &$1.48~(0.26)\times10^{27}$ &$1.67\pm0.05$  &$2.530\pm0.010$  &$(2.611\pm0.072)\times10^{46}$\\
OS 562 		&$0.091\pm0.002$ &$0.245\pm0.011$     &$0.208\pm0.011$     &$0.389\pm0.076$  &$+0.043\pm0.001~(-0.246,+0.429)$  &$2.21~(0.21)\times10^{27}$   &$2.47~(0.67)\times10^{27}$ &$3.03\pm0.03$  &$3.010\pm0.008$  &$(4.642\pm0.051)\times10^{46}$\\
NRAO 512 	&$0.153\pm0.003$ &$0.193\pm0.029$     &$0.339\pm0.019$     &$0.303\pm0.089$  &$-0.186\pm0.005~(-0.366,-0.032)$ &$8.63~(1.32)\times10^{27}$   &$7.13~(1.76)\times10^{27}$ &$4.67\pm0.10$  &$2.440\pm0.012$  &$(7.043\pm0.152)\times10^{47}$\\
\hline
\end{tabular}}
%\tablecomments{}
\end{table*}
\end{longrotatetable}

\begin{longrotatetable}
\addtocounter{table}{-1}
\begin{table*}[h!]
\centering
\caption{Continued.}
\resizebox{20cm}{!}{
\begin{tabular}{lcccccccccc}\hline\hline
SMMAN source   &\multicolumn{2}{c}{${F}_{var}$}  &\multicolumn{2}{c}{${V}_{S}$} &$\alpha_{4.8-23.6}$  &$P_{4.8}$ &$P_{23.6}$ &Flux (0.1--100\,GeV) &LAT $\Gamma$ &$L_\gamma$ (0.1--100\,GeV)\\
  &4.8 GHz &23.6 GHz &4.8 GHz &23.6 GHz & &W Hz$^{-1}$ &W Hz$^{-1}$ &($10^{-8}\,\mathrm{ph}\,\mathrm{cm}^{-2}\,\mathrm{s}^{-1}$) & (photon index) & ($\mathrm{erg}\,\mathrm{s}^{-1}$)\\
  (1) &(2)     &(3)     &(4)   &(5)  &(6)   &(7)  &(8) &(9)  &(10) &(11)\\
\hline
3C 345 		&$0.077\pm0.001$ &$0.069\pm0.007$  &$0.186\pm0.027$     &$0.324\pm0.068$   &$+0.004\pm0.001~(-0.211,+0.128)$  &$5.30~(0.43)\times10^{27}$	&$5.69~(0.69)\times10^{27}$ &$13.00\pm0.17$ &$2.440\pm0.001$  &$(1.224\pm0.016)\times10^{47}$\\
Mkn 501 	&$0.058\pm0.003$ &$0.230\pm0.029$  &$0.116\pm0.024$    &$0.357\pm0.110$  &$-0.212\pm0.004~(-0.383,+0.052)$ &$3.84~(0.24)\times10^{24}$     &$2.74~(0.78)\times10^{24}$ &$4.94\pm0.03$  &$1.800\pm0.001$  &$(2.525\pm0.013)\times10^{44}$\\
S4 1716+68 	&$0.082\pm0.005$ &...  &$0.147\pm0.023$     &...   &...  &... &... &$1.85\pm0.09$  &$2.370\pm0.004$  &$(3.718\pm0.182)\times10^{46}$\\
S4 1726+45 	&$0.286\pm0.003$ &$0.212\pm0.024$  &$0.547\pm0.006$     &$0.347\pm0.093$ &$-0.055\pm0.002~(-0.344,+0.314)$  &$1.64~(0.47)\times10^{27}$     &$1.51~(0.40)\times10^{27}$ &$7.98\pm0.09$  &$2.390\pm0.002$  &$(1.280\pm0.015)\times10^{47}$\\
B2 1732+38A  &$0.201\pm0.002$ &$0.348\pm0.009$ &$0.395\pm0.018$     &$0.595\pm0.049$   &$+0.209\pm0.001~(-0.085,+0.603)$ &$2.21~(0.46)\times10^{27}$     &$3.40~(1.28)\times10^{27}$  &$6.25\pm0.03$  &$2.260\pm0.003$  &$(2.513\pm0.012)\times10^{47}$\\
S4 1738+47 	 &$0.112\pm0.003$ &$0.310\pm0.037$ &$0.233\pm0.015$     &$0.390\pm0.067$   &$-0.049\pm0.006~(-0.277,+0.496)$ &... &... &$1.13\pm0.10$ &$1.92\pm0.008$ &...\\
4C +51.37 	 &$0.233\pm0.003$ &$0.423\pm0.021$ &$0.407\pm0.011$     &$0.581\pm0.035$   &$-0.097\pm0.004~(-0.343,+0.294)$ &$6.21~(1.45)\times10^{27}$     &$6.94~(3.08)\times10^{27}$ &$6.00\pm0.06$  &$2.500\pm0.002$  &$(5.277\pm0.052)\times10^{47}$\\
S4 1749+70 	 &$0.257\pm0.003$ &$0.346\pm0.027$ &$0.478\pm0.010$     &$0.520\pm0.057$   &$-0.105\pm0.003~(-0.236,+0.252)$ &$1.43~(0.37)\times10^{27}$     &$1.48~(0.57)\times10^{27}$ &$4.01\pm0.11$  &$1.950\pm0.001$  &$(1.423\pm0.039)\times10^{47}$\\
OT 081 		 &$0.228\pm0.003$ &$0.190\pm0.012$ &$0.423\pm0.013$     &$0.378\pm0.074$    &$+0.228\pm0.001~(+0.023,+0.418)$ &$5.18~(1.18)\times10^{26}$     &$7.36~(1.64)\times10^{26}$ &$8.69\pm0.07$  &$2.220\pm0.005$  &$(2.319\pm0.020)\times10^{46}$\\
S5 1803+784  &$0.078\pm0.002$ &$0.028\pm0.023$ &$0.170\pm0.016$     &$0.186\pm0.080$    &$-0.039\pm0.001~(-0.162,+0.197)$ &$3.42~(0.28)\times10^{27}$     &$3.13~(0.32)\times10^{27}$   &$15.10\pm0.13$ &$2.200\pm0.001$  &$(2.691\pm0.022)\times10^{47}$\\
S4 1800+44 	 &$0.209\pm0.003$ &$0.322\pm0.027$ &$0.378\pm0.013$     &$0.519\pm0.042$    &$+0.035\pm0.005~(-0.100,+0.251)$ &$1.13~(0.24)\times10^{27}$     &$1.36~(0.48)\times10^{27}$ &$3.58\pm0.06$  &$2.360\pm0.009$  &$(4.808\pm0.082)\times10^{46}$\\
3C 371 		&$0.119\pm0.002$ &$0.170\pm0.008$  &$0.235\pm0.014$     &$0.551\pm0.049$   &$-0.045\pm0.001~(-0.229,+0.122)$  &$1.19~(0.14)\times10^{25}$     &$1.17~(0.24)\times10^{25}$ &$4.86\pm0.05$  &$2.320\pm0.002$  &$(1.758\pm0.016)\times10^{44}$\\
4C +56.27 	&$0.097\pm0.002$ &$0.121\pm0.012$  &$0.193\pm0.021$     &$0.350\pm0.078$   &$+0.015\pm0.001~(-0.129,+0.225)$ &$1.51~(0.16)\times10^{27}$     &$1.52~(0.25)\times10^{27}$ &$3.33\pm0.27$  &$2.280\pm0.000$  &$(4.867\pm0.390)\times10^{46}$\\
3C 380 		&$0.014\pm0.003$ &$0.071\pm0.014$  &$0.033\pm0.022$     &$0.160\pm0.099$    &$-0.528\pm0.001~(-0.635,-0.392)$ &$9.18~(0.24)\times10^{27}$     &$3.84~(0.49)\times10^{27}$ &$2.89\pm0.01$  &$2.280\pm0.005$  &$(4.693\pm0.011)\times10^{46}$\\
S4 1849+67 	&$0.203\pm0.002$ &$0.199\pm0.013$  &$0.399\pm0.015$     &$0.384\pm0.076$    &$+0.222\pm0.002~(+0.022,+0.467)$ &$1.12~(0.22)\times10^{27}$     &$1.81~(0.44)\times10^{27}$ &$2.14\pm0.04$  &$2.210\pm0.005$  &$(3.315\pm0.056)\times10^{46}$\\
S4 1926+61 	 &$0.195\pm0.002$ &$0.290\pm0.012$ &$0.343\pm0.026$     &$0.549\pm0.056$    &$-0.022\pm0.001~(-0.237,+0.300)$ &$9.16~(1.85)\times10^{26}$     &$9.65~(3.05)\times10^{26}$ &$1.56\pm0.02$ &$1.99\pm0.004$ &$(2.163\pm0.030)\times10^{46}$\\
S5 2007+77 	 &$0.170\pm0.002$ &$0.184\pm0.021$ &$0.351\pm0.013$     &$0.323\pm0.078$   &$+0.052\pm0.003~(-0.134,+0.242)$ &$3.17~(0.55)\times10^{26}$     &$3.14~(0.76)\times10^{26}$ &$2.32\pm0.02$  &$2.120\pm0.003$  &$(8.422\pm0.086)\times10^{45}$\\
TXS 2007+659 &$0.245\pm0.003$ &$0.163\pm0.031$	  &$0.436\pm0.009$     &$0.326\pm0.079$   &$+0.065\pm0.006~(-0.060,+0.386)$ &$3.87~(0.95)\times10^{27}$     &$3.33~(0.77)\times10^{27}$ &$4.30\pm0.07$  &$2.550\pm0.006$  &$(3.338\pm0.051)\times10^{47}$\\
4C +72.28 	&$0.069\pm0.002$ &$0.104\pm0.013$  &$0.192\pm0.012$     &$0.303\pm0.066$   &$-0.081\pm0.001~(-0.219,+0.083)$ &... &... &$2.28\pm0.03$ &$2.23\pm0.008$ &... \\
S5 2023+760 &$0.103\pm0.003$ &$0.078\pm0.034$  &$0.202\pm0.031$     &$0.254\pm0.084$   &$+0.042\pm0.005~(-0.030,+0.121)$  &$6.27~(0.66)\times10^{26}$     &$6.74~(1.03)\times10^{26}$ &$2.81\pm0.09$  &$2.320\pm0.002$  &$(2.912\pm0.091)\times10^{46}$\\
4C +31.56 	&$0.101\pm0.002$ &$0.116\pm0.031$  &$0.209\pm0.037$     &$0.340\pm0.062$   &$-0.227\pm0.005~(-0.377,-0.026)$  &$5.34~(0.54)\times10^{26}$     &$4.08~(0.76)\times10^{26}$ &... &... &... \\
B2 2023+33 	&$0.192\pm0.002$ &$0.243\pm0.011$  &$0.353\pm0.014$     &$0.544\pm0.030$   &$+0.141\pm0.001~(-0.149,+0.548)$ &$2.58~(0.50)\times10^{26}$     &$3.45~(0.92)\times10^{26}$ &$7.27\pm0.03$  &$2.860\pm0.014$  &$(4.469\pm0.020)\times10^{45}$\\
PKS 2029+121 &$0.095\pm0.002$ &$0.088\pm0.024$	  &$0.179\pm0.022$     &$0.125\pm0.103$  &$-0.210\pm0.006~(-0.325,-0.080)$  &$5.67~(0.57)\times10^{27}$     &$4.19~(0.62)\times10^{27}$  &$3.08\pm0.01$  &$2.480\pm0.011$  &$(1.918\pm0.003)\times10^{47}$ \\
3C 418 	&$0.056\pm0.002$ &$0.187\pm0.017$	  &$0.123\pm0.019$     &$0.322\pm0.038$  &$-0.085\pm0.001~(-0.262,+0.120)$ &$3.47~(0.21)\times10^{28}$     &$2.97~(0.68)\times10^{28}$ &$4.15\pm0.06$  &$2.540\pm0.009$  &$(6.313\pm0.089)\times10^{47}$ \\
B2 2107+35A  &$0.265\pm0.003$ &$0.126\pm0.012$	  &$0.454\pm0.010$     &$0.314\pm0.088$  &$+0.013\pm0.001~(-0.133,+0.265)$ &$2.34~(0.62)\times10^{26}$     &$2.59~(0.43)\times10^{26}$ &... &... &...\\
OX 036 	&$0.163\pm0.002$ &$0.200\pm0.012$	  &$0.329\pm0.015$     &$0.361\pm0.086$   &$-0.027\pm0.001~(-0.255,+0.319)$ &$2.02~(0.34)\times10^{28}$     &$1.97~(0.45)\times10^{28}$ &$2.91\pm0.04$  &$2.130\pm0.018$  &$(7.800\pm0.101)\times10^{47}$\\
NRAO 676 &$0.088\pm0.003$ &...	  &$0.256\pm0.022$     &...  &...  &...  &... &$5.25\pm0.01$  &$2.710\pm0.010$  &$(1.147\pm0.001)\times10^{48}$\\
BL Lacertae  &$0.335\pm0.002$ &$0.439\pm0.004$	  &$0.585\pm0.011$     &$0.886\pm0.008$ &$+0.313\pm0.001~(-0.145,+0.87)$  &$2.73~(0.92)\times10^{25}$     &$5.81~(2.60)\times10^{25}$  &$64.00\pm0.58$ &$2.070\pm0.000$  &$(7.010\pm0.063)\times10^{45}$\\
4C +31.63 	&$0.145\pm0.003$ &...  &$0.253\pm0.017$     &...  &...  &...  &... &... &... &...\\
PKS 2209+236  &$0.217\pm0.004$ &$0.120\pm0.043$	  &$0.519\pm0.007$     &$0.198\pm0.083$  &$-0.212\pm0.037~(-0.308,-0.115)$ &$3.27~(0.70)\times10^{27}$     &$2.41~(0.45)\times10^{27}$ &... &... &... \\
CTA 102  &$0.072\pm0.002$ &$0.254\pm0.005$	  &$0.166\pm0.034$     &$0.568\pm0.027$  &$-0.071\pm0.001~(-0.440,+0.328)$ &$1.30~(0.10)\times10^{28}$     &$1.25~(0.34)\times10^{28}$ &$71.40\pm0.56$ &$2.320\pm0.000$  &$(3.146\pm0.025)\times10^{48}$\\
B2 2234+28A &$0.125\pm0.002$ &$0.209\pm0.010$   &$0.260\pm0.015$     &$0.542\pm0.034$   &$+0.074\pm0.001~(-0.109,+0.476)$ &$2.60~(0.33)\times10^{27}$     &$2.95~(0.68)\times10^{27}$ &$8.72\pm0.13$  &$2.160\pm0.001$  &$(2.300\pm0.034)\times10^{47}$\\
3C 454.3  &$0.137\pm0.001$ &$0.254\pm0.006$ &$0.205\pm0.030$     &$0.355\pm0.091$   &$-0.147\pm0.001~(-0.310,+0.058)$  &$3.82~(0.53)\times10^{28}$     &$2.62~(0.73)\times10^{28}$ &$49.60\pm0.50$ &$2.180\pm0.001$  &$(1.575\pm0.016)\times10^{48}$\\
4C +27.50 &$0.095\pm0.002$ &$0.389\pm0.013$	  &$0.199\pm0.017$     &$0.574\pm0.051$   &$+0.055\pm0.001~(-0.263,+0.482)$ &$5.04~(0.49)\times10^{27}$     &$5.91~(2.42)\times10^{27}$ &... &... &... \\
4C +10.73 &$0.074\pm0.003$ &$0.058\pm0.063$	  &$0.130\pm0.022$     &$0.172\pm0.093$  &$-0.243\pm0.006~(-0.337,-0.126)$ &$7.81~(0.60)\times10^{27}$     &$5.56~(0.90)\times10^{27}$ &$3.56\pm0.08$  &$2.500\pm0.013$  &$(3.940\pm0.090)\times10^{47}$\\
4C +45.51 &$0.065\pm0.003$ &...	  &$0.151\pm0.016$     &...   &...  &... &... &... &... &... \\
\hline
\end{tabular}}
%\tablecomments{}
\end{table*}
\end{longrotatetable}

\clearpage
\section{Radio Light curves of Individual Sources Monitored in SMMAN Program} \label{Appendix B:SMMAN_lcs}
Figure~\ref{fig:SMMAN_sources_lc} of Appendix~\ref{Appendix B:SMMAN_lcs} presents the light curves for all the 131 SMMAN sources and the three calibrator sources in the program at 4.8 and 23.6 GHz. The data points shown here are unbinned. For some sources, either 4.8 GHz or 23.6 GHz light curves are available; accordingly, those are presented.\\

%%%%%%% Figure Sets for Figure 19. %%%%%%%%%%%%%%%%%%%%%%%%%%%%%%%%%%%%%%%%%
\begin{figure*}[p]
\centering
\begin{tabular}{cc}
\includegraphics[width=0.49\textwidth]{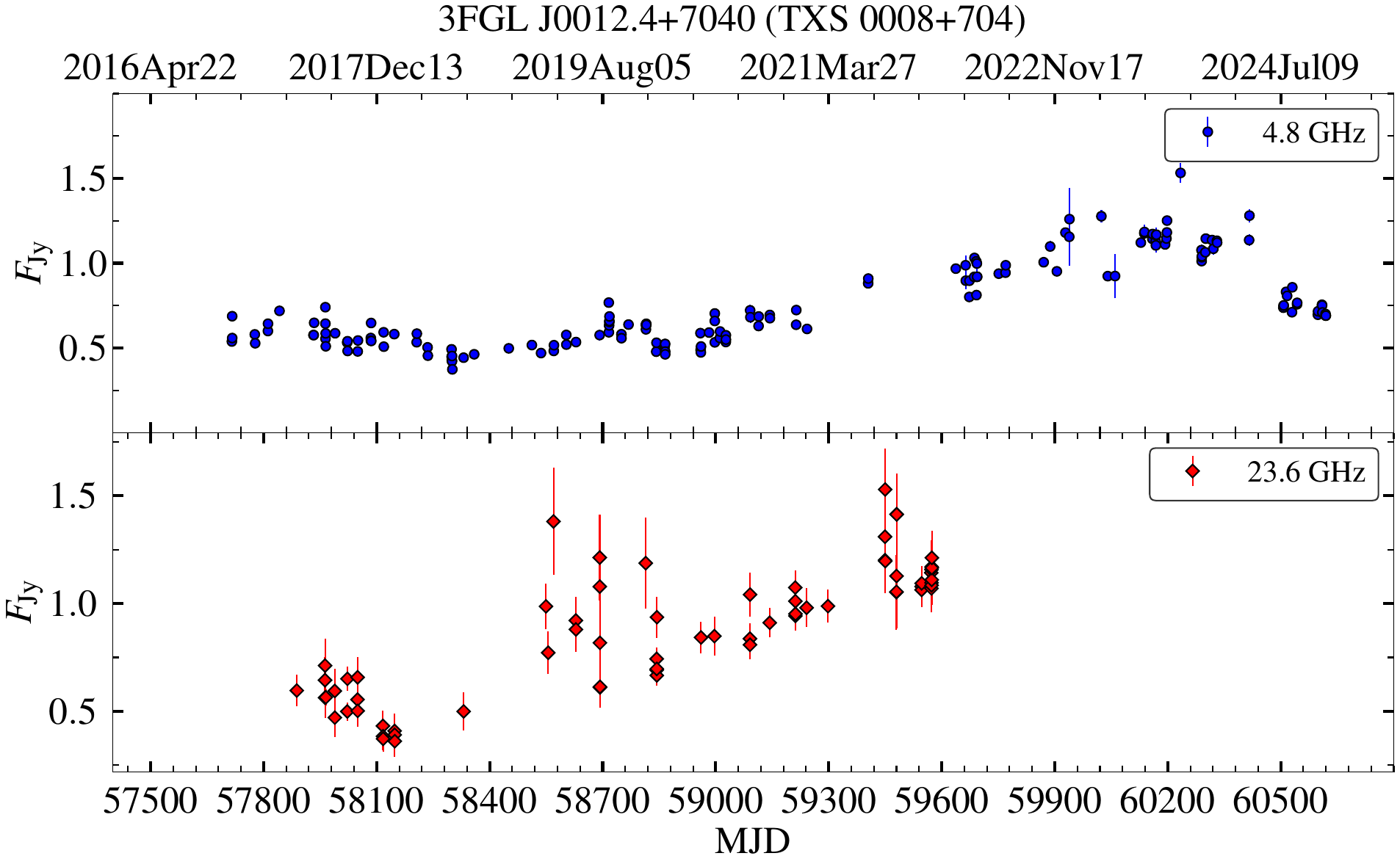}
\includegraphics[width=0.49\textwidth]{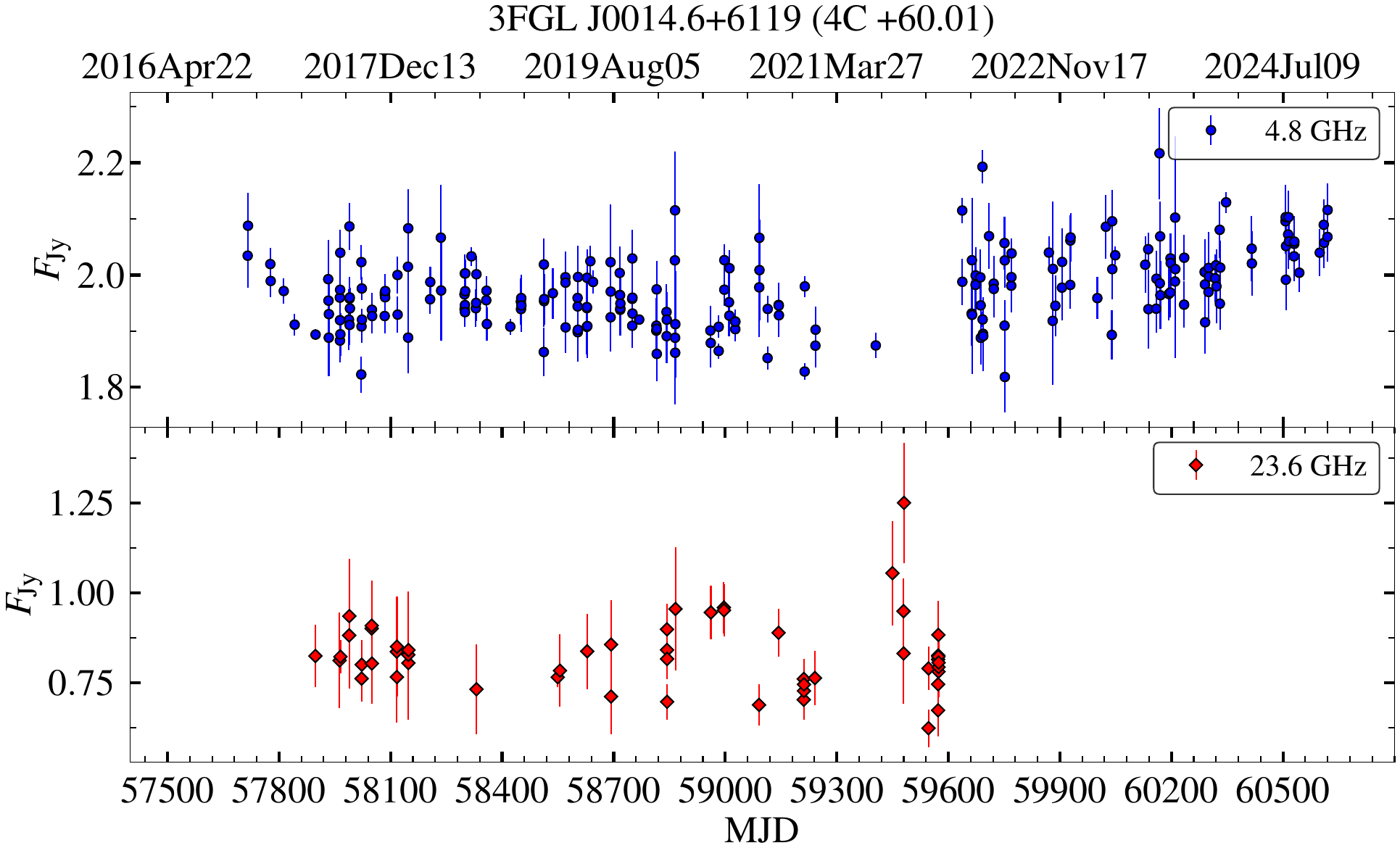}\\
\includegraphics[width=0.49\textwidth]{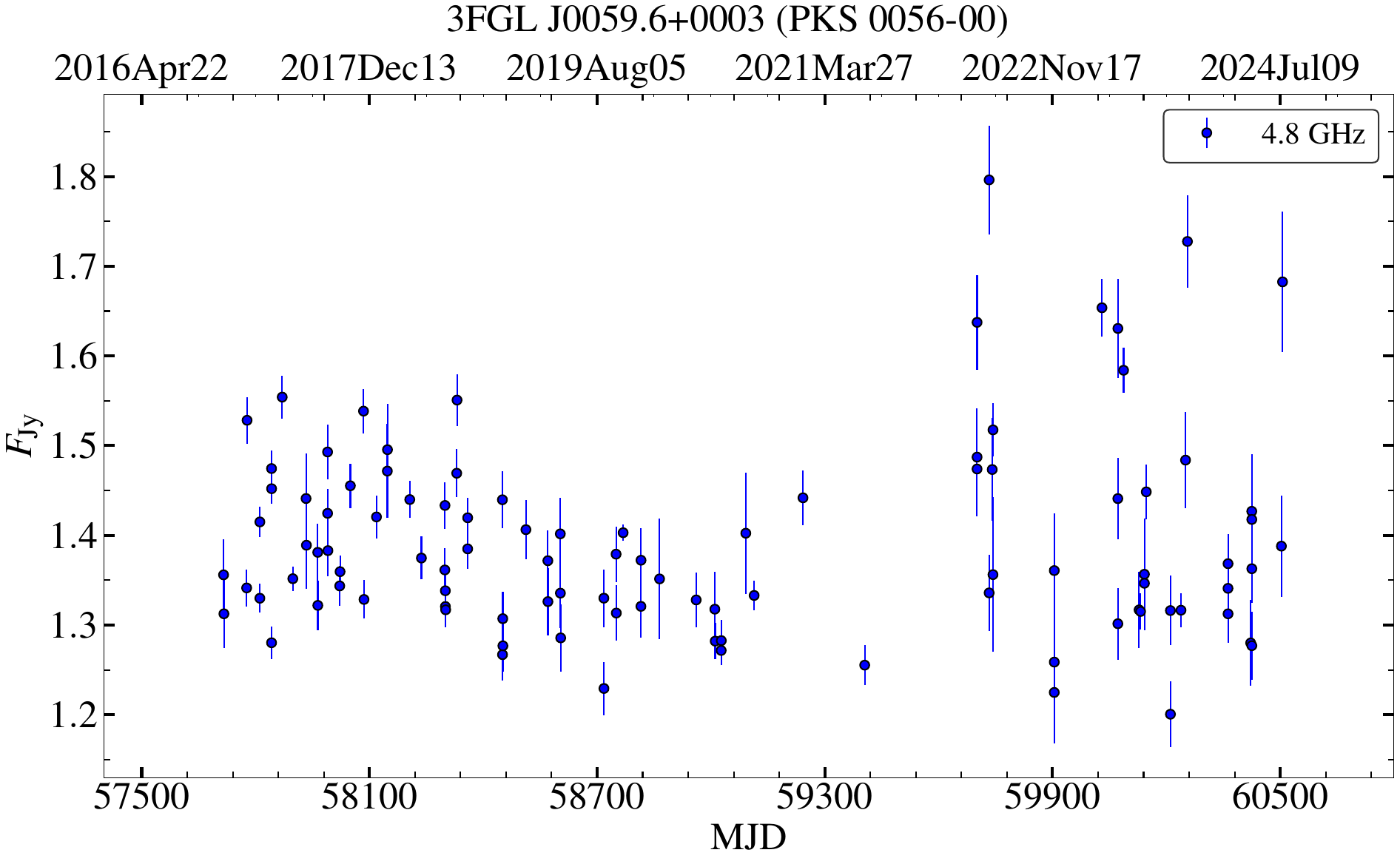}
\includegraphics[width=0.49\textwidth]{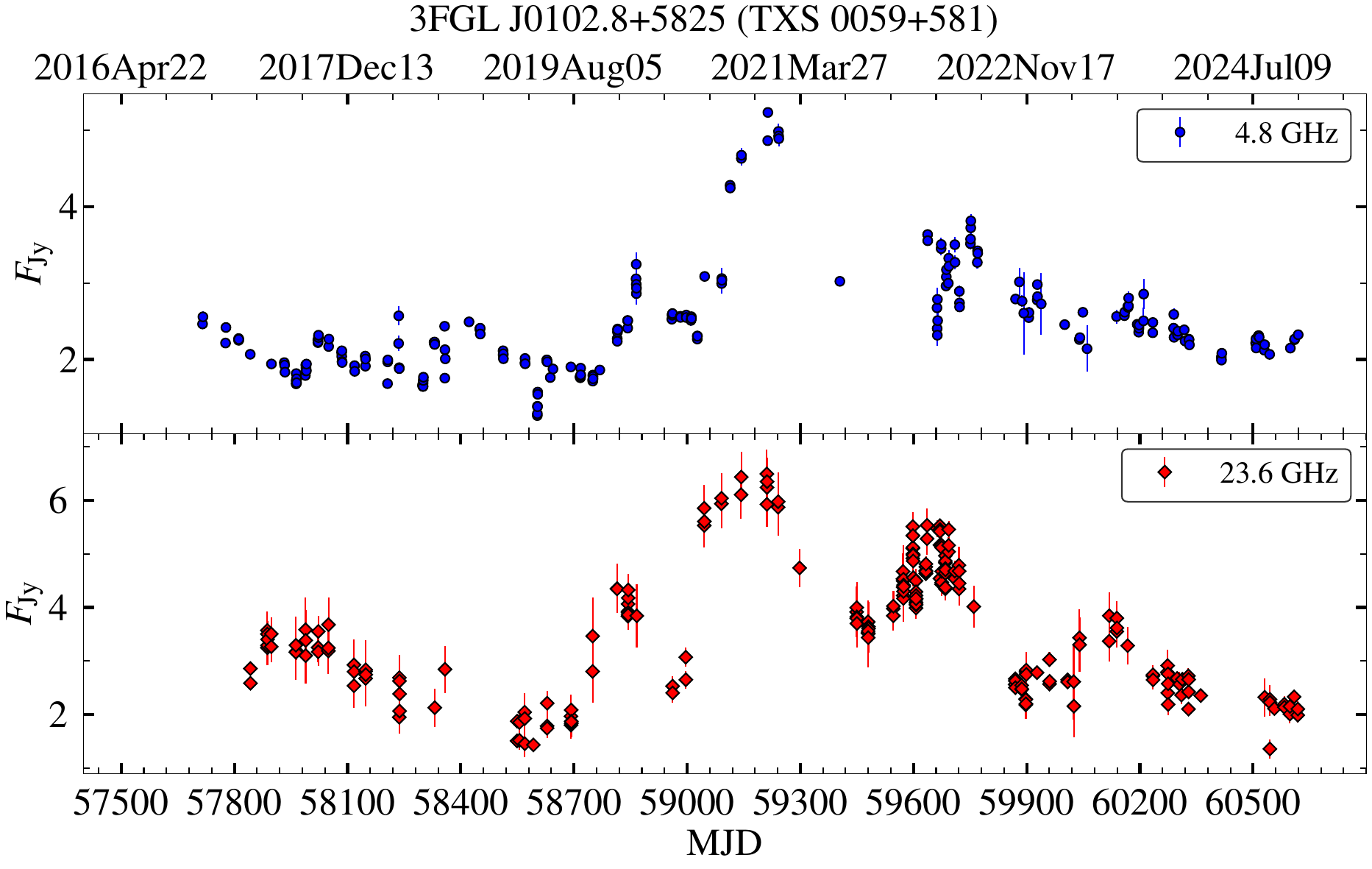}\\
\includegraphics[width=0.49\textwidth]{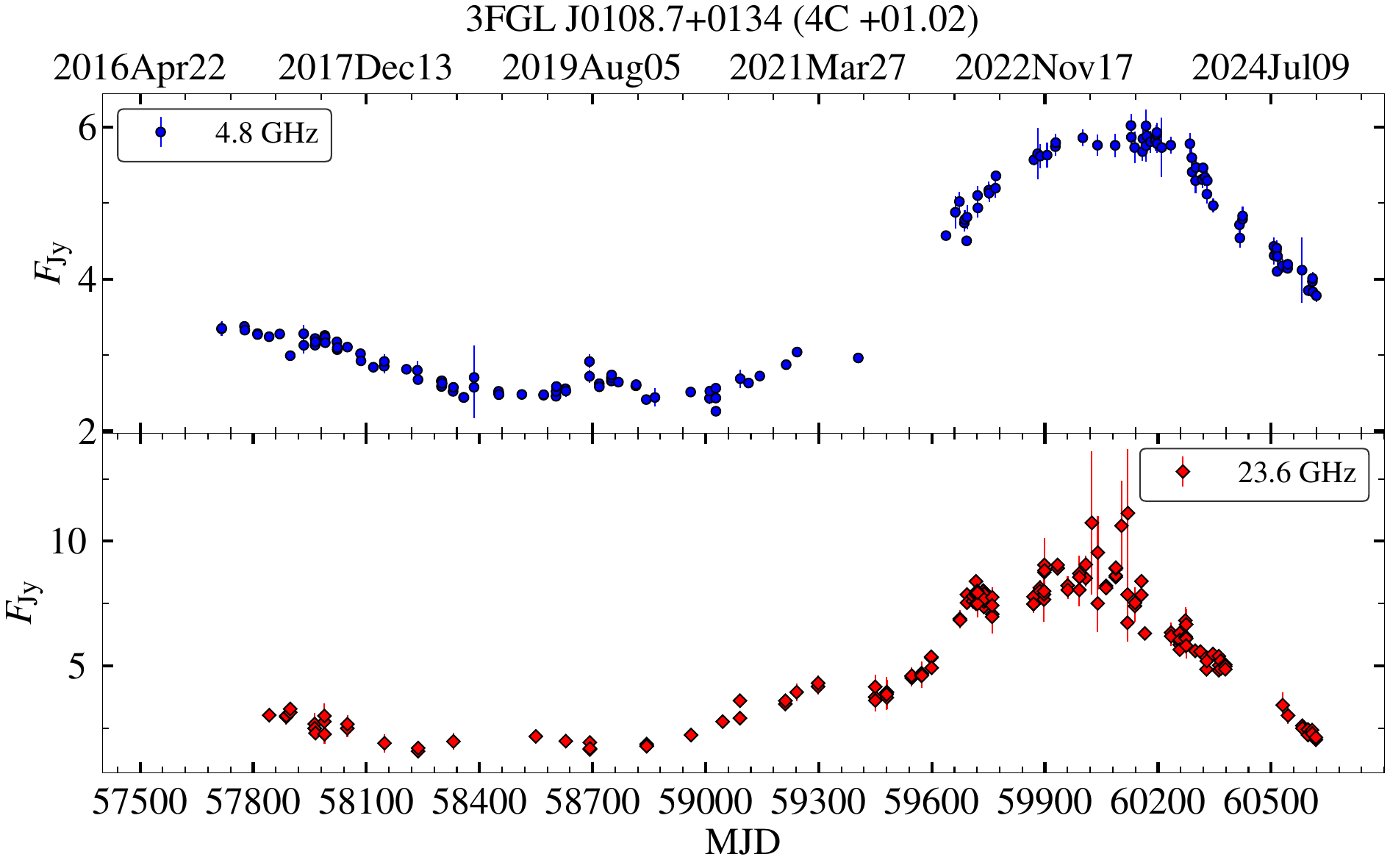}
\includegraphics[width=0.49\textwidth]{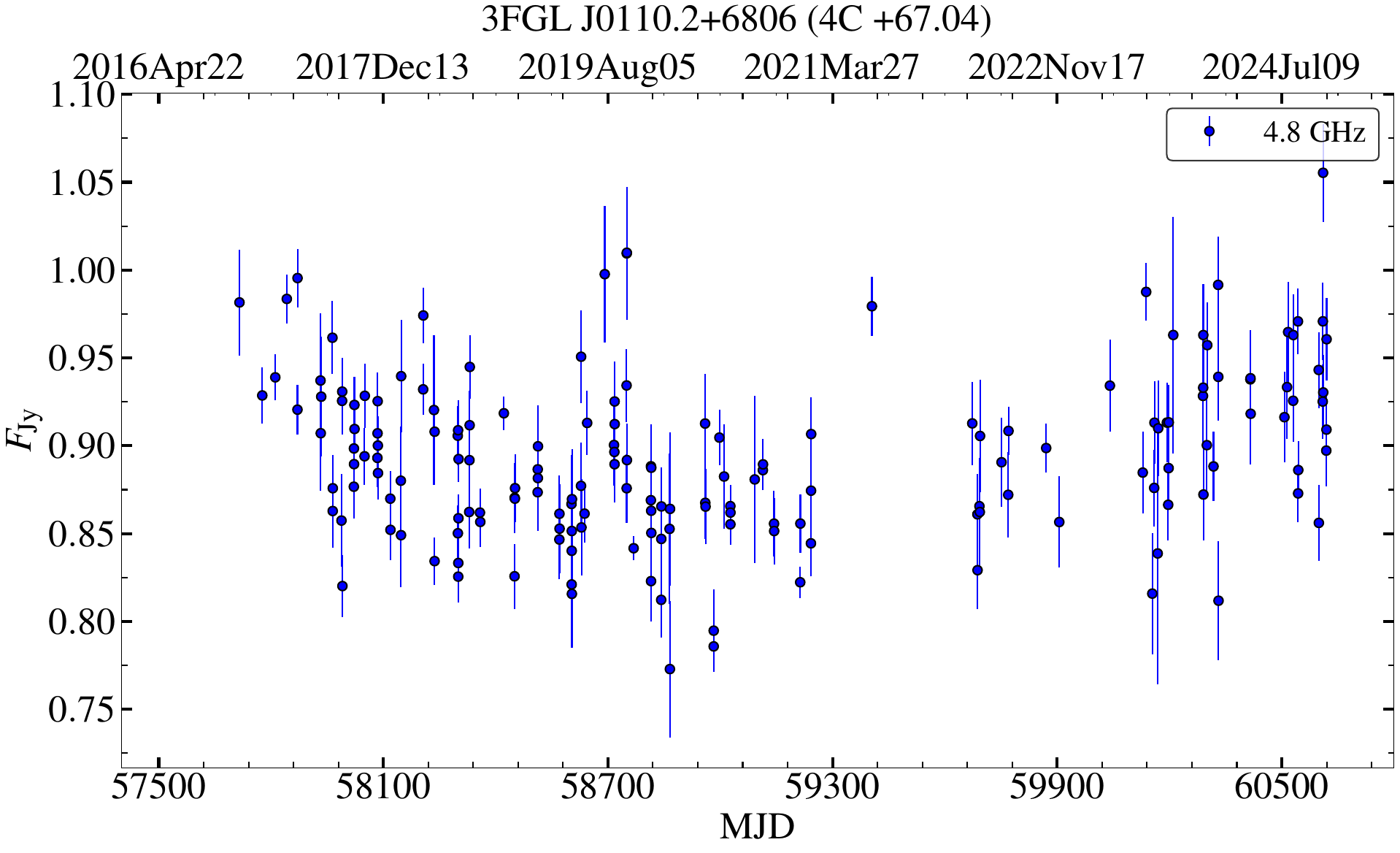}\\
\includegraphics[width=0.49\textwidth]{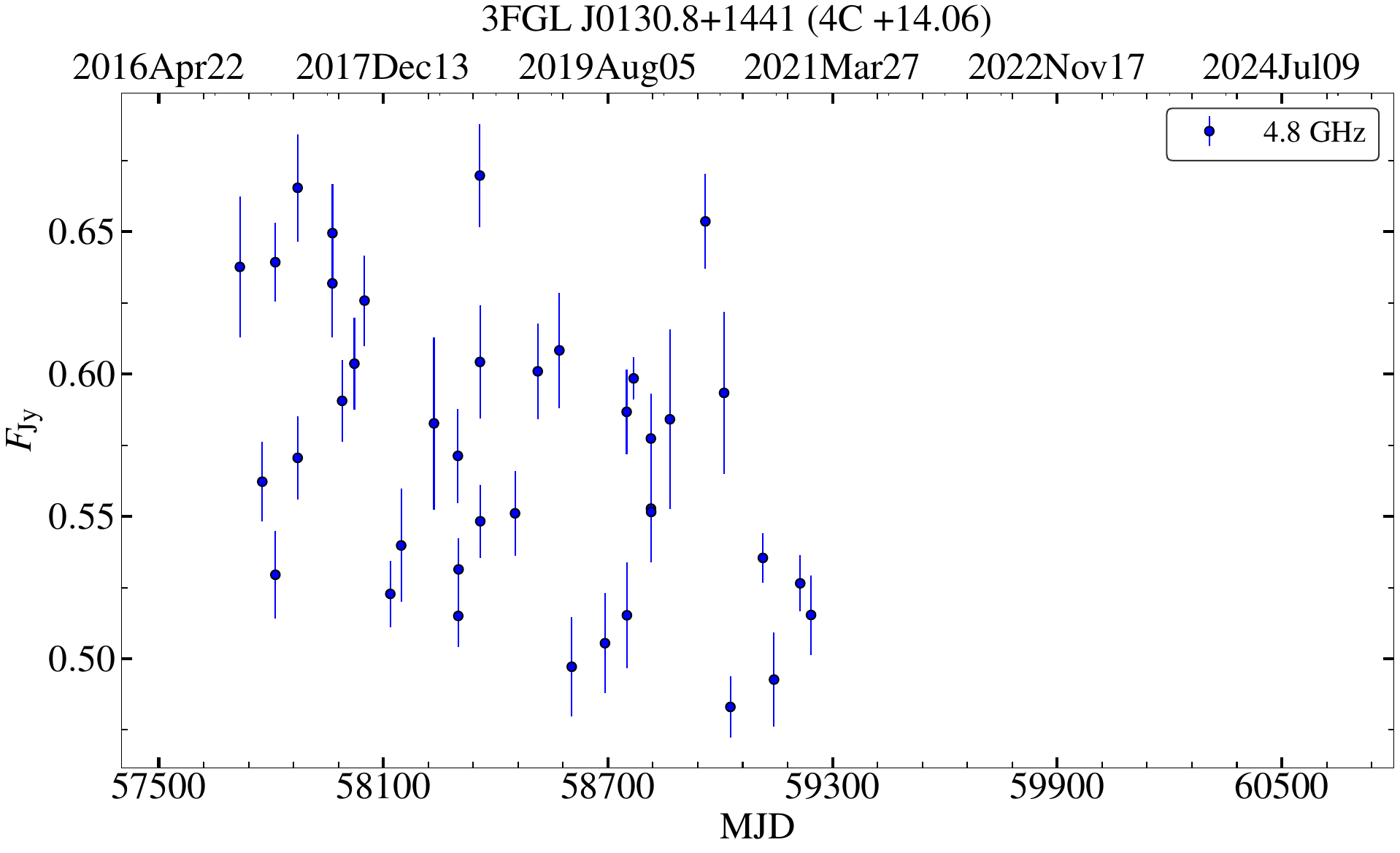}
\includegraphics[width=0.49\textwidth]{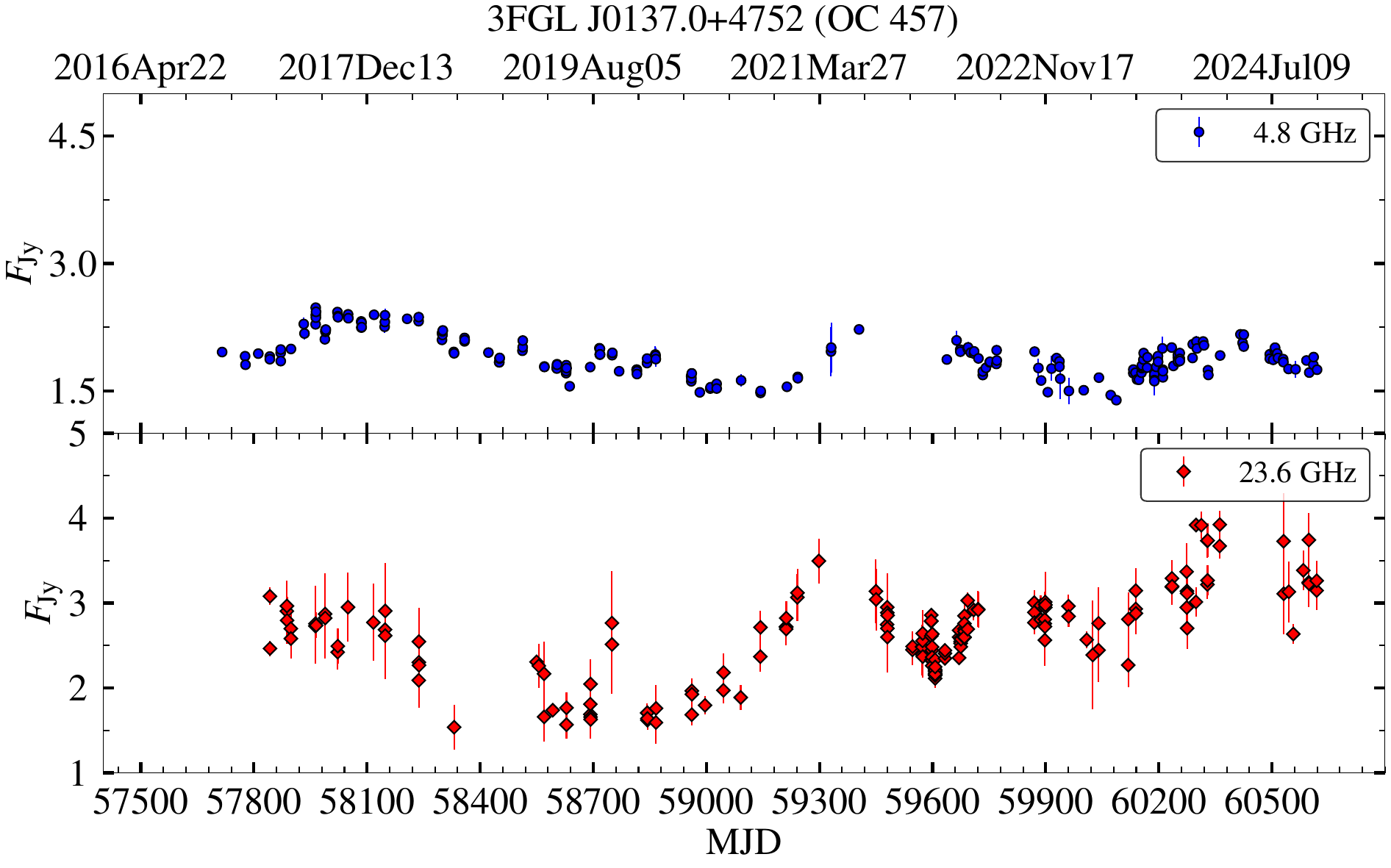}\\

\end{tabular}
\caption{Light curves of the sources monitored by the SMMAN program at 4.8 GHz and 23.6 GHz. }
\label{fig:SMMAN_sources_lc}
\end{figure*}

\addtocounter{figure}{-1}
\begin{figure*}[p]
\centering
\begin{tabular}{cc}
\includegraphics[width=0.49\textwidth]{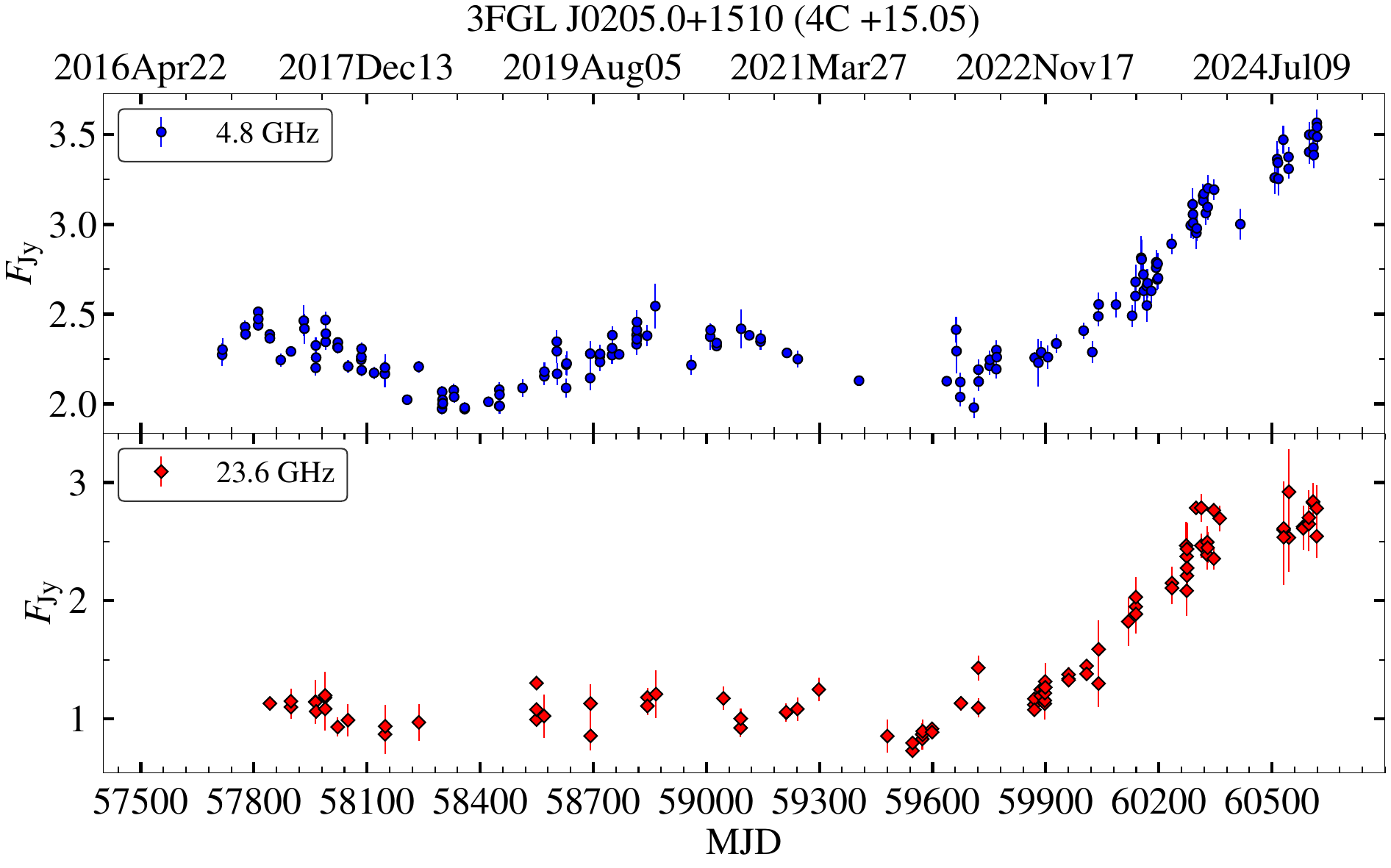}
\includegraphics[width=0.49\textwidth]{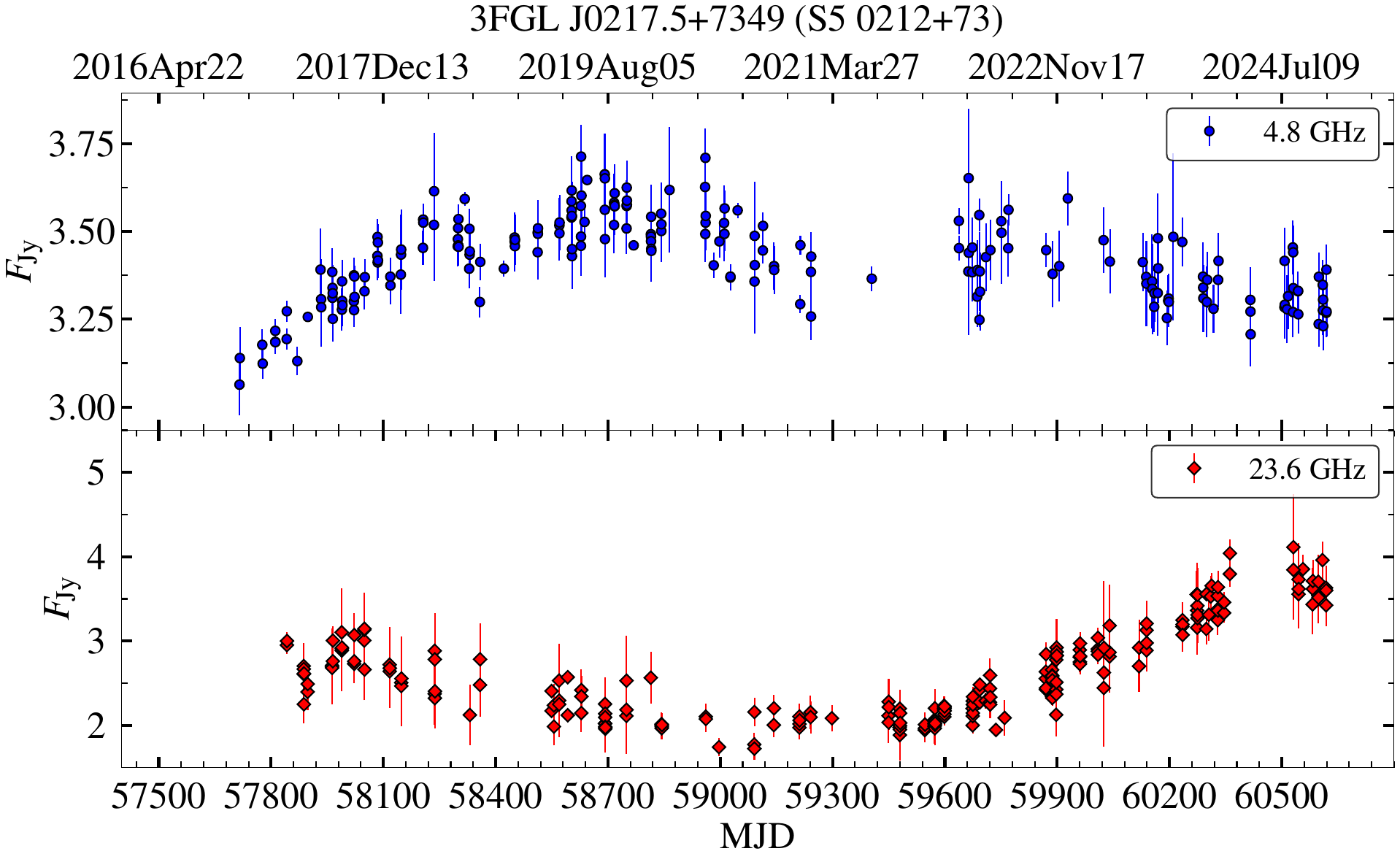}\\
\includegraphics[width=0.49\textwidth]{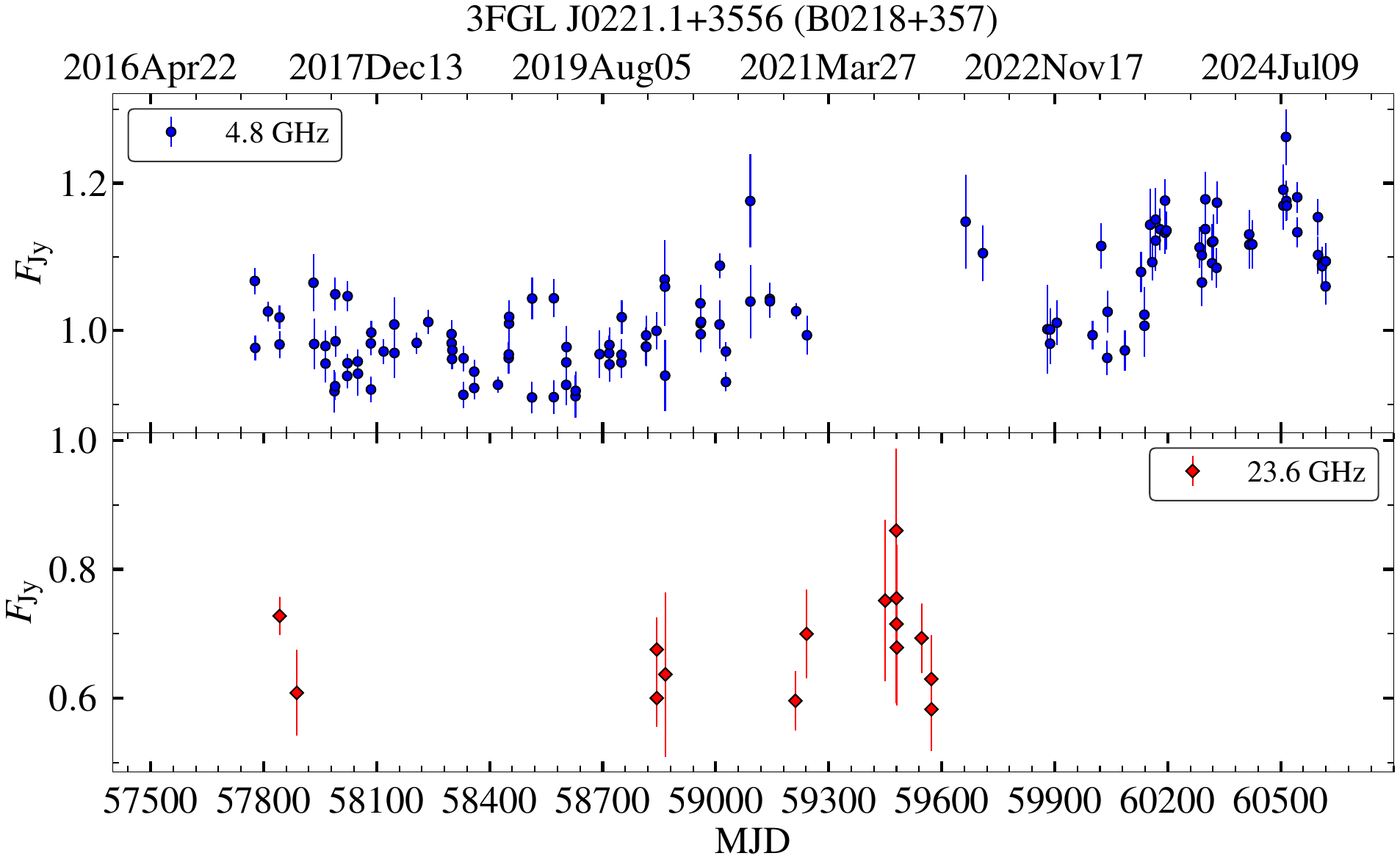}
\includegraphics[width=0.49\textwidth]{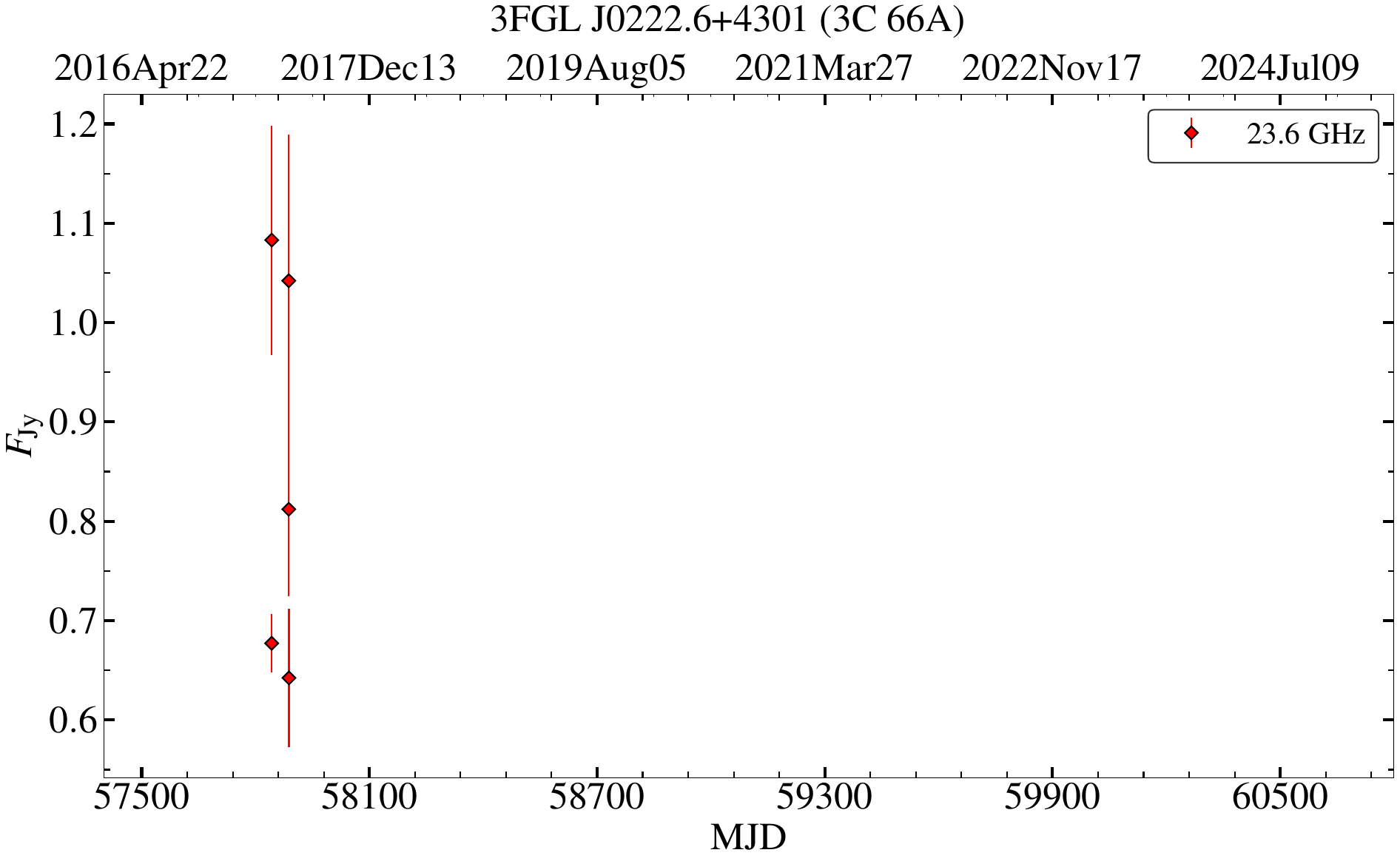}\\
\includegraphics[width=0.49\textwidth]{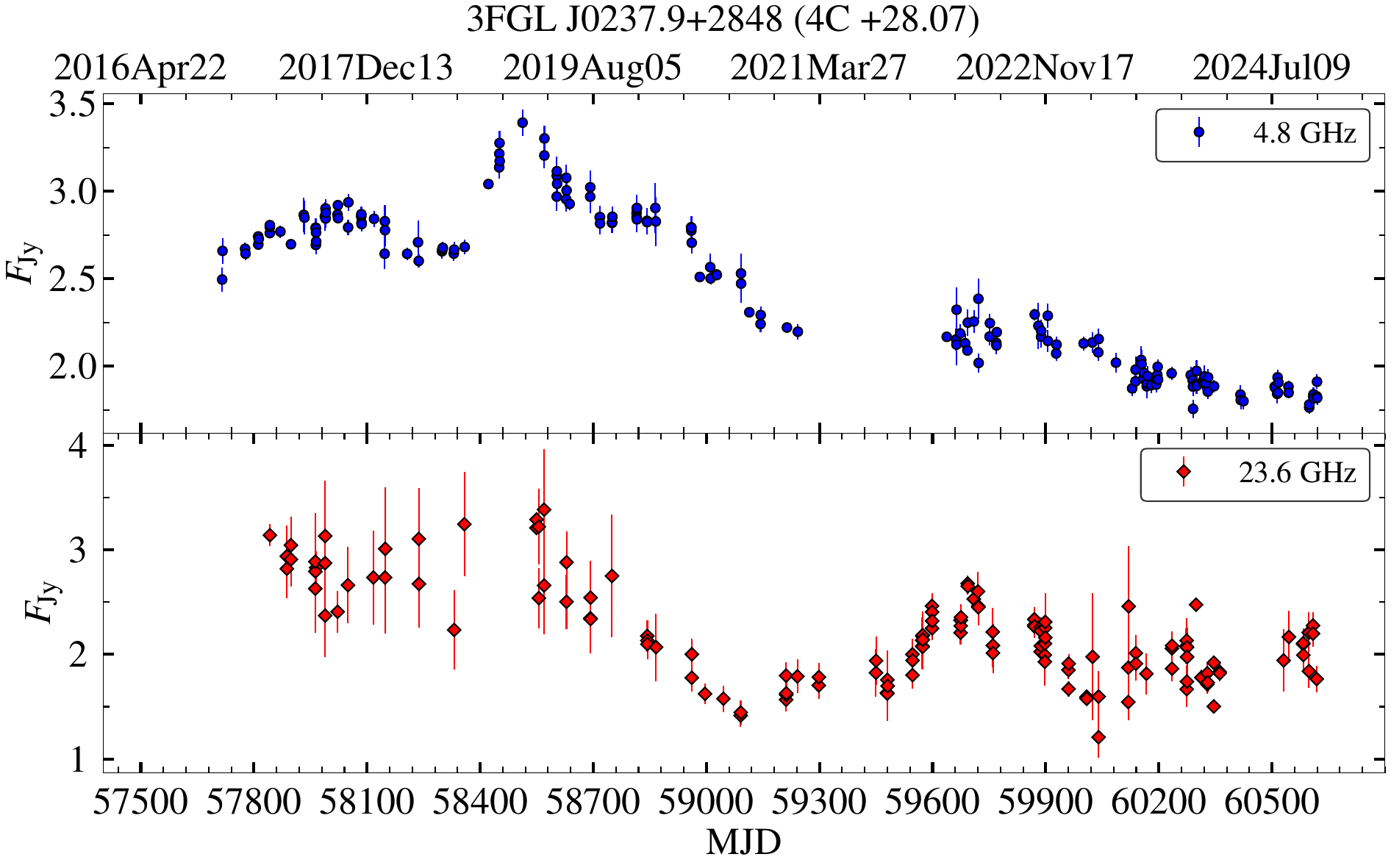}
\includegraphics[width=0.49\textwidth]{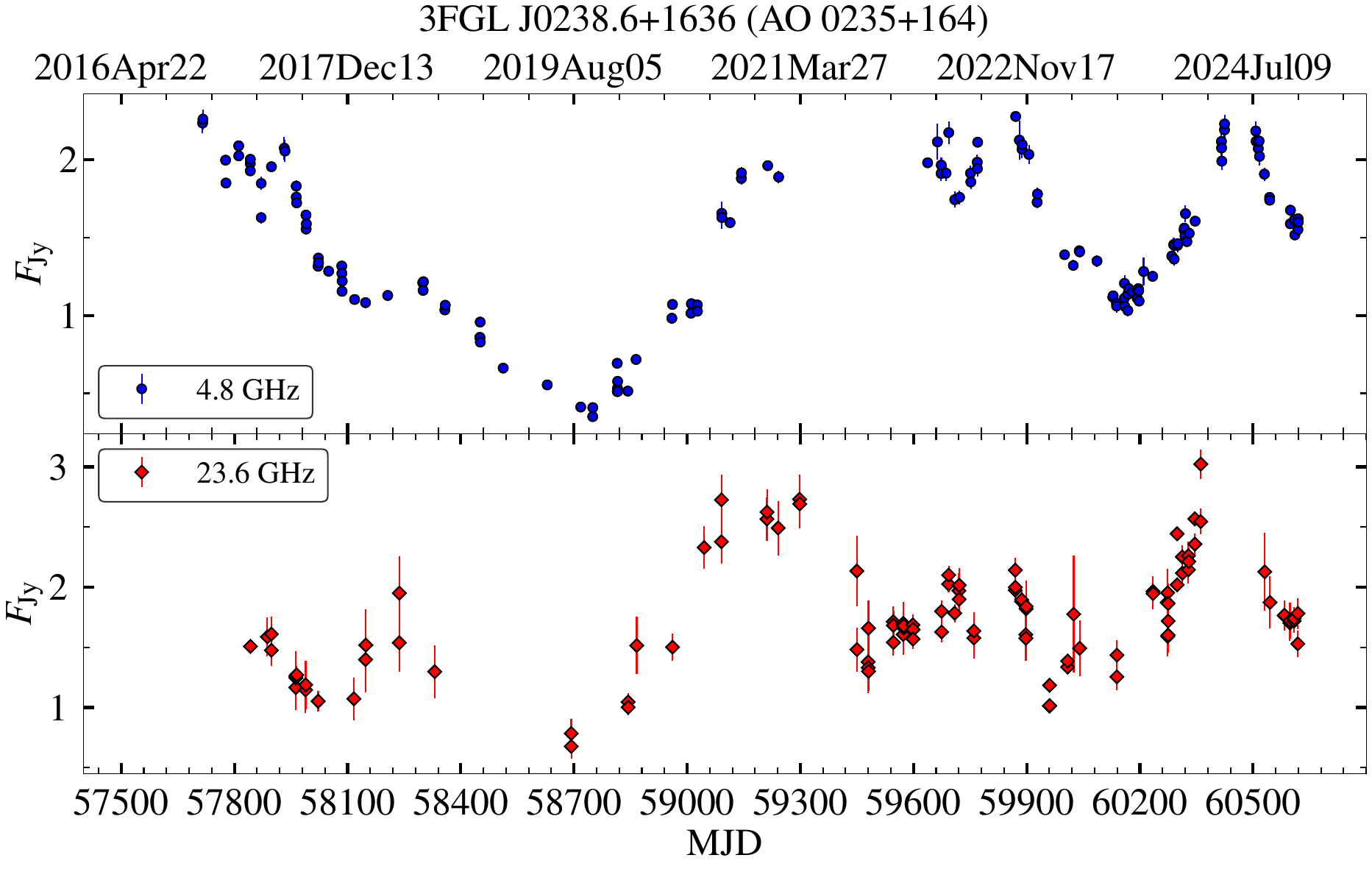}\\
\includegraphics[width=0.49\textwidth]{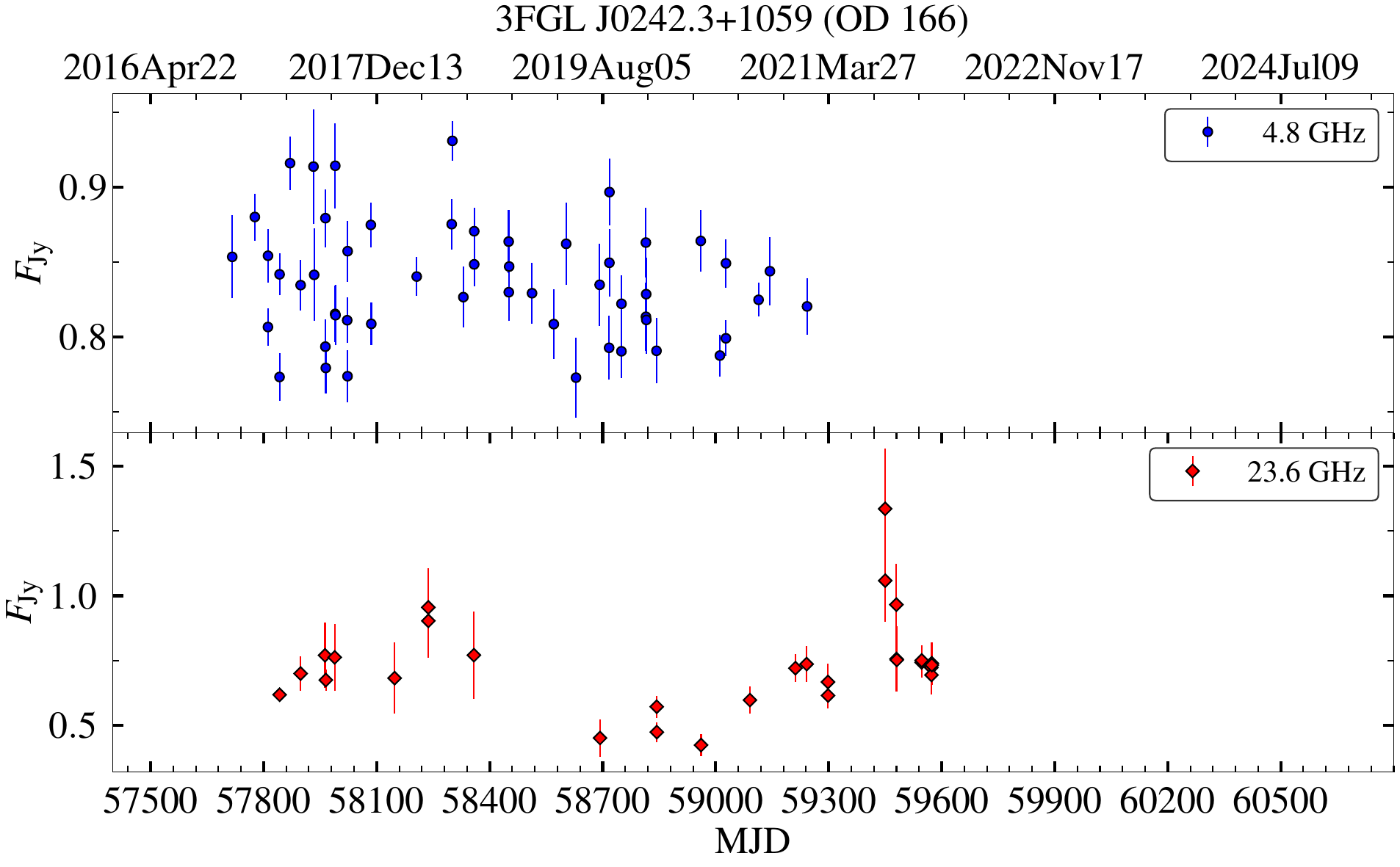}
\includegraphics[width=0.49\textwidth]{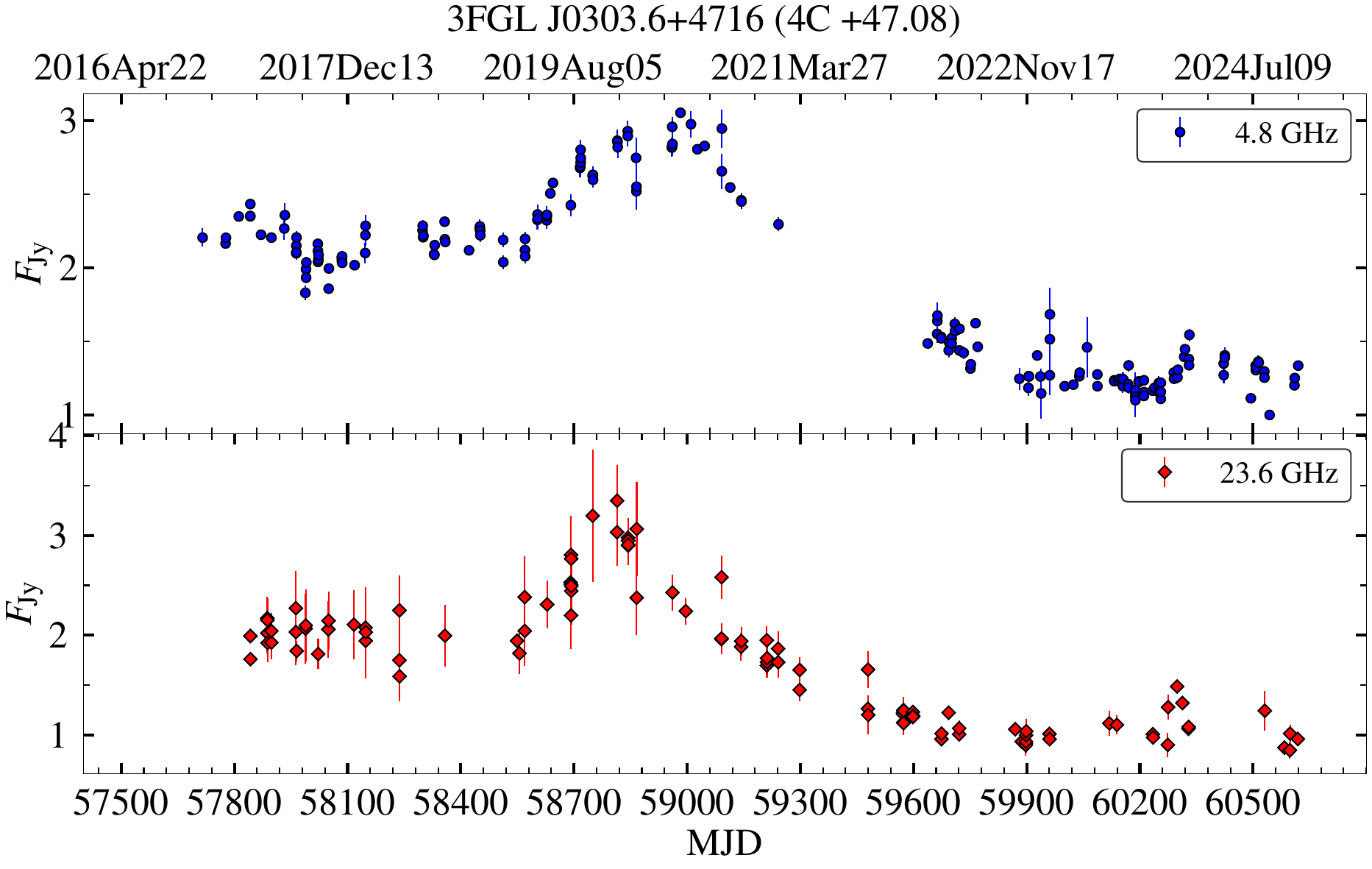}\\
\end{tabular}
\caption{Continued.}
\end{figure*}

\begin{figure*}[p]
\centering
\addtocounter{figure}{-1}
\caption{Continued.}
\begin{tabular}{cc}
\includegraphics[width=0.49\textwidth]{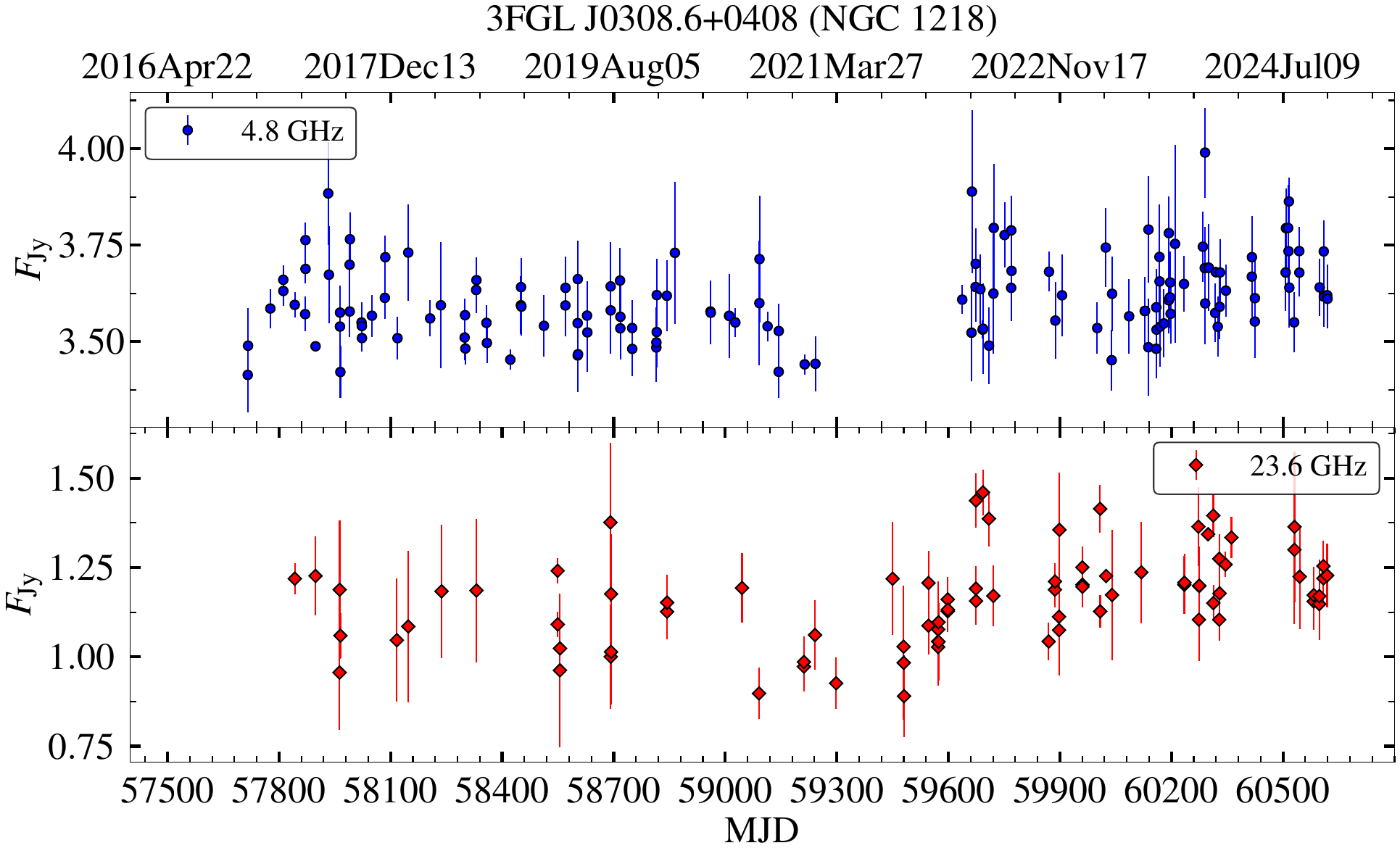}
\includegraphics[width=0.49\textwidth]{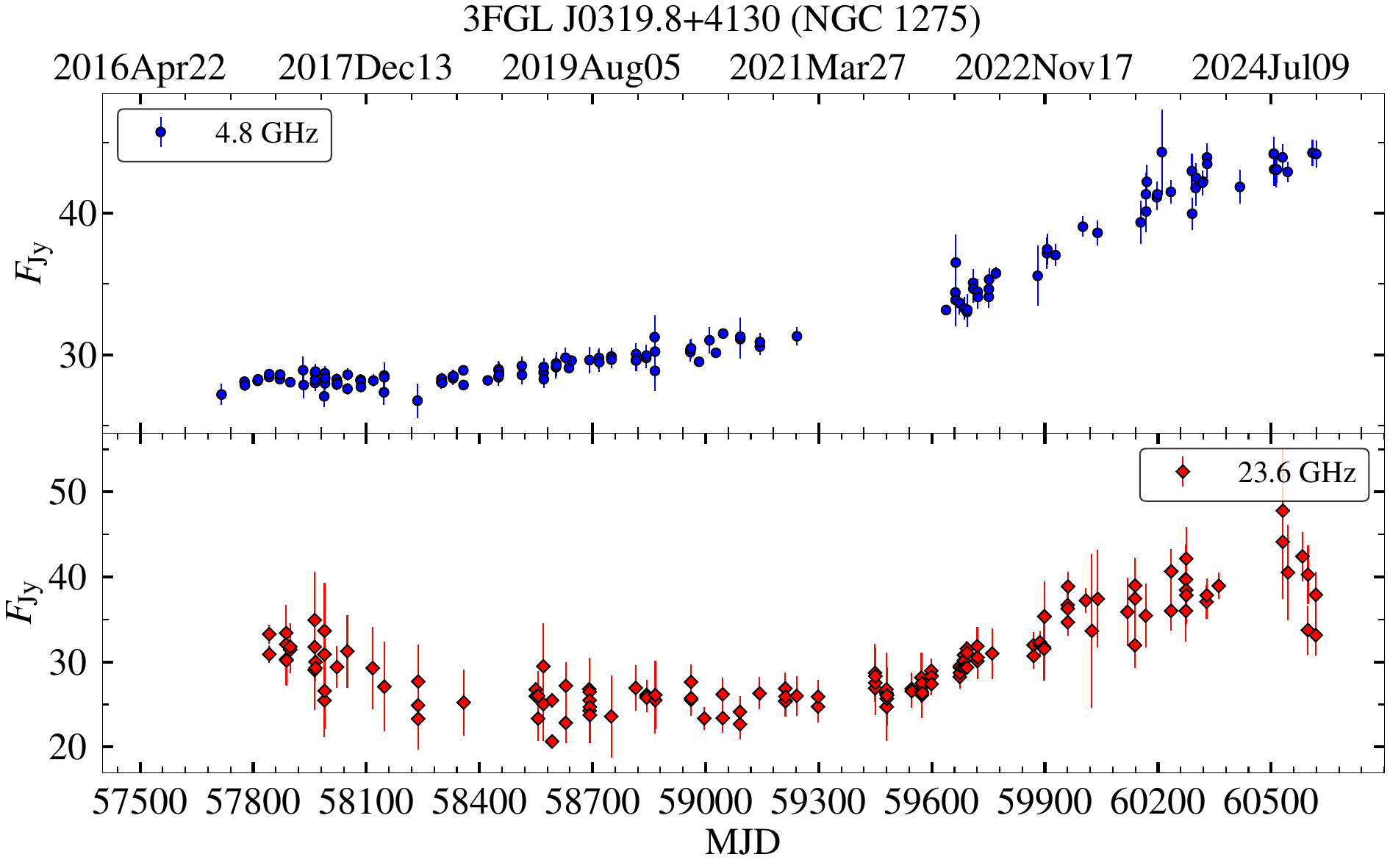}\\
\includegraphics[width=0.49\textwidth]{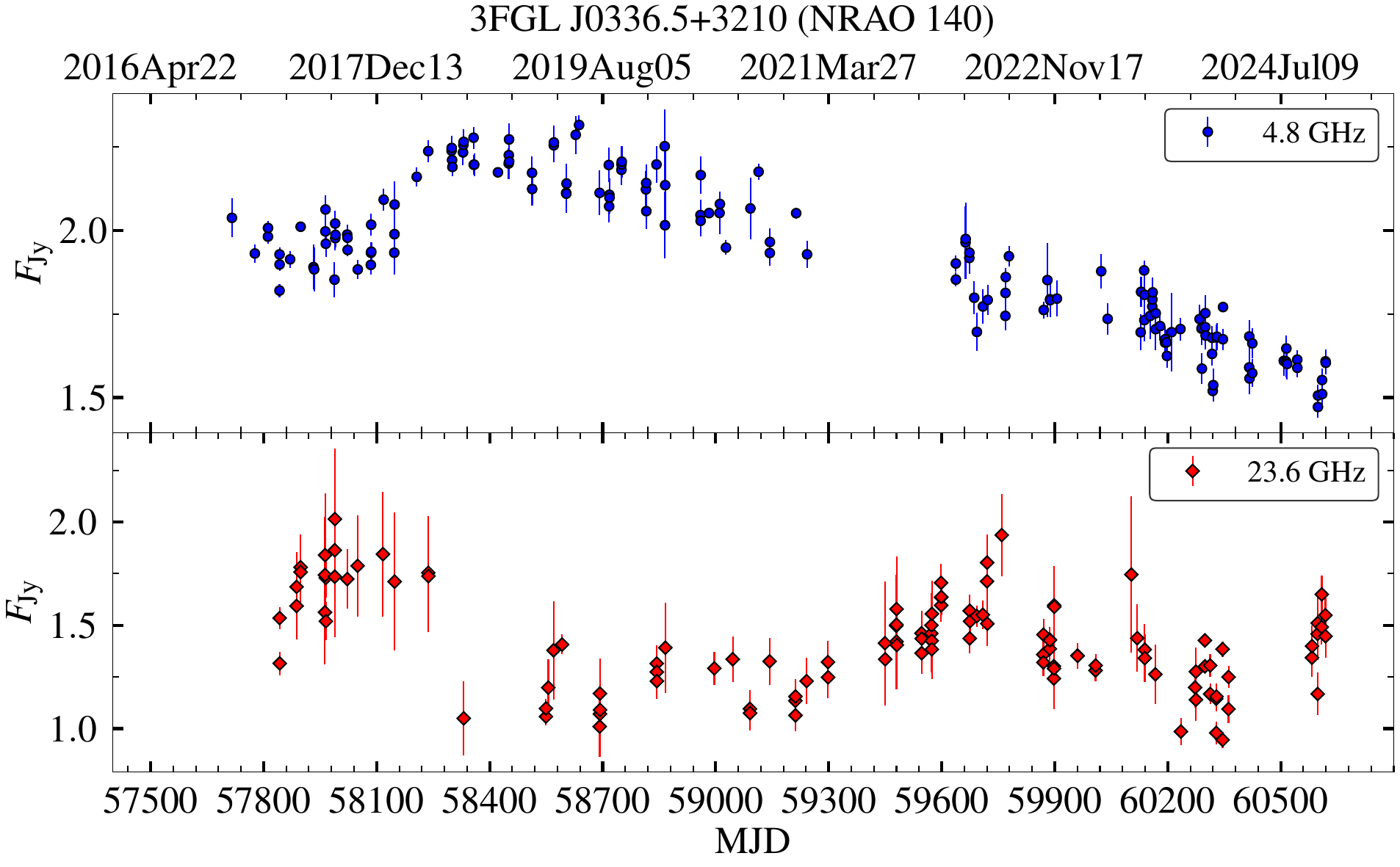}
\includegraphics[width=0.49\textwidth]{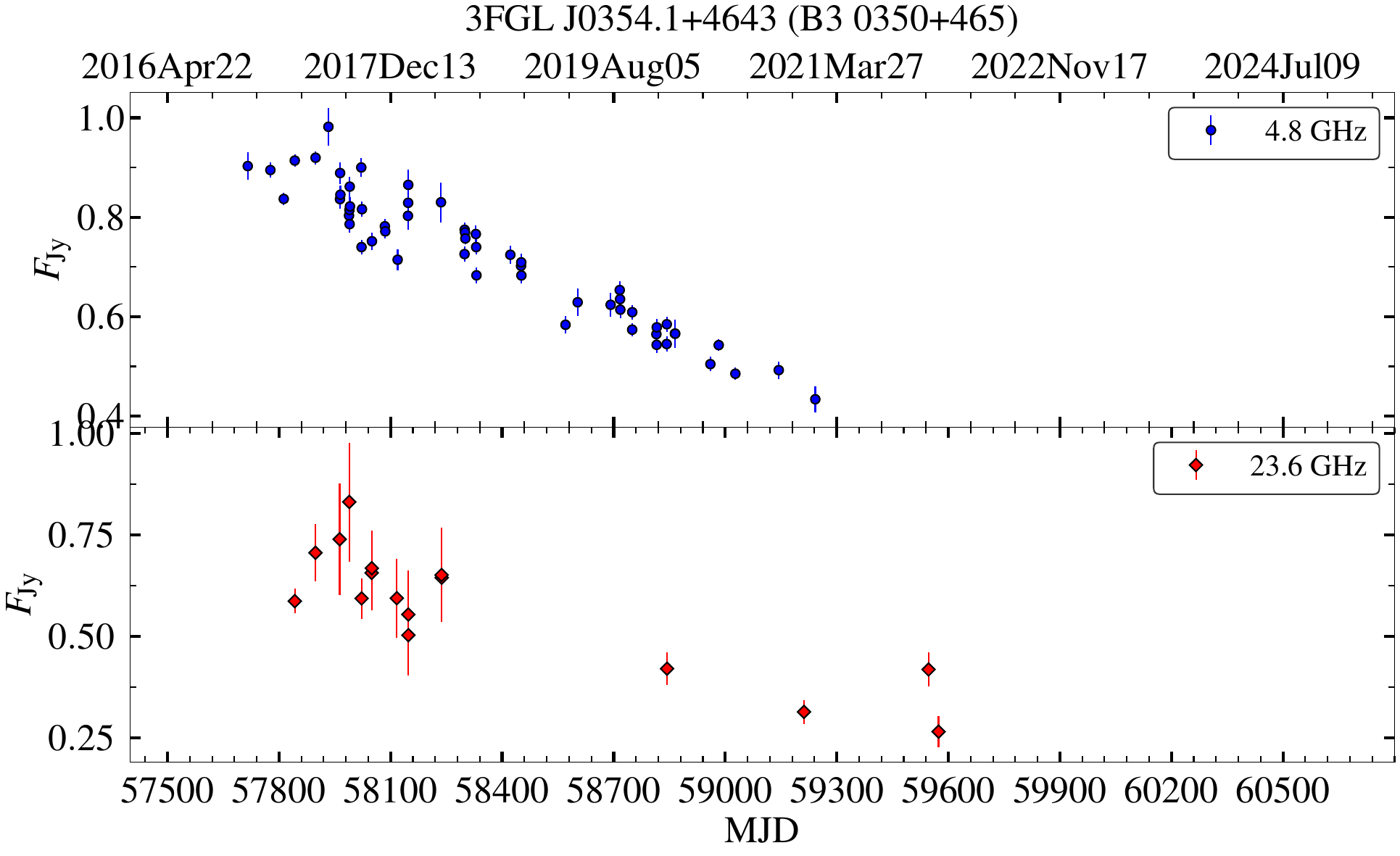}\\
\includegraphics[width=0.49\textwidth]{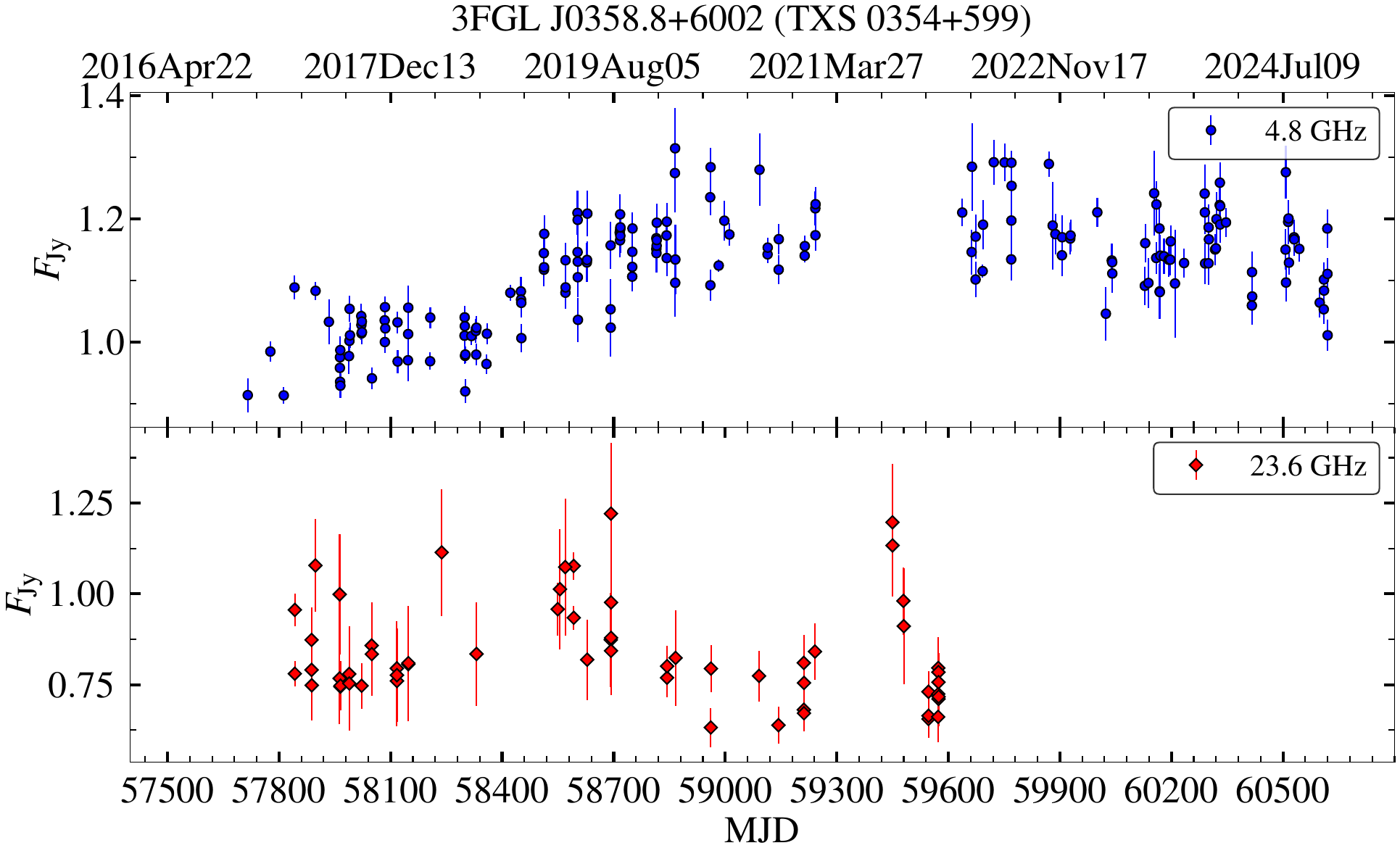}
\includegraphics[width=0.49\textwidth]{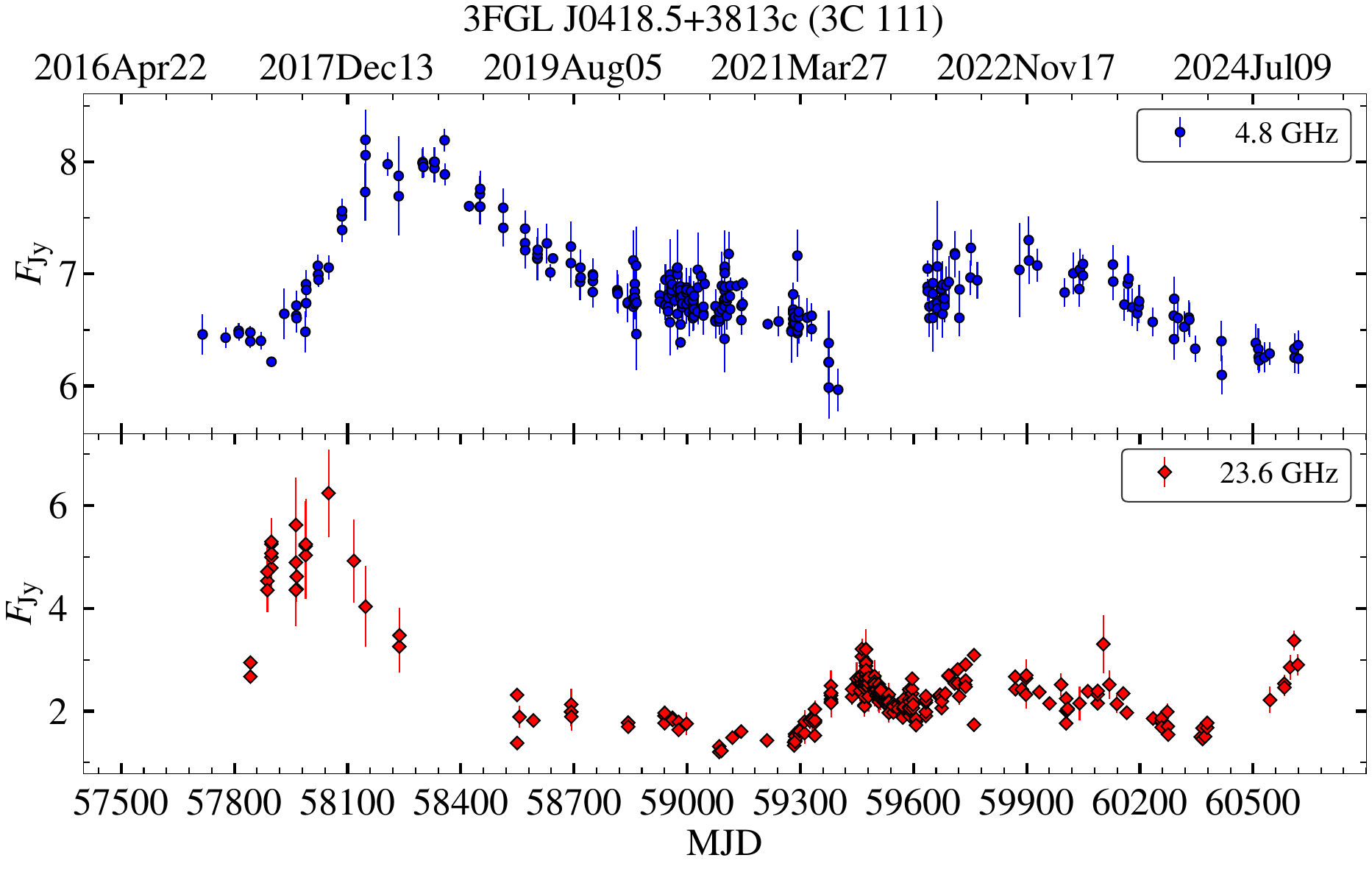}\\
\includegraphics[width=0.49\textwidth]{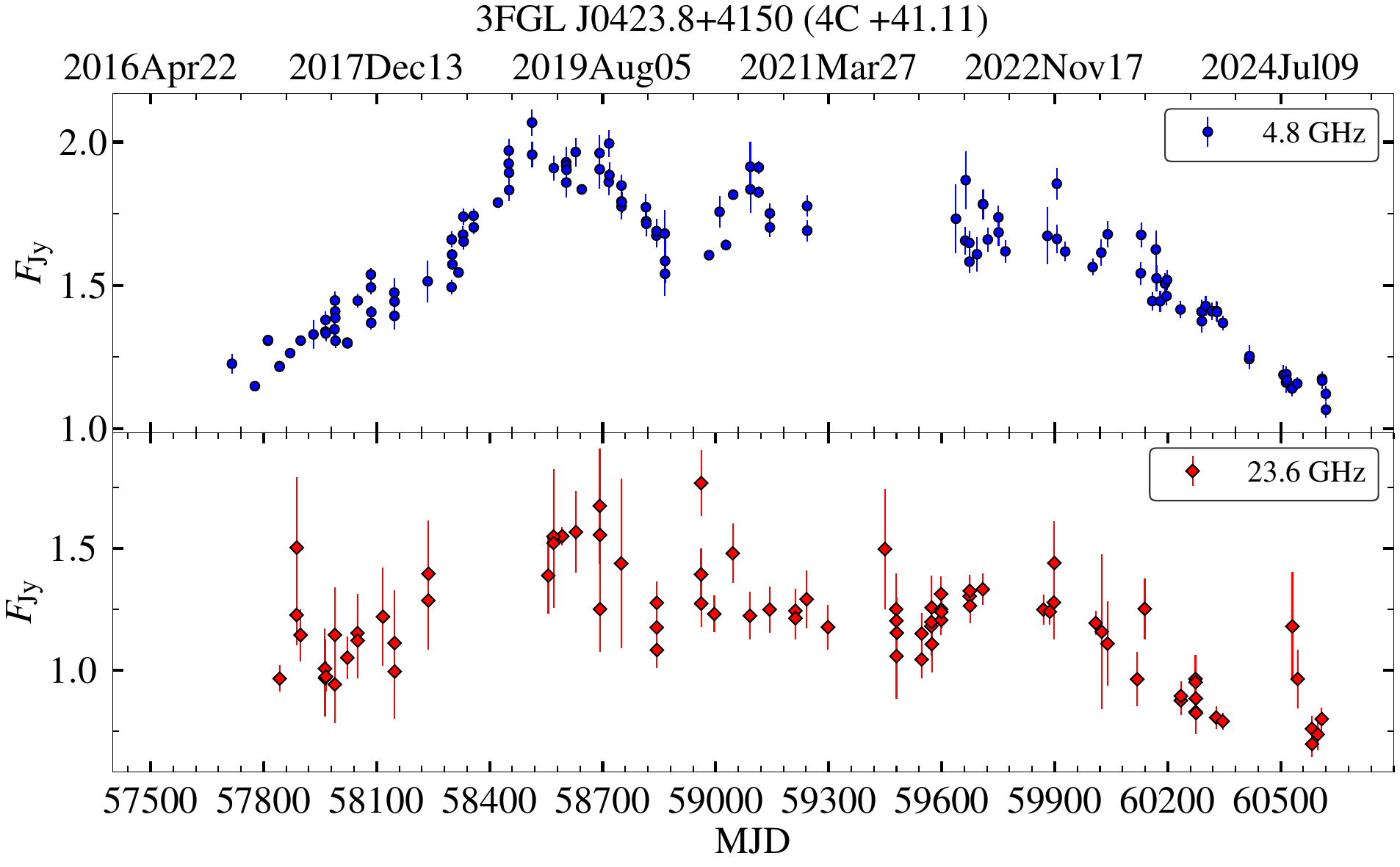}
\includegraphics[width=0.49\textwidth]{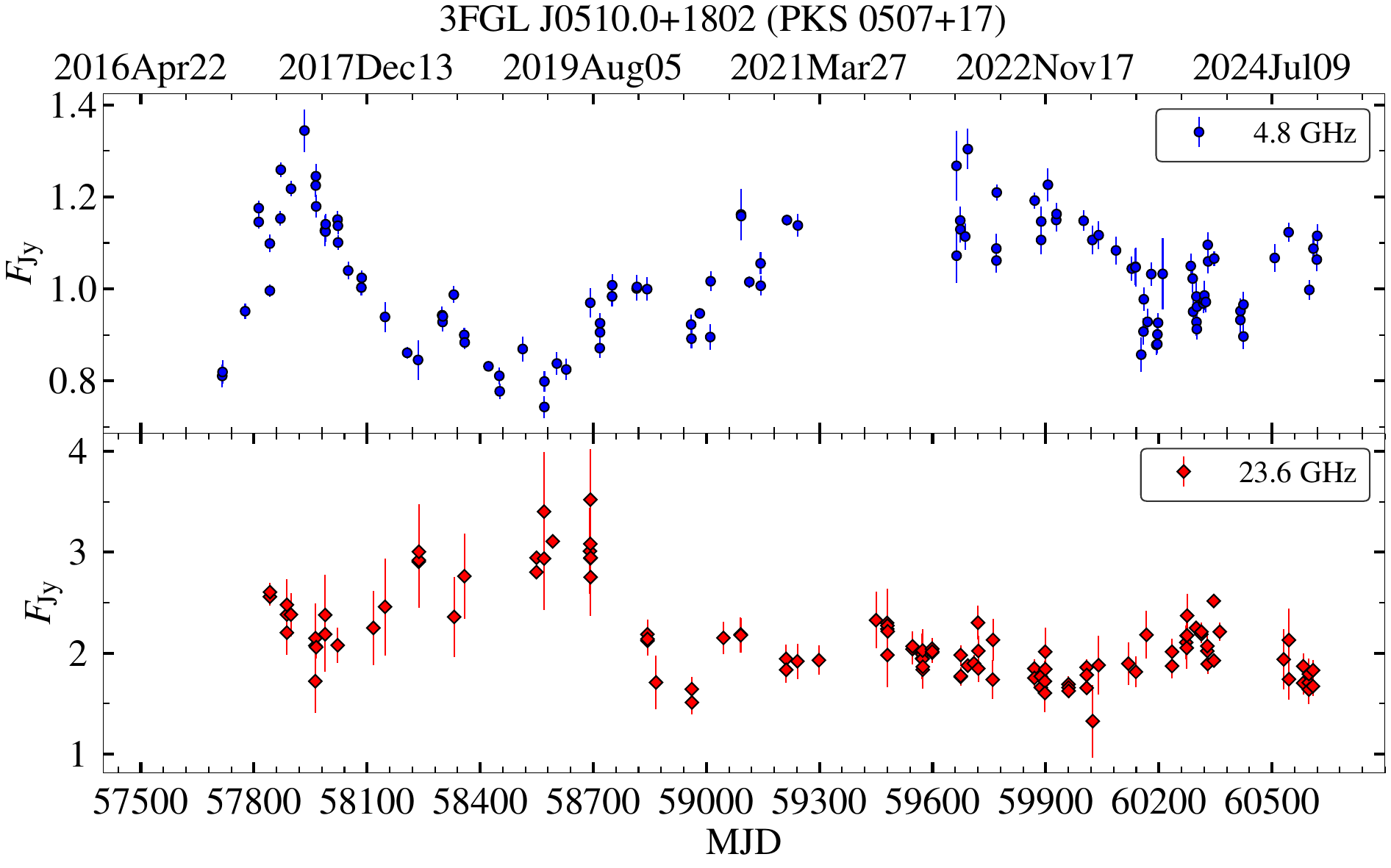}\\
\end{tabular}
\end{figure*}

\begin{figure*}[p]
\centering
\addtocounter{figure}{-1}
\caption{Continued.}
\begin{tabular}{cc}
\includegraphics[width=0.49\textwidth]{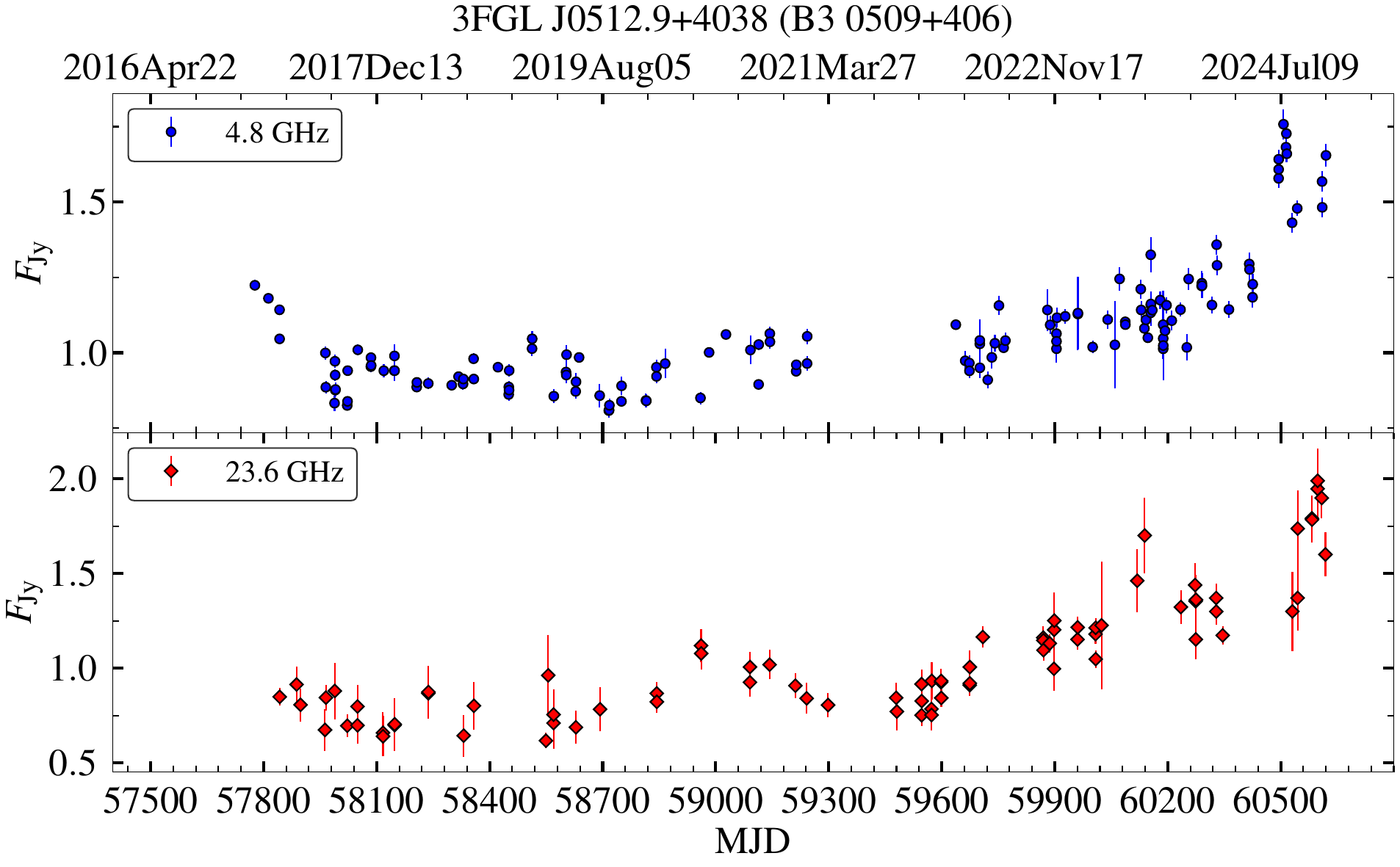}
\includegraphics[width=0.49\textwidth]{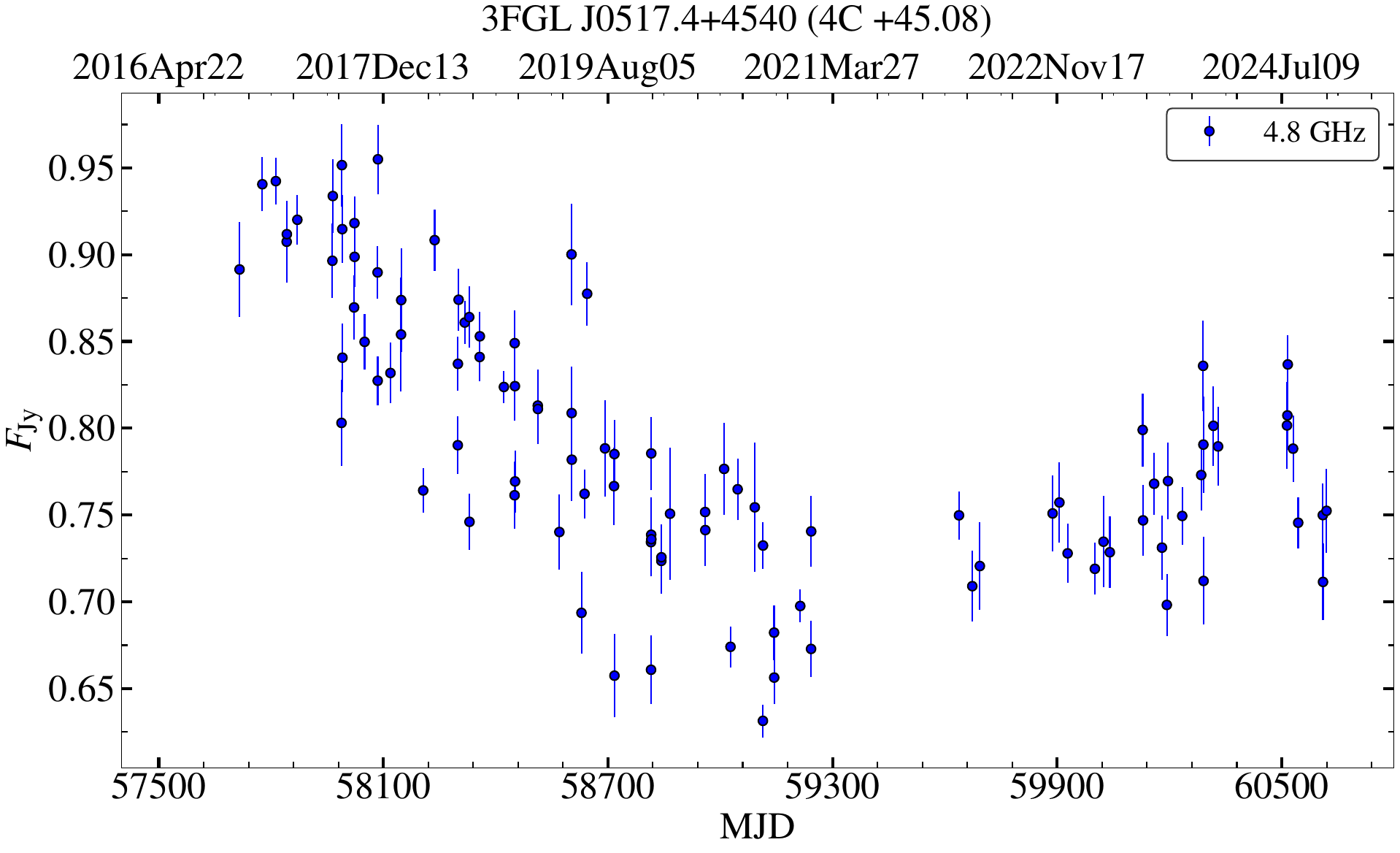}\\
\includegraphics[width=0.49\textwidth]{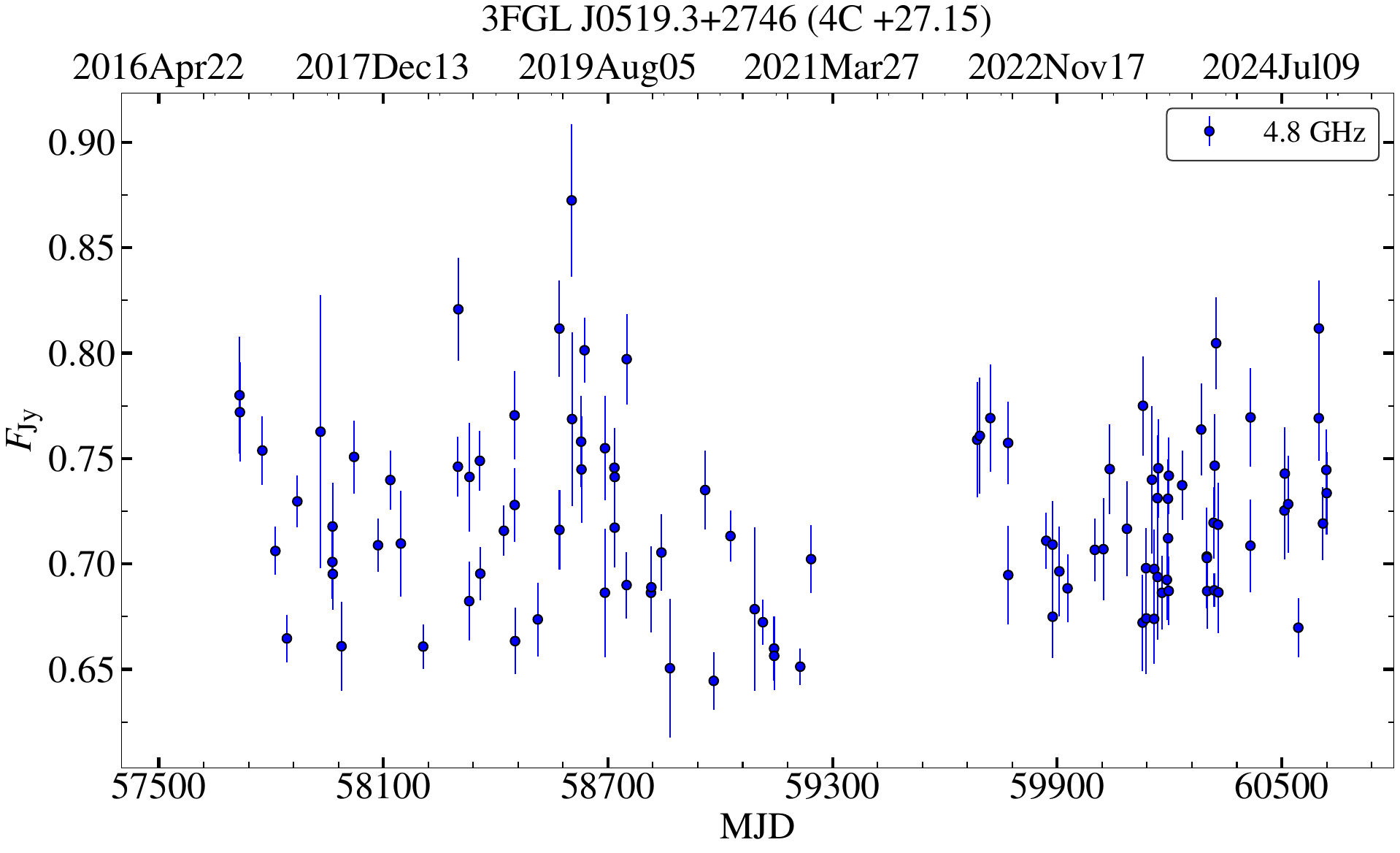}
\includegraphics[width=0.49\textwidth]{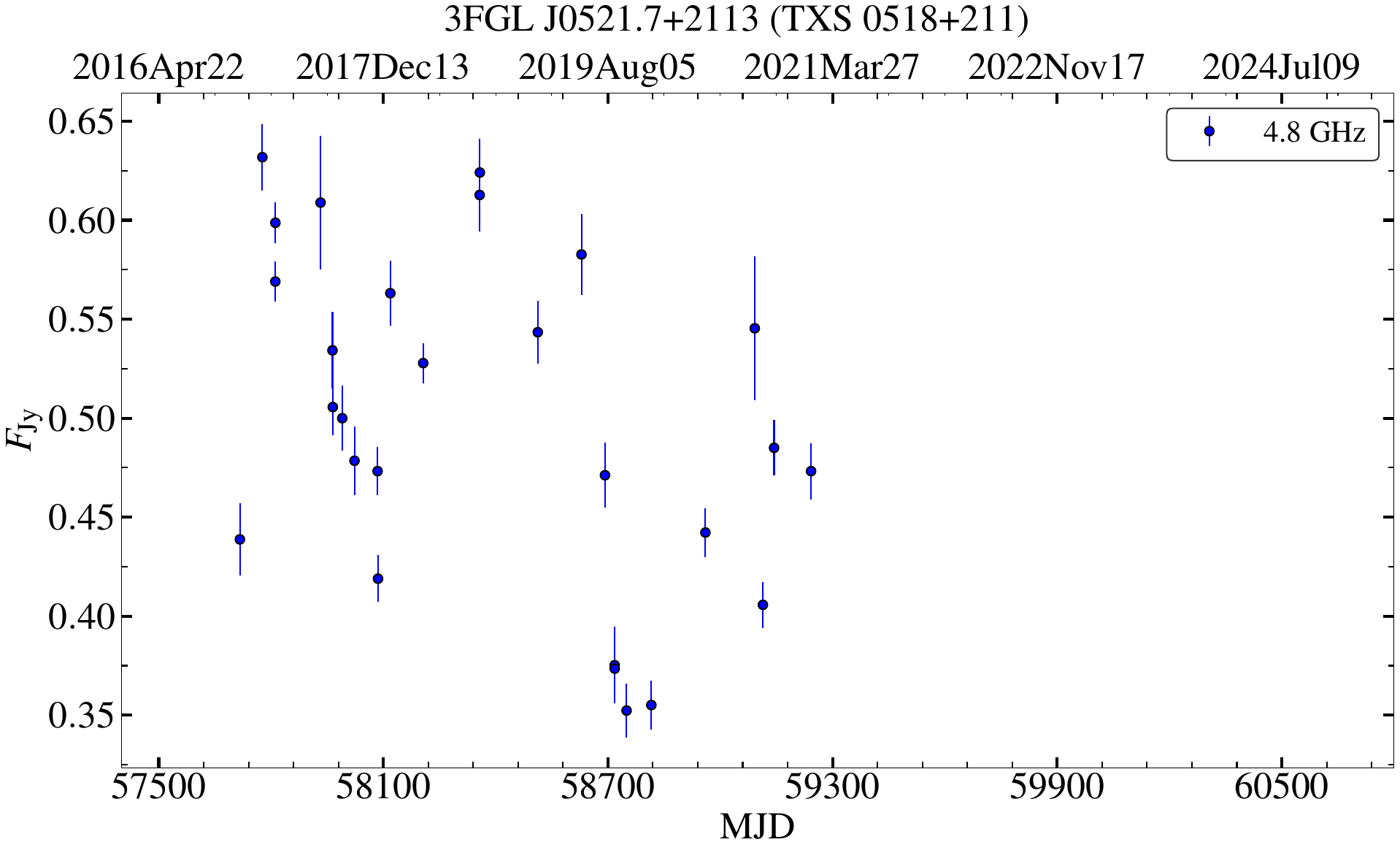}\\
\includegraphics[width=0.49\textwidth]{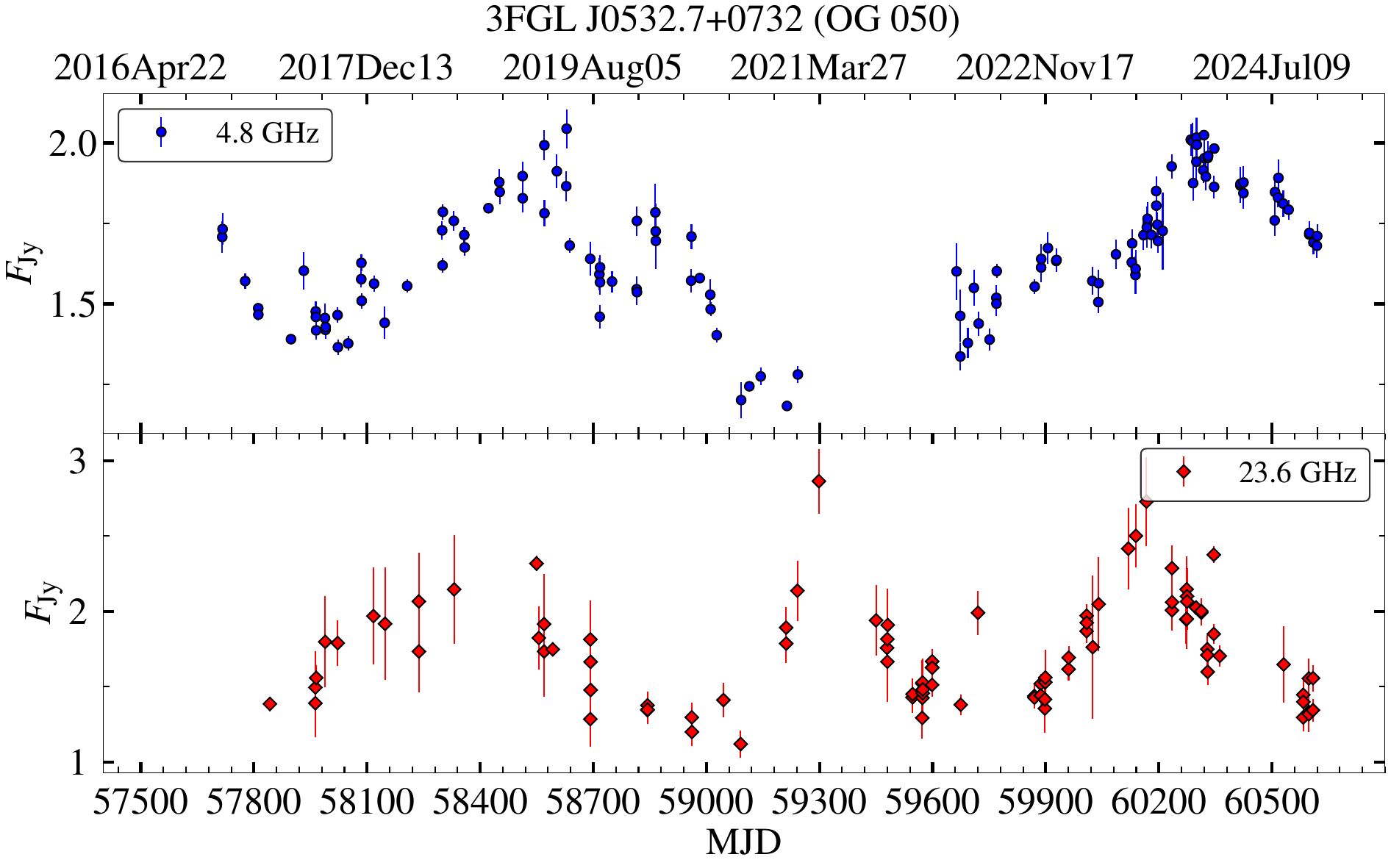}
\includegraphics[width=0.49\textwidth]{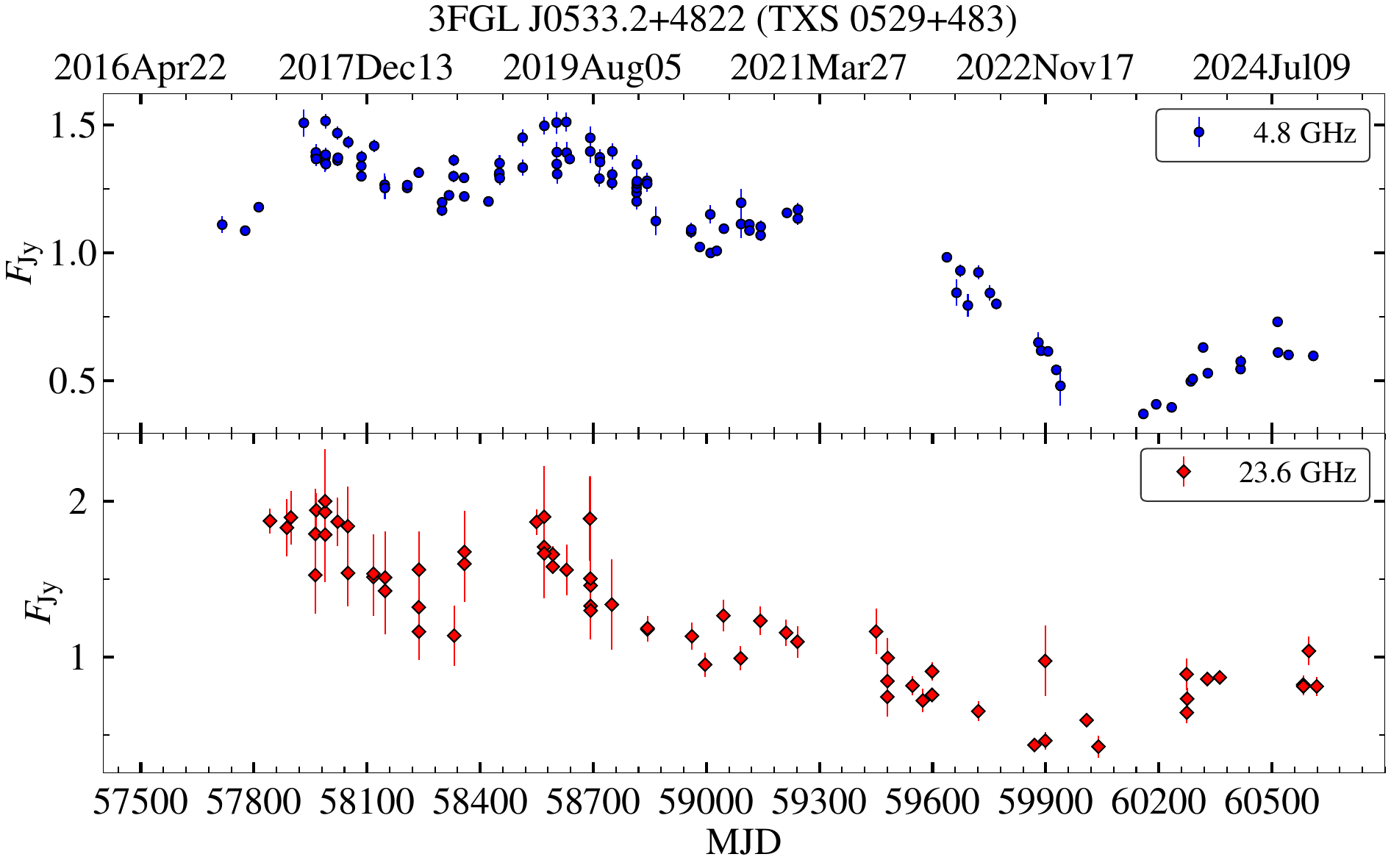}\\
\includegraphics[width=0.49\textwidth]{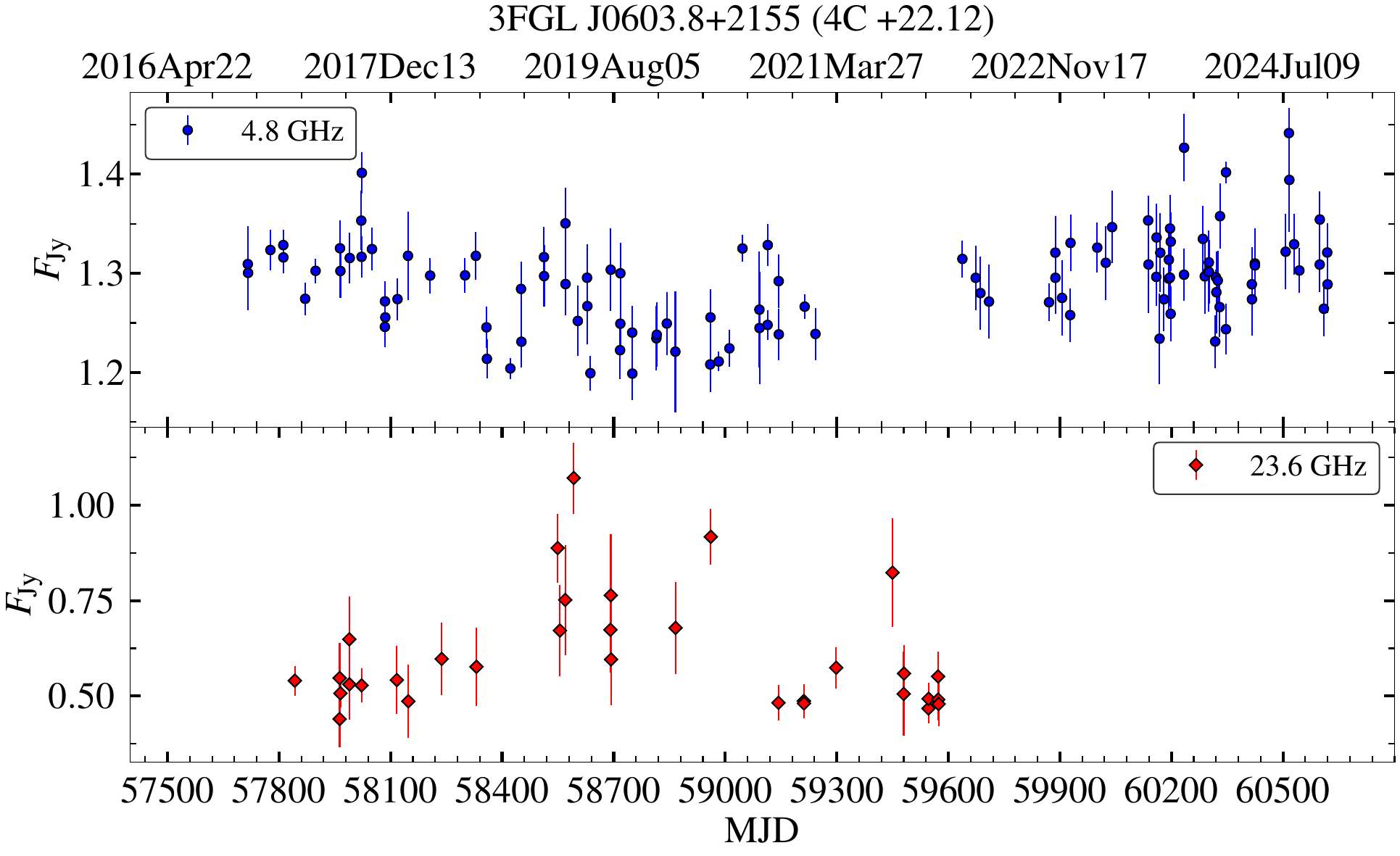}
\includegraphics[width=0.49\textwidth]{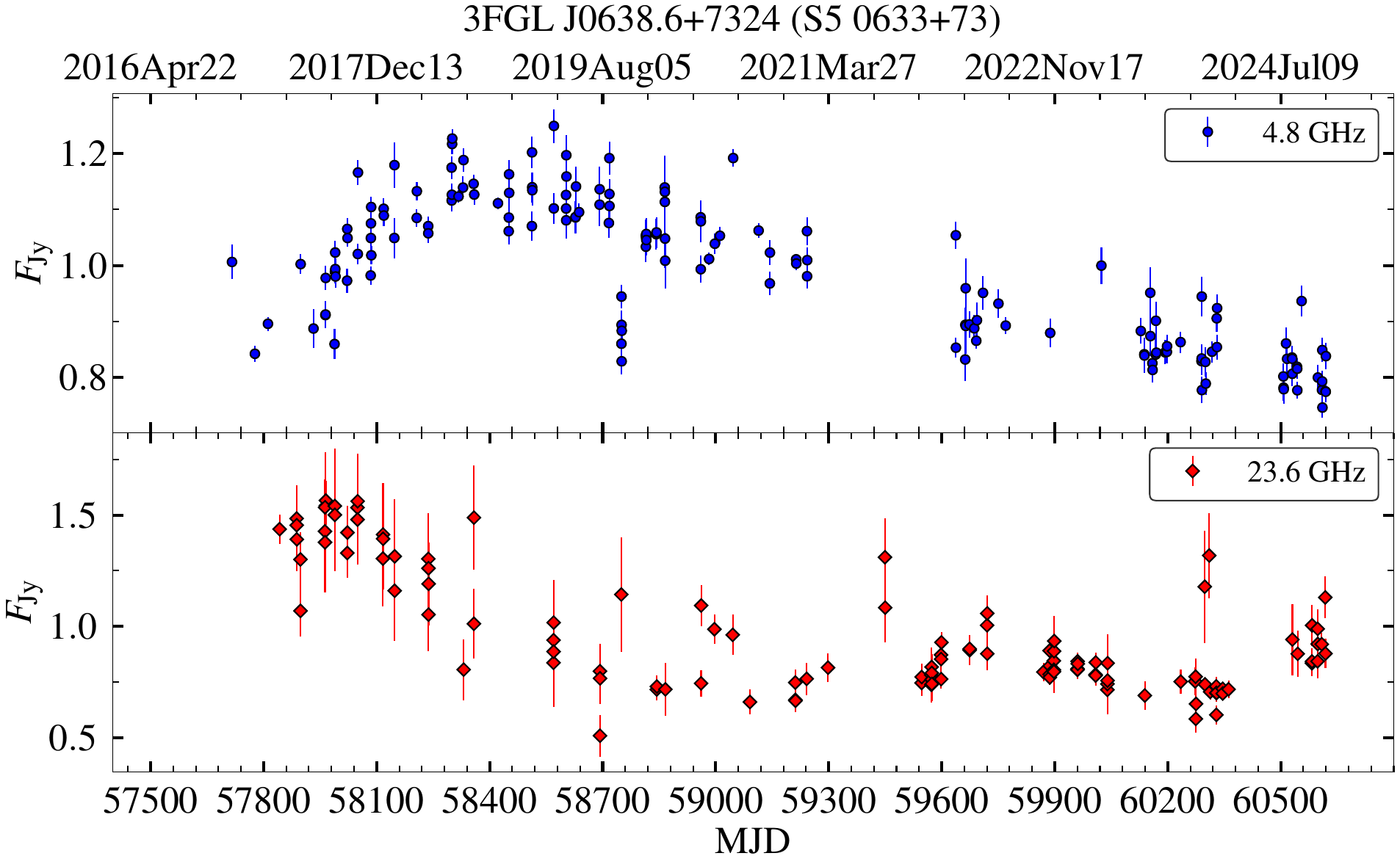}\\
\end{tabular}
\end{figure*}

\begin{figure*}[p]
\centering
\addtocounter{figure}{-1}
\caption{Continued.}
\begin{tabular}{cc}
\includegraphics[width=0.49\textwidth]{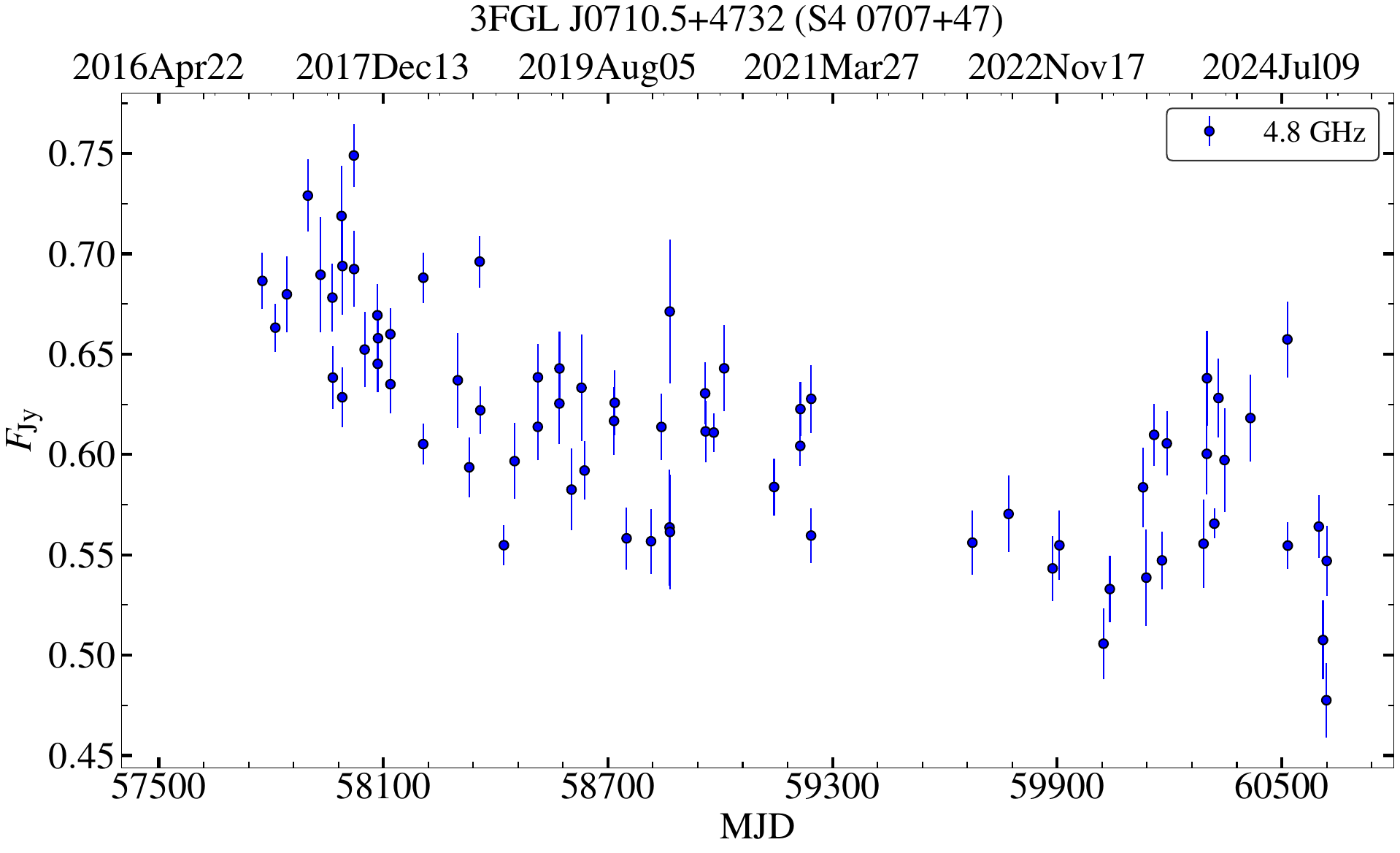}
\includegraphics[width=0.49\textwidth]{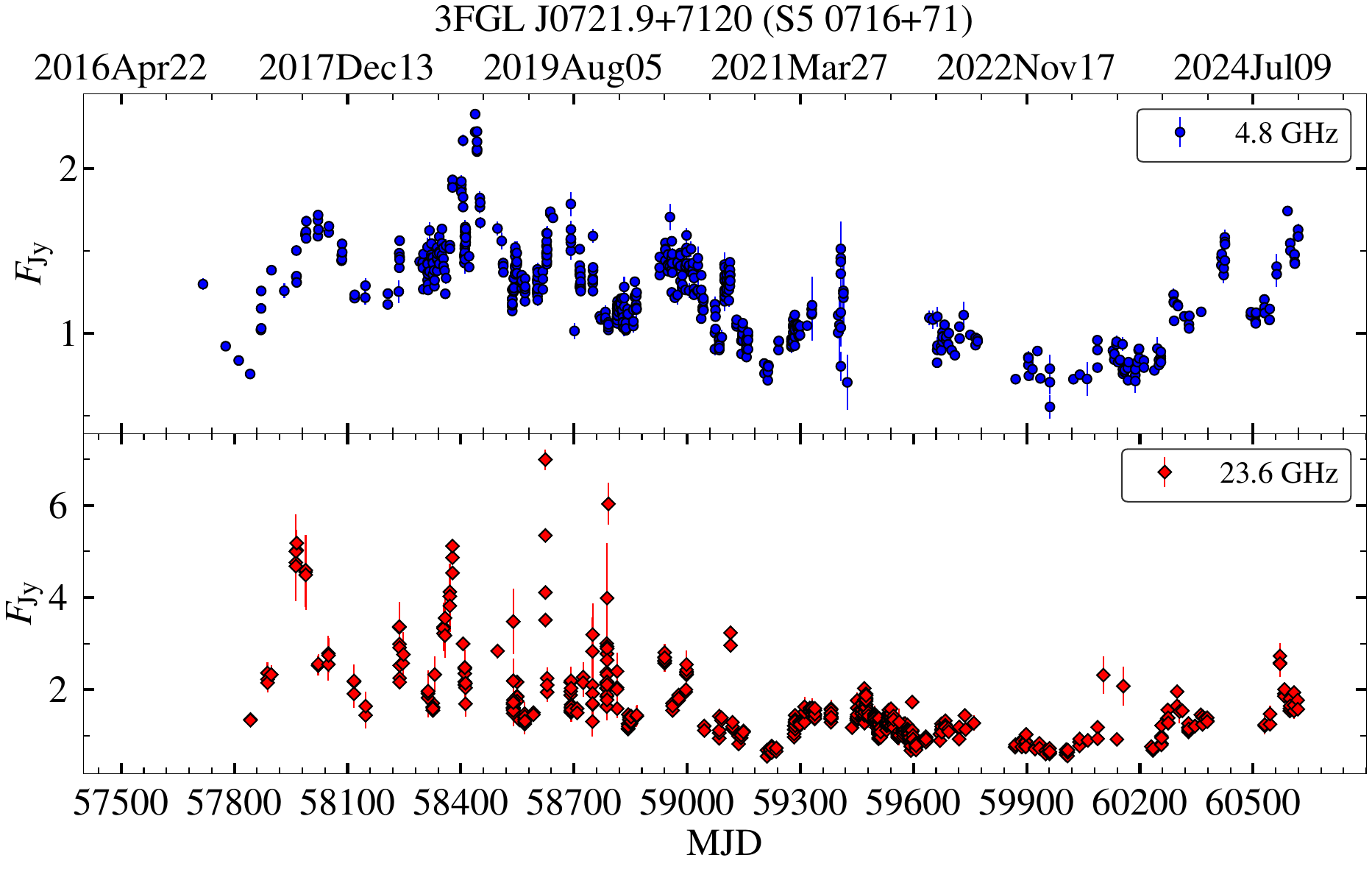}\\
\includegraphics[width=0.49\textwidth]{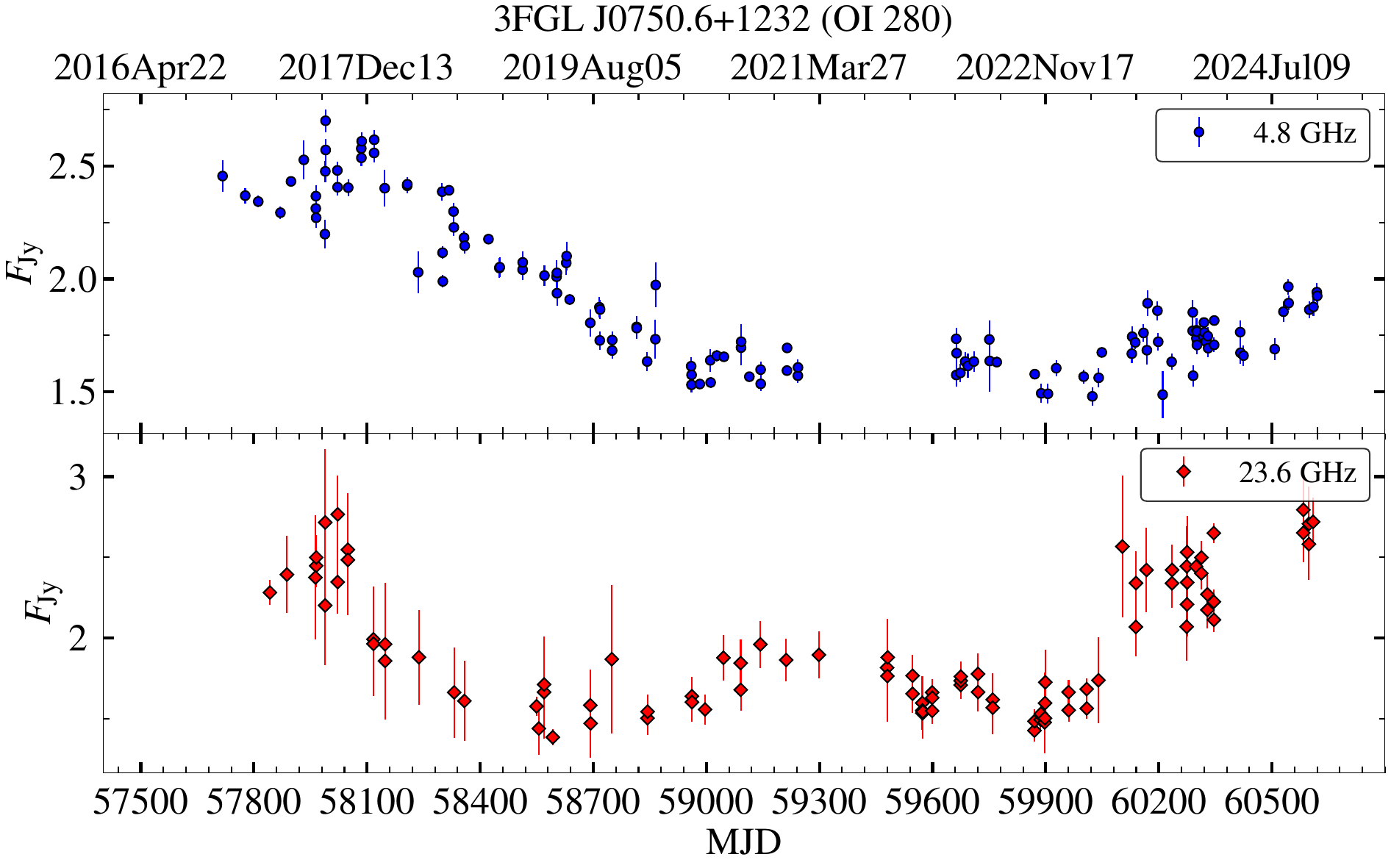}
\includegraphics[width=0.49\textwidth]{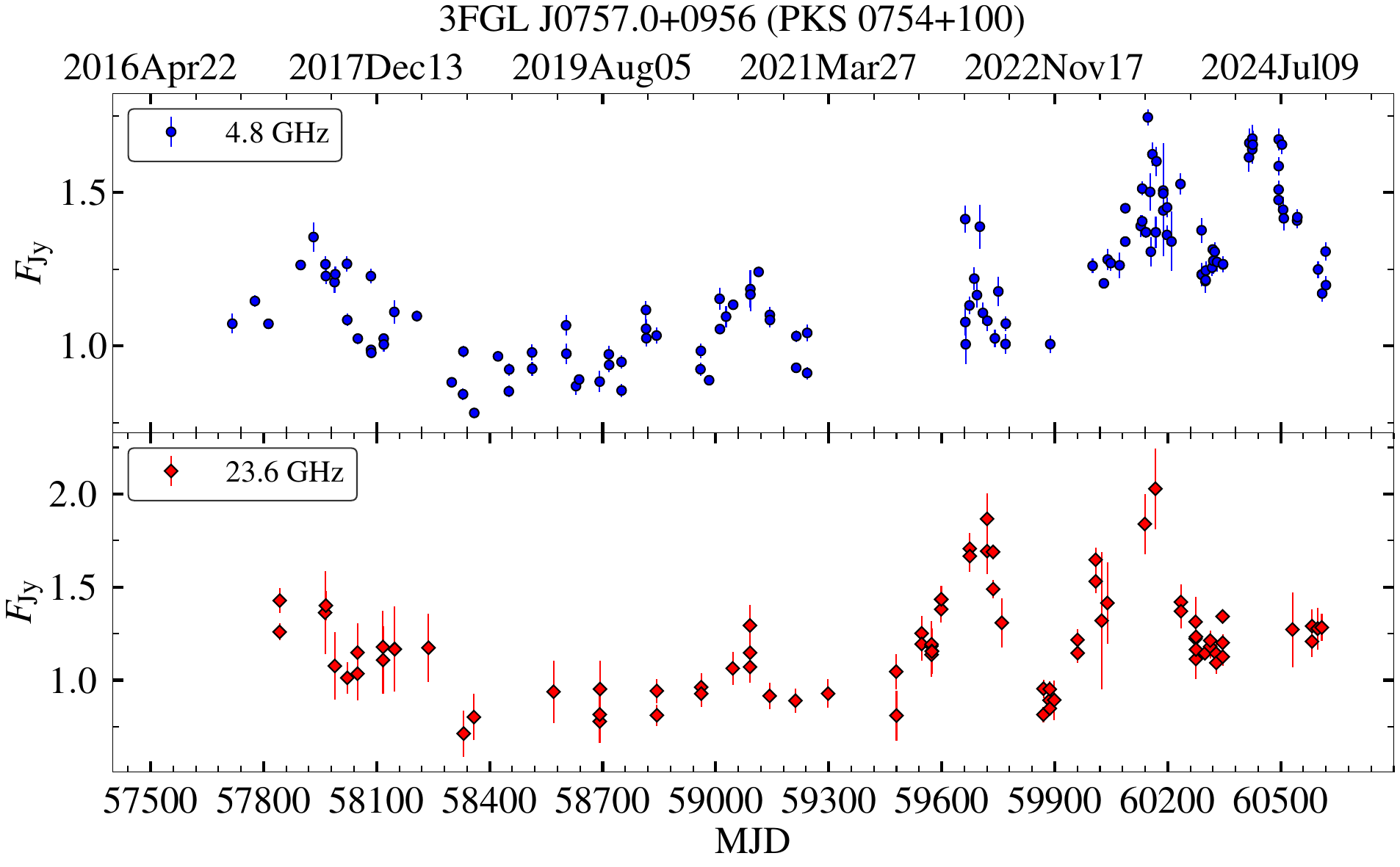}\\
\includegraphics[width=0.49\textwidth]{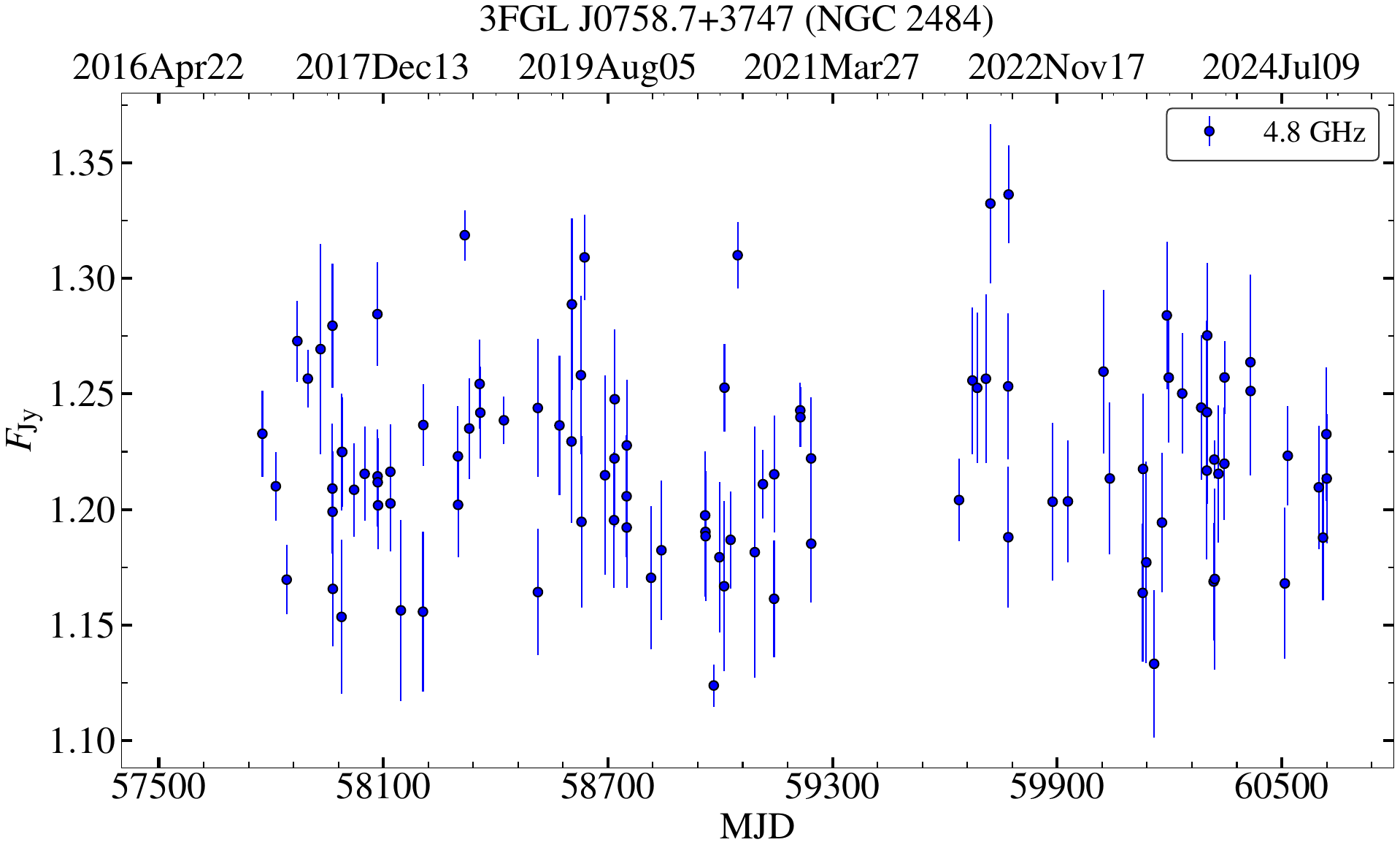}
\includegraphics[width=0.49\textwidth]{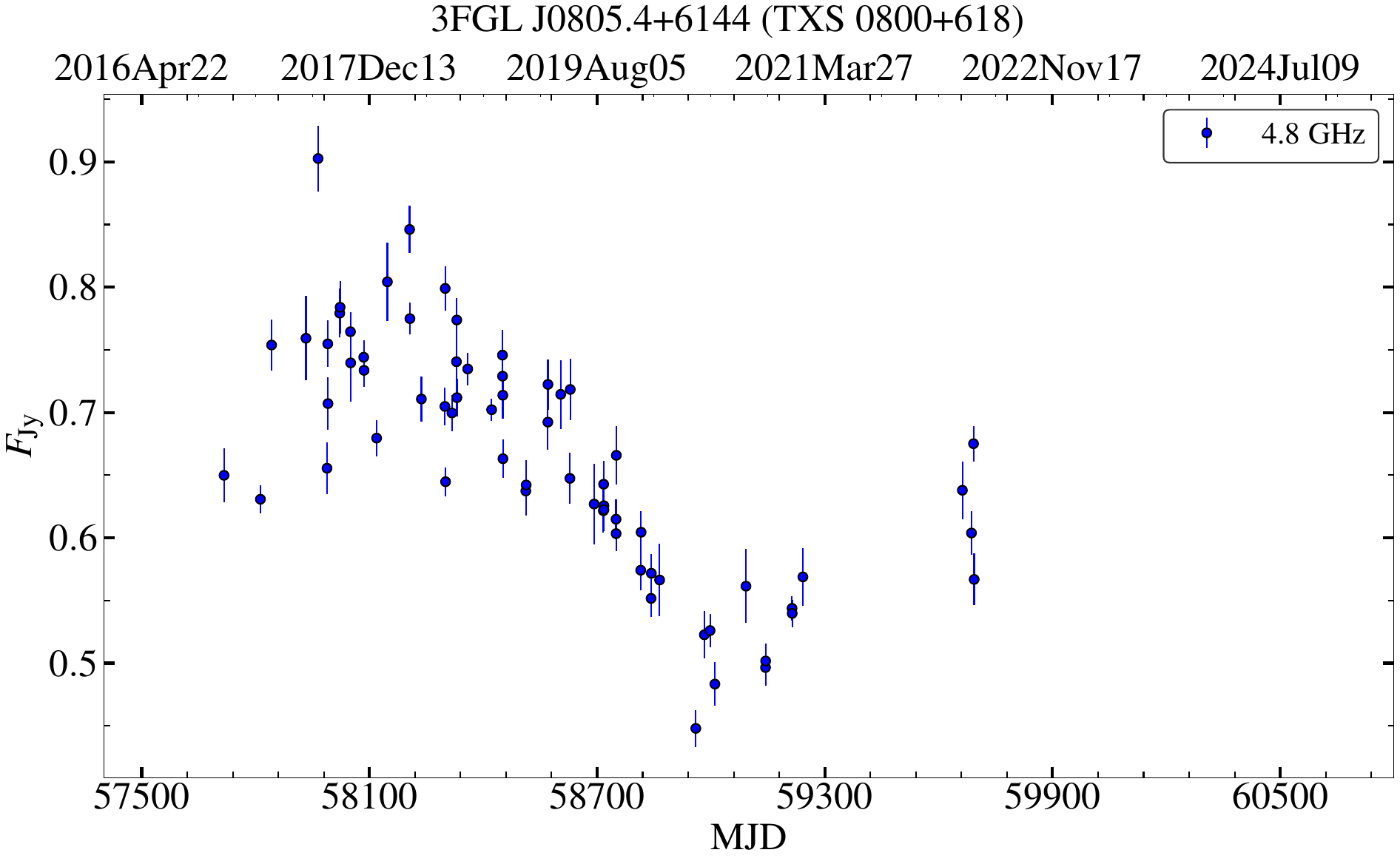}\\
\includegraphics[width=0.49\textwidth]{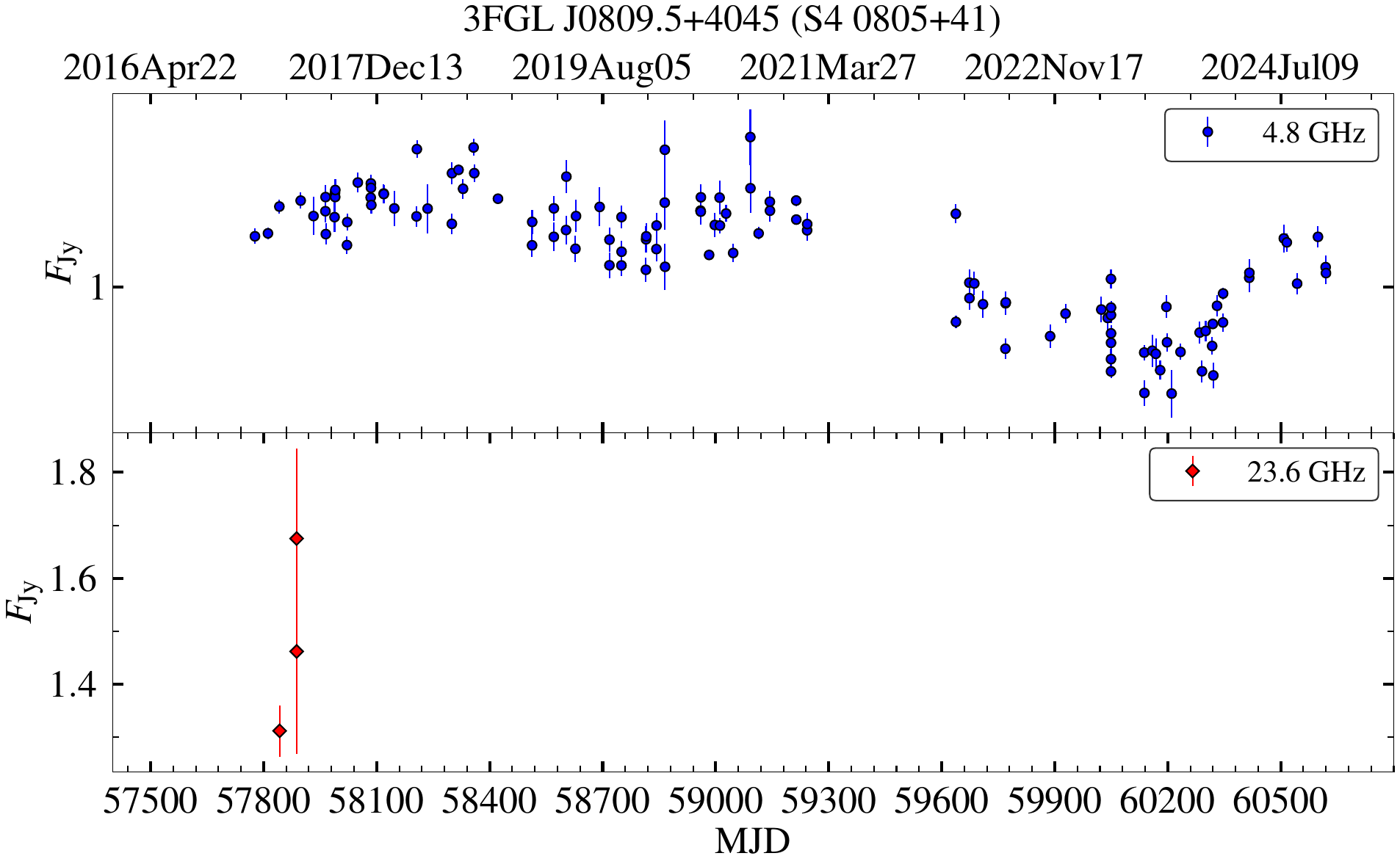}
\includegraphics[width=0.49\textwidth]{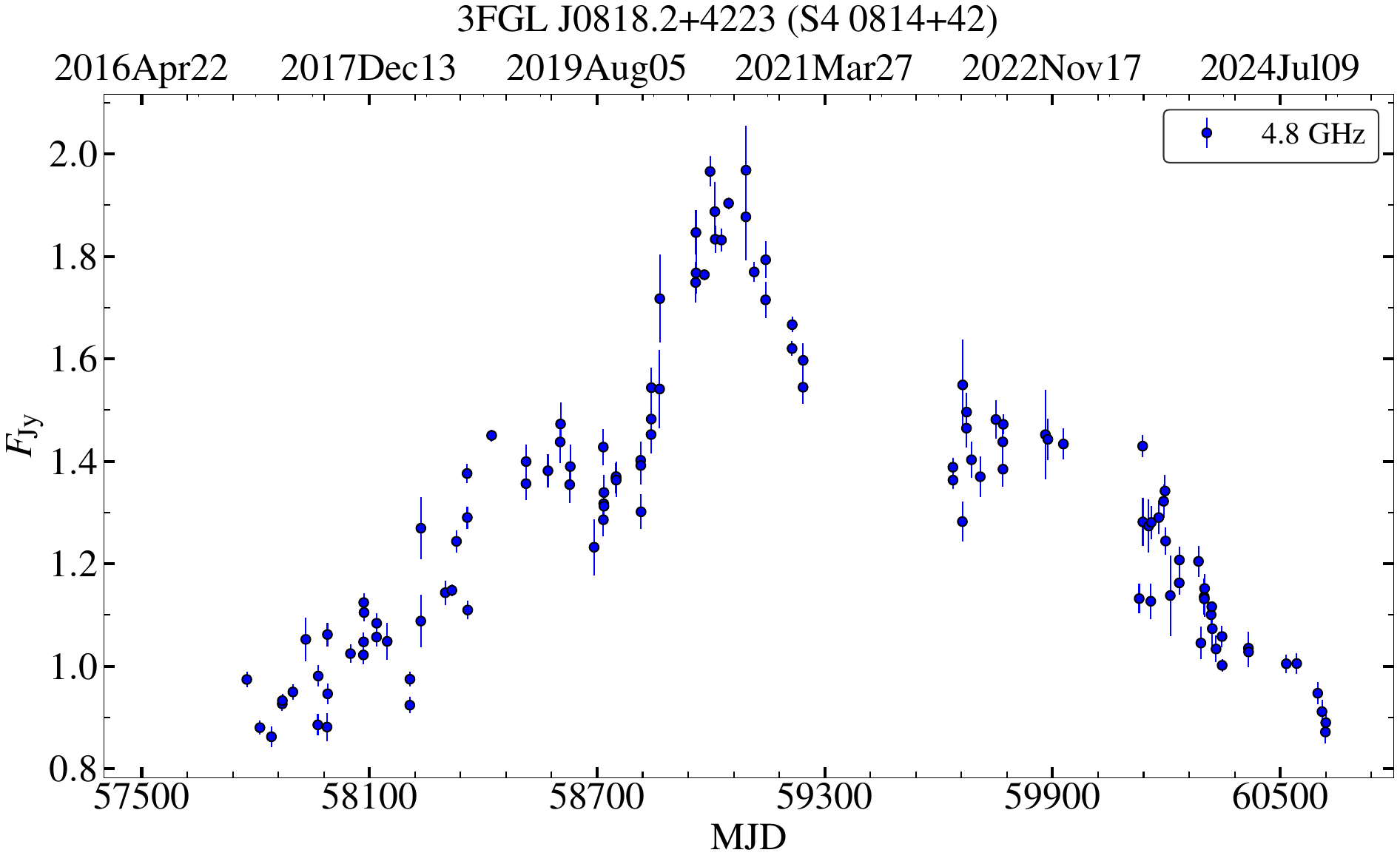}\\
\end{tabular}
\end{figure*}

\begin{figure*}[p]
\centering
\addtocounter{figure}{-1}
\caption{Continued.}
\begin{tabular}{cc}
\includegraphics[width=0.49\textwidth]{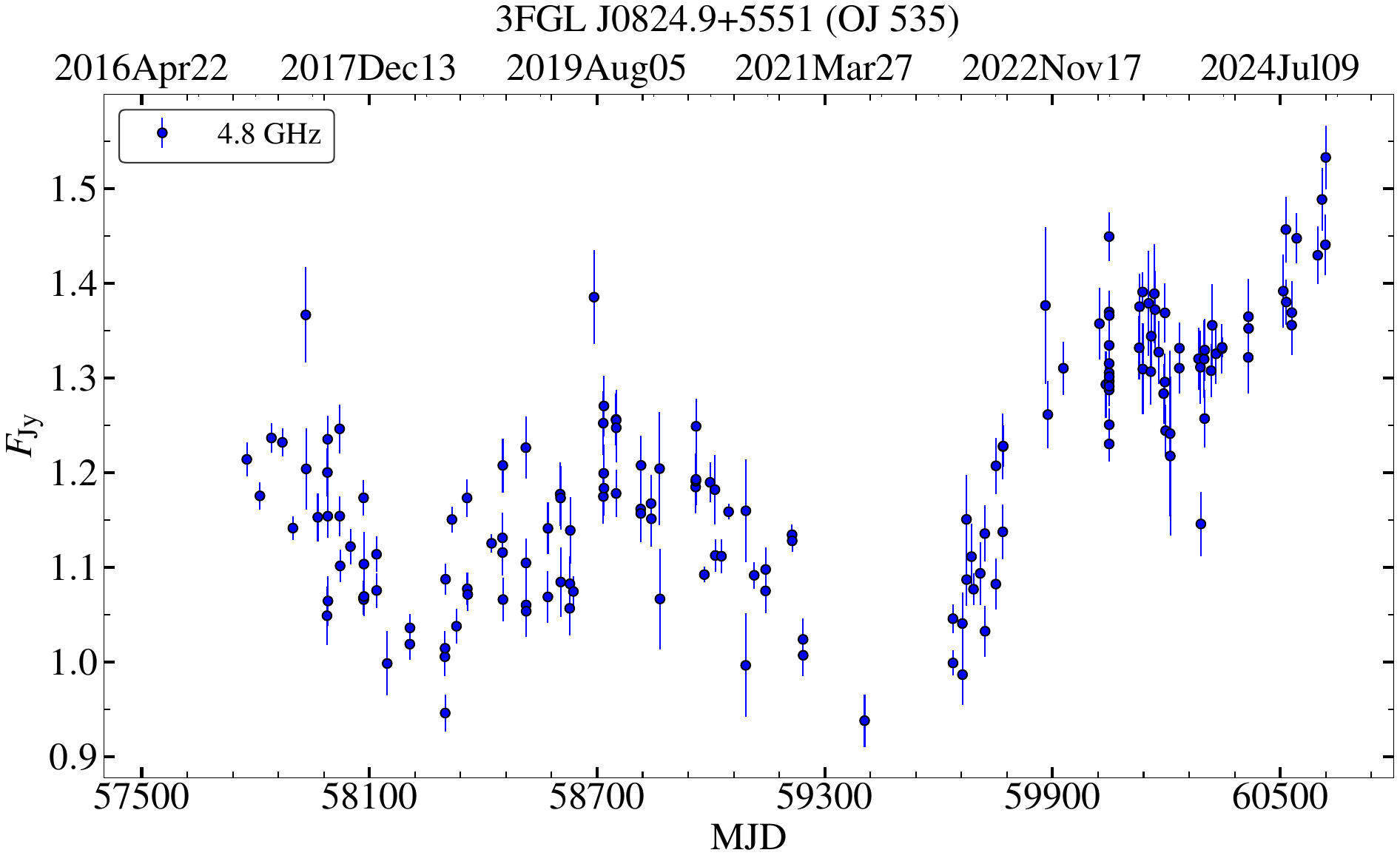}
\includegraphics[width=0.49\textwidth]{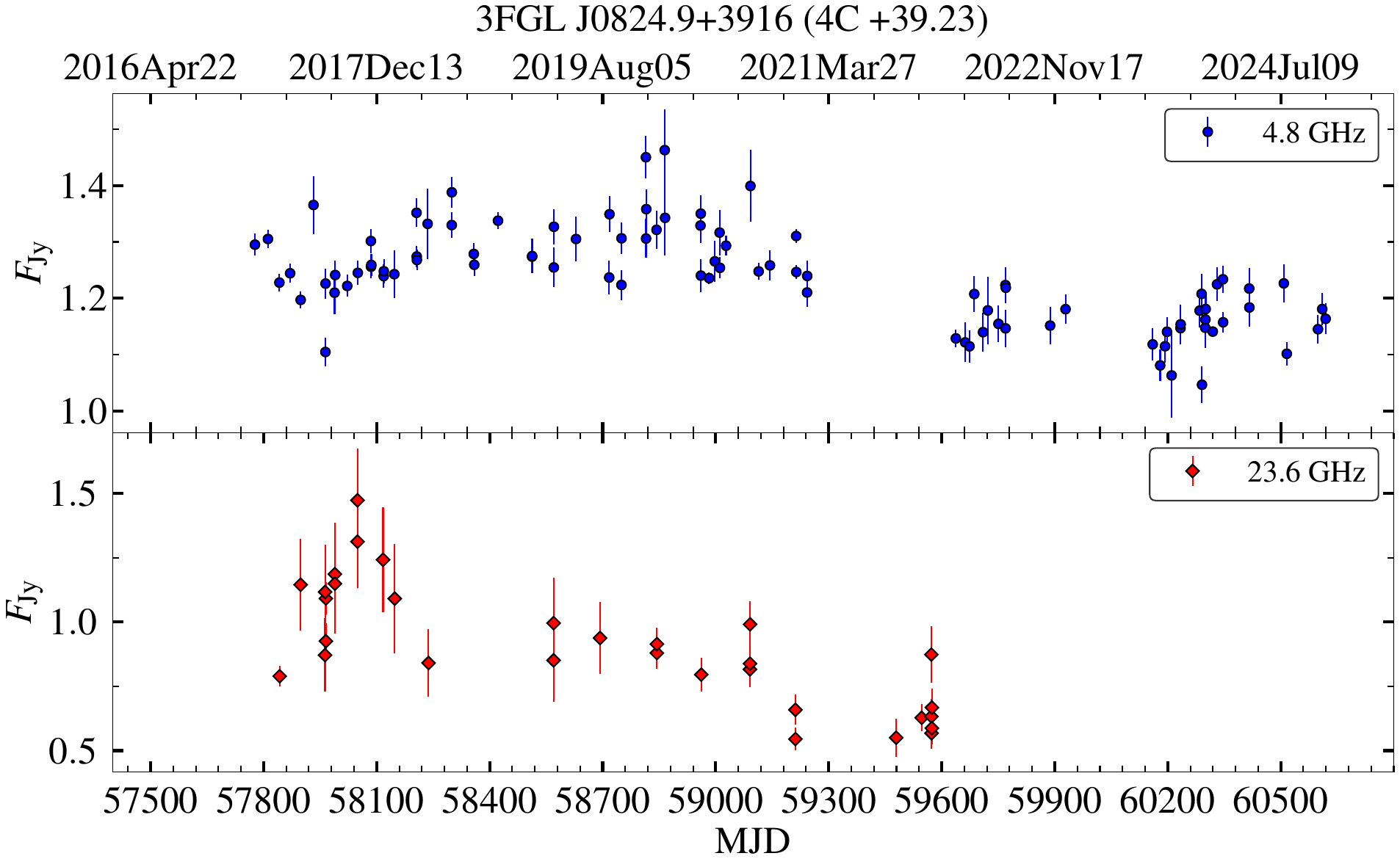}\\
\includegraphics[width=0.49\textwidth]{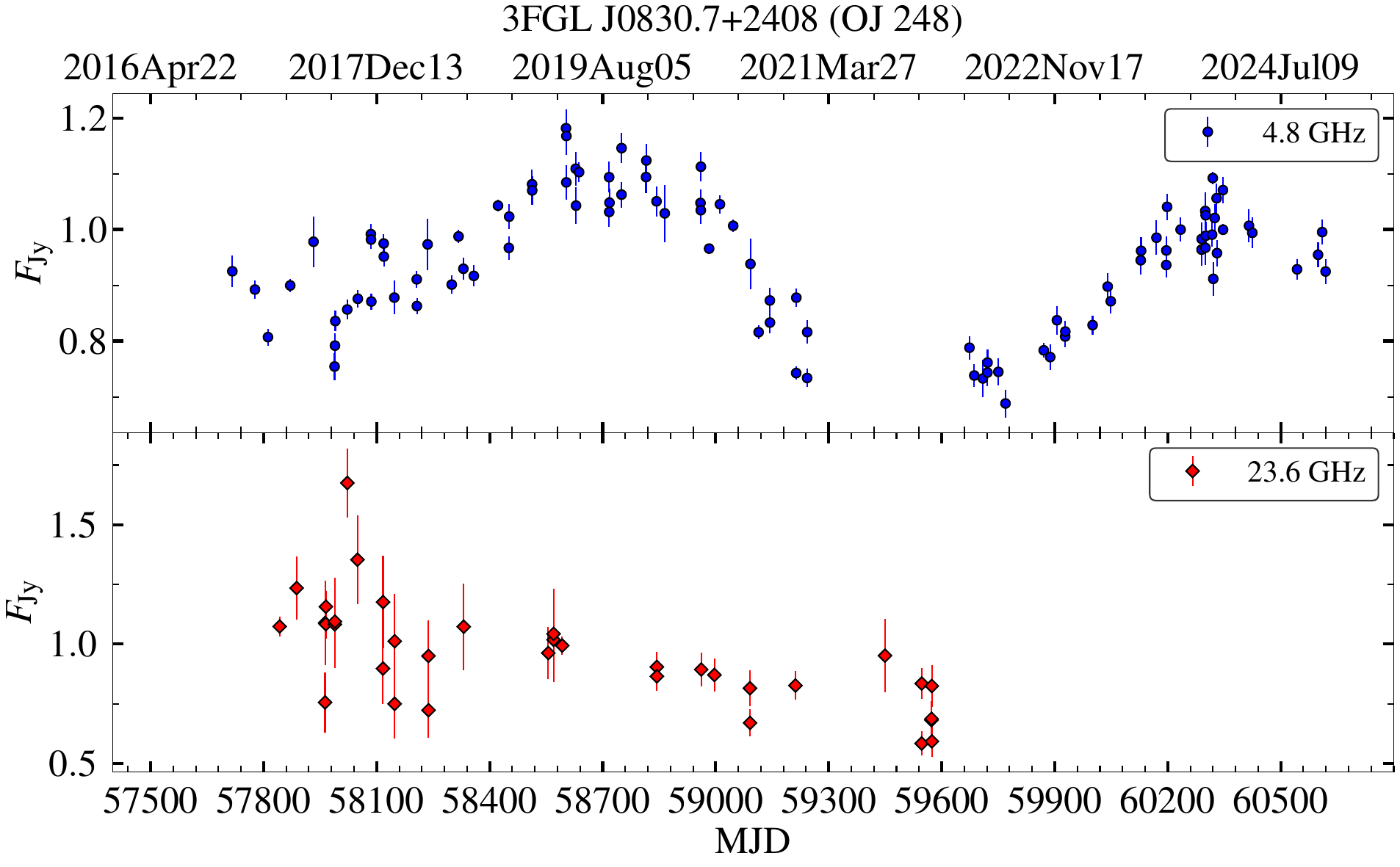}
\includegraphics[width=0.49\textwidth]{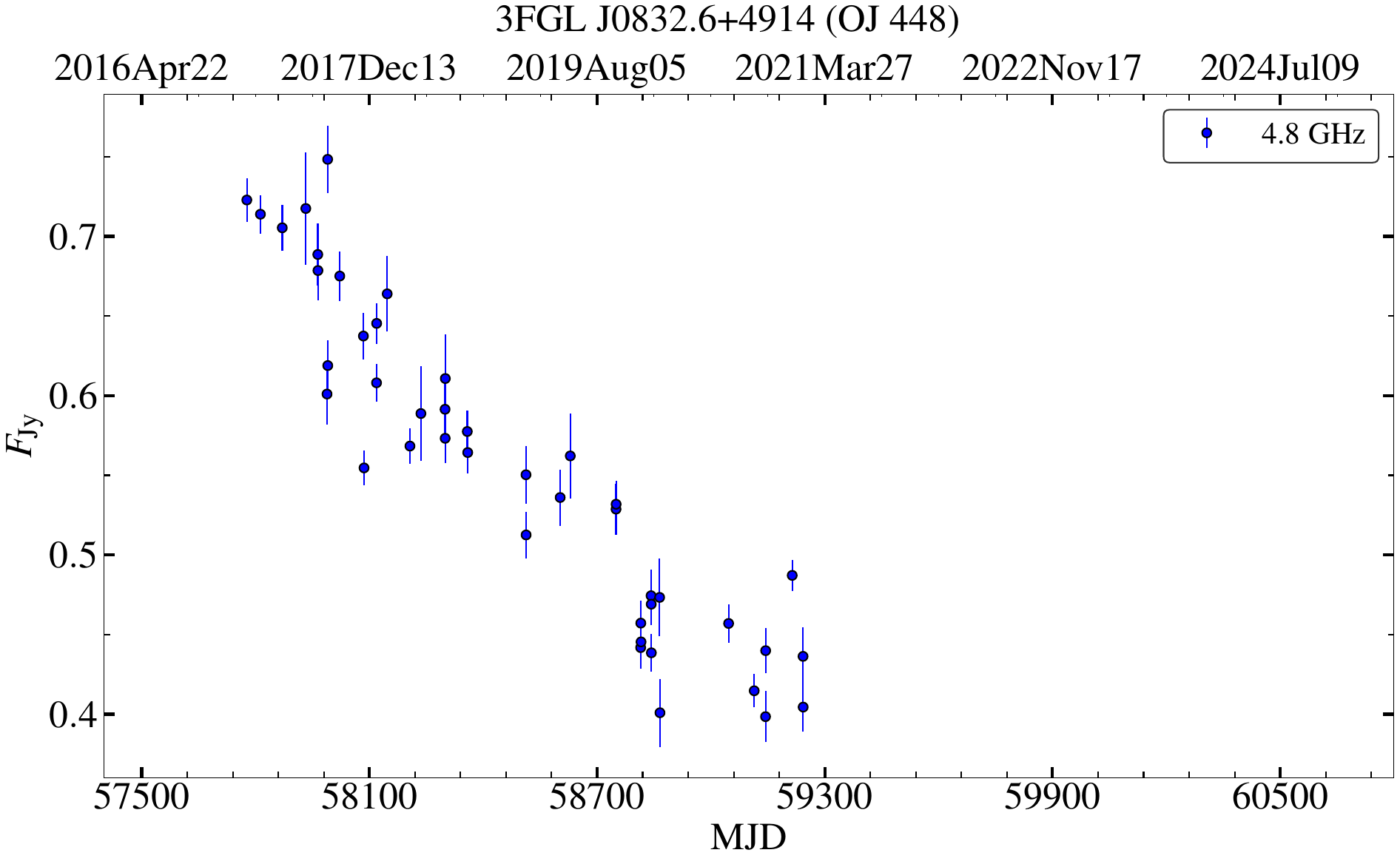}\\
\includegraphics[width=0.49\textwidth]{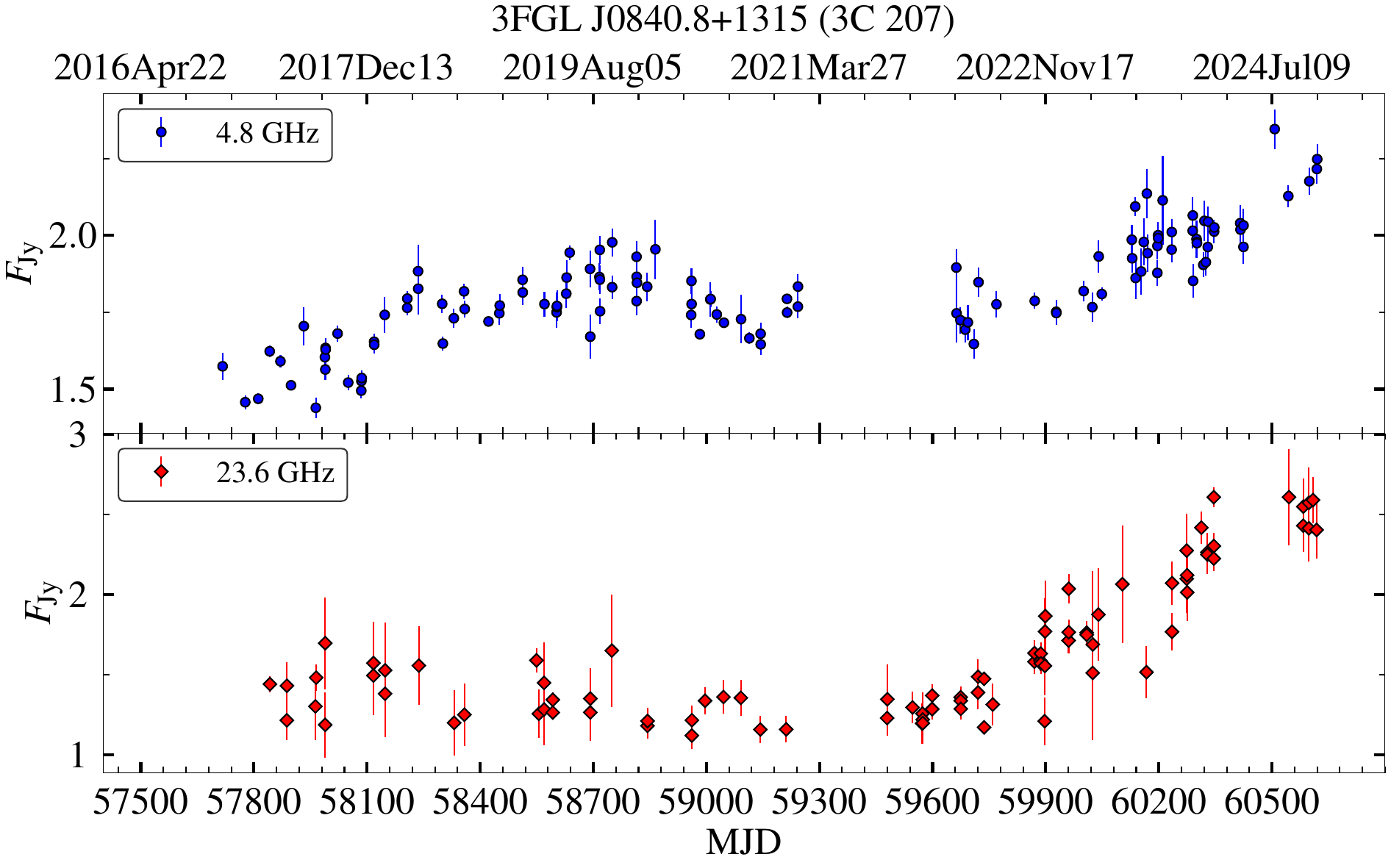}
\includegraphics[width=0.49\textwidth]{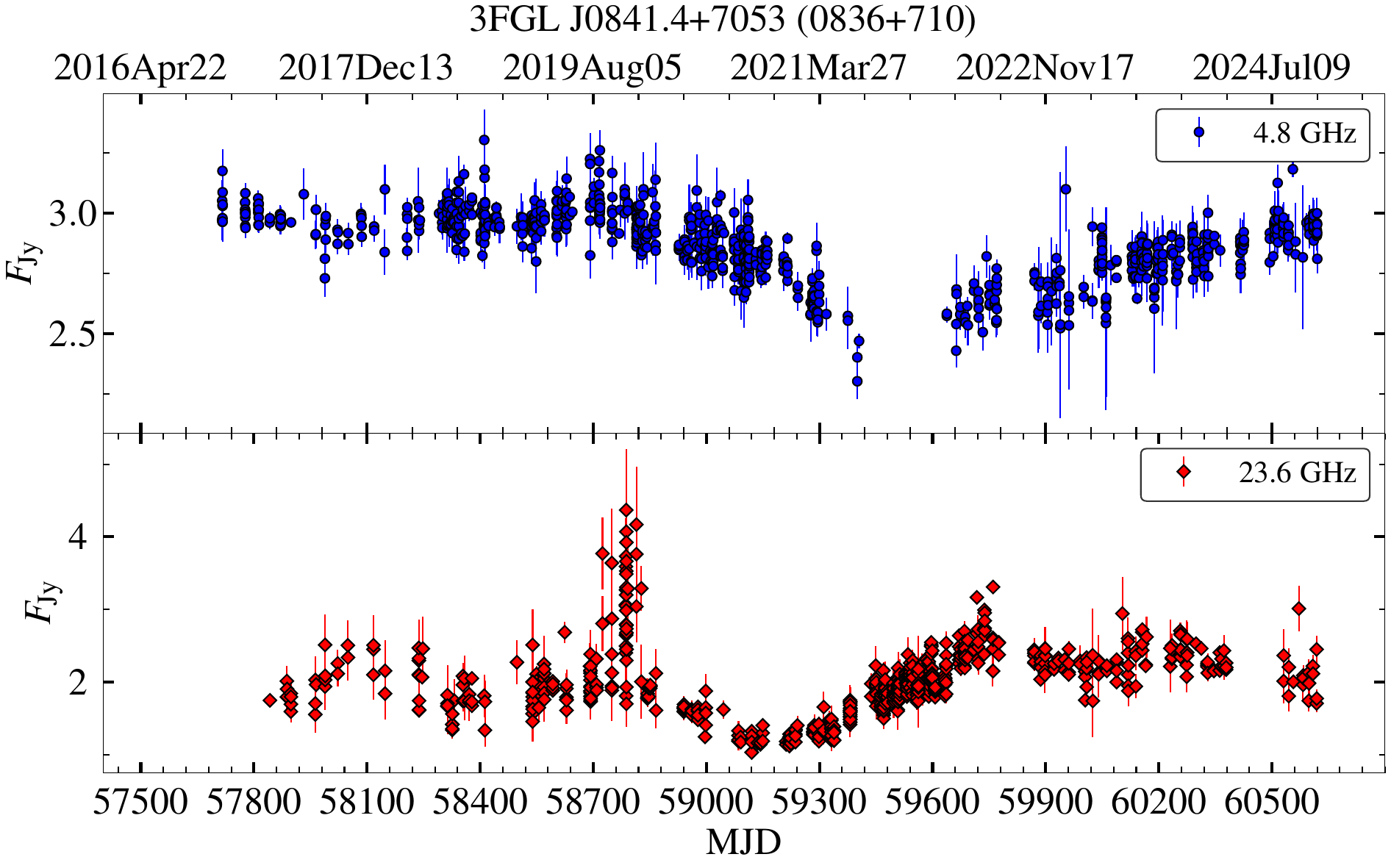}\\
\includegraphics[width=0.49\textwidth]{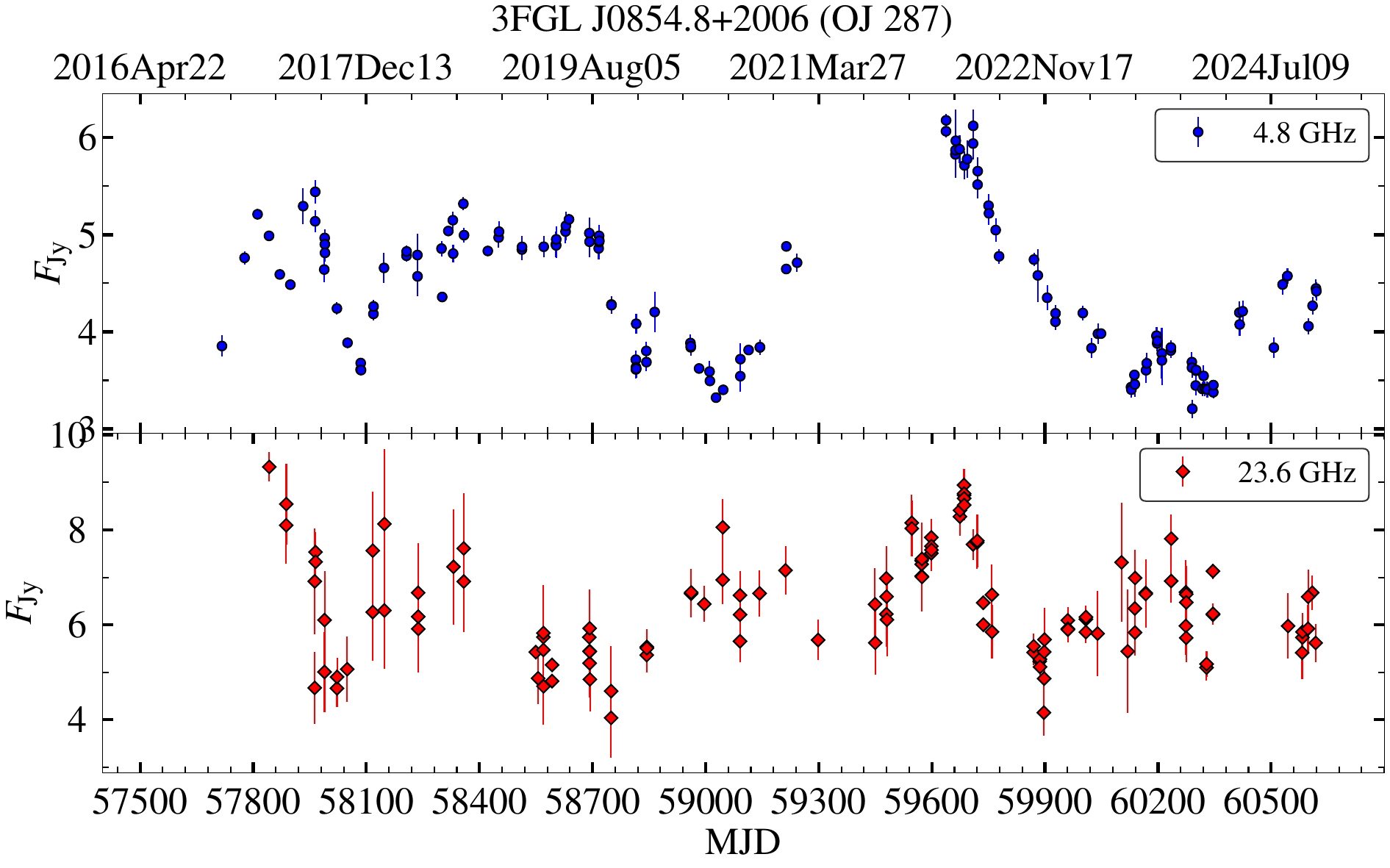}
\includegraphics[width=0.49\textwidth]{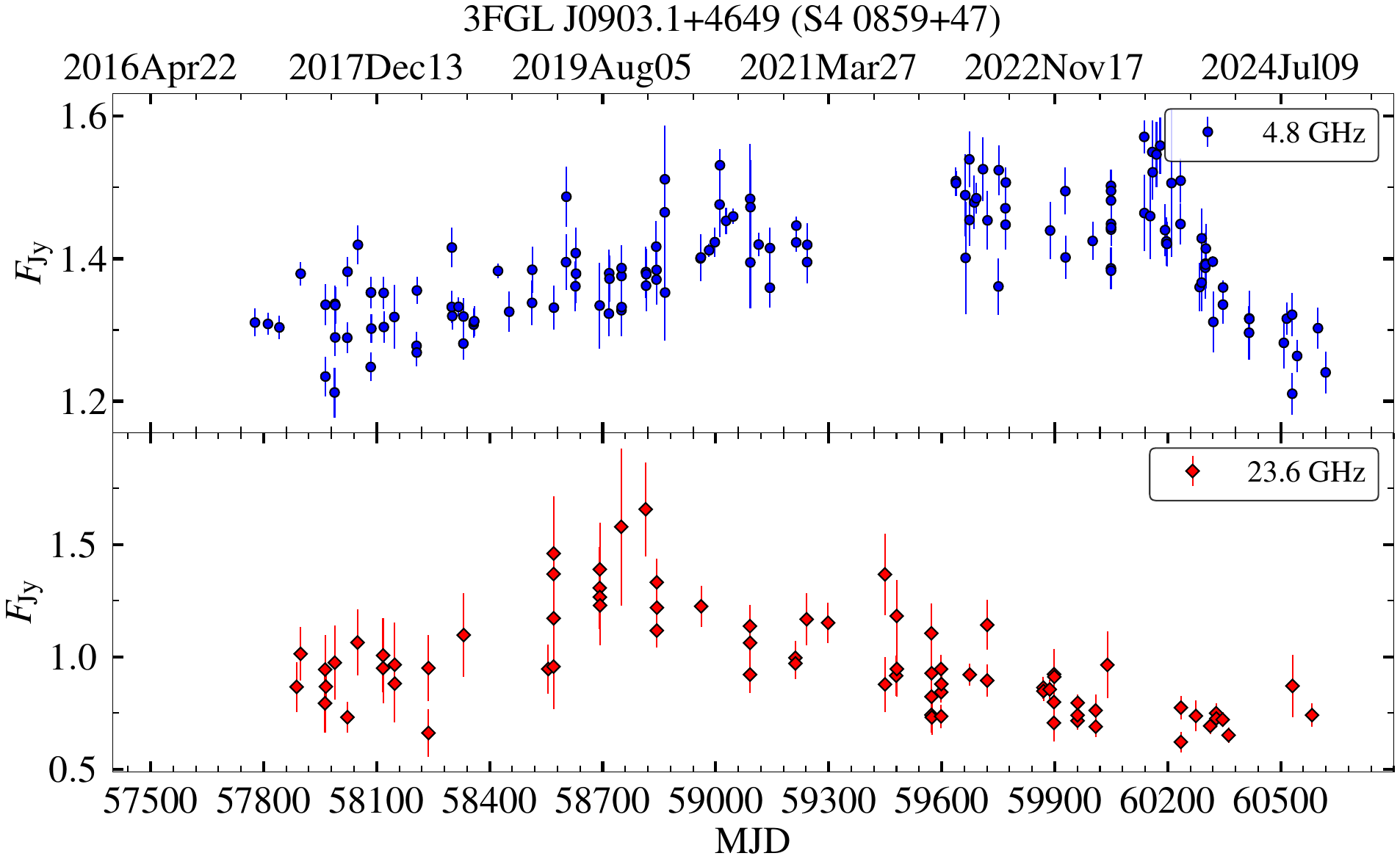}\\
\end{tabular}
\end{figure*}

\begin{figure*}[p]
\centering
\addtocounter{figure}{-1}
\caption{Continued.}
\begin{tabular}{cc}
\includegraphics[width=0.49\textwidth]{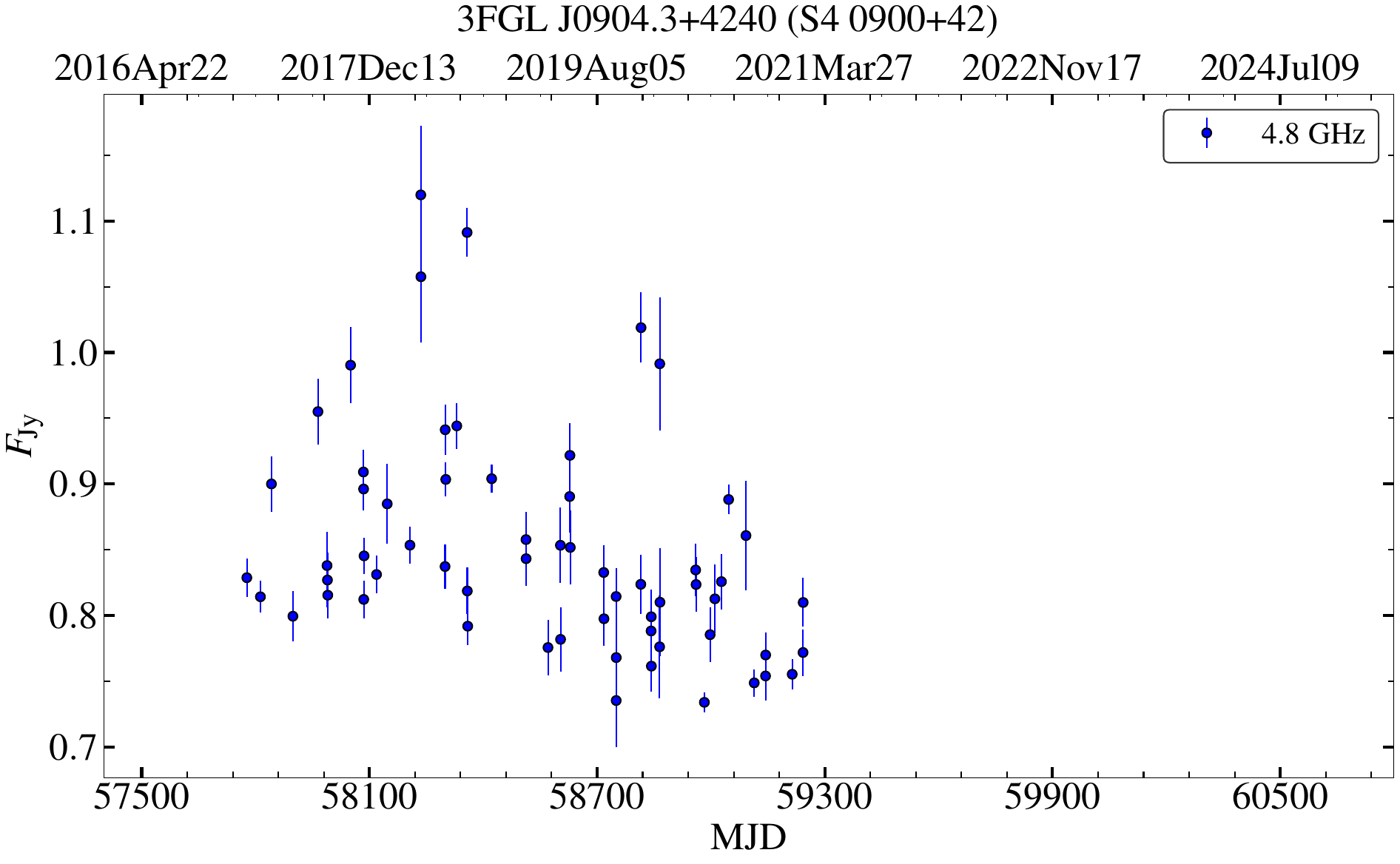}
\includegraphics[width=0.49\textwidth]{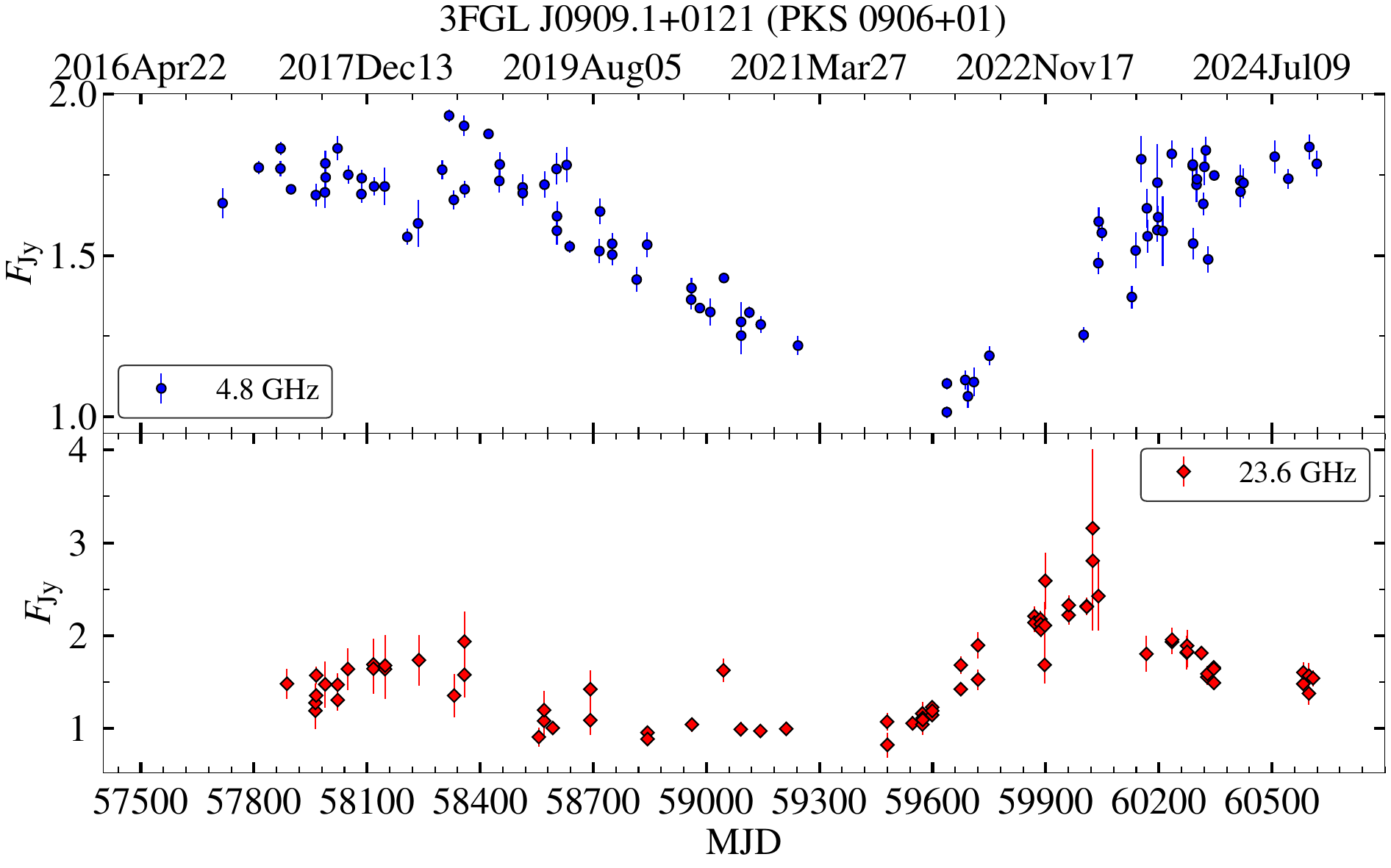}\\
\includegraphics[width=0.49\textwidth]{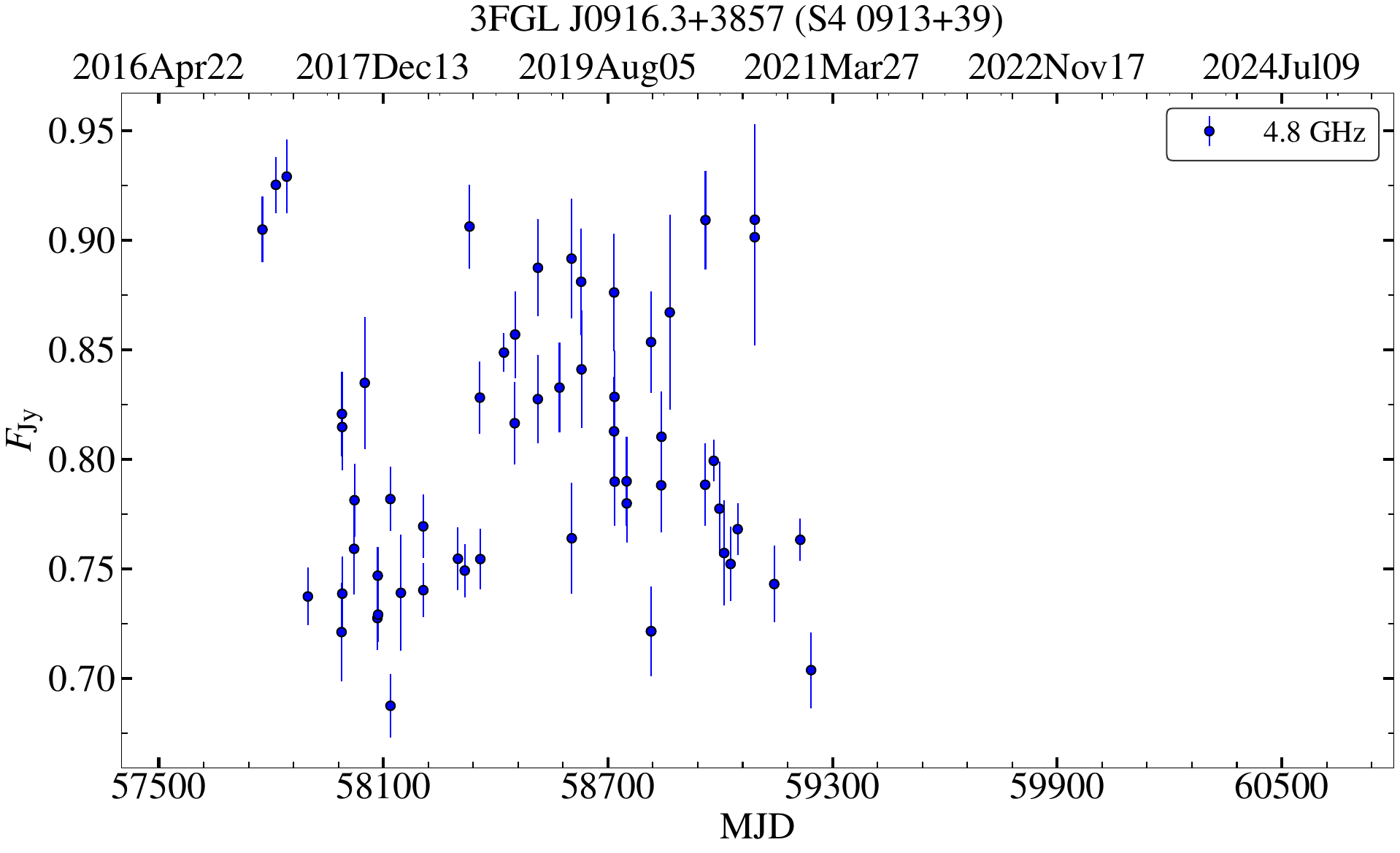}
\includegraphics[width=0.49\textwidth]{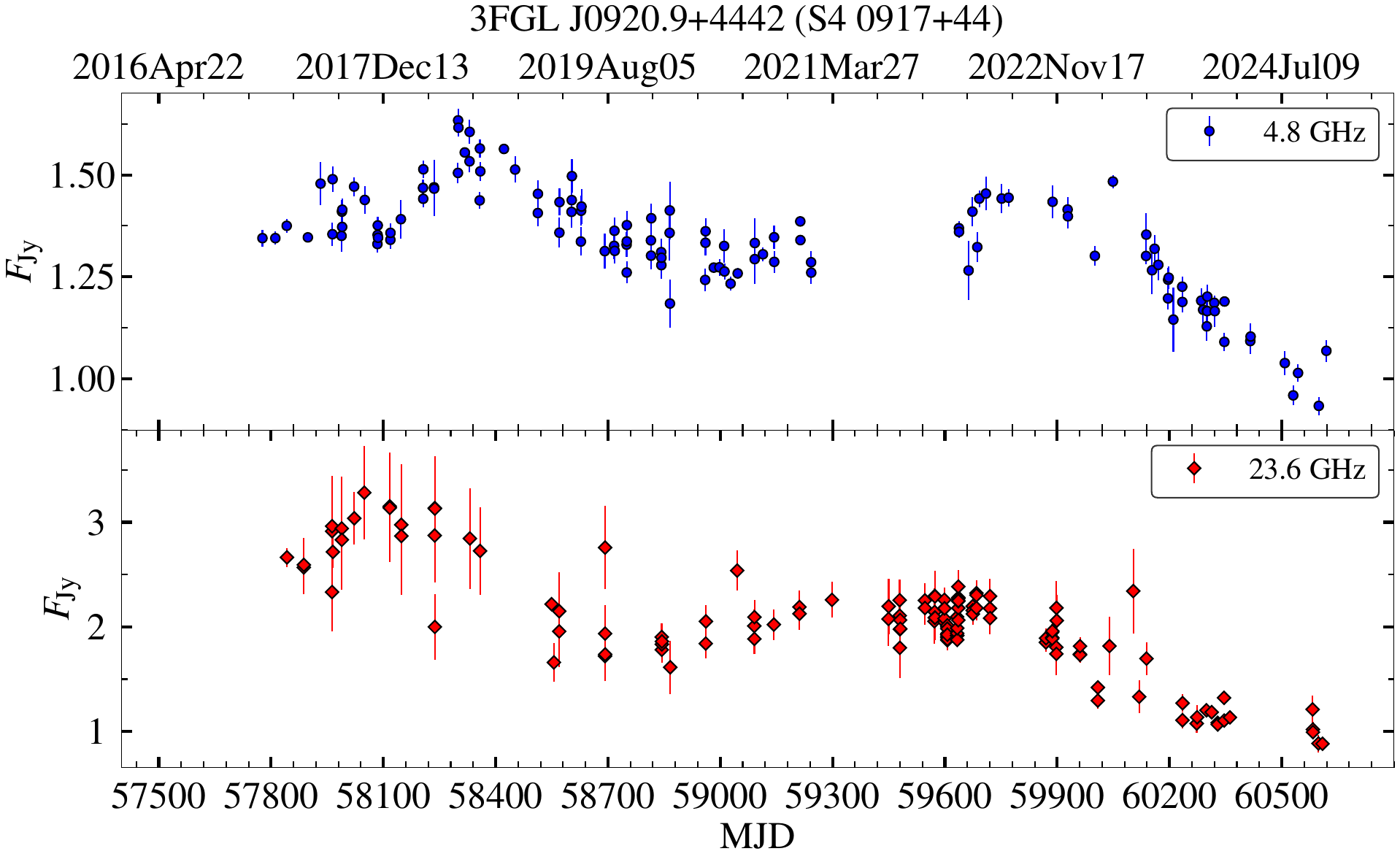}\\
\includegraphics[width=0.49\textwidth]{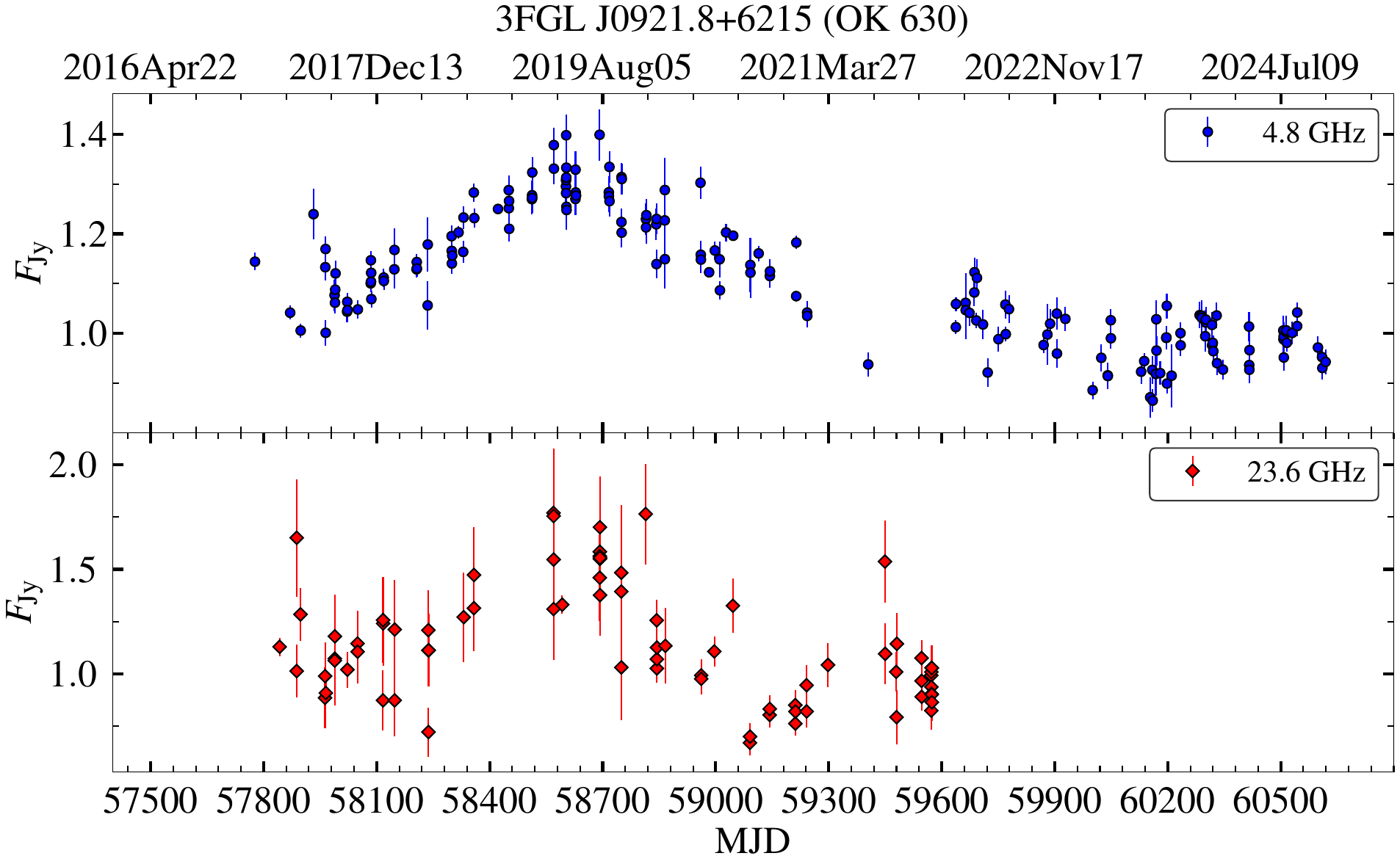}
\includegraphics[width=0.49\textwidth]{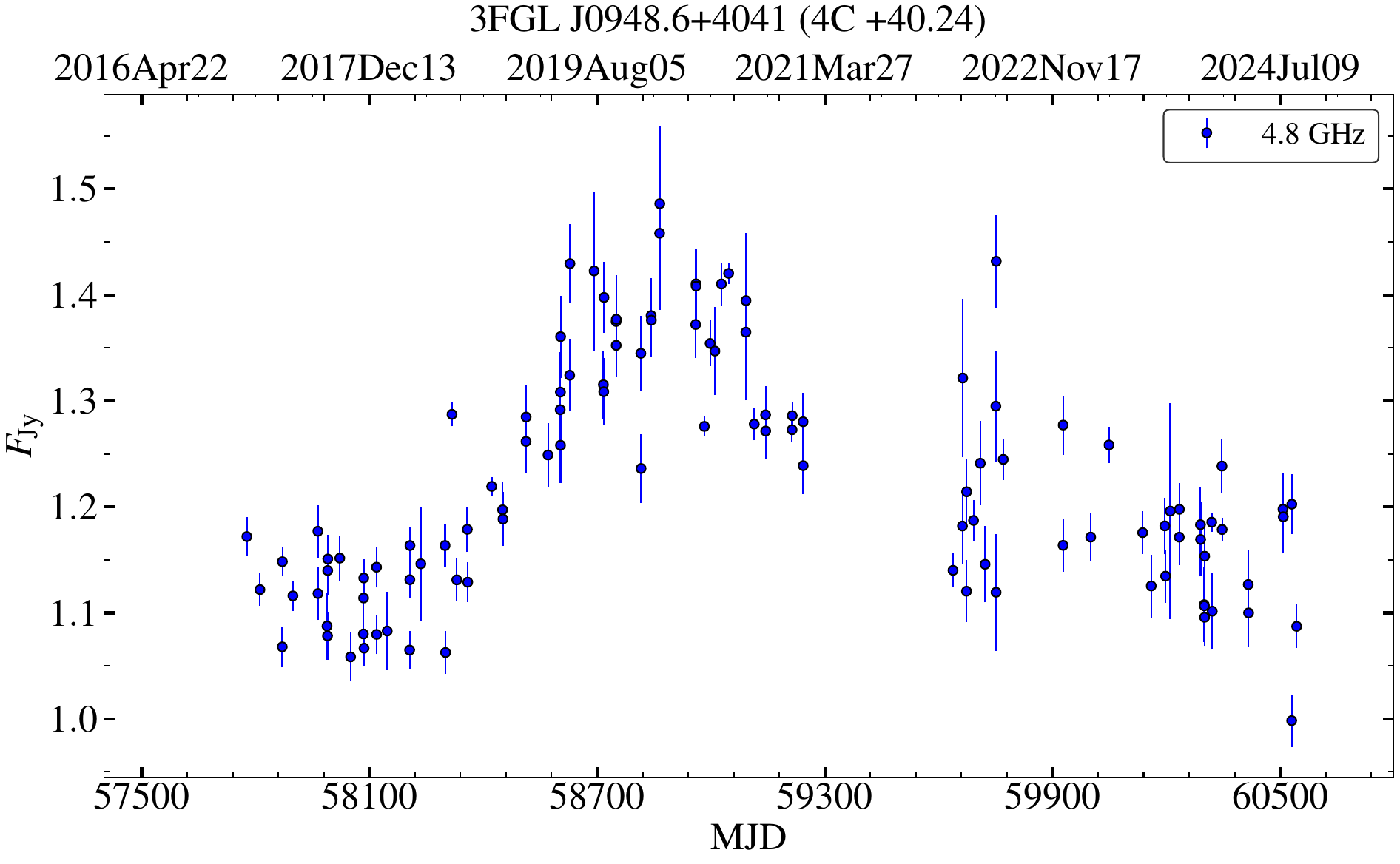}\\
\includegraphics[width=0.49\textwidth]{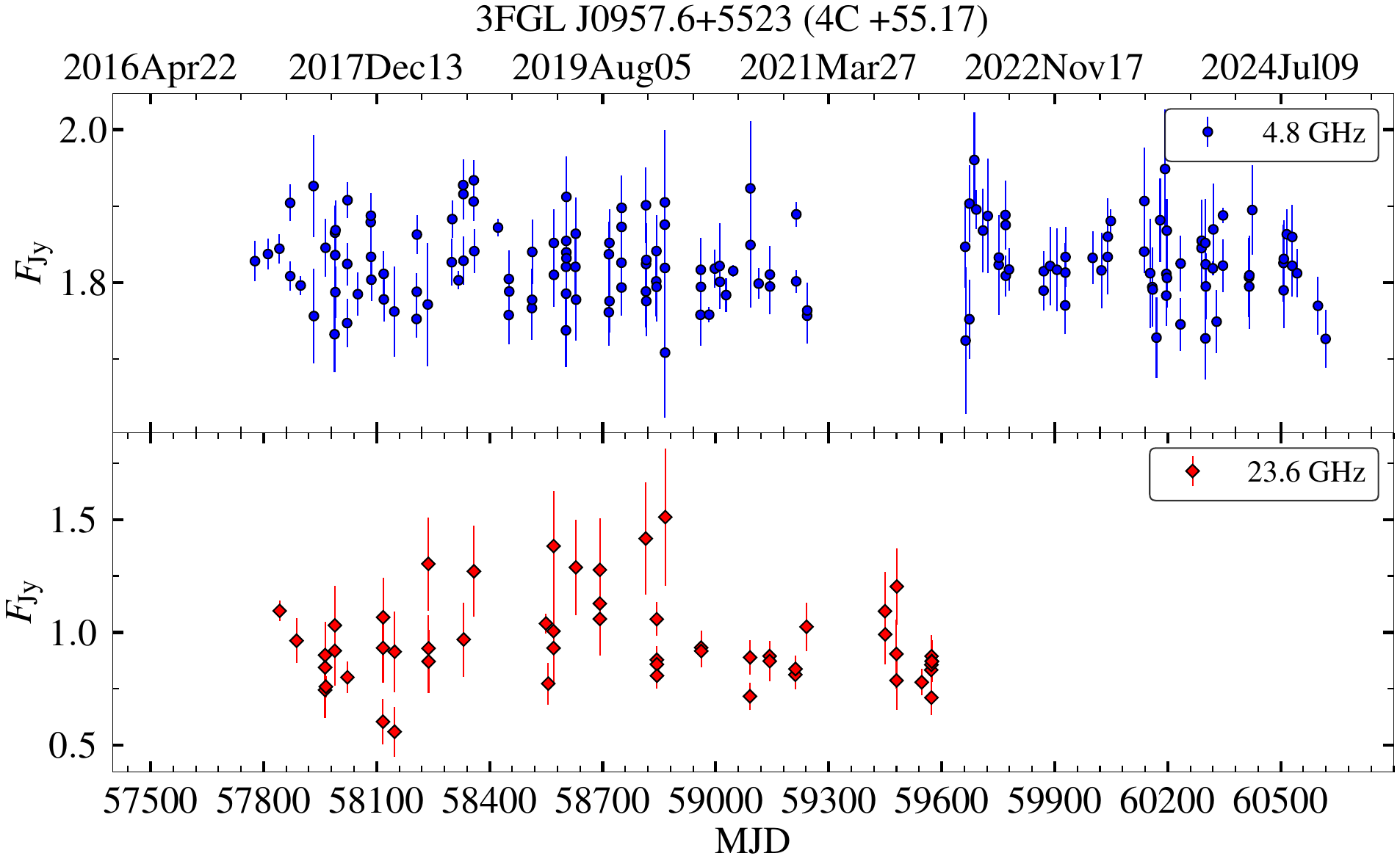}
\includegraphics[width=0.49\textwidth]{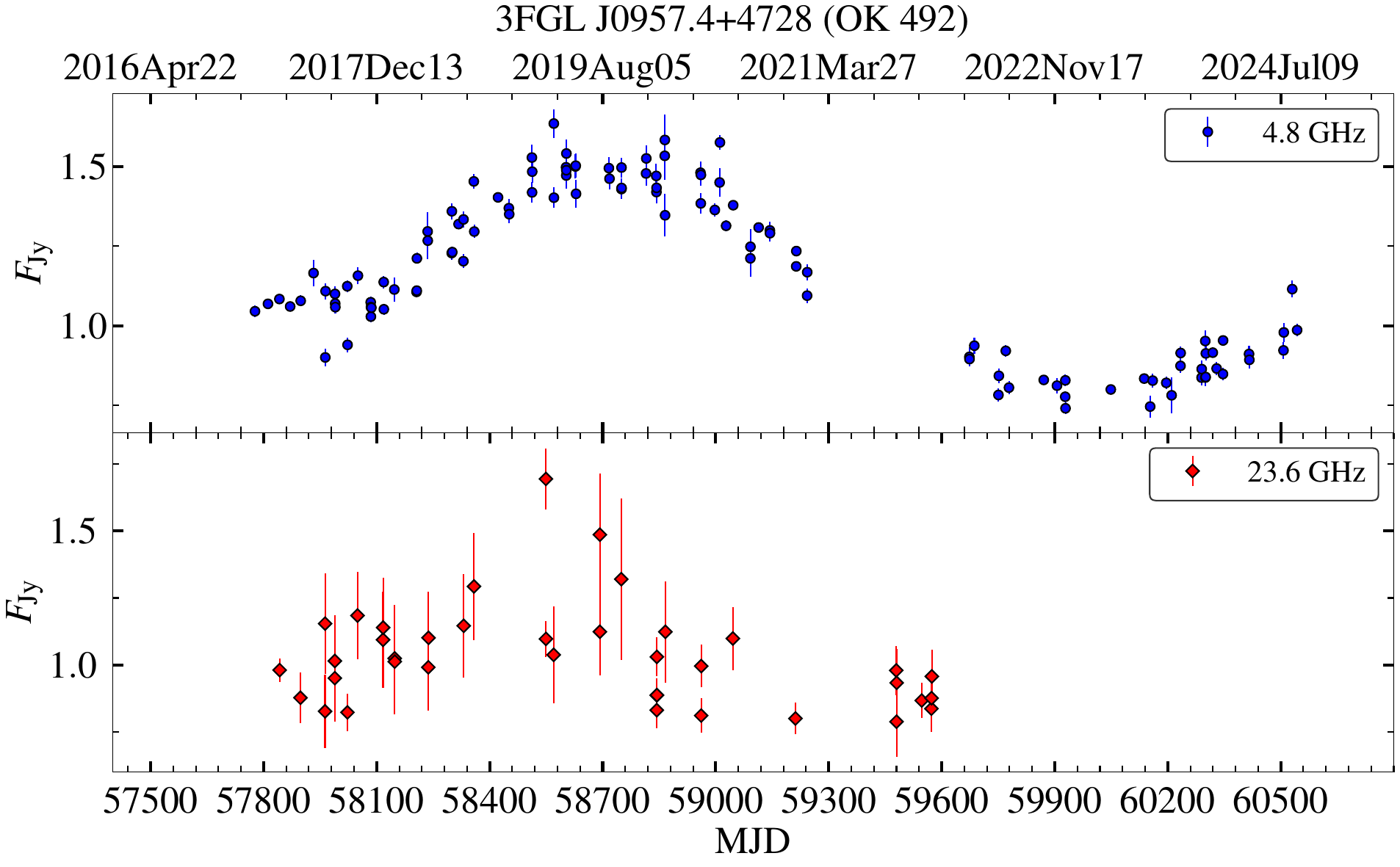}\\
\end{tabular}
\end{figure*}

\begin{figure*}[p]
\centering
\addtocounter{figure}{-1}
\caption{Continued.}
\begin{tabular}{cc}
\includegraphics[width=0.49\textwidth]{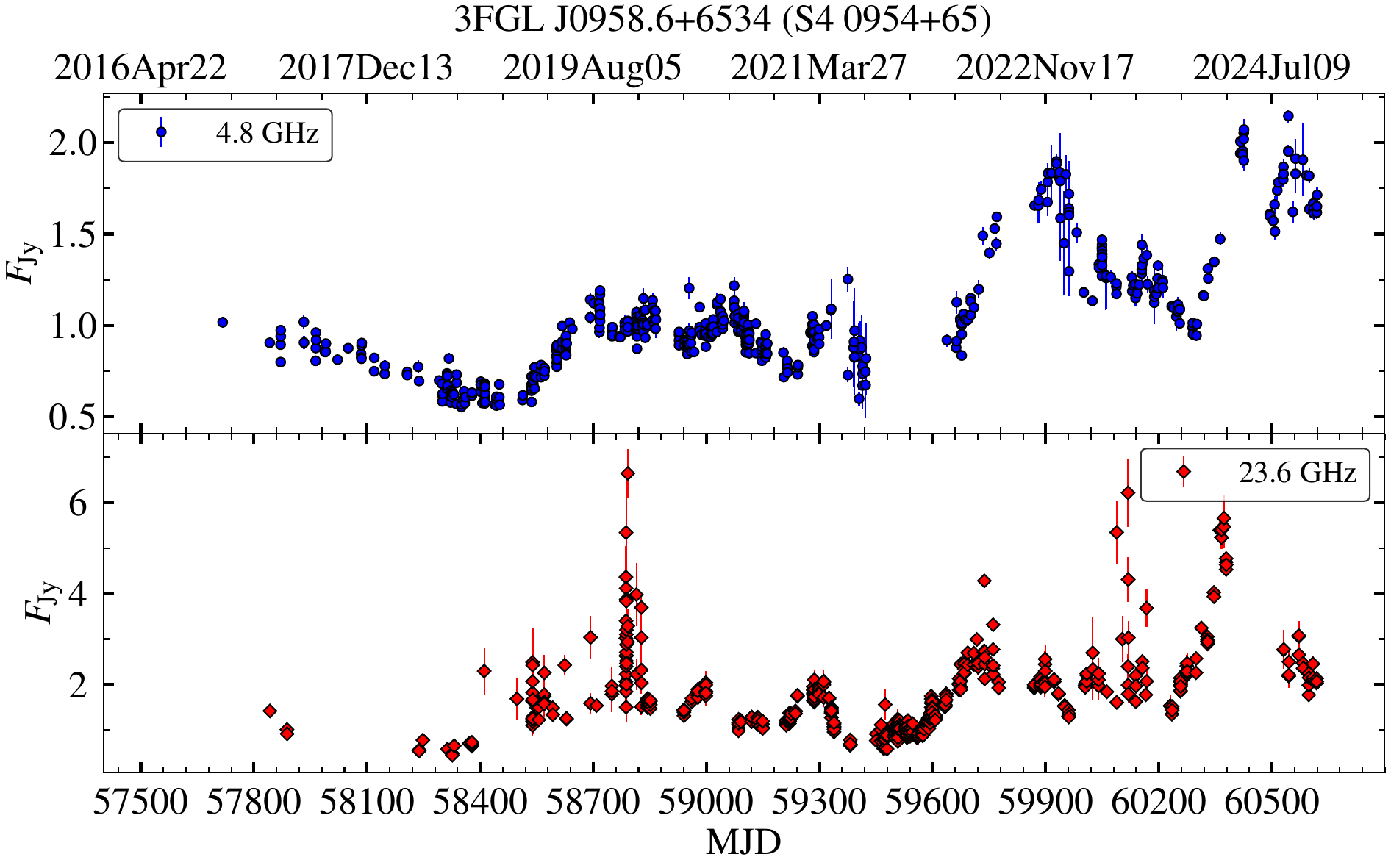}
\includegraphics[width=0.49\textwidth]{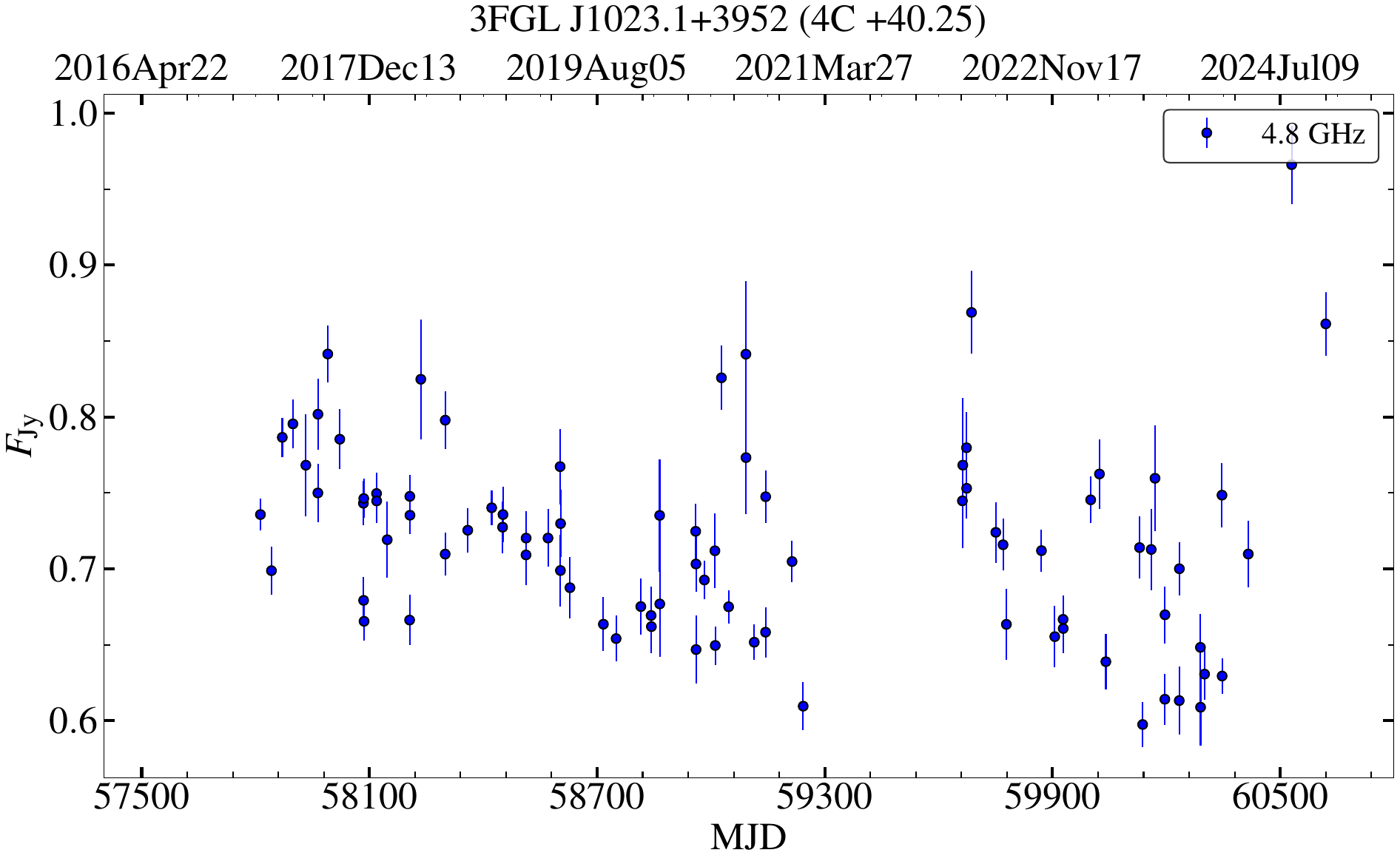}\\
\includegraphics[width=0.49\textwidth]{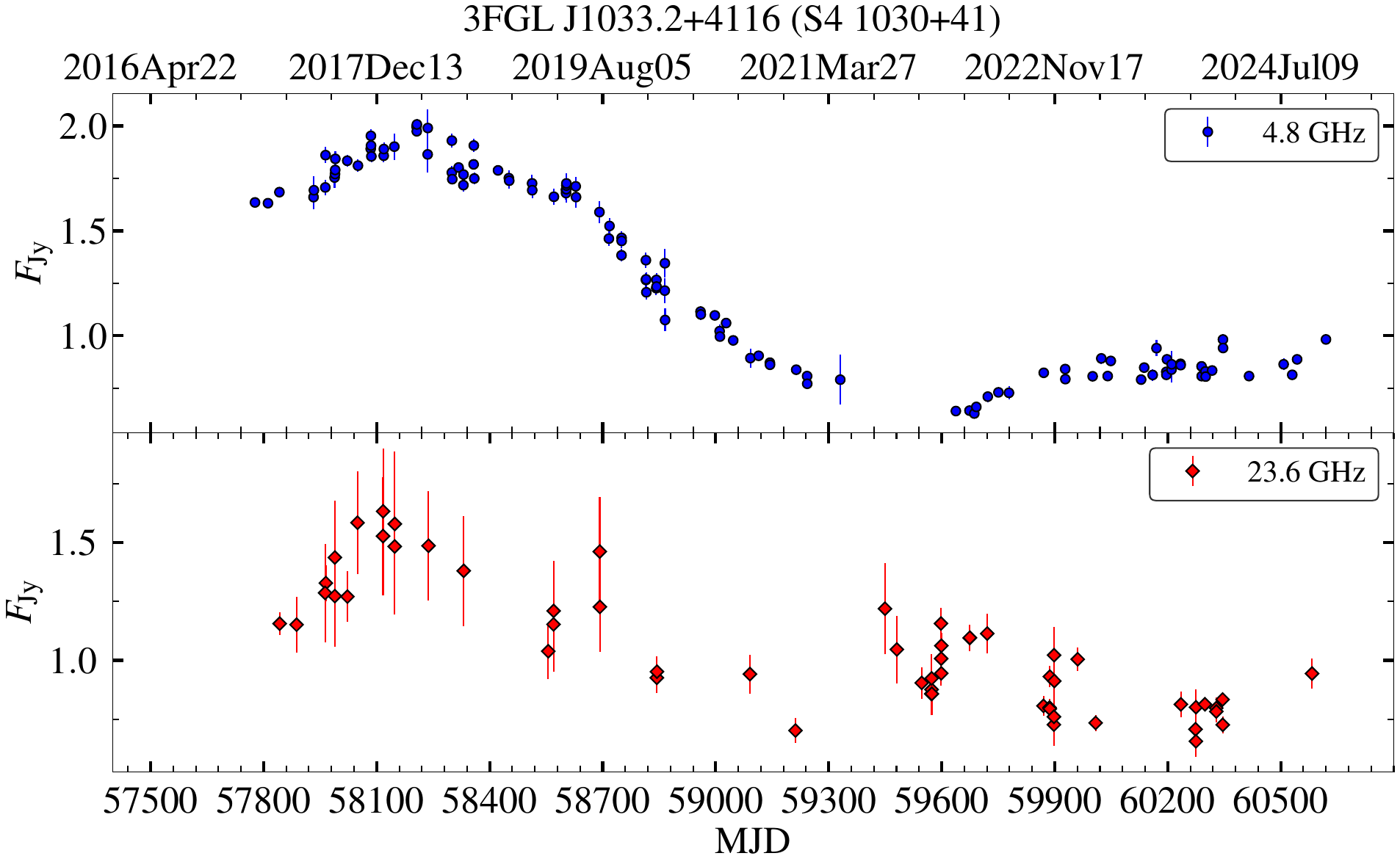}
\includegraphics[width=0.49\textwidth]{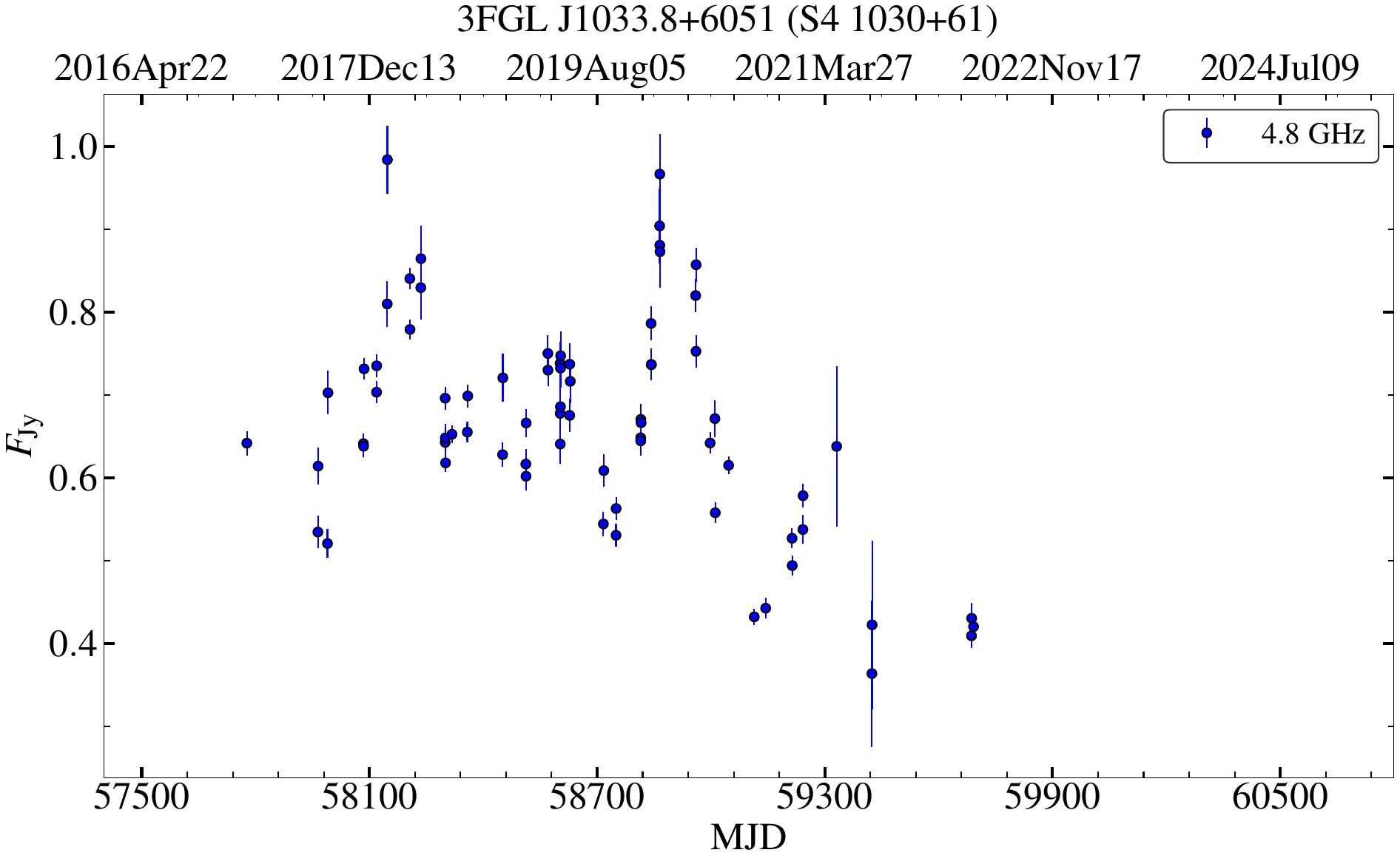}\\
\includegraphics[width=0.49\textwidth]{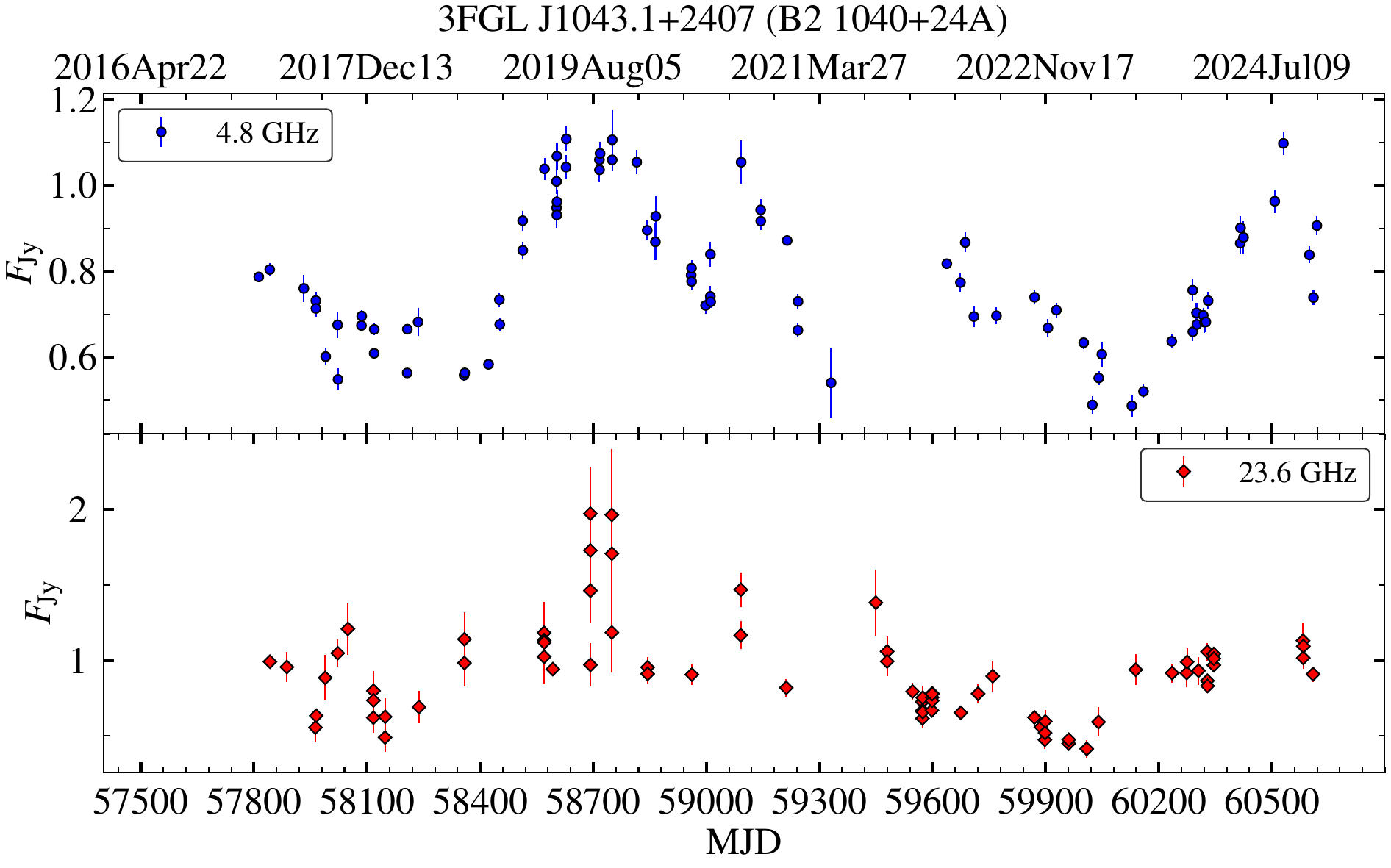}
\includegraphics[width=0.49\textwidth]{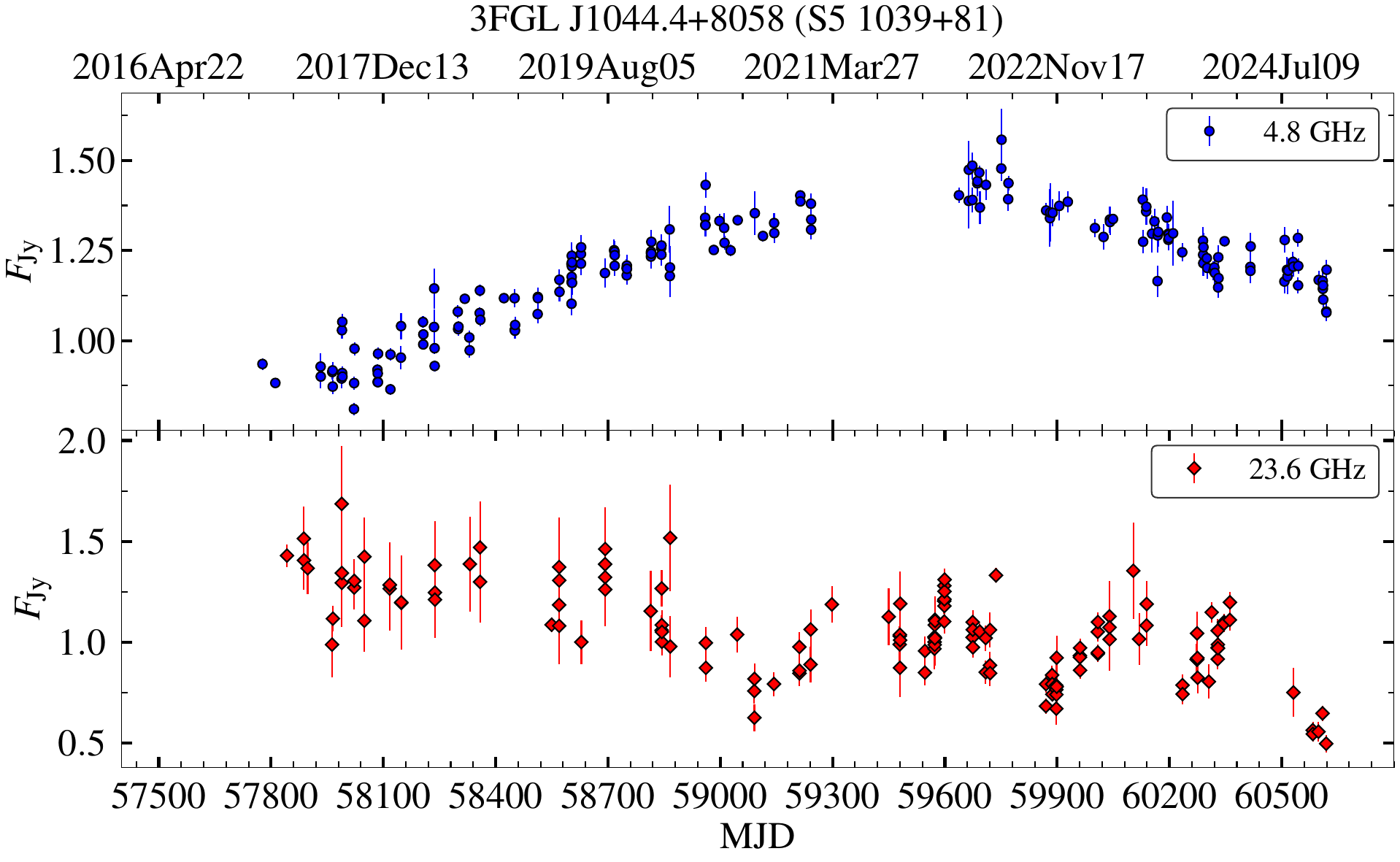}\\
\includegraphics[width=0.49\textwidth]{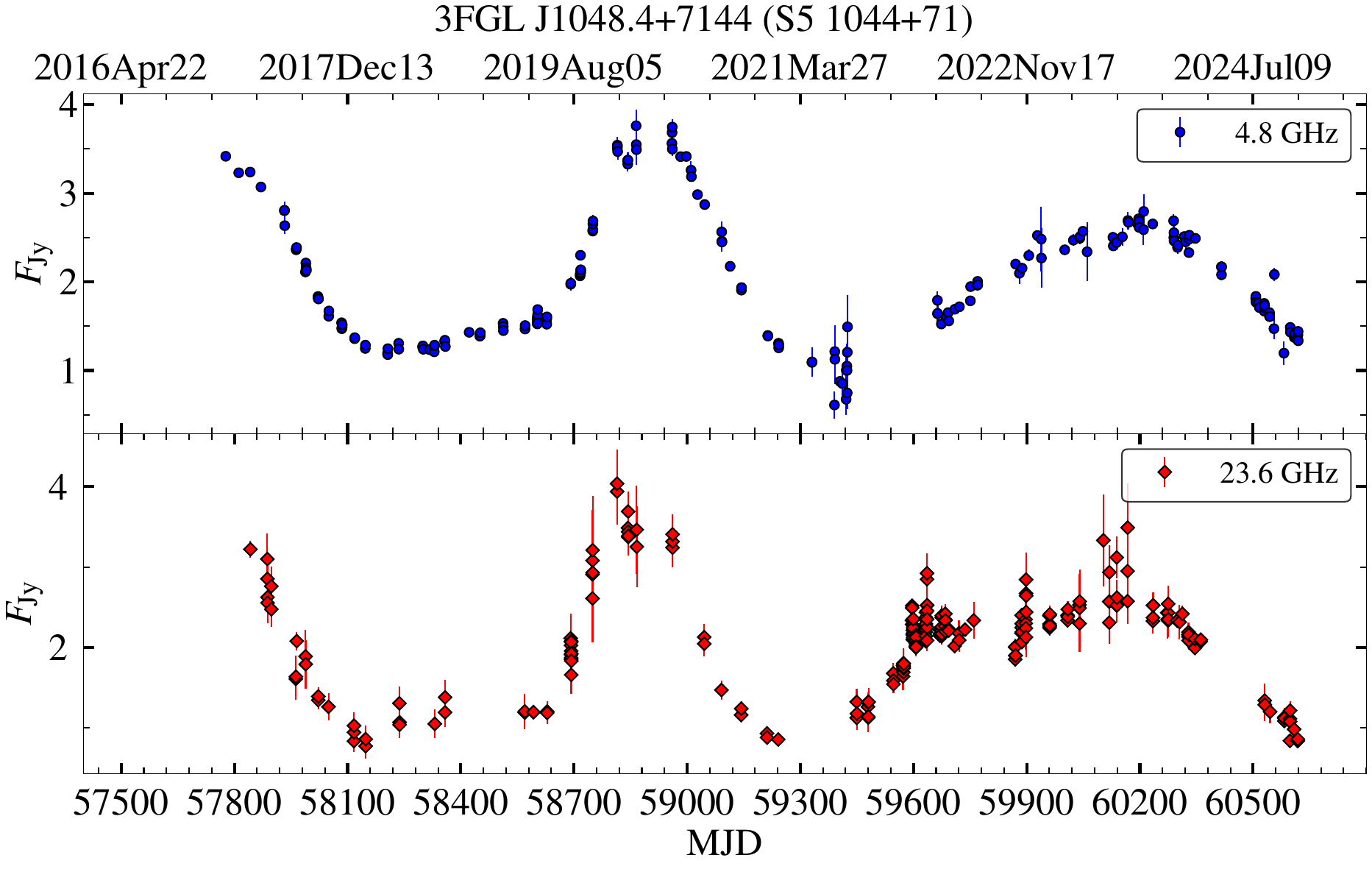}
\includegraphics[width=0.49\textwidth]{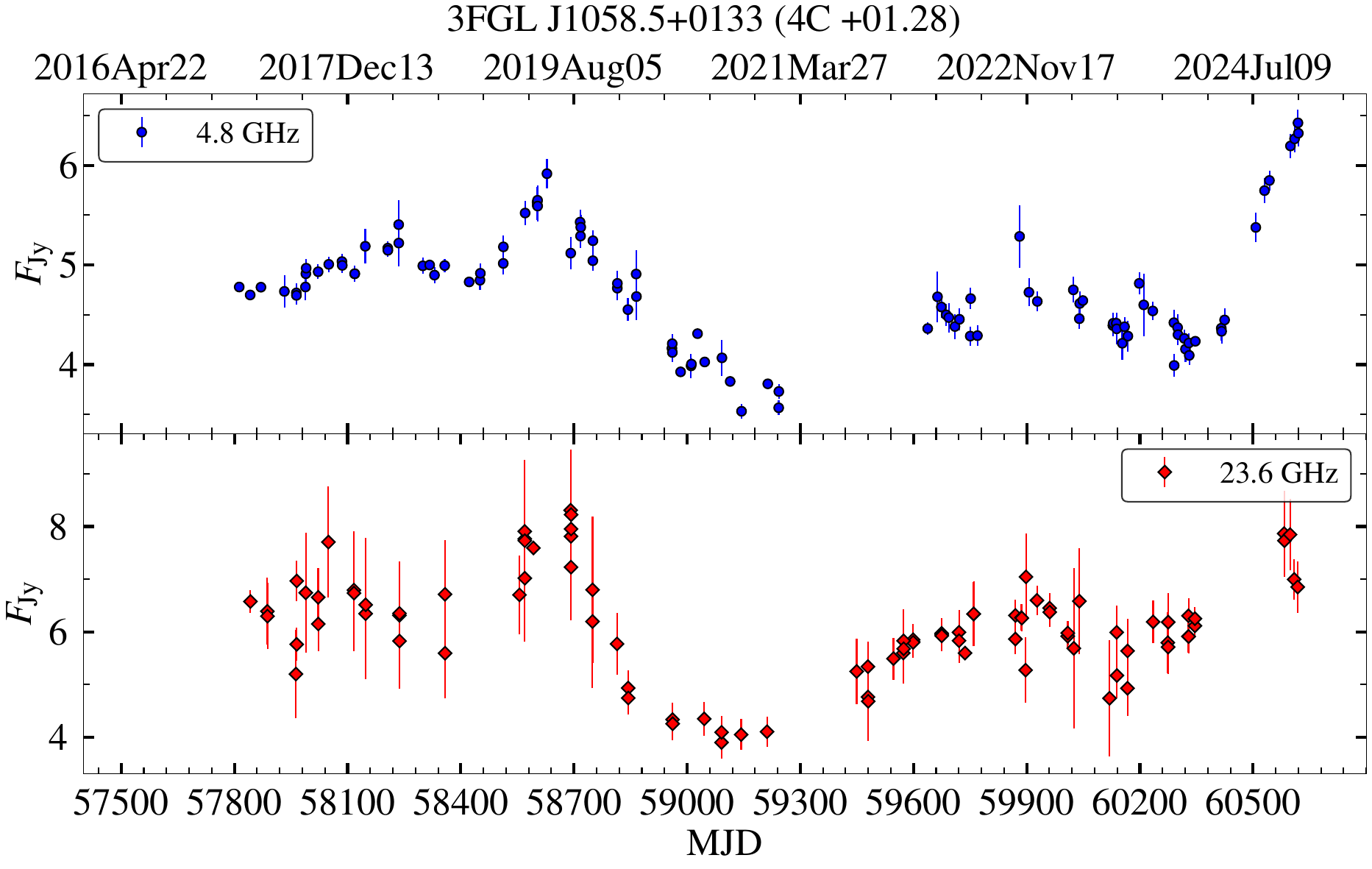}\\
\end{tabular}
\end{figure*}

\begin{figure*}[p]
\centering
\addtocounter{figure}{-1}
\caption{Continued.}
\begin{tabular}{cc}
\includegraphics[width=0.49\textwidth]{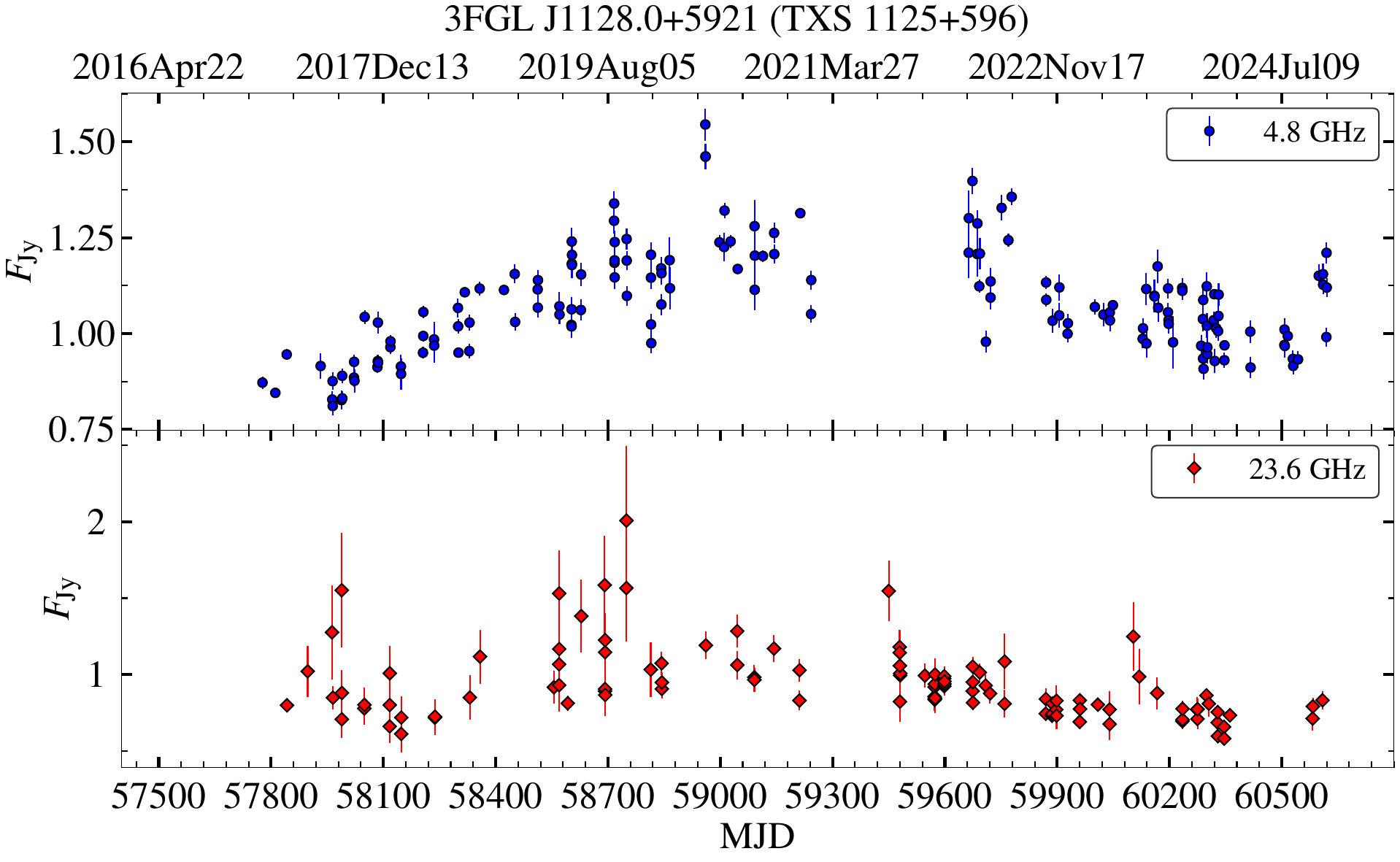}
\includegraphics[width=0.49\textwidth]{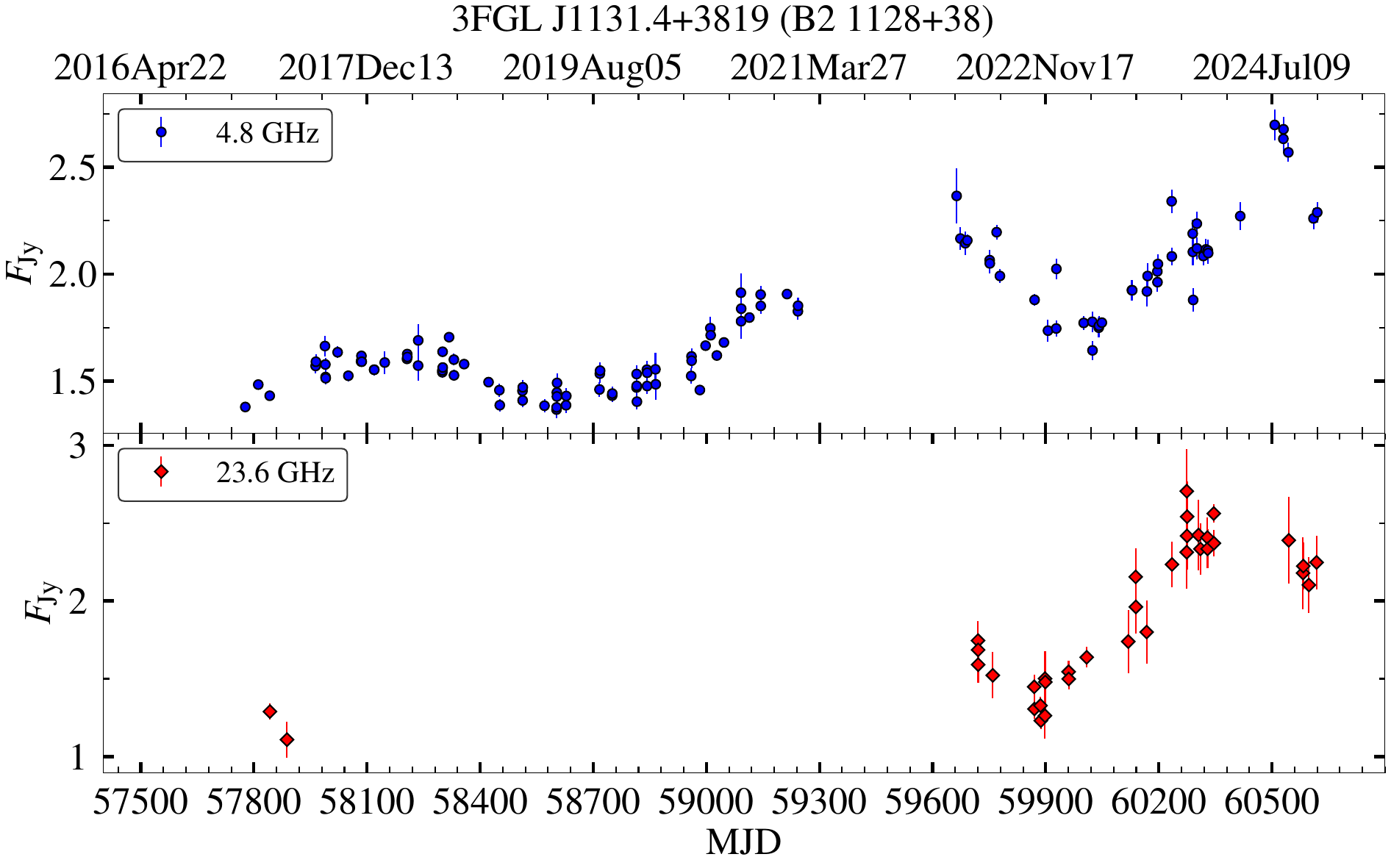}\\
\includegraphics[width=0.49\textwidth]{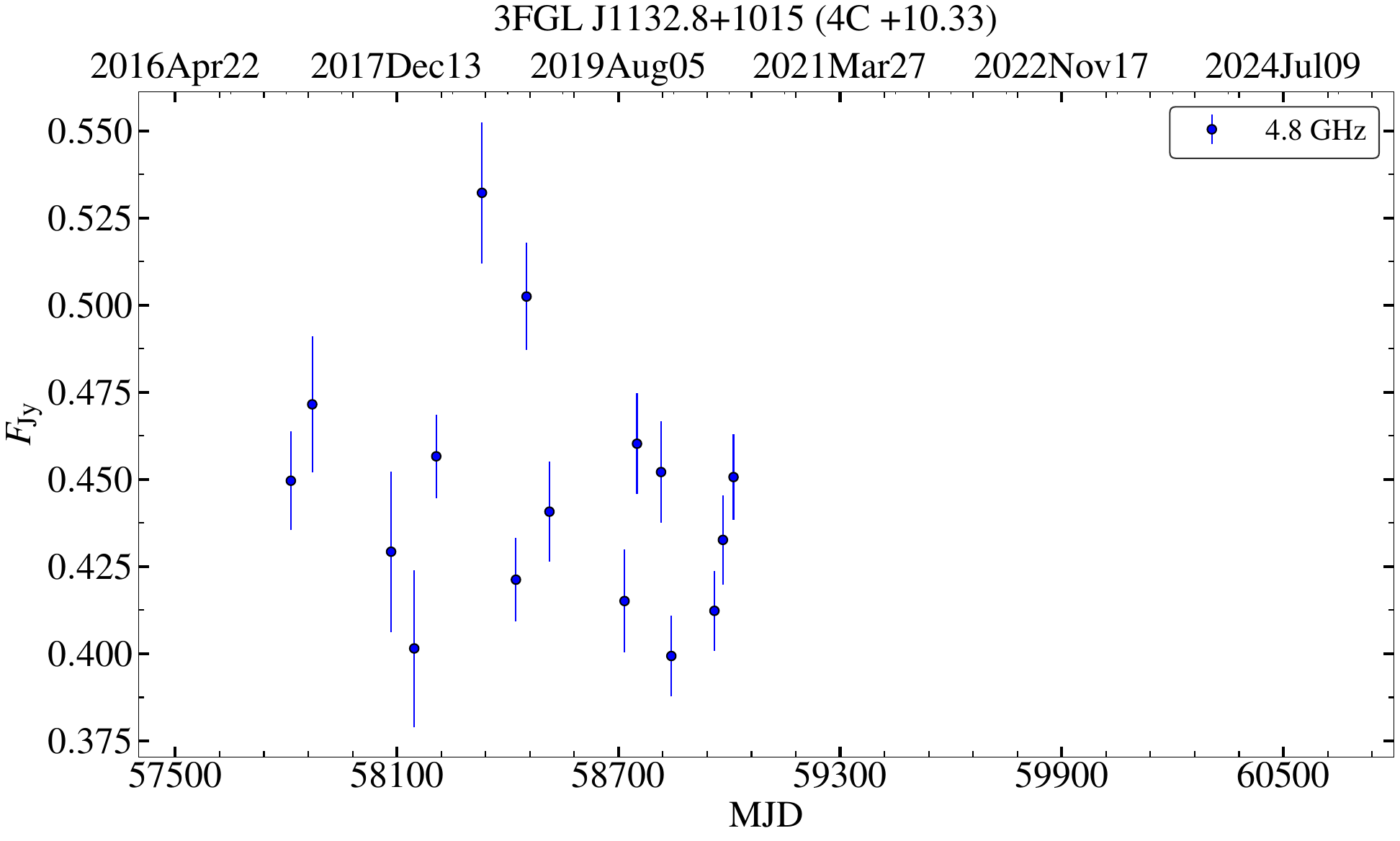}
\includegraphics[width=0.49\textwidth]{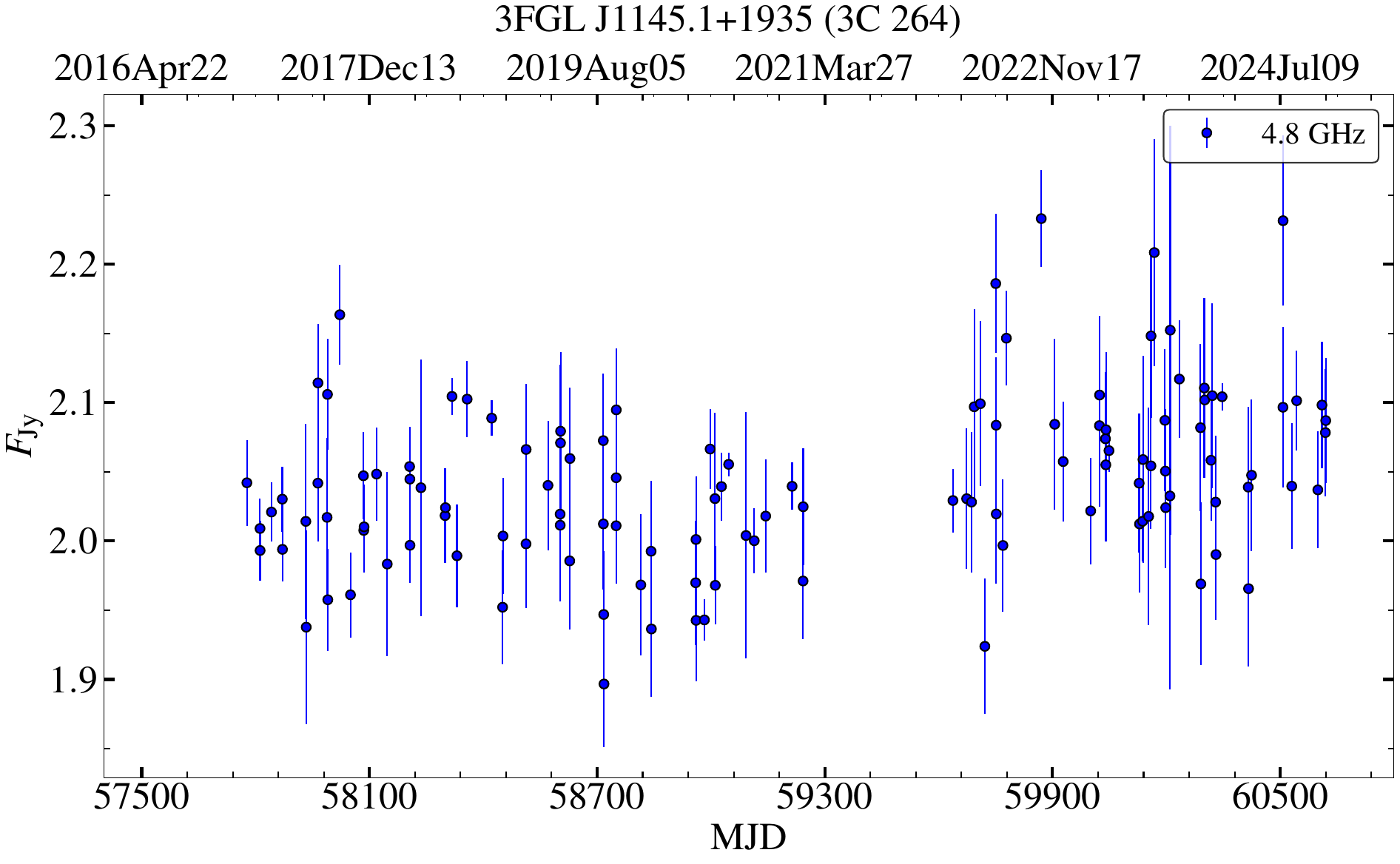}\\
\includegraphics[width=0.49\textwidth]{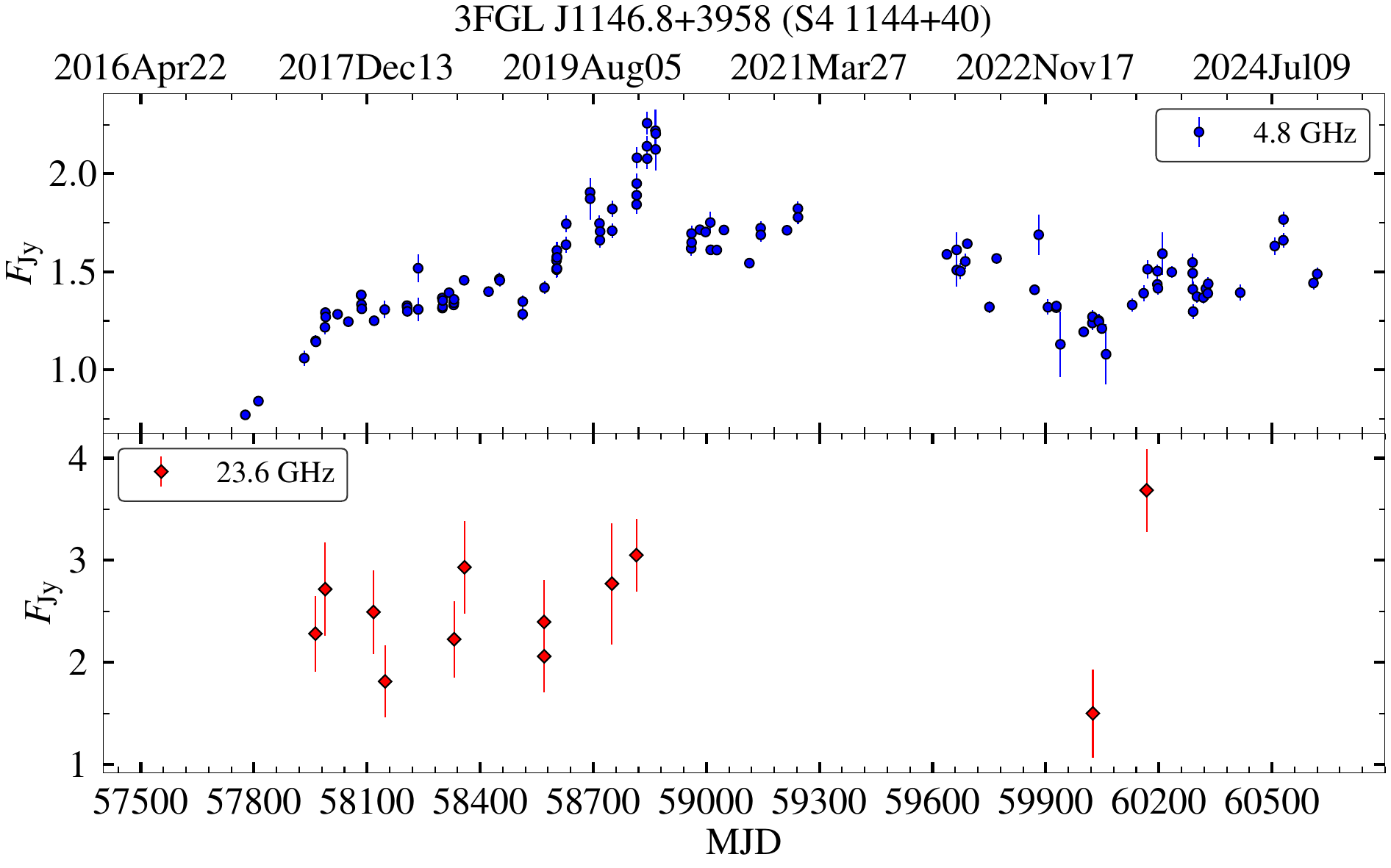}
\includegraphics[width=0.49\textwidth]{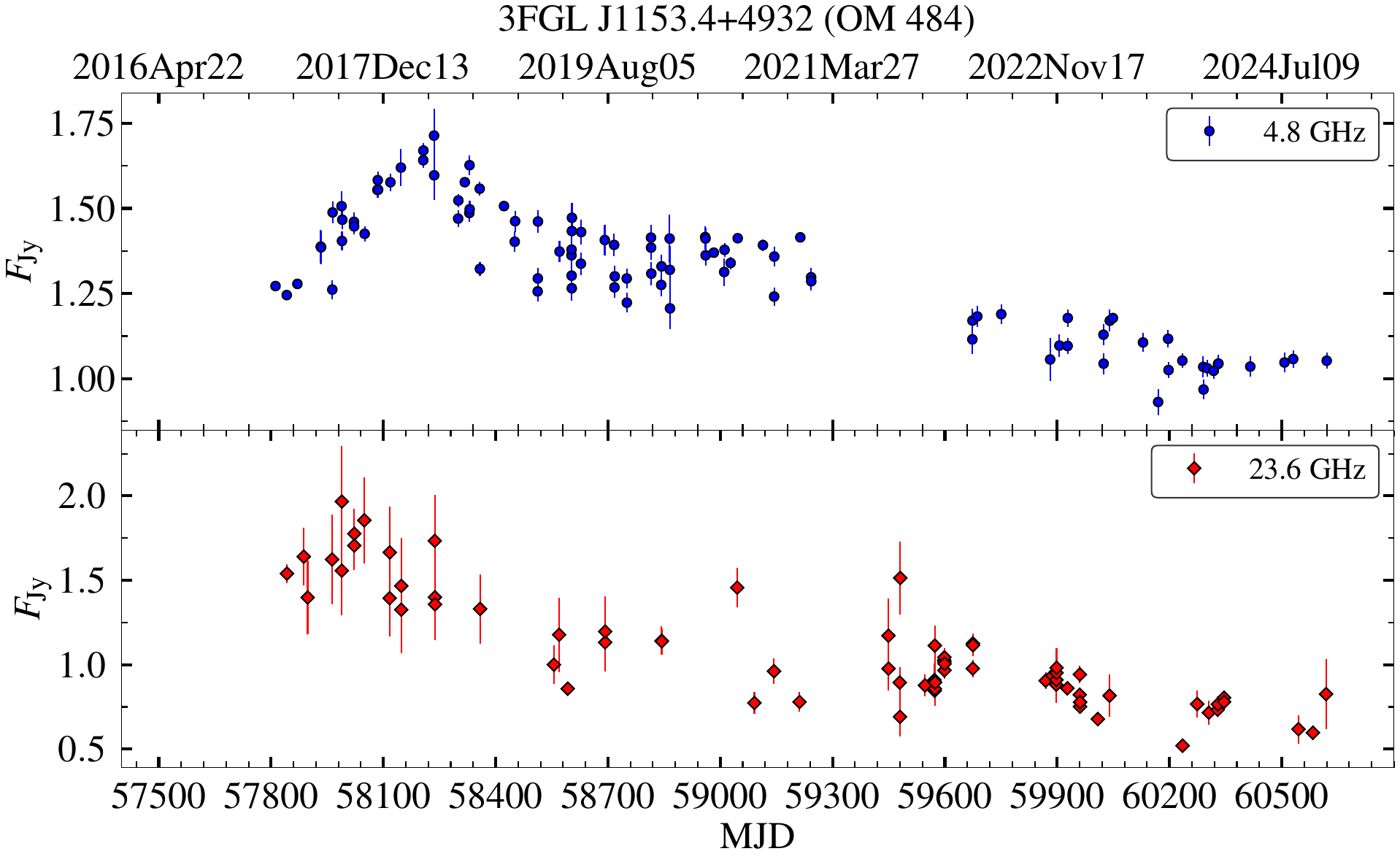}\\
\includegraphics[width=0.49\textwidth]{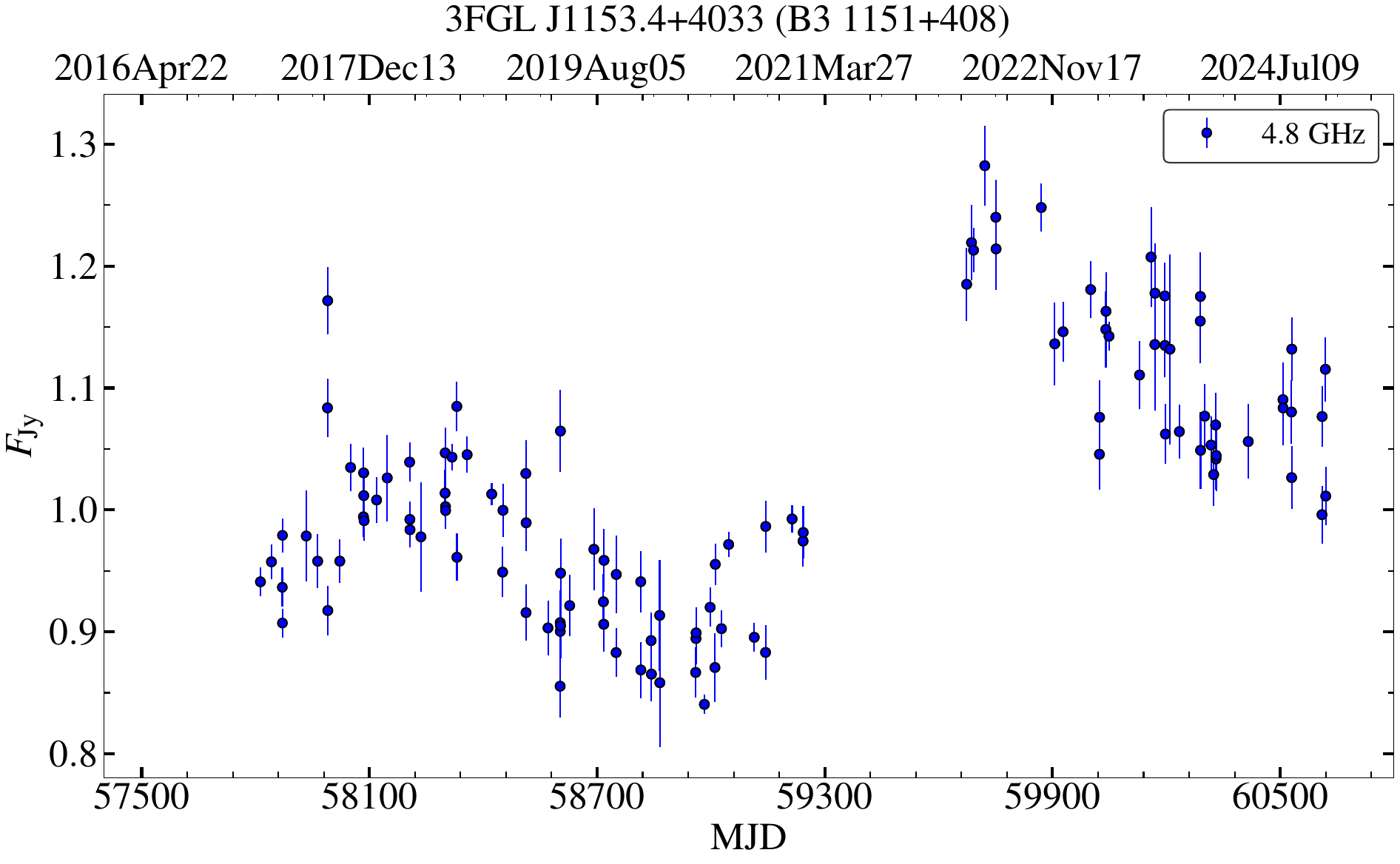}
\includegraphics[width=0.49\textwidth]{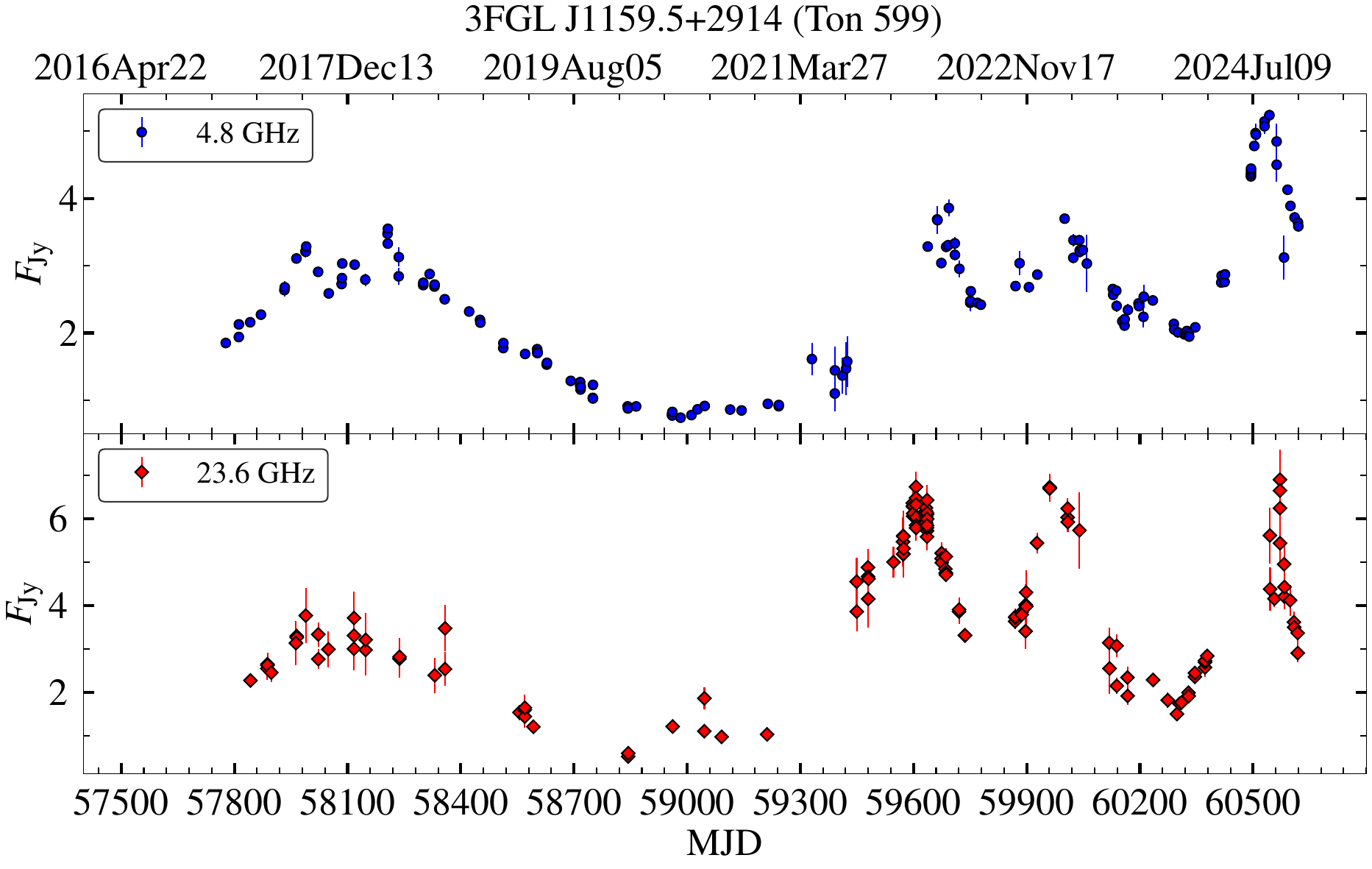}\\
\end{tabular}
\end{figure*}

\begin{figure*}[p]
\centering
\addtocounter{figure}{-1}
\caption{Continued.}
\begin{tabular}{cc}
\includegraphics[width=0.49\textwidth]{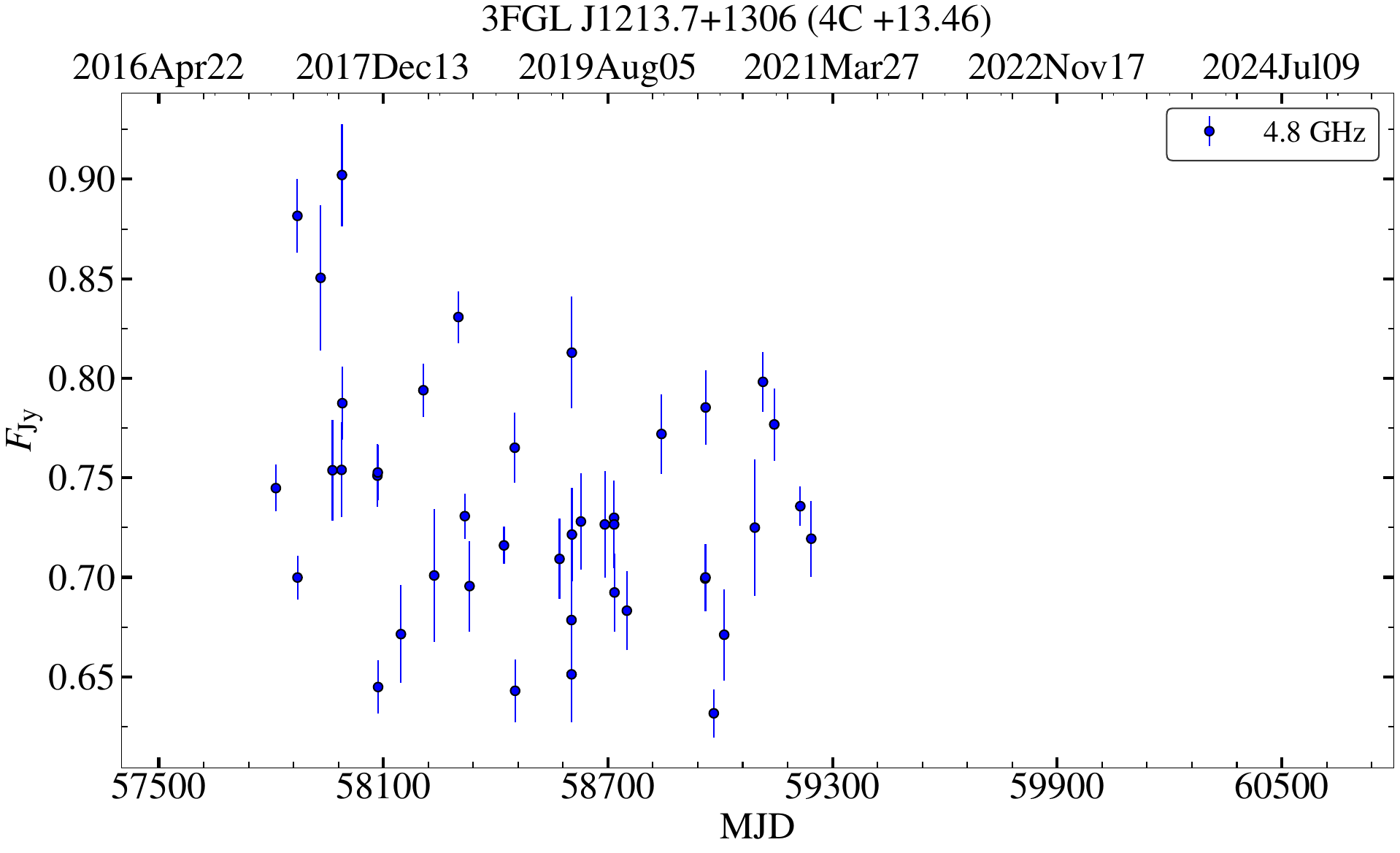}
\includegraphics[width=0.49\textwidth]{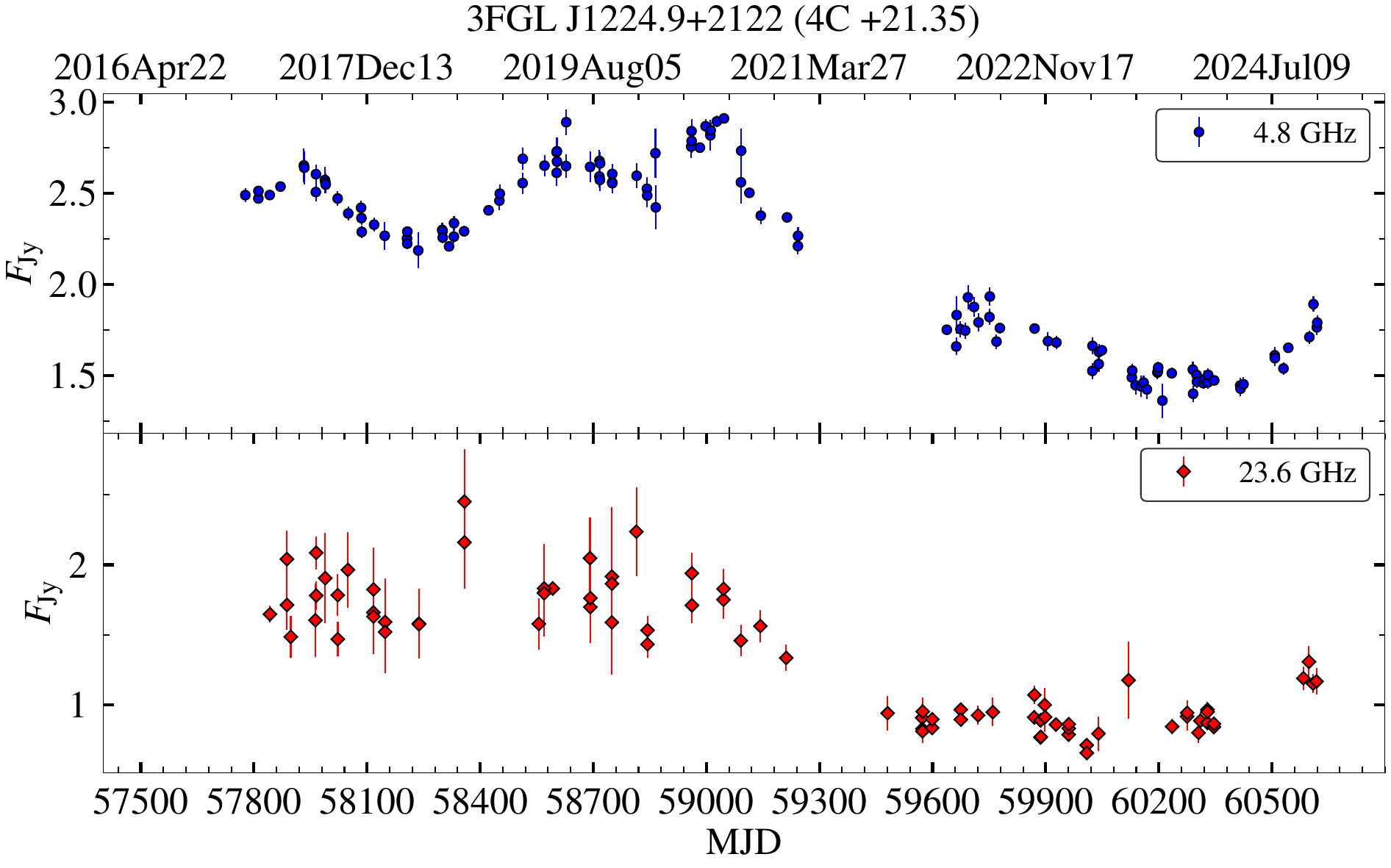}\\
\includegraphics[width=0.49\textwidth]{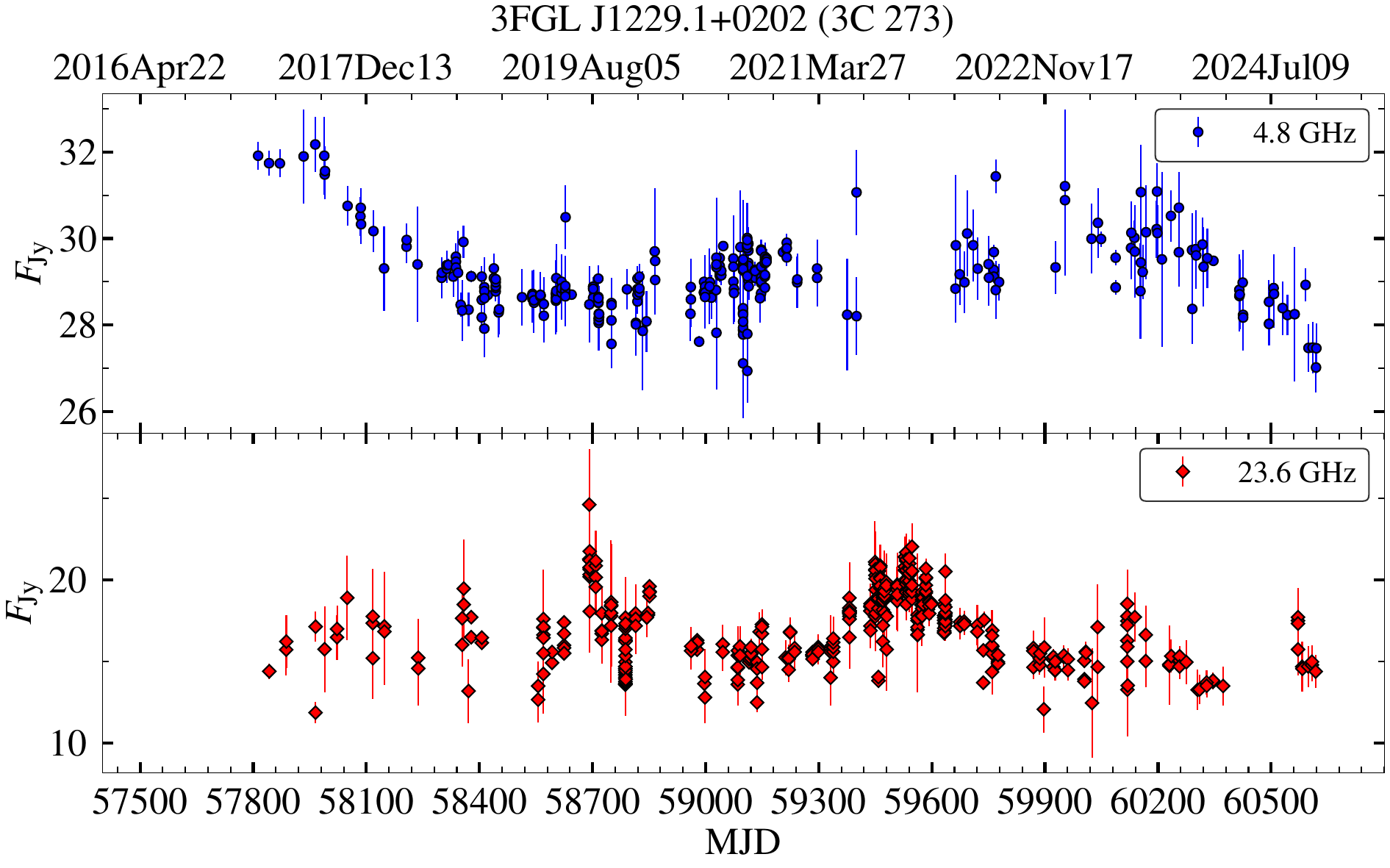}
\includegraphics[width=0.49\textwidth]{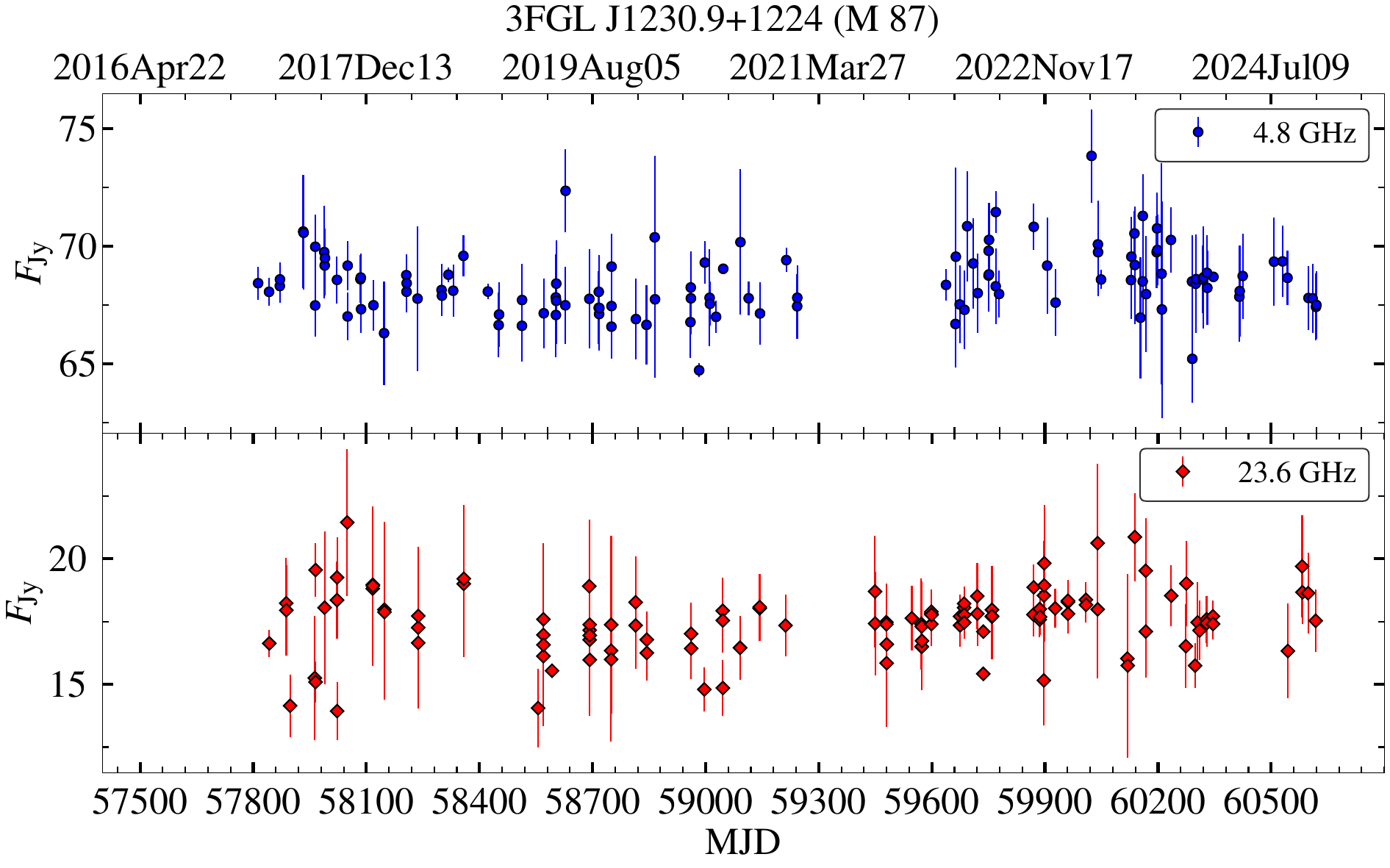}\\
\includegraphics[width=0.49\textwidth]{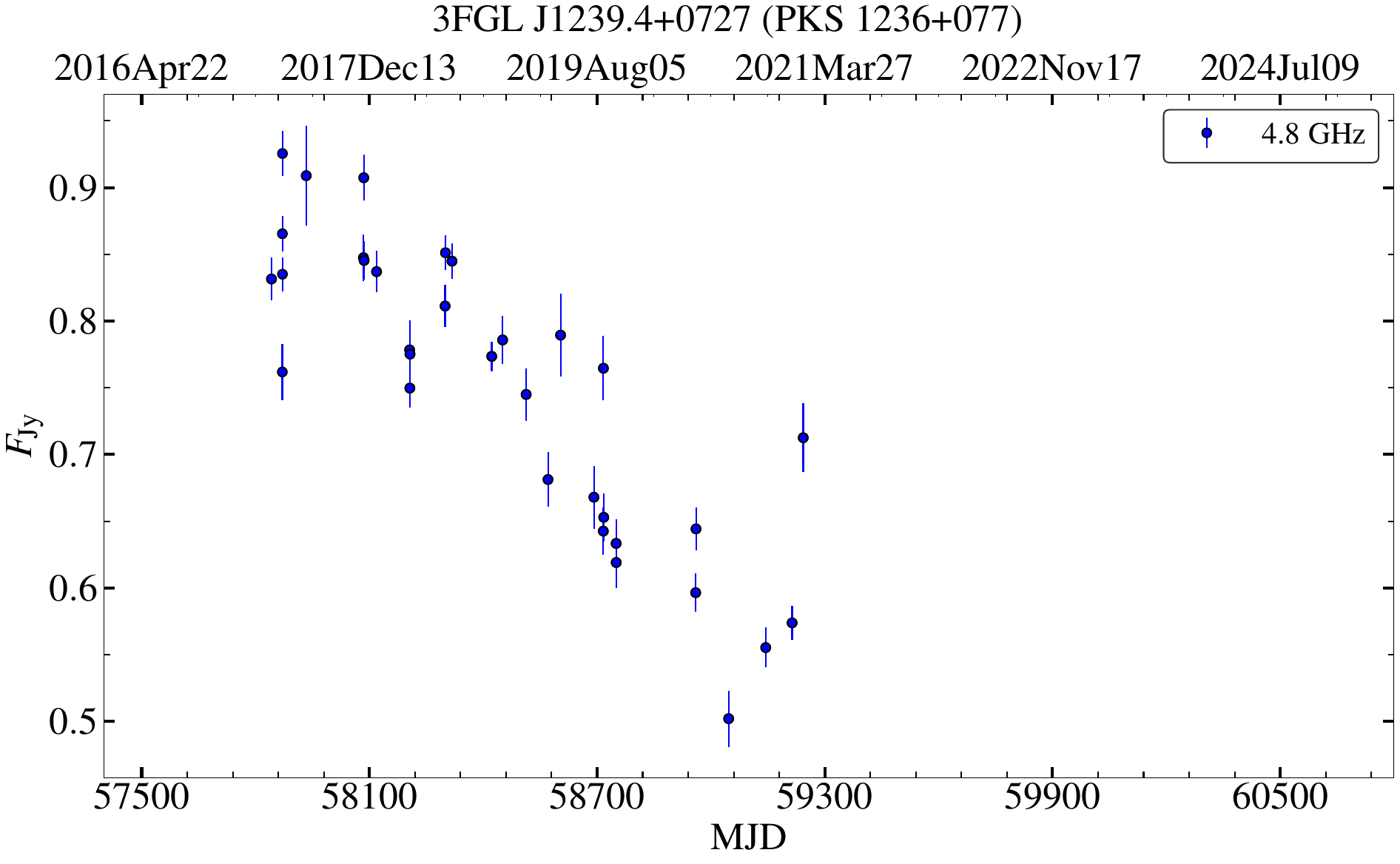}
\includegraphics[width=0.49\textwidth]{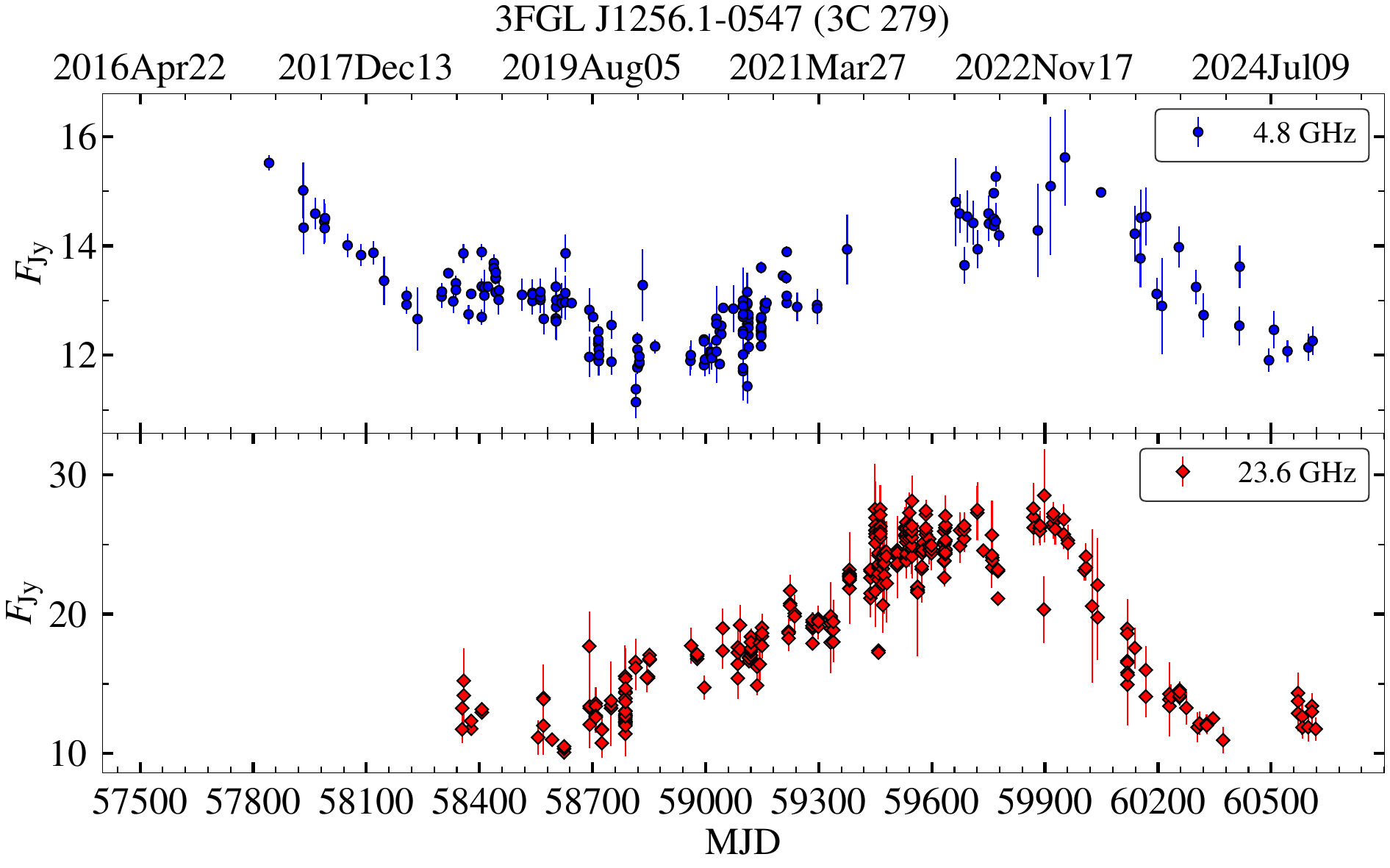}\\
\includegraphics[width=0.49\textwidth]{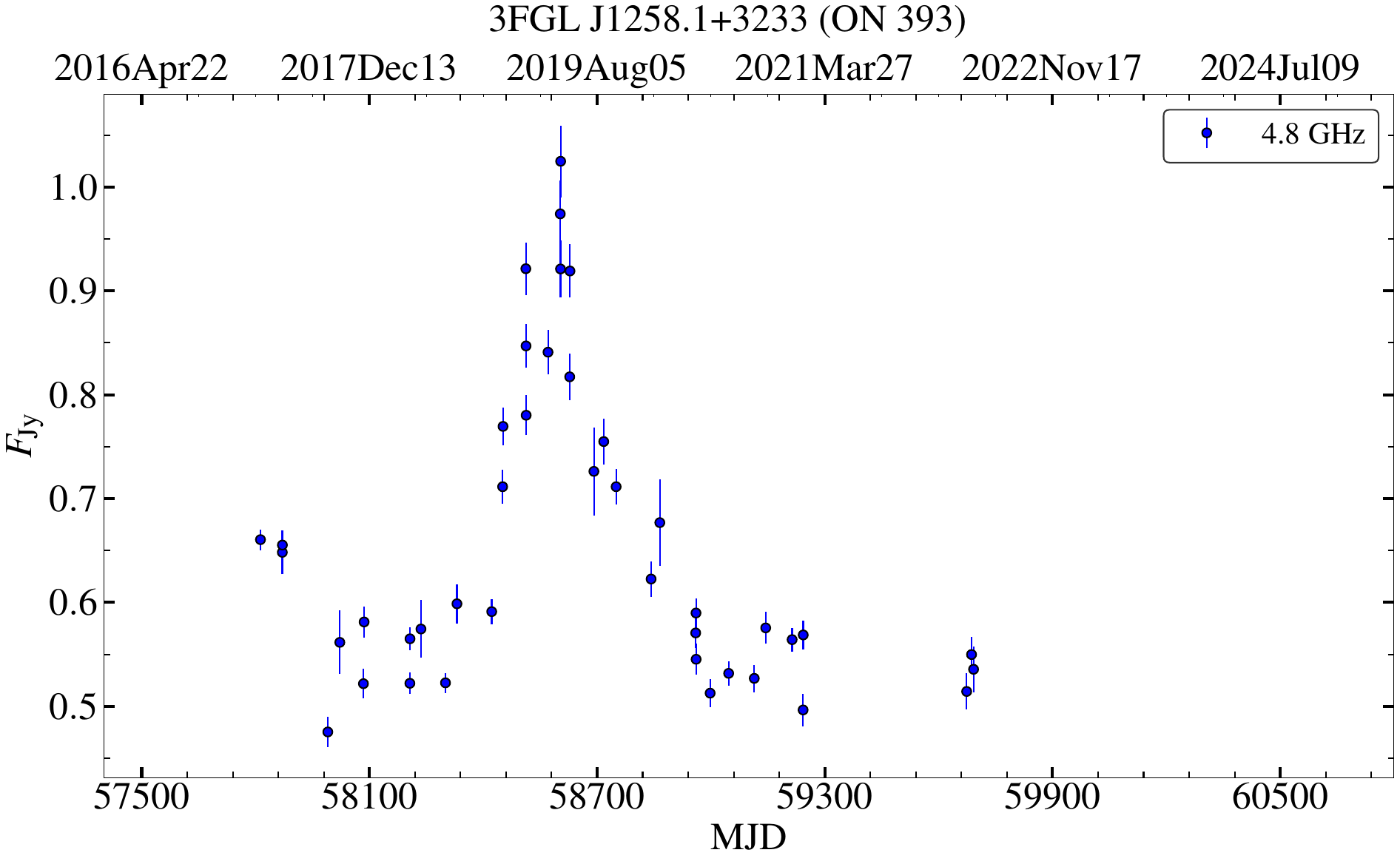}
\includegraphics[width=0.49\textwidth]{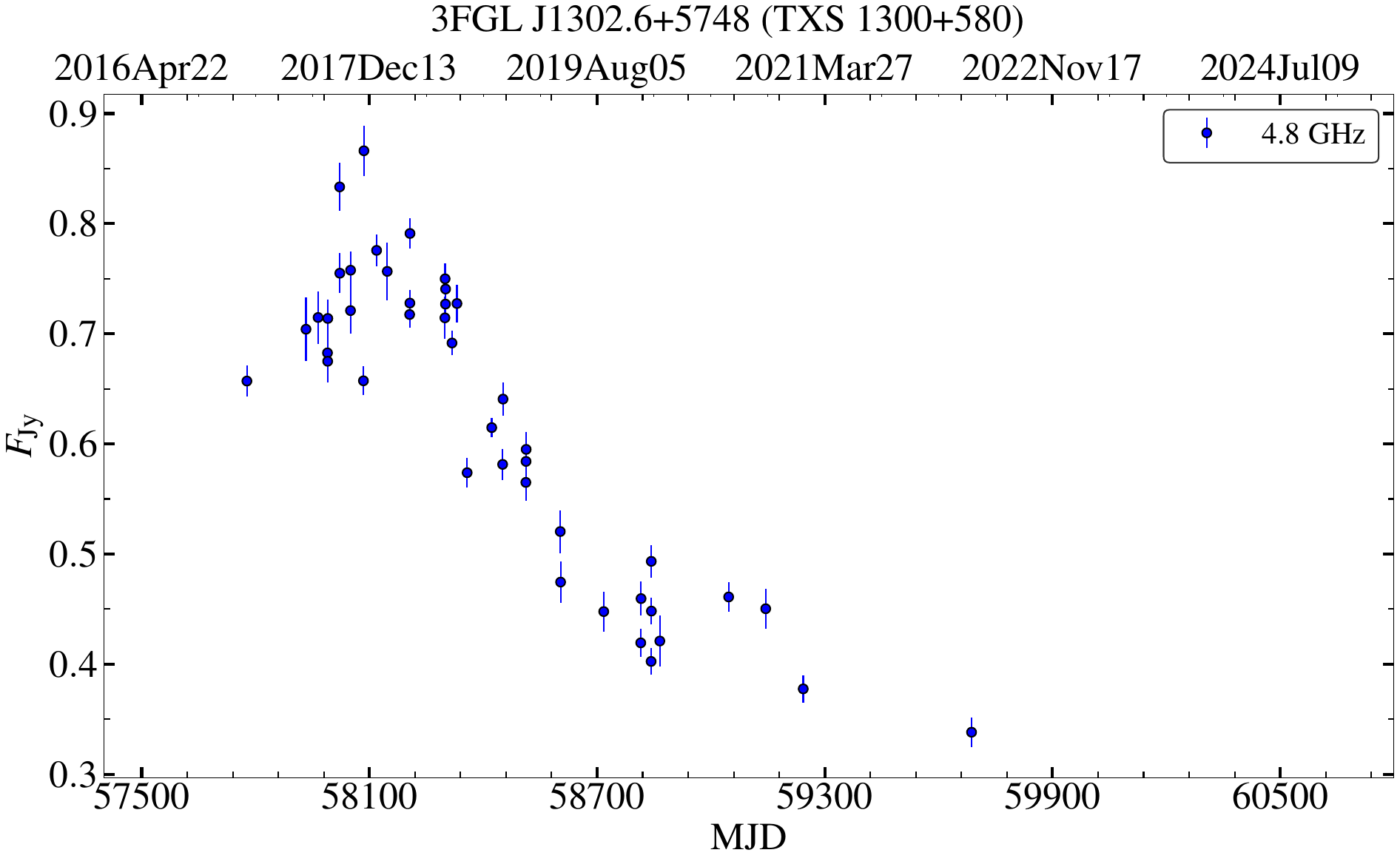}\\
\end{tabular}
\end{figure*}

\begin{figure*}[p]
\centering
\addtocounter{figure}{-1}
\caption{Continued.}
\begin{tabular}{cc}
\includegraphics[width=0.49\textwidth]{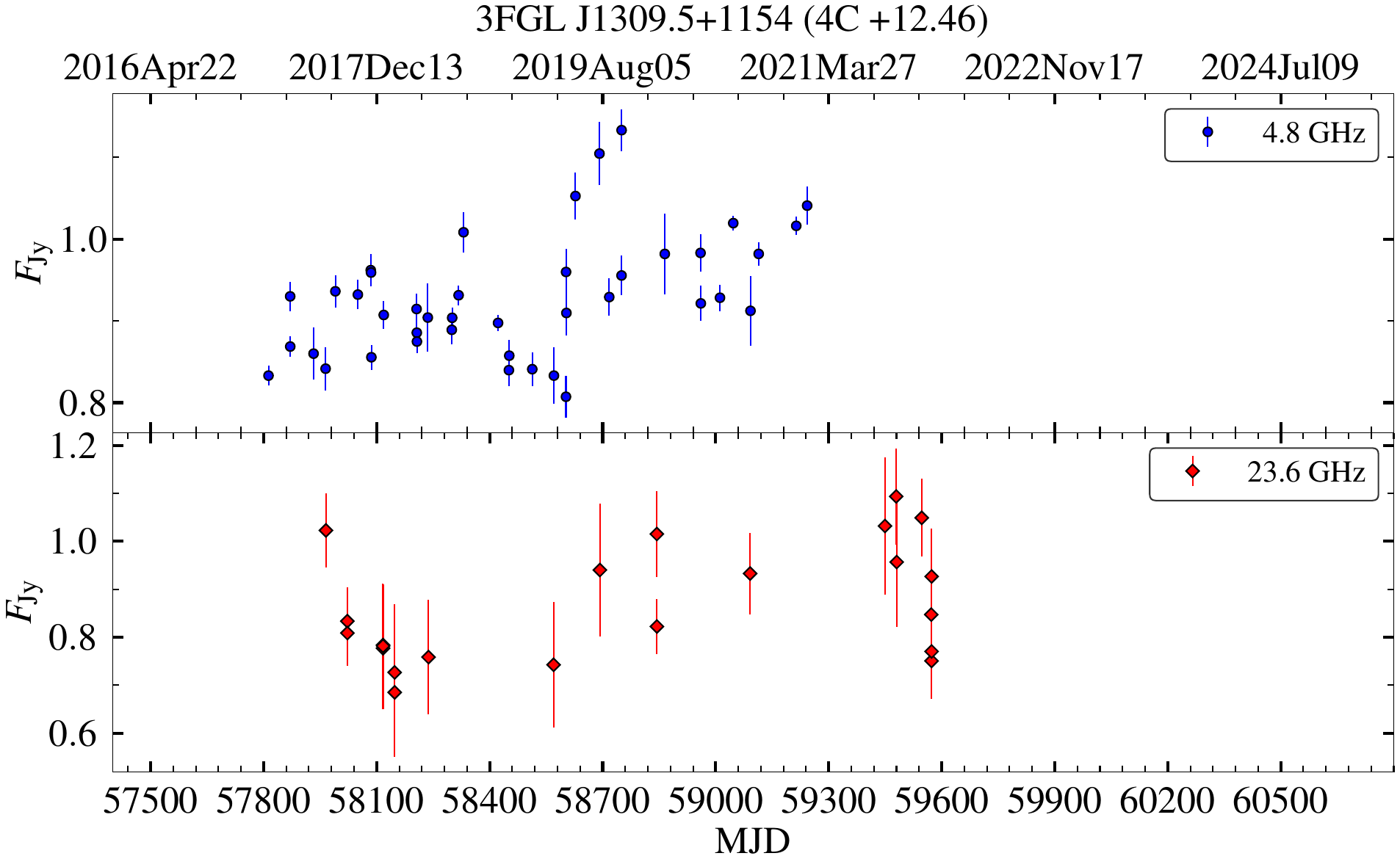}
\includegraphics[width=0.49\textwidth]{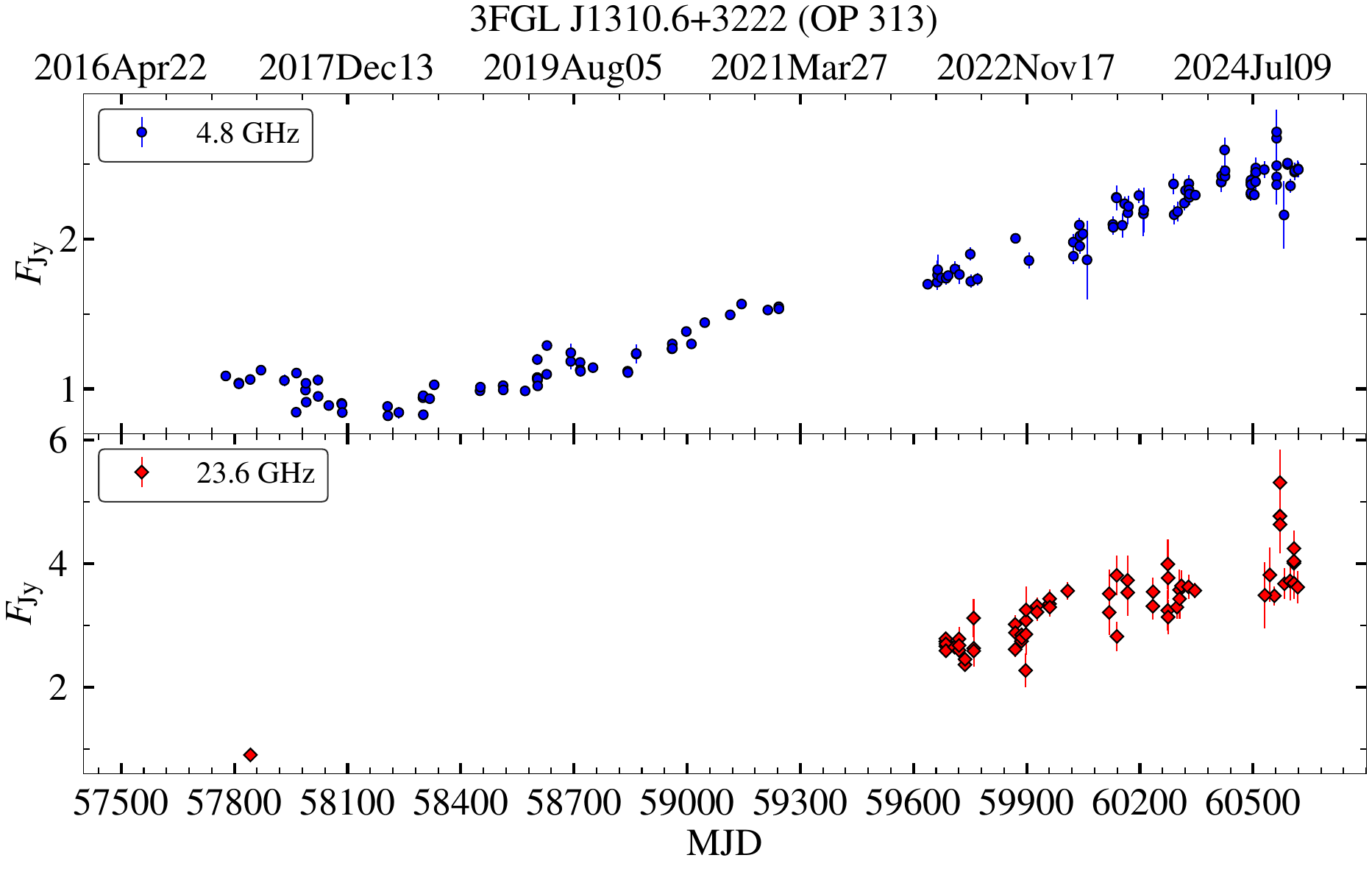}\\
\includegraphics[width=0.49\textwidth]{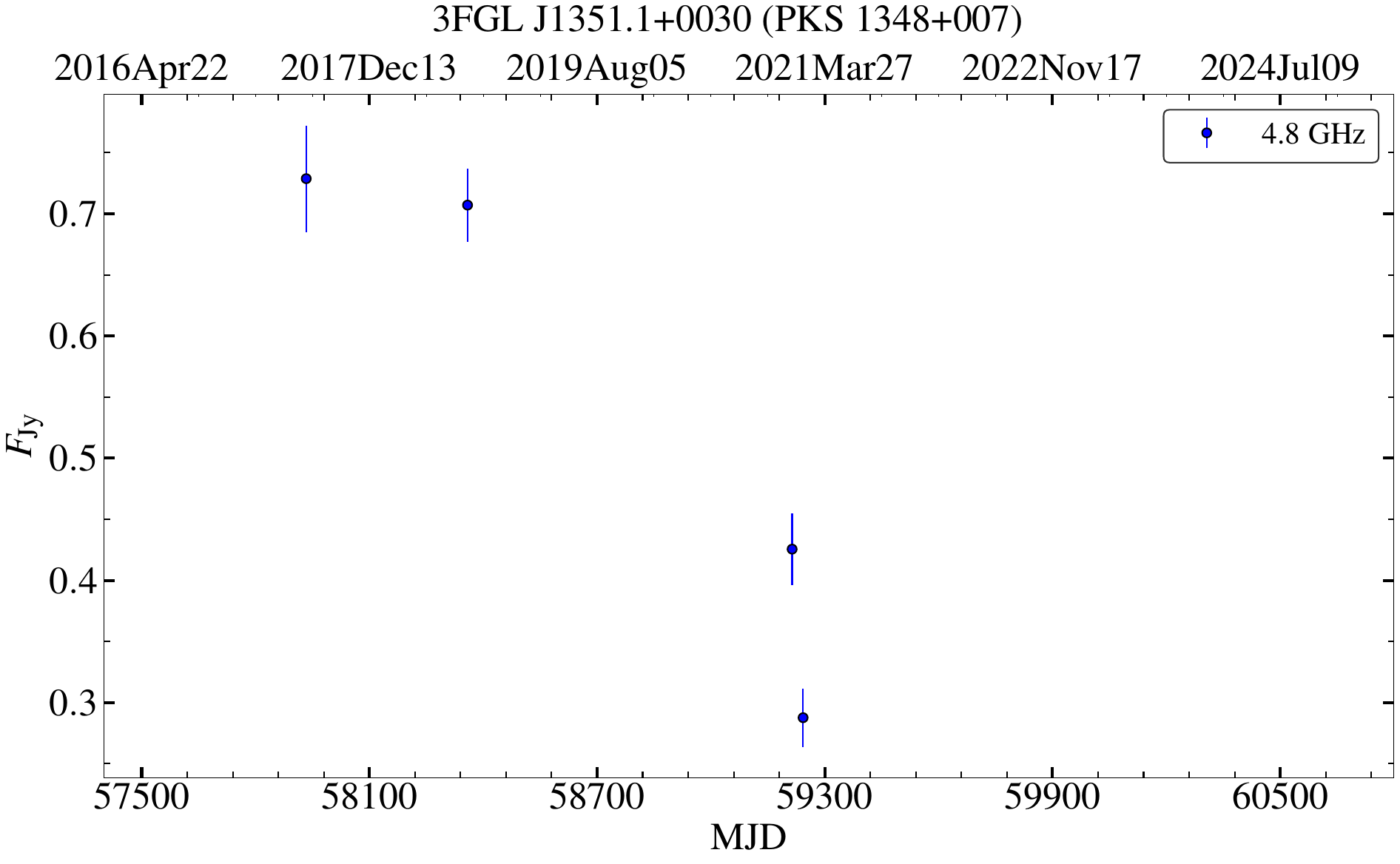}
\includegraphics[width=0.49\textwidth]{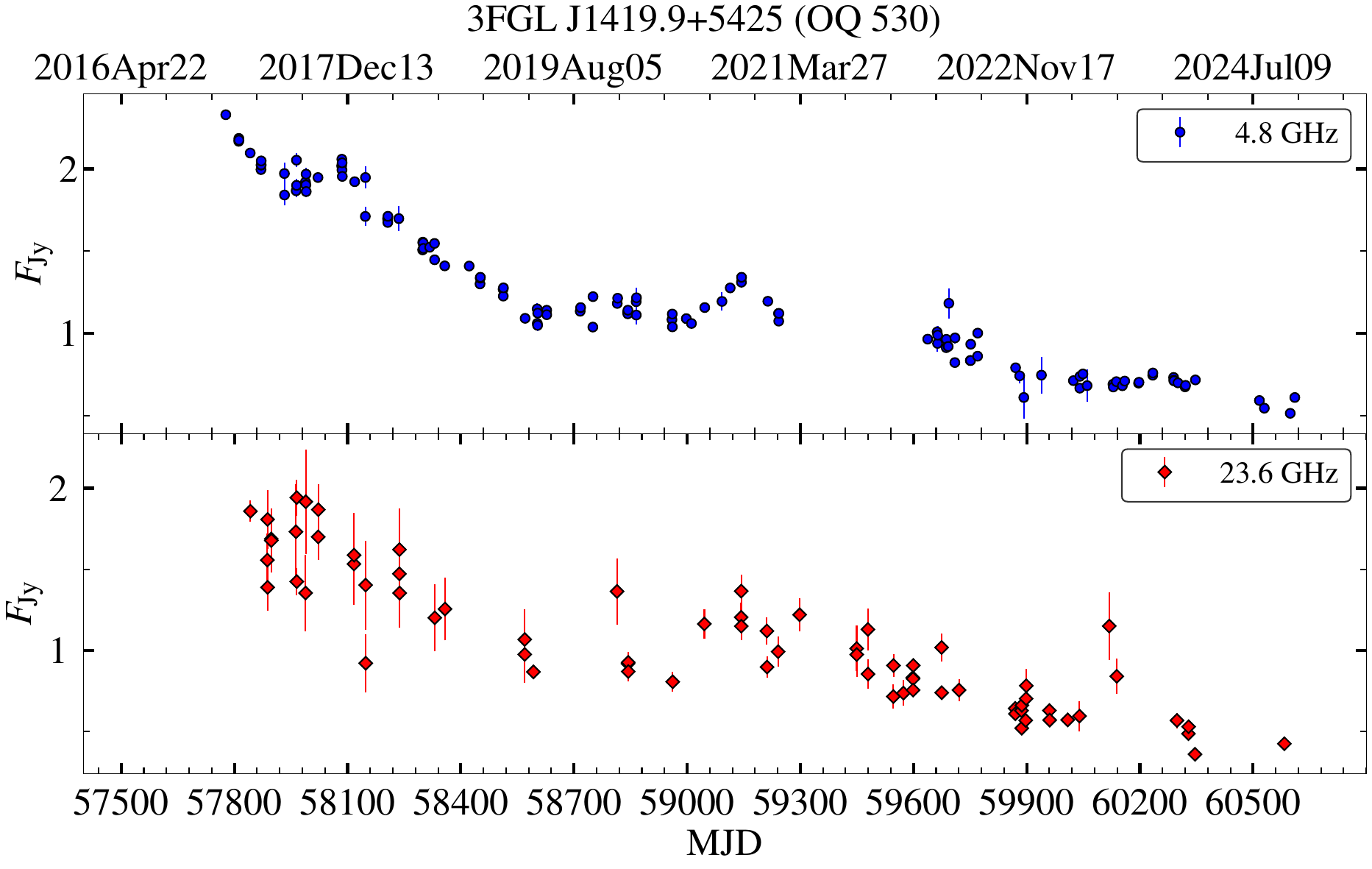}\\
\includegraphics[width=0.49\textwidth]{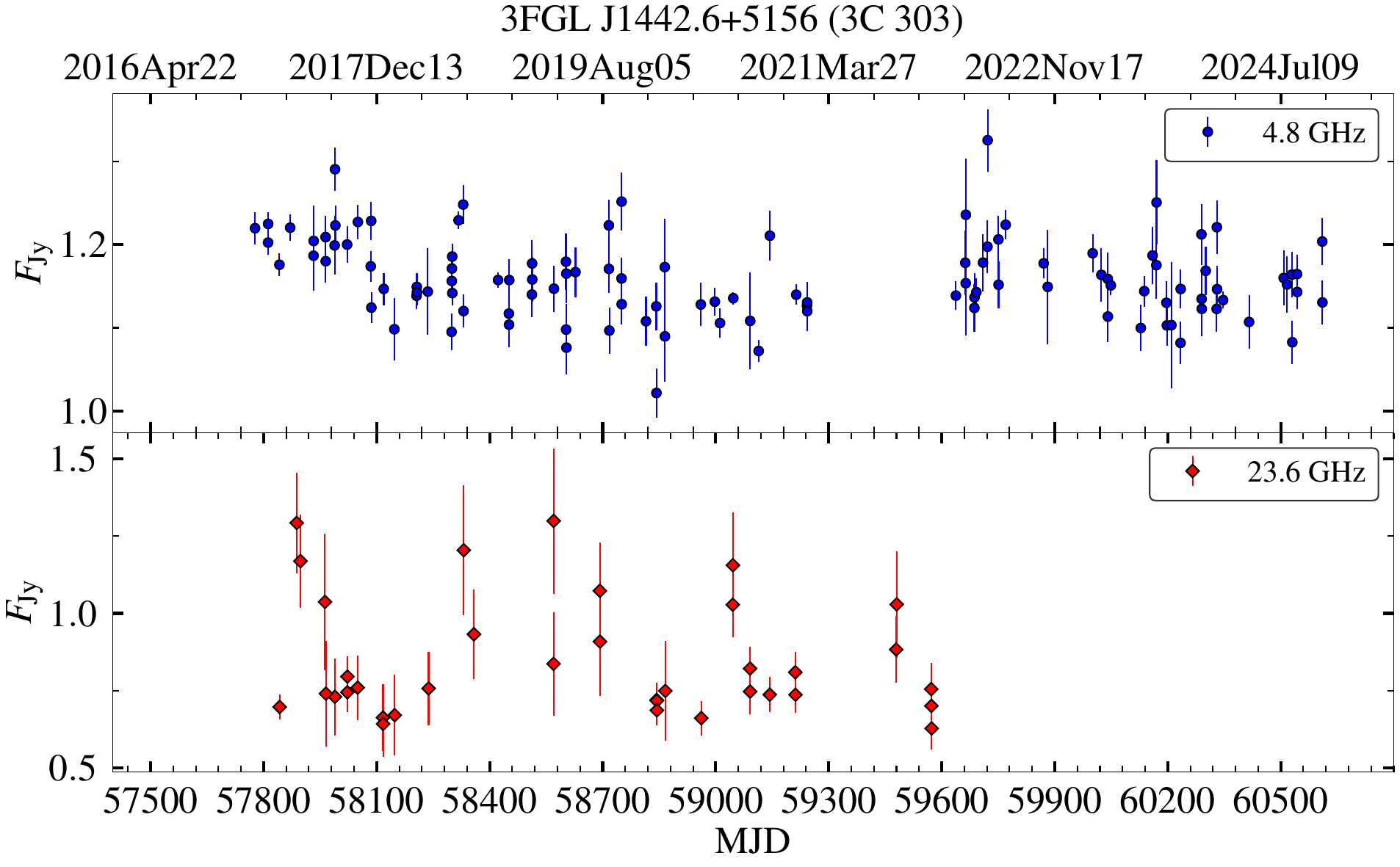}
\includegraphics[width=0.49\textwidth]{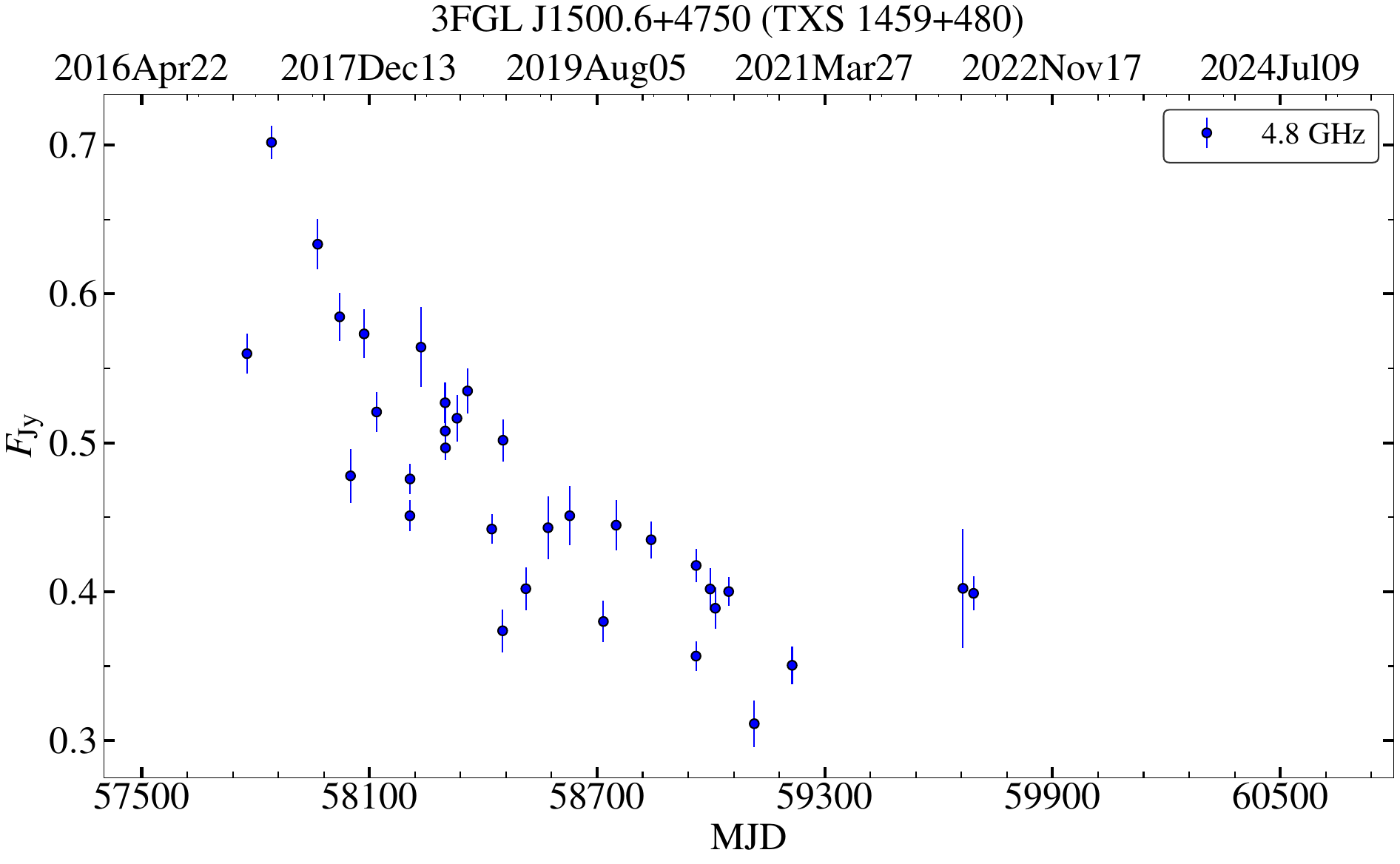}\\
\includegraphics[width=0.49\textwidth]{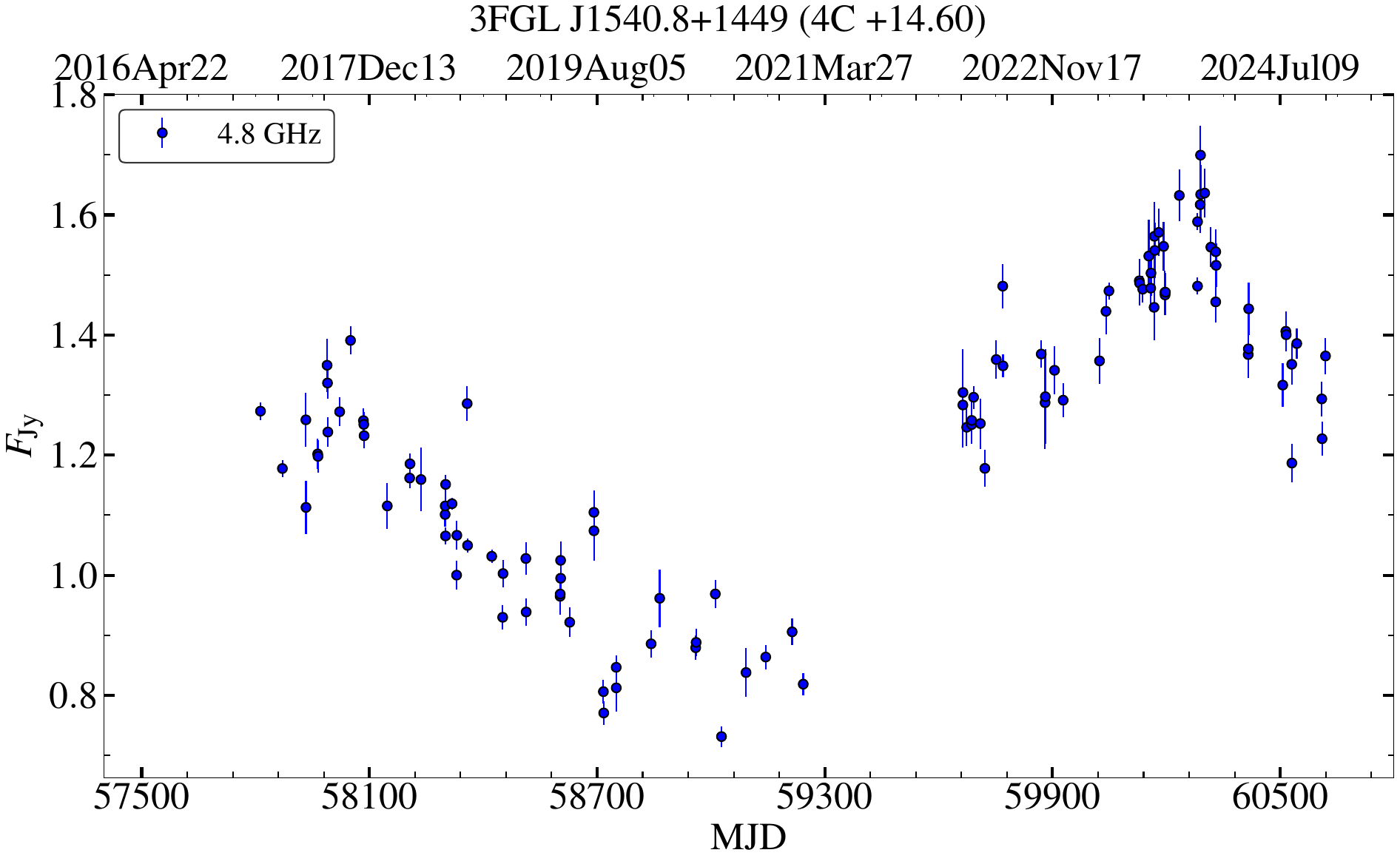}
\includegraphics[width=0.49\textwidth]{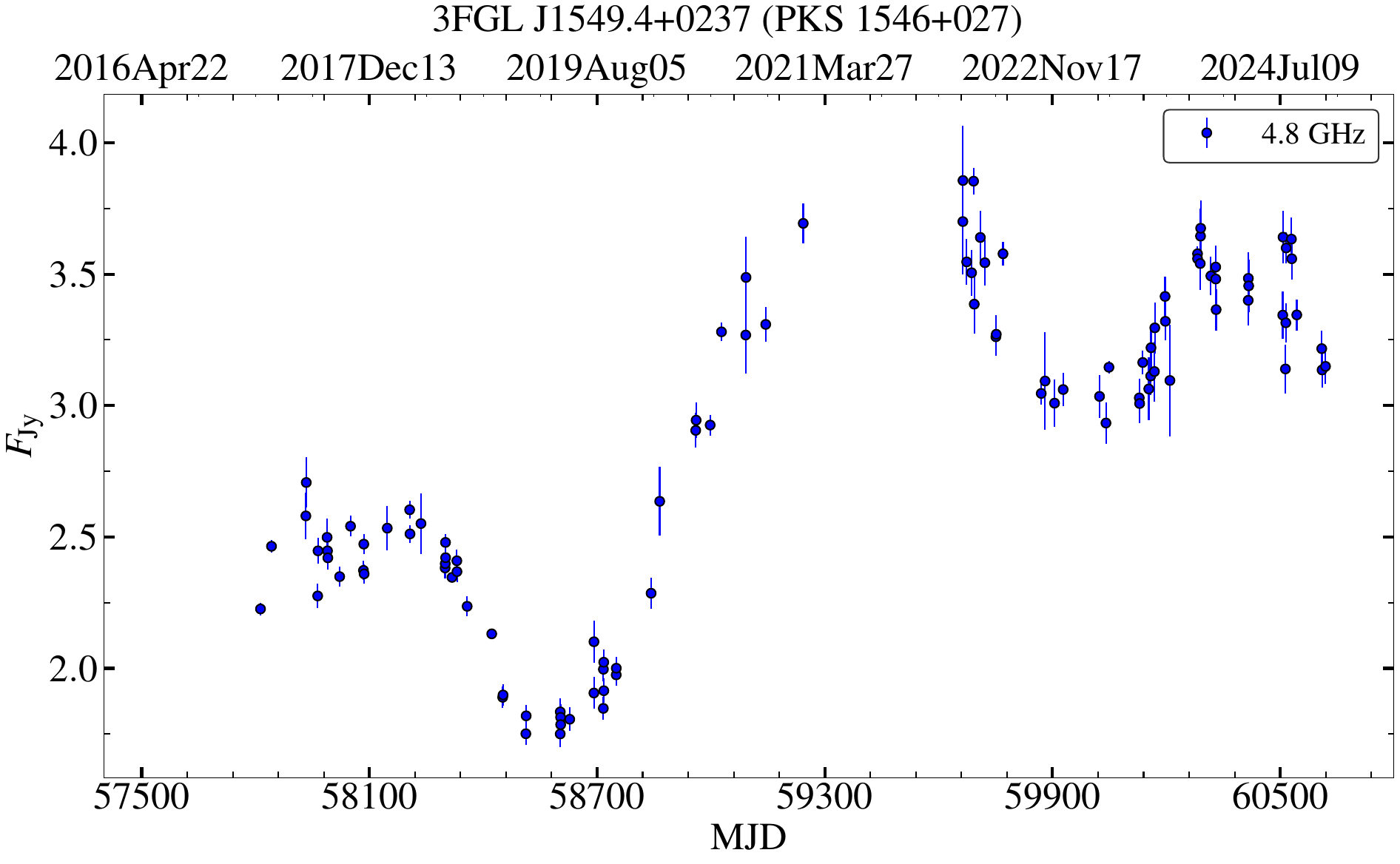}\\
\end{tabular}
\end{figure*}

\begin{figure*}[p]
\centering
\addtocounter{figure}{-1}
\caption{Continued.}
\begin{tabular}{cc}
\includegraphics[width=0.49\textwidth]{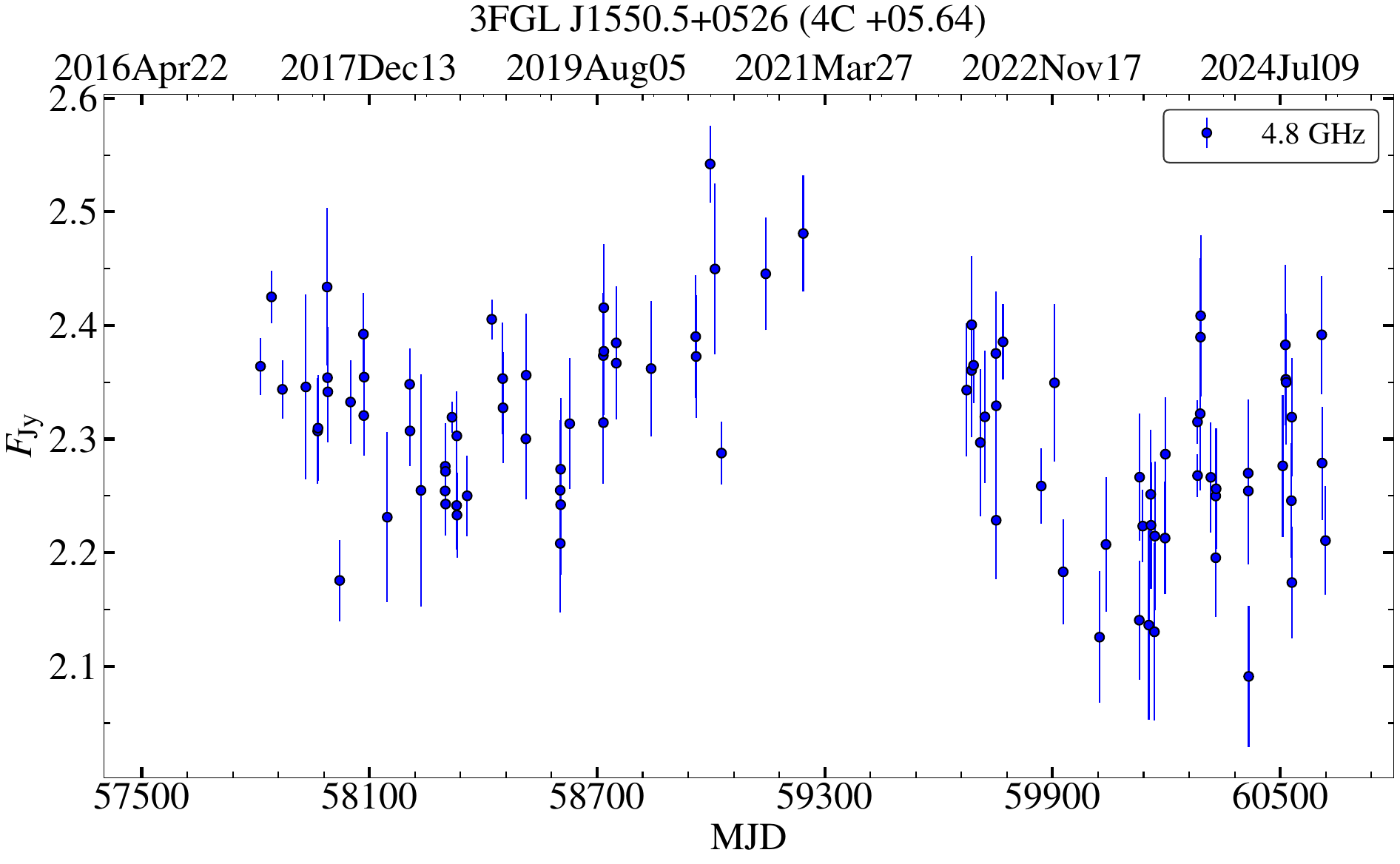}
\includegraphics[width=0.49\textwidth]{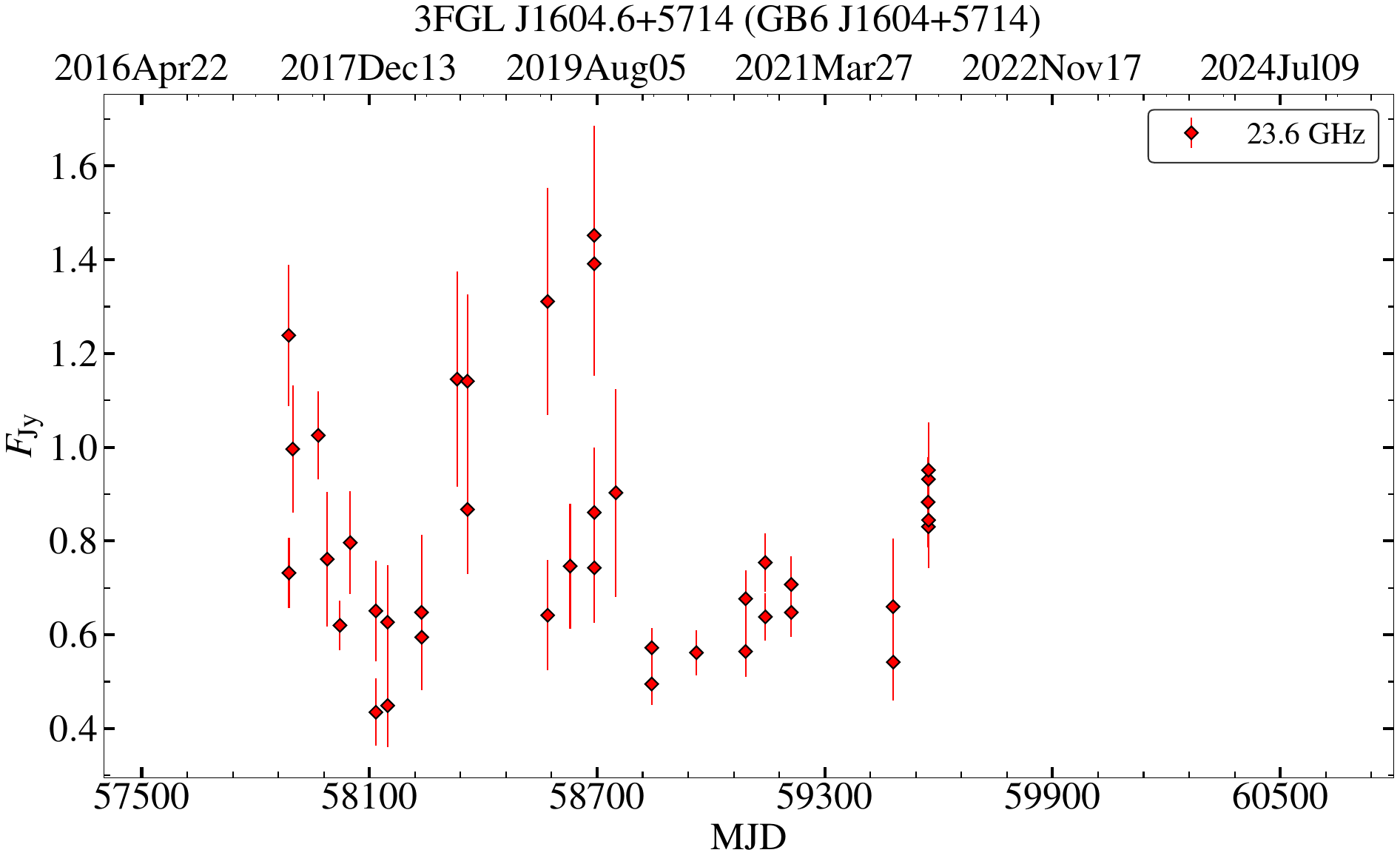}\\
\includegraphics[width=0.49\textwidth]{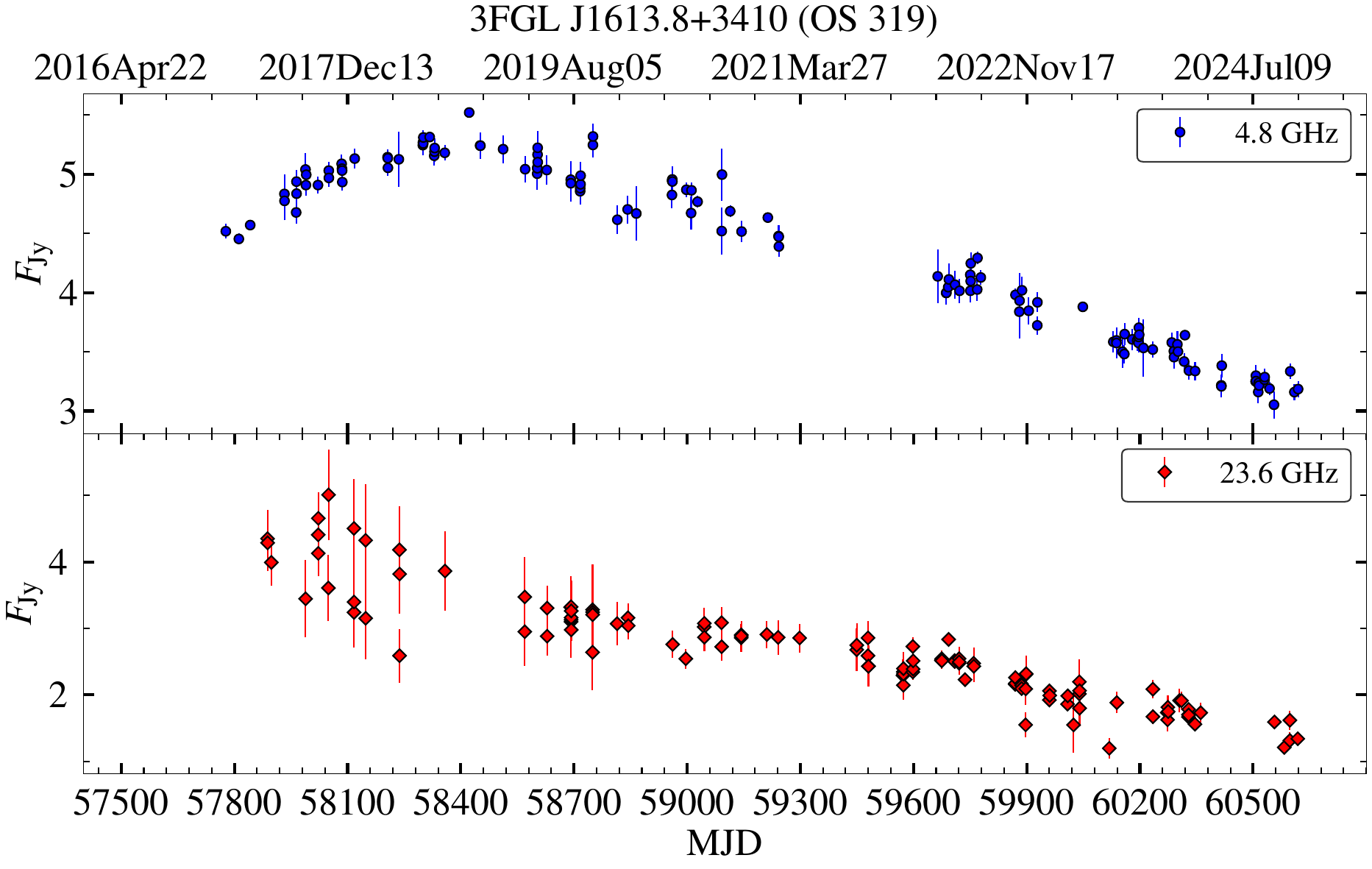}
\includegraphics[width=0.49\textwidth]{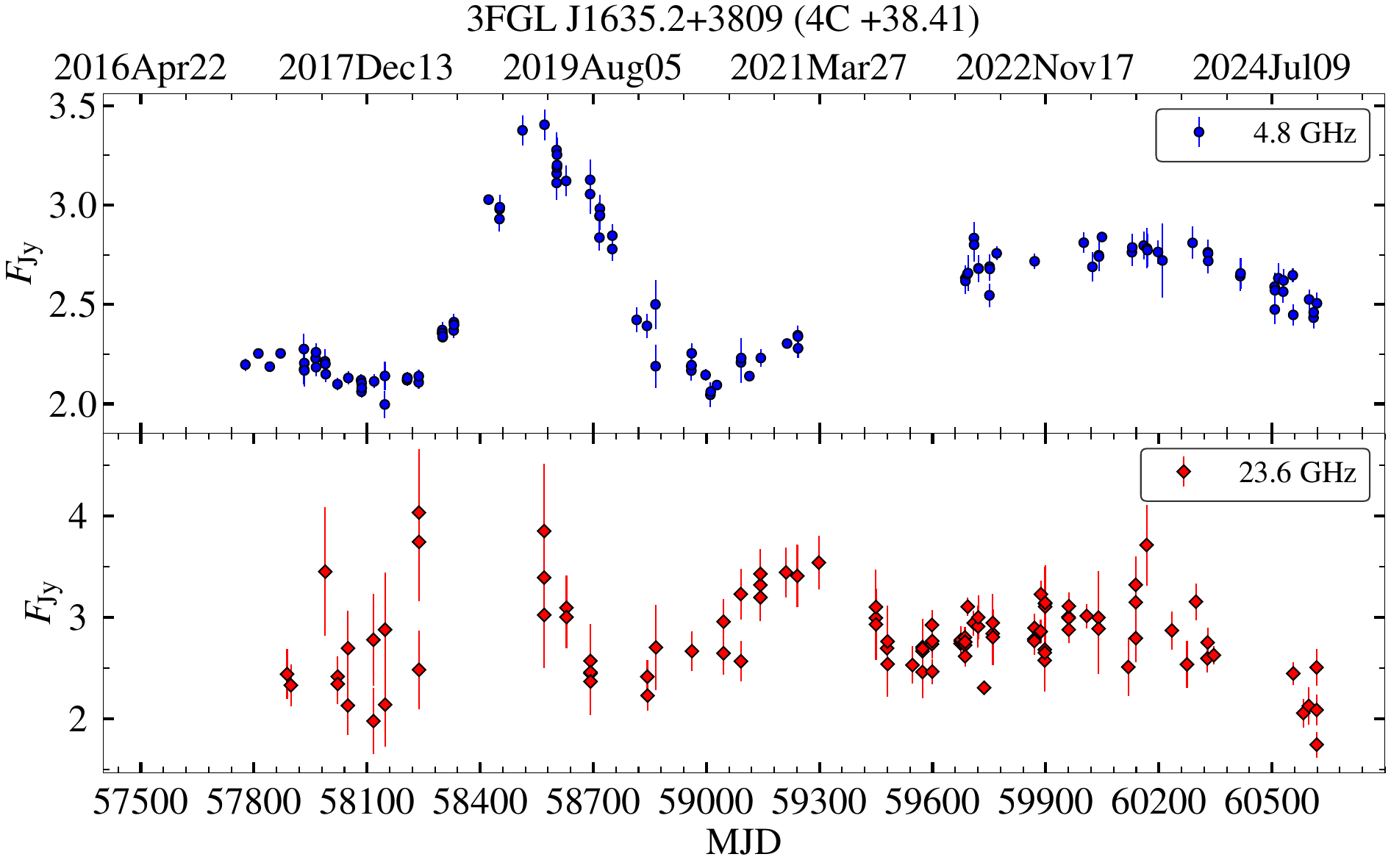}\\
\includegraphics[width=0.49\textwidth]{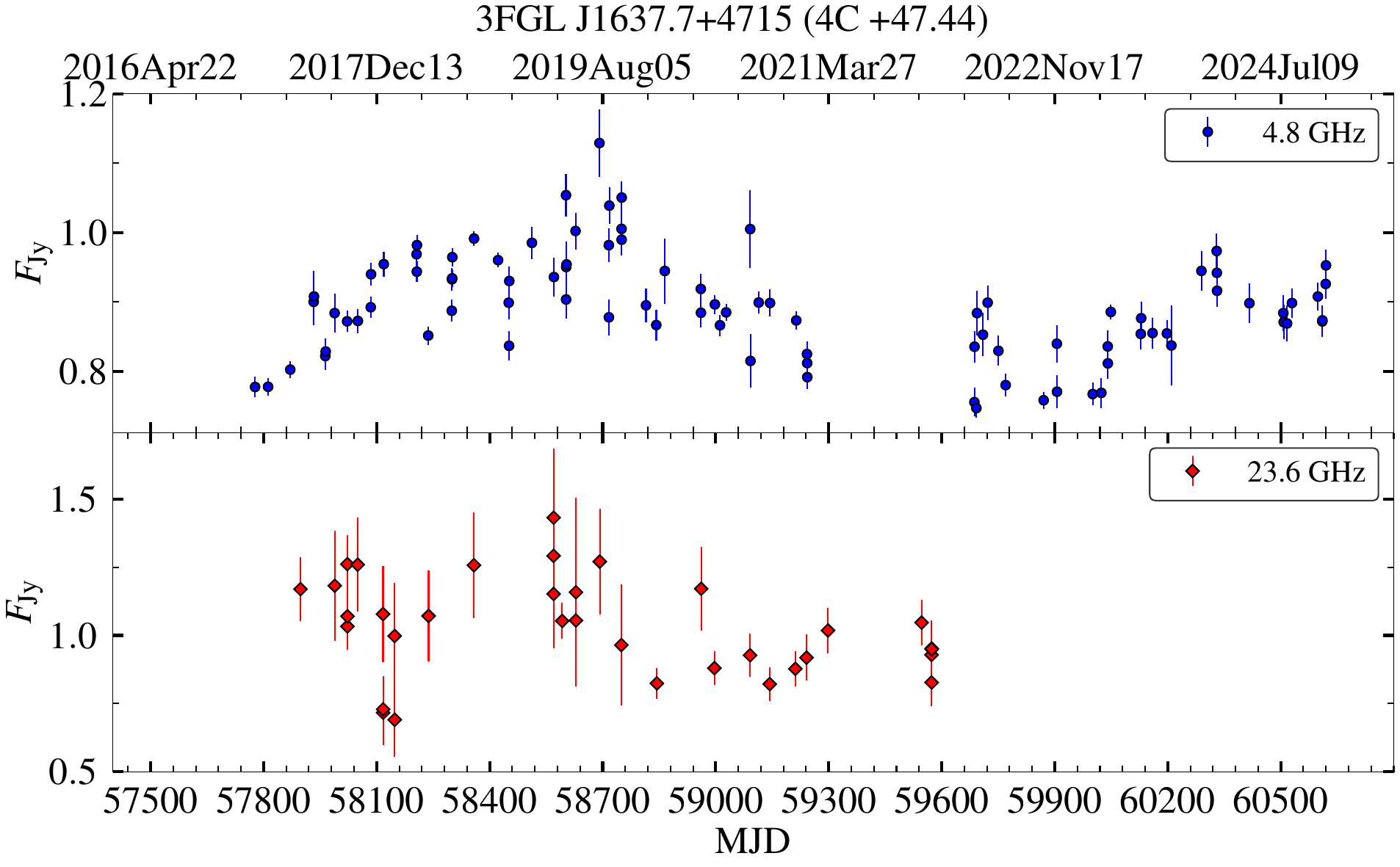}
\includegraphics[width=0.49\textwidth]{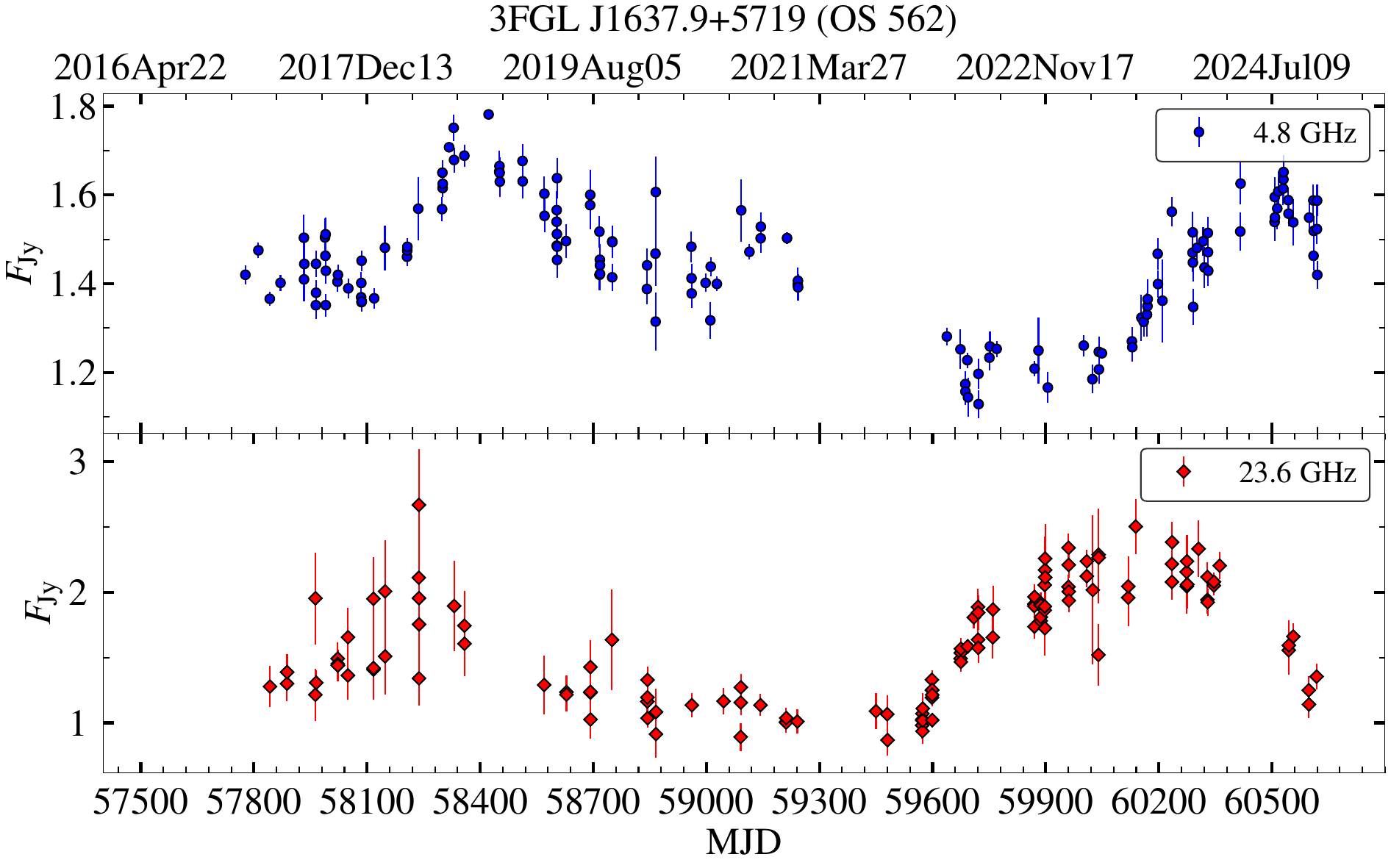}\\
\includegraphics[width=0.49\textwidth]{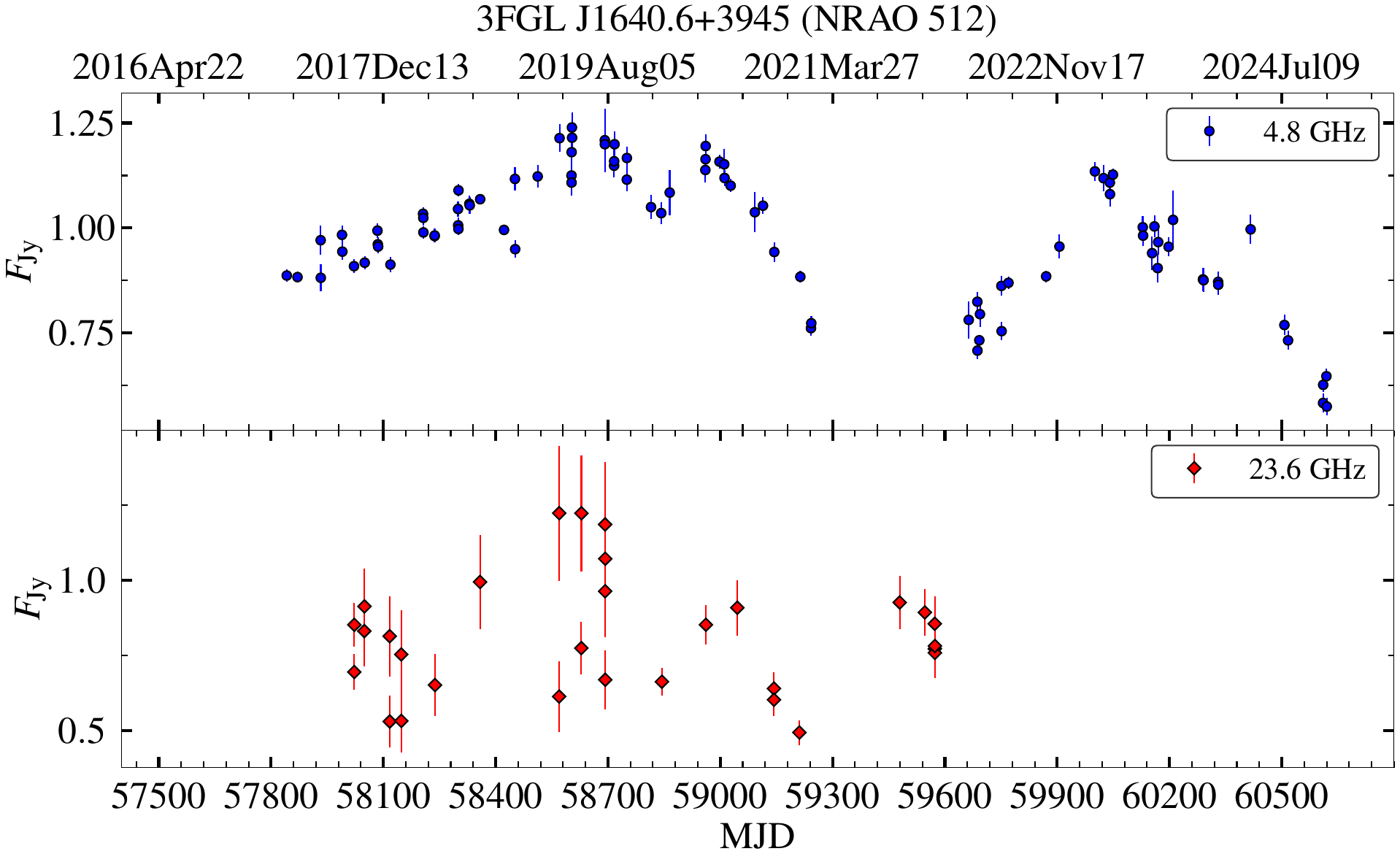}
\includegraphics[width=0.49\textwidth]{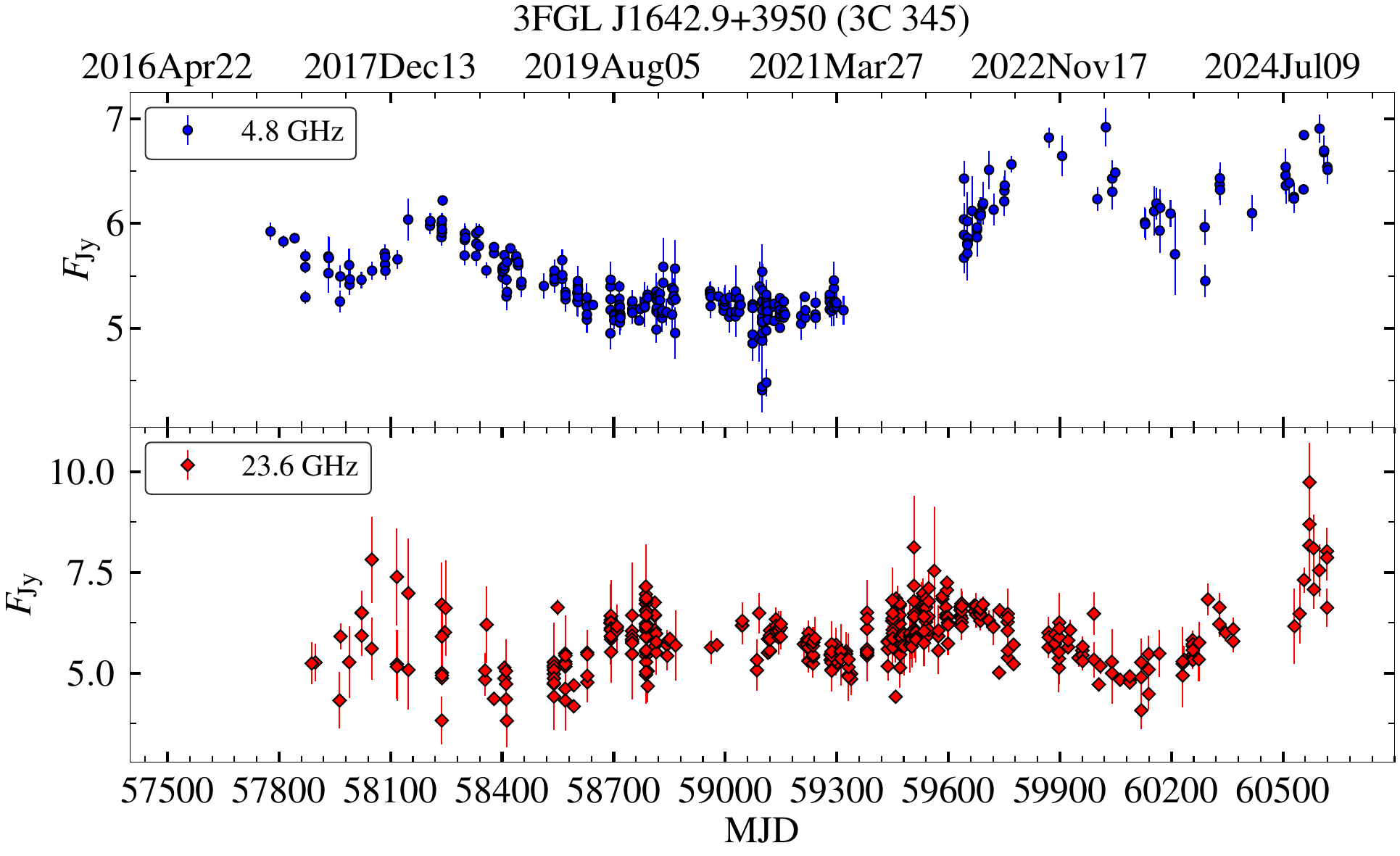}\\
\end{tabular}
\end{figure*}

\begin{figure*}[p]
\centering
\addtocounter{figure}{-1}
\caption{Continued.}
\begin{tabular}{cc}
\includegraphics[width=0.49\textwidth]{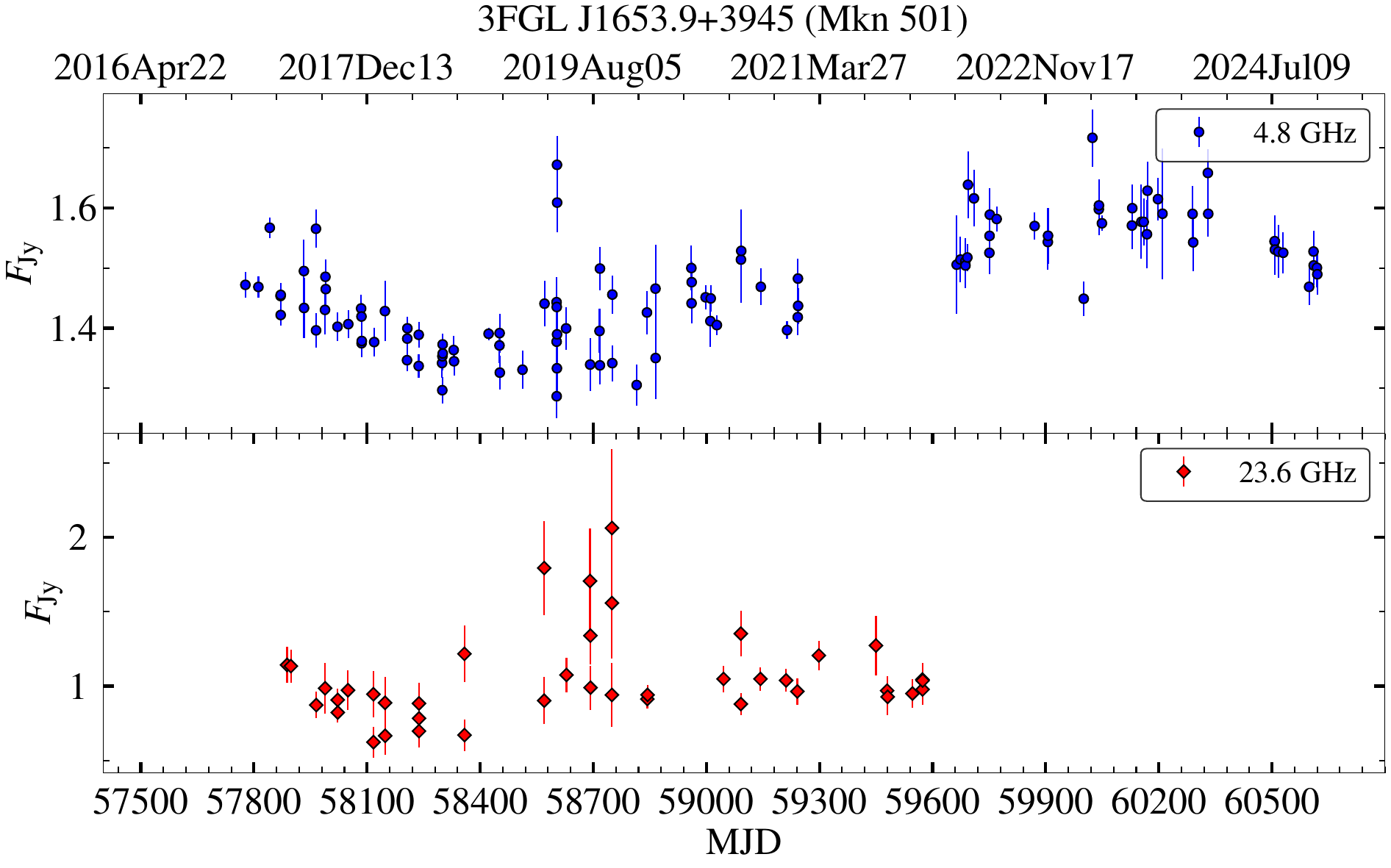}
\includegraphics[width=0.49\textwidth]{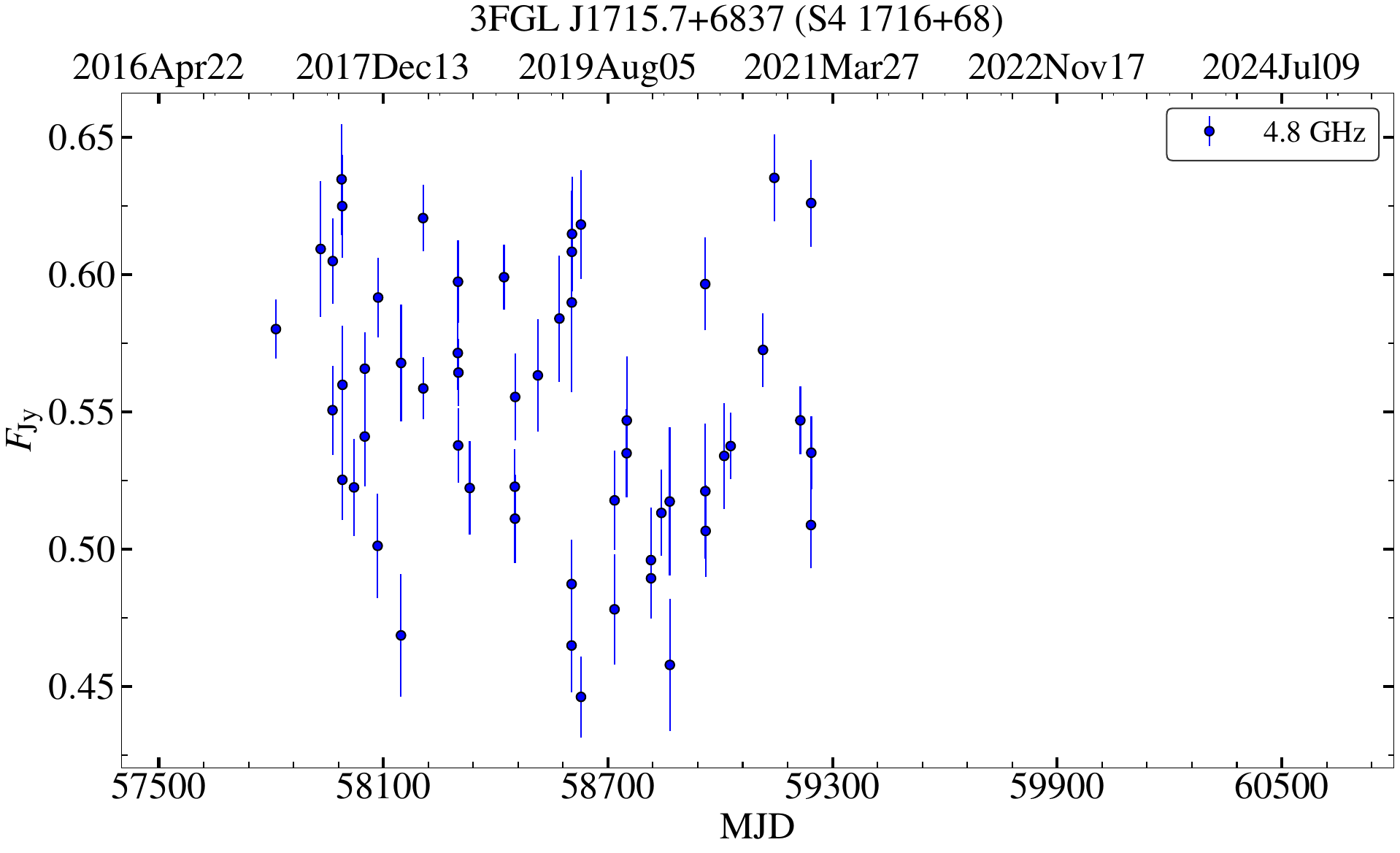}\\
\includegraphics[width=0.49\textwidth]{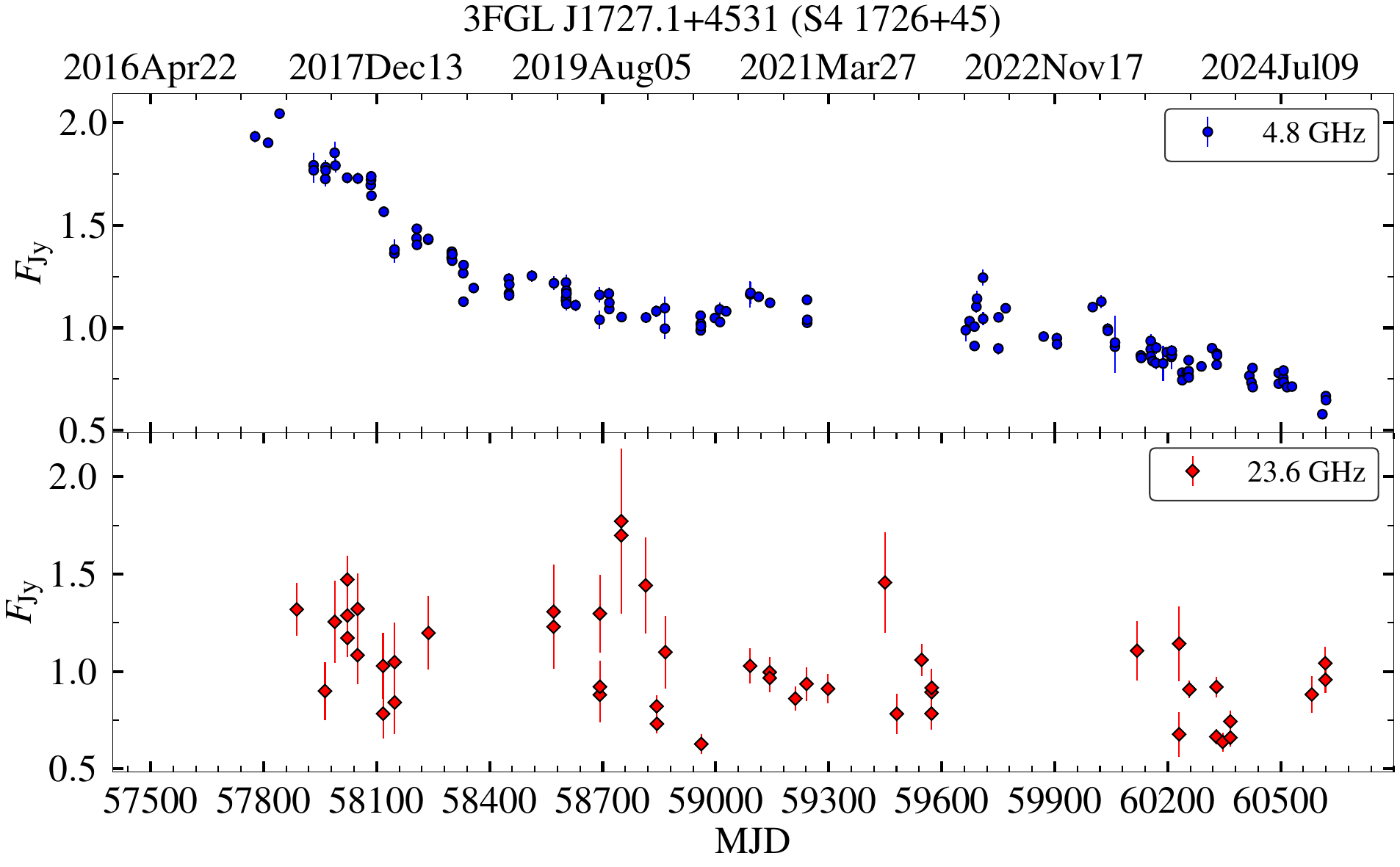}
\includegraphics[width=0.49\textwidth]{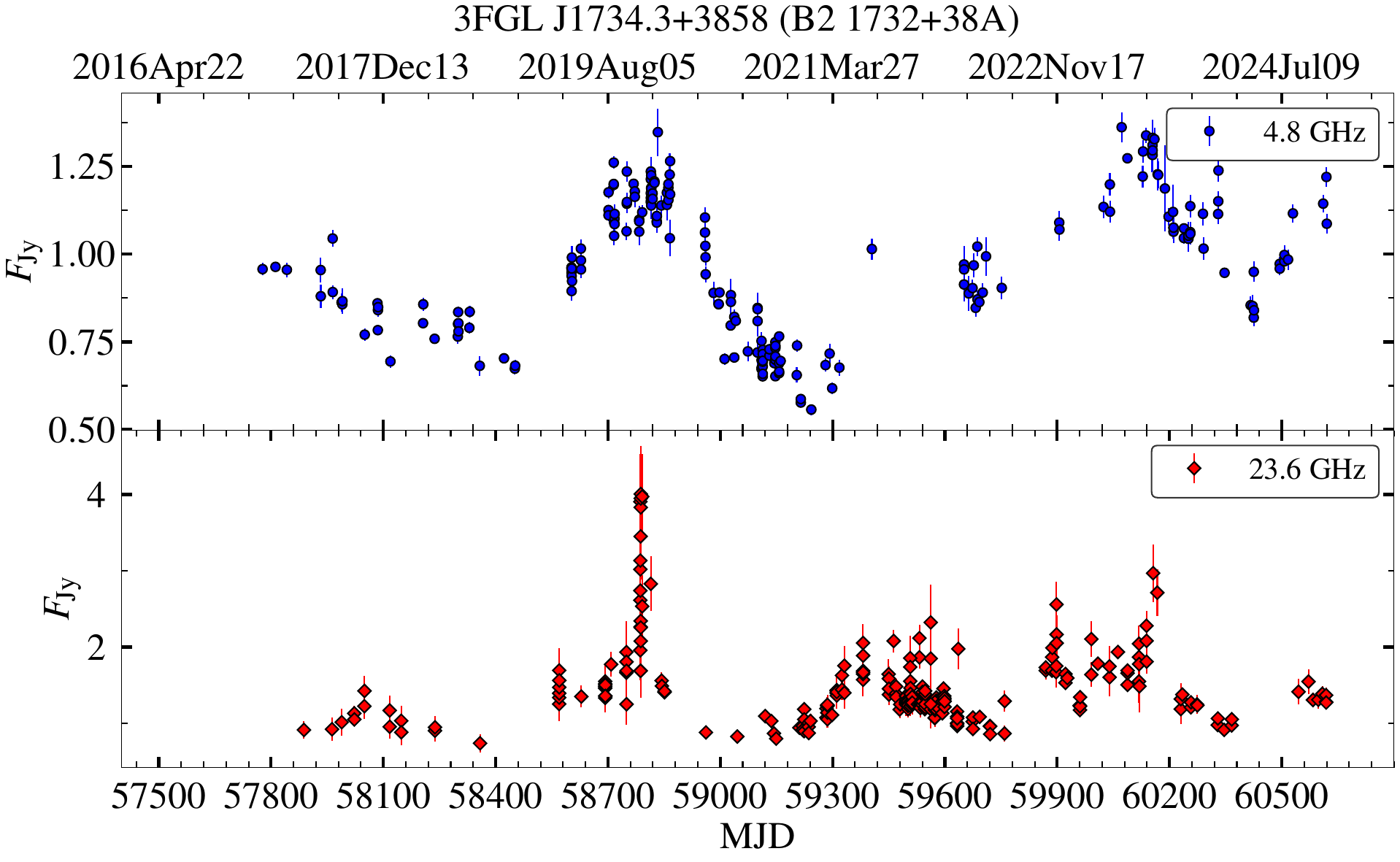}\\
\includegraphics[width=0.49\textwidth]{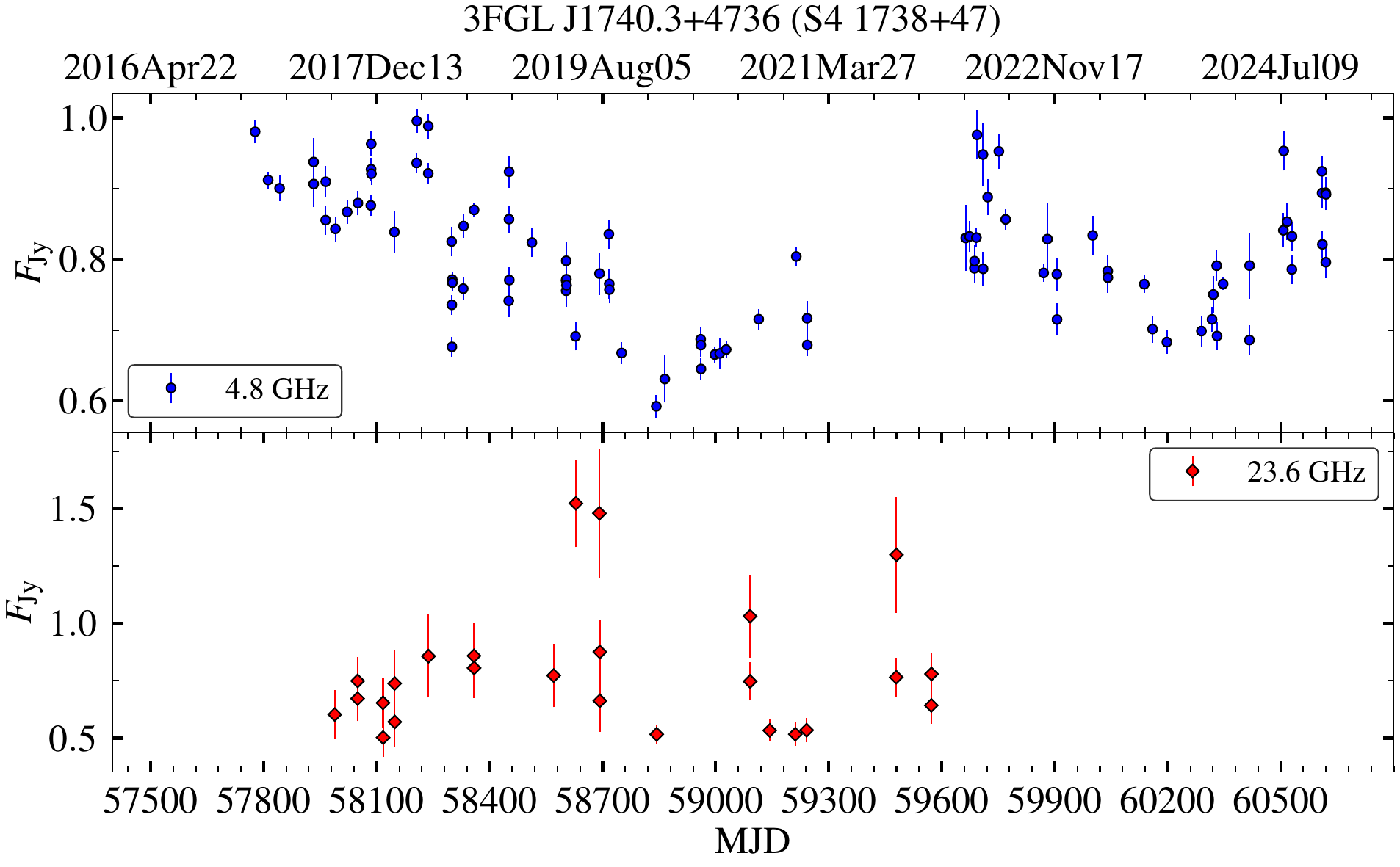}
\includegraphics[width=0.49\textwidth]{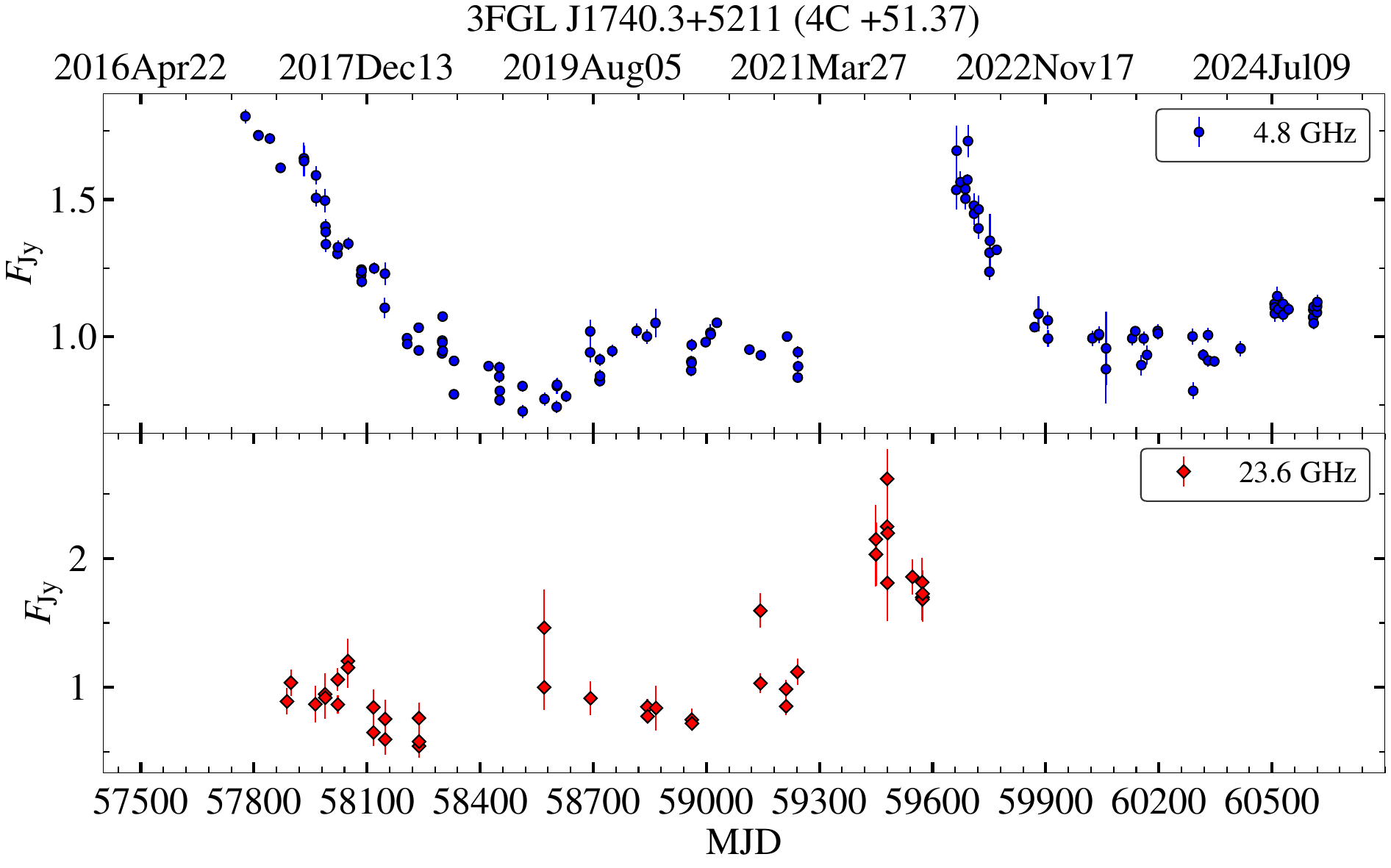}\\
\includegraphics[width=0.49\textwidth]{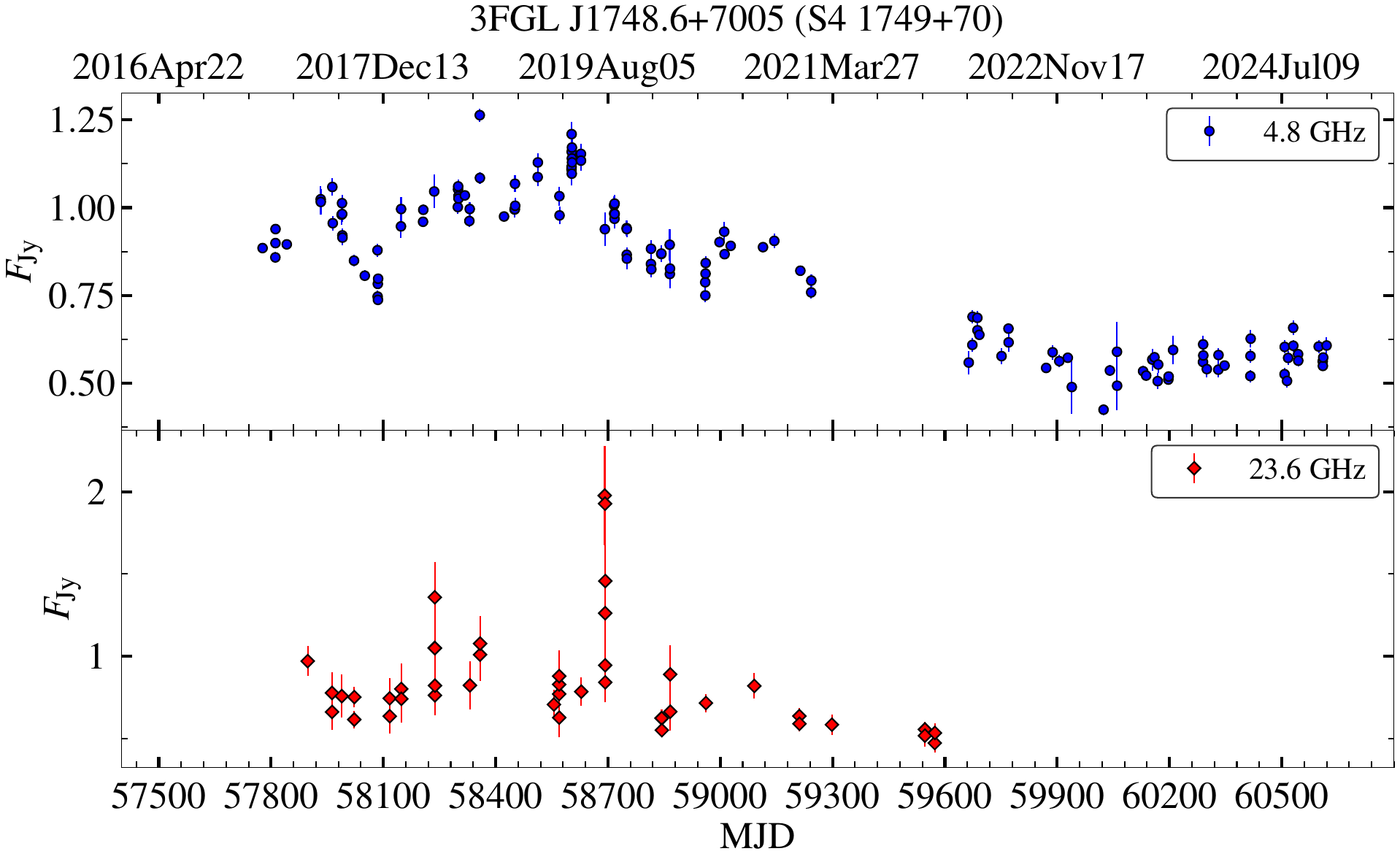}
\includegraphics[width=0.49\textwidth]{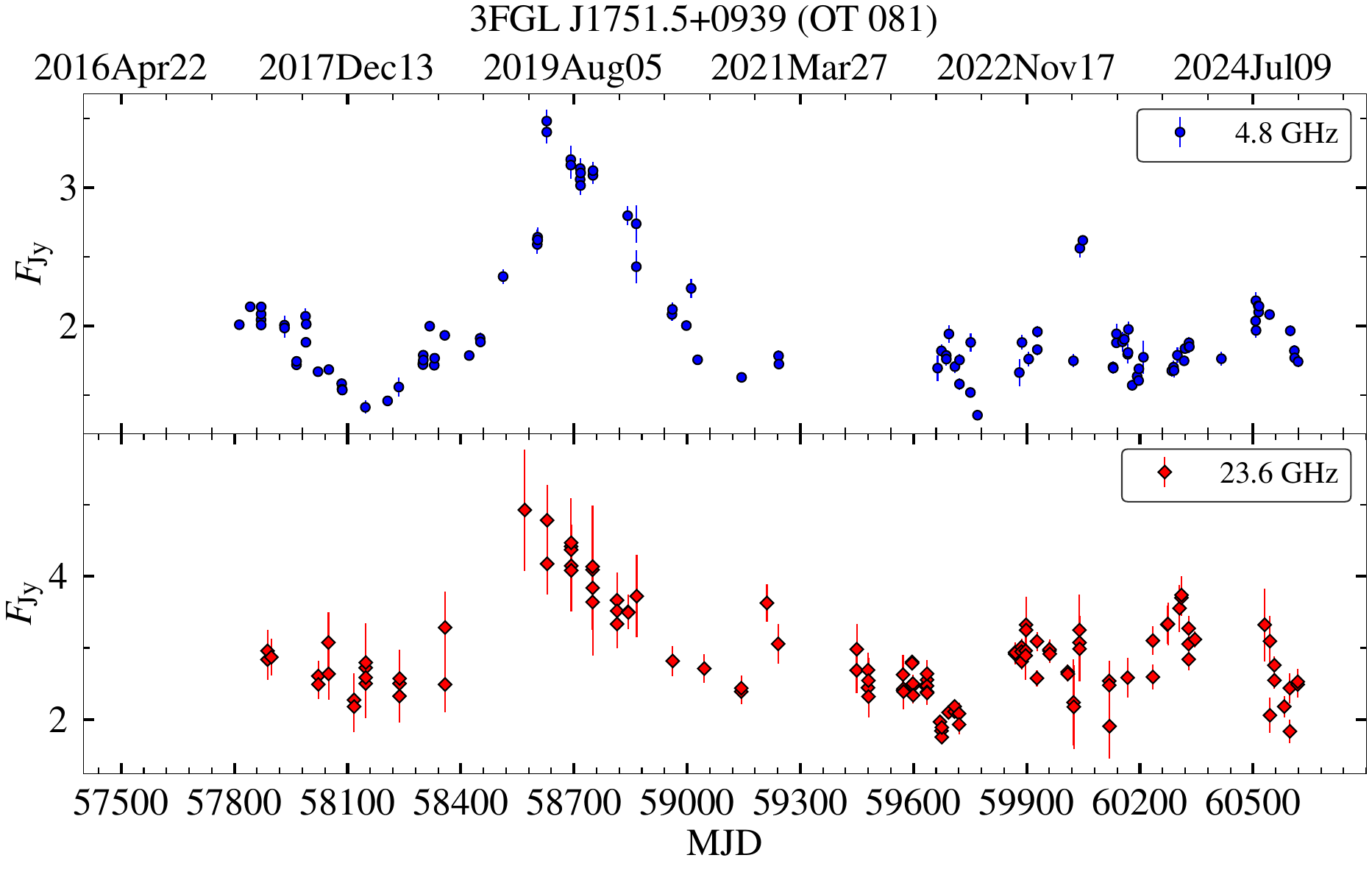}\\
\end{tabular}
\end{figure*}

\begin{figure*}[p]
\centering
\addtocounter{figure}{-1}
\caption{Continued.}
\begin{tabular}{cc}
\includegraphics[width=0.49\textwidth]{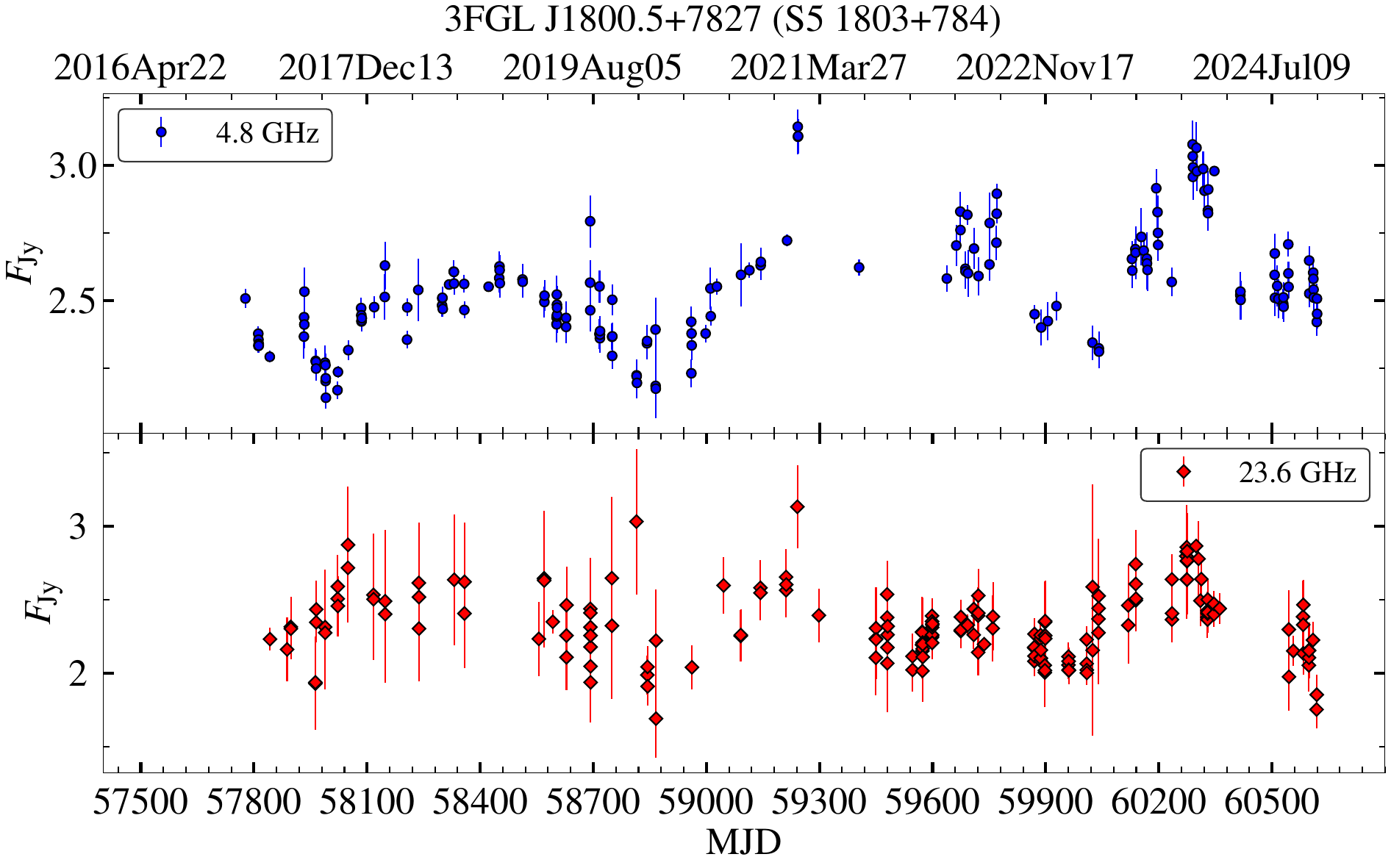}
\includegraphics[width=0.49\textwidth]{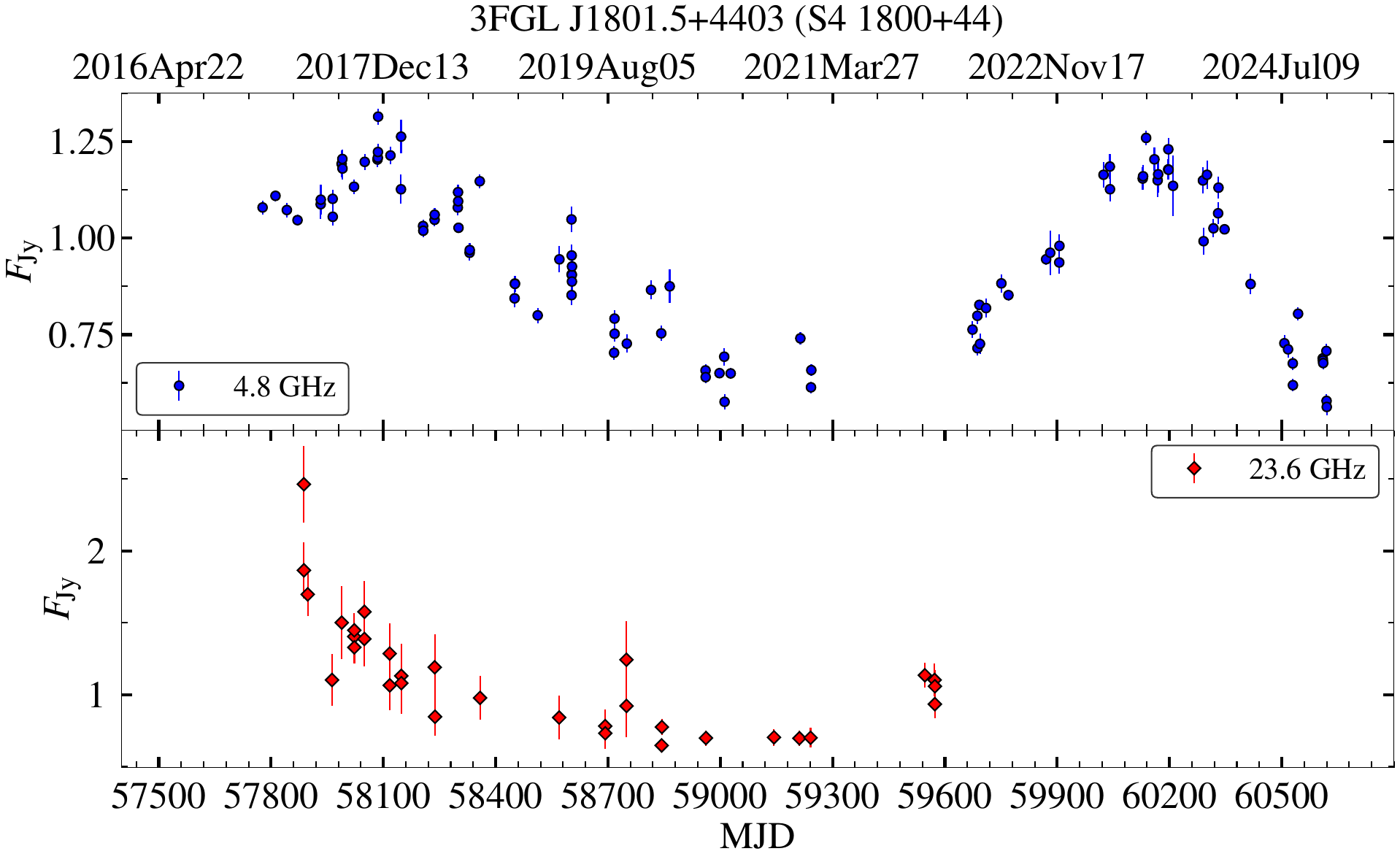}\\
\includegraphics[width=0.49\textwidth]{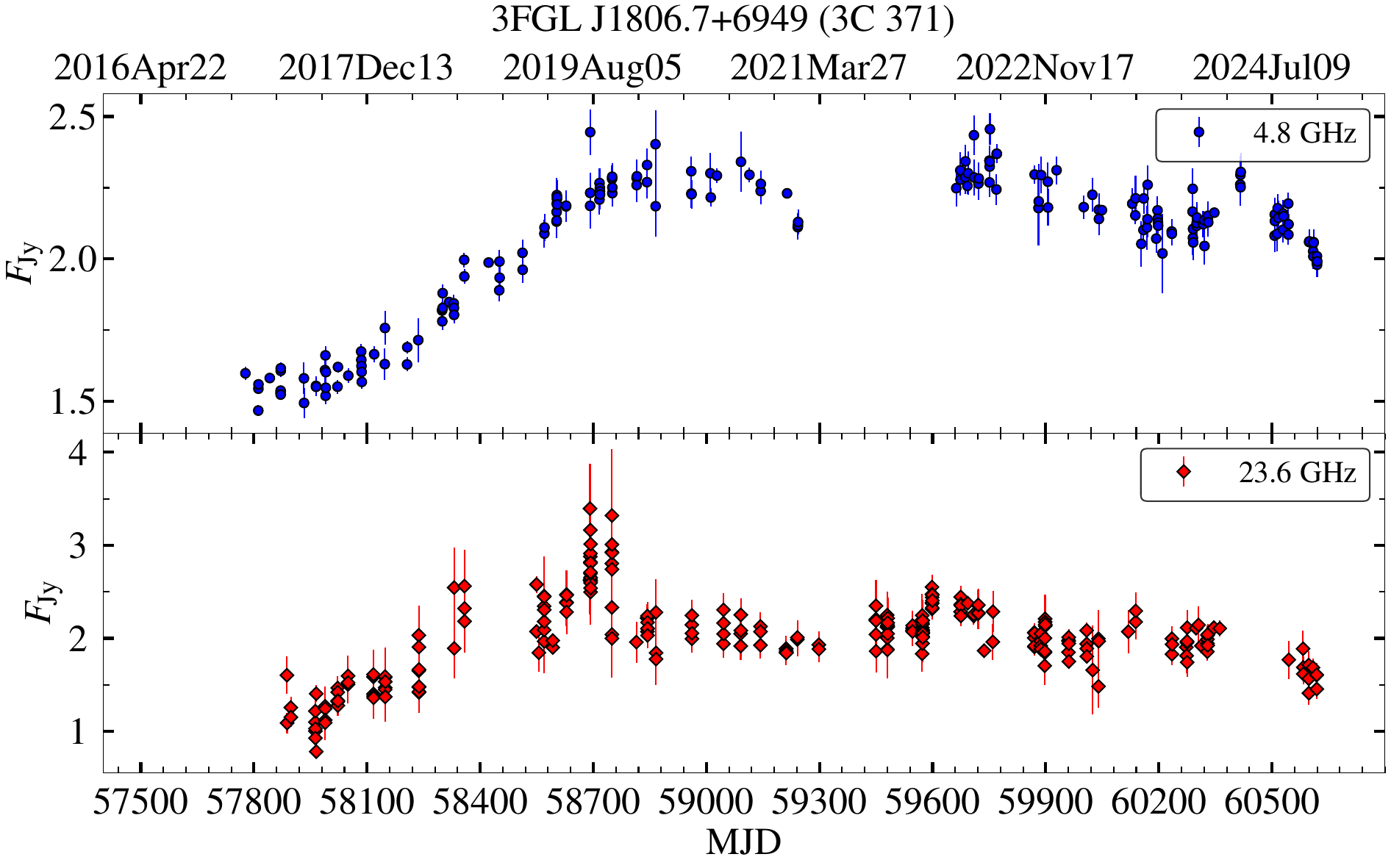}
\includegraphics[width=0.49\textwidth]{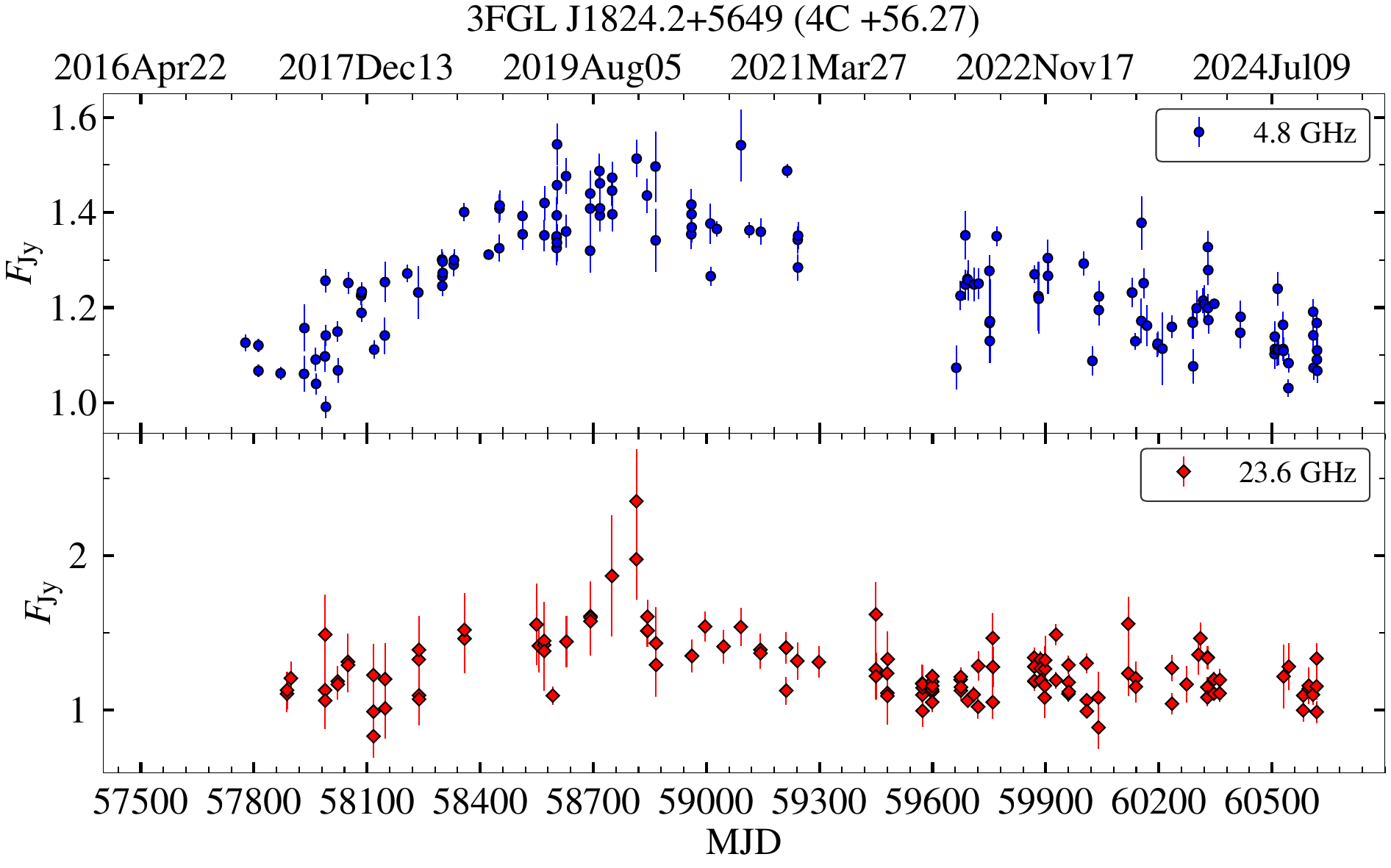}\\
\includegraphics[width=0.49\textwidth]{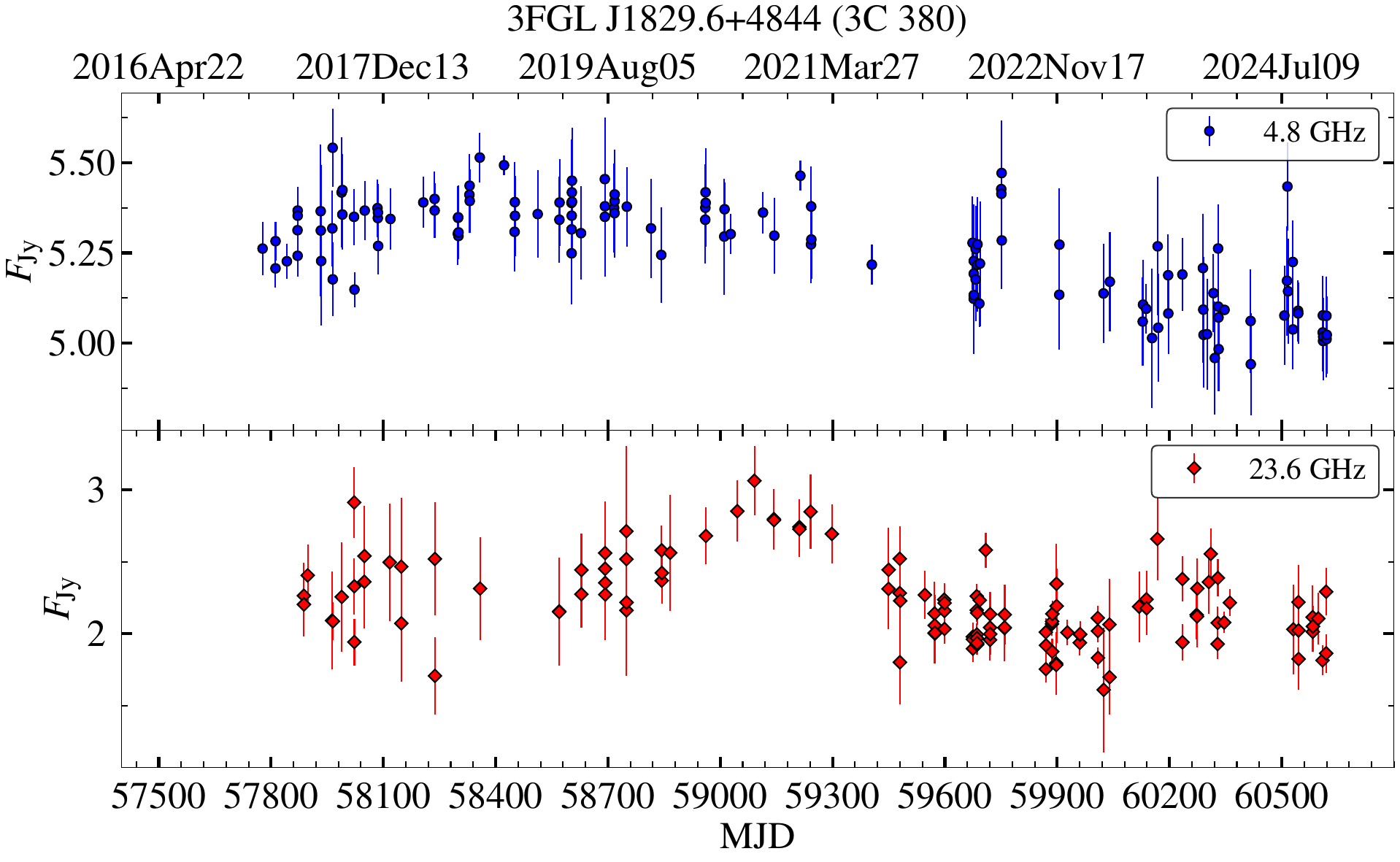}
\includegraphics[width=0.49\textwidth]{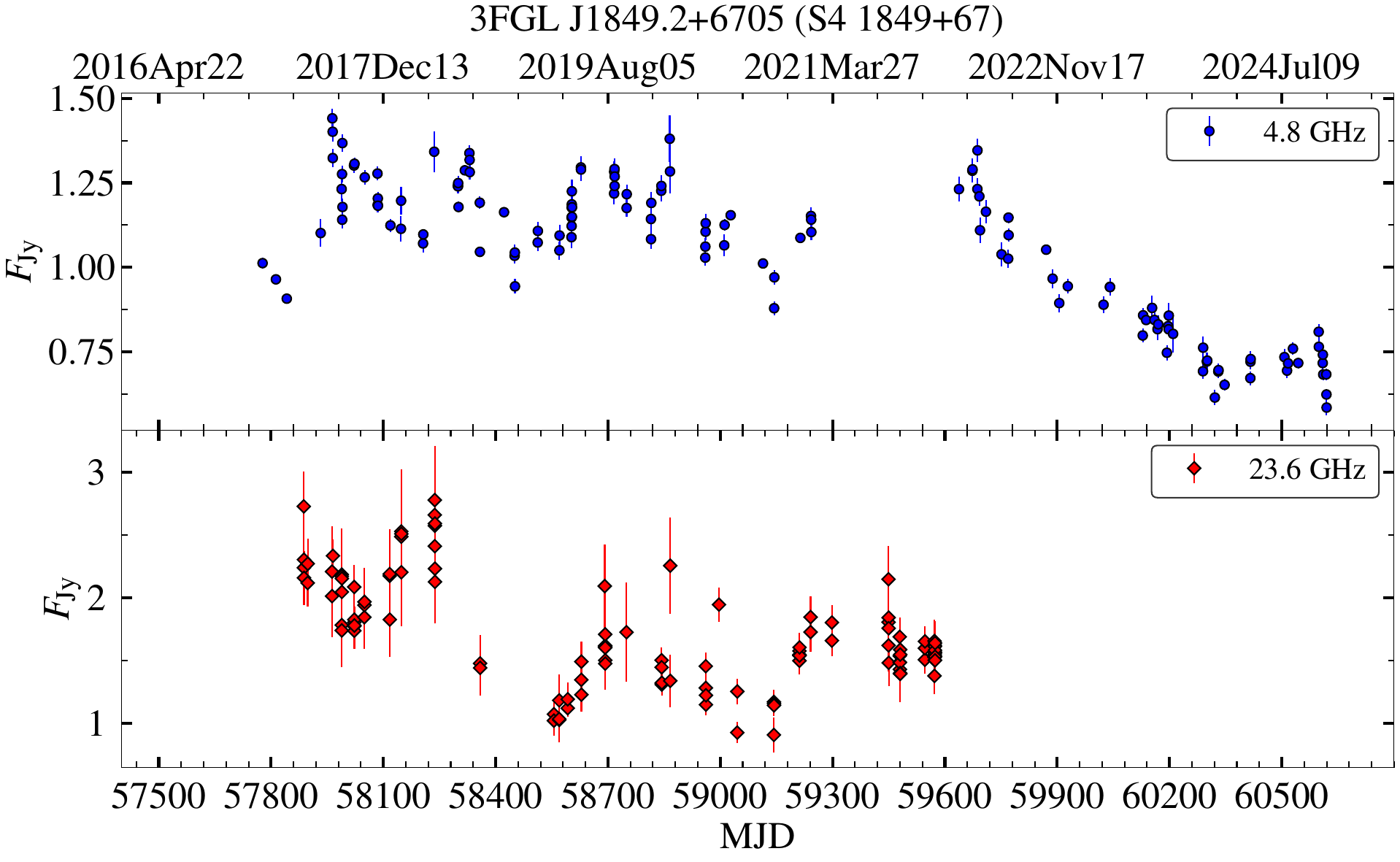}\\
\includegraphics[width=0.49\textwidth]{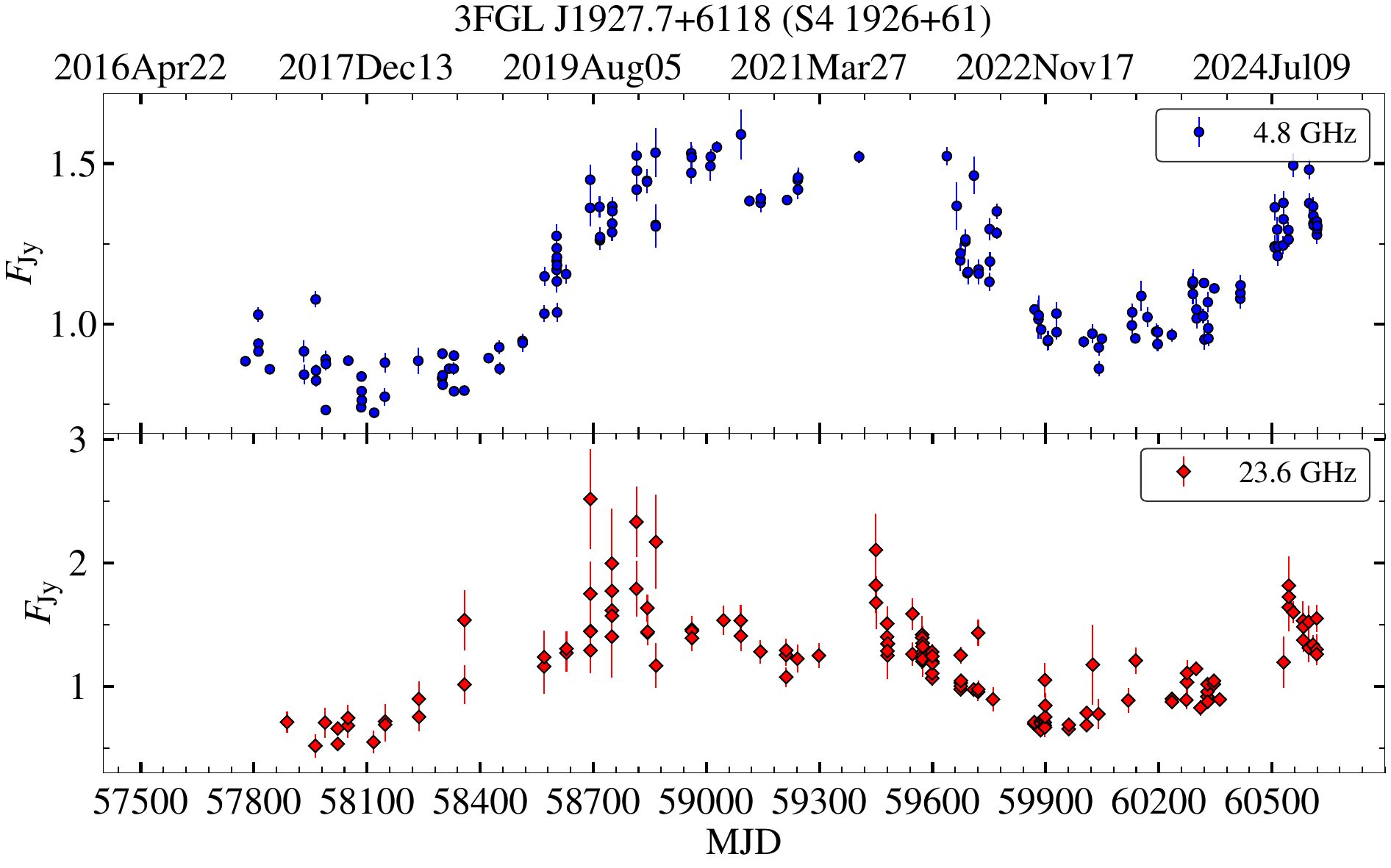}
\includegraphics[width=0.49\textwidth]{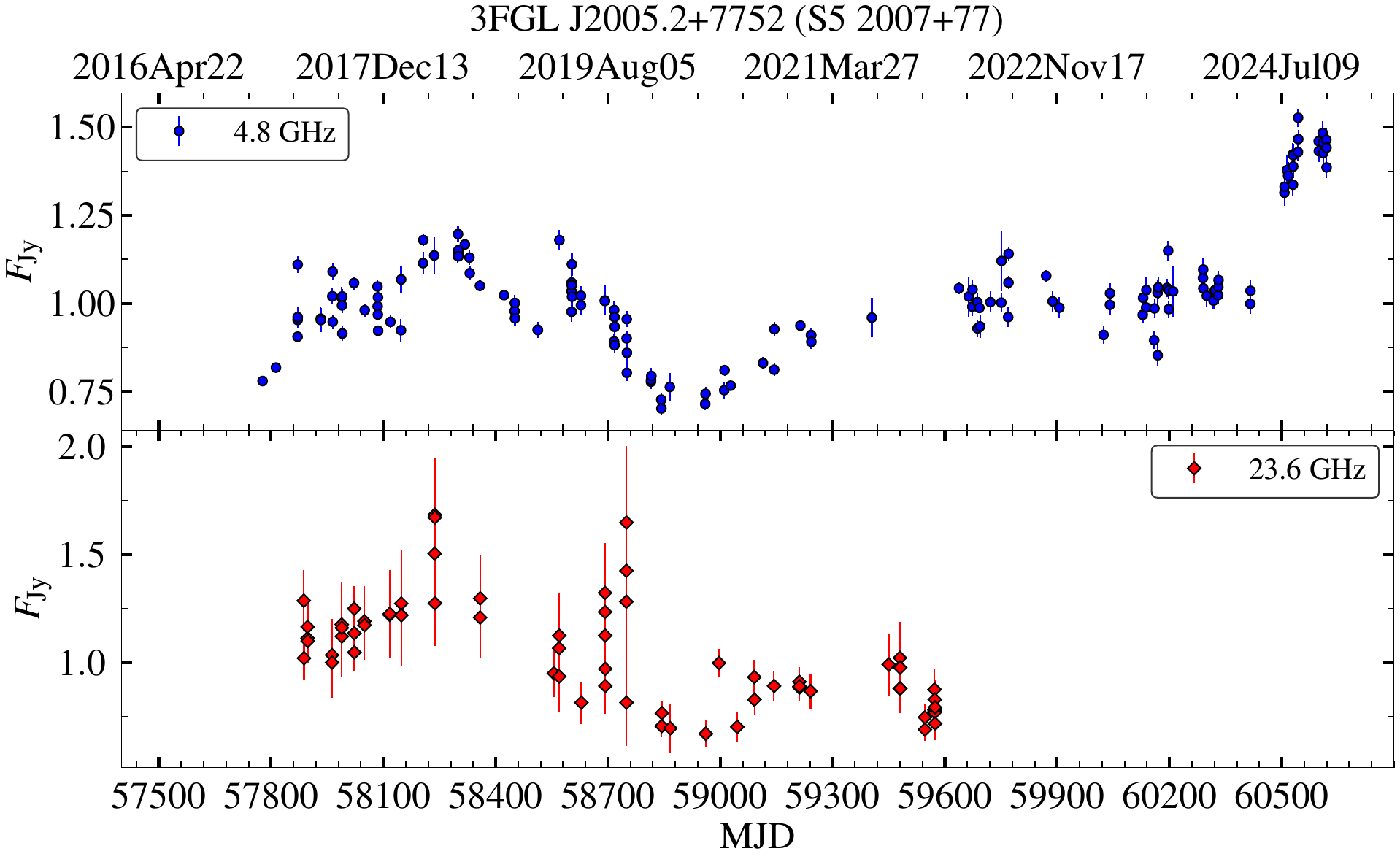}\\
\end{tabular}
\end{figure*}

\begin{figure*}[p]
\centering
\addtocounter{figure}{-1}
\caption{Continued.}
\begin{tabular}{cc}
\includegraphics[width=0.49\textwidth]{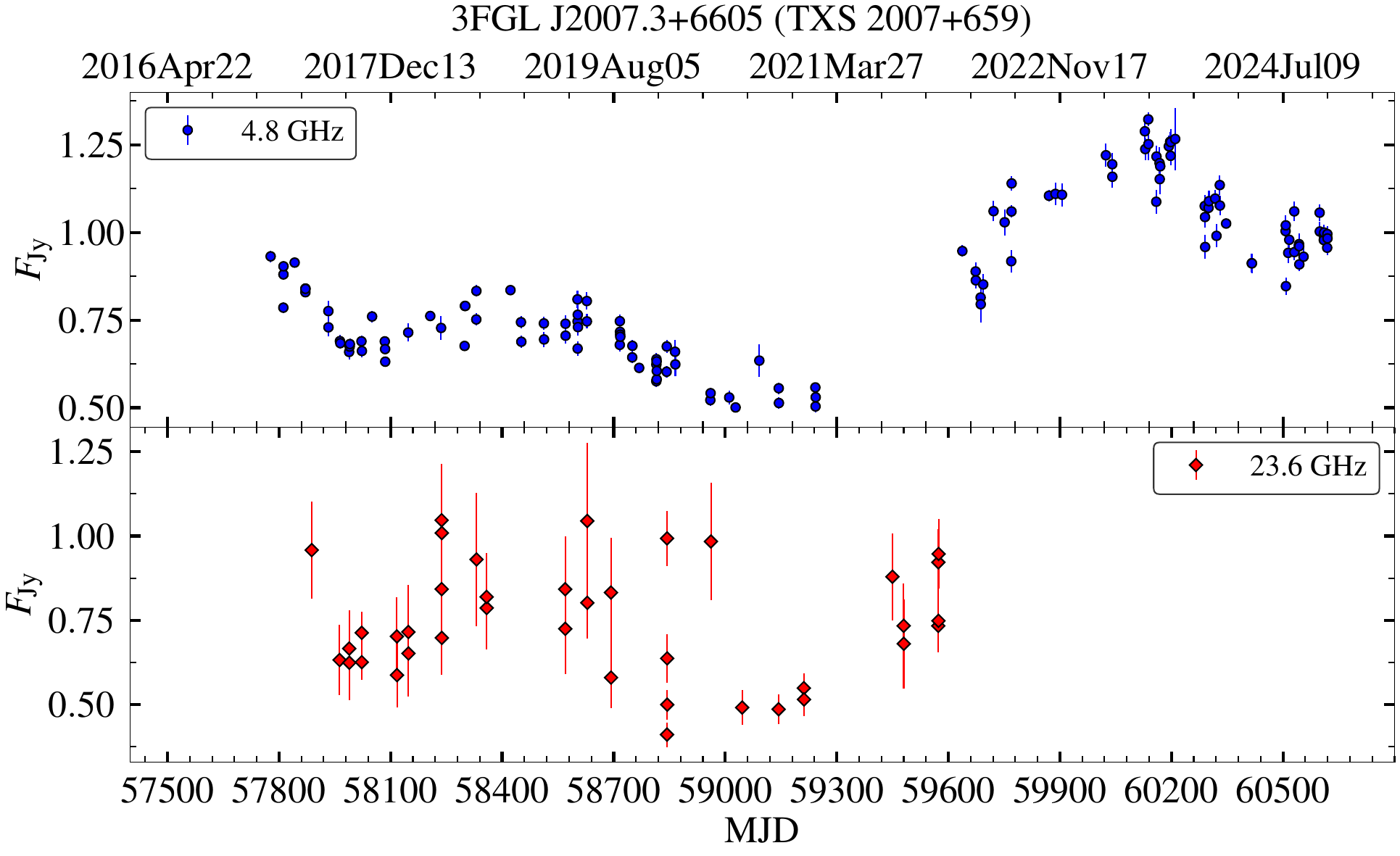}
\includegraphics[width=0.49\textwidth]{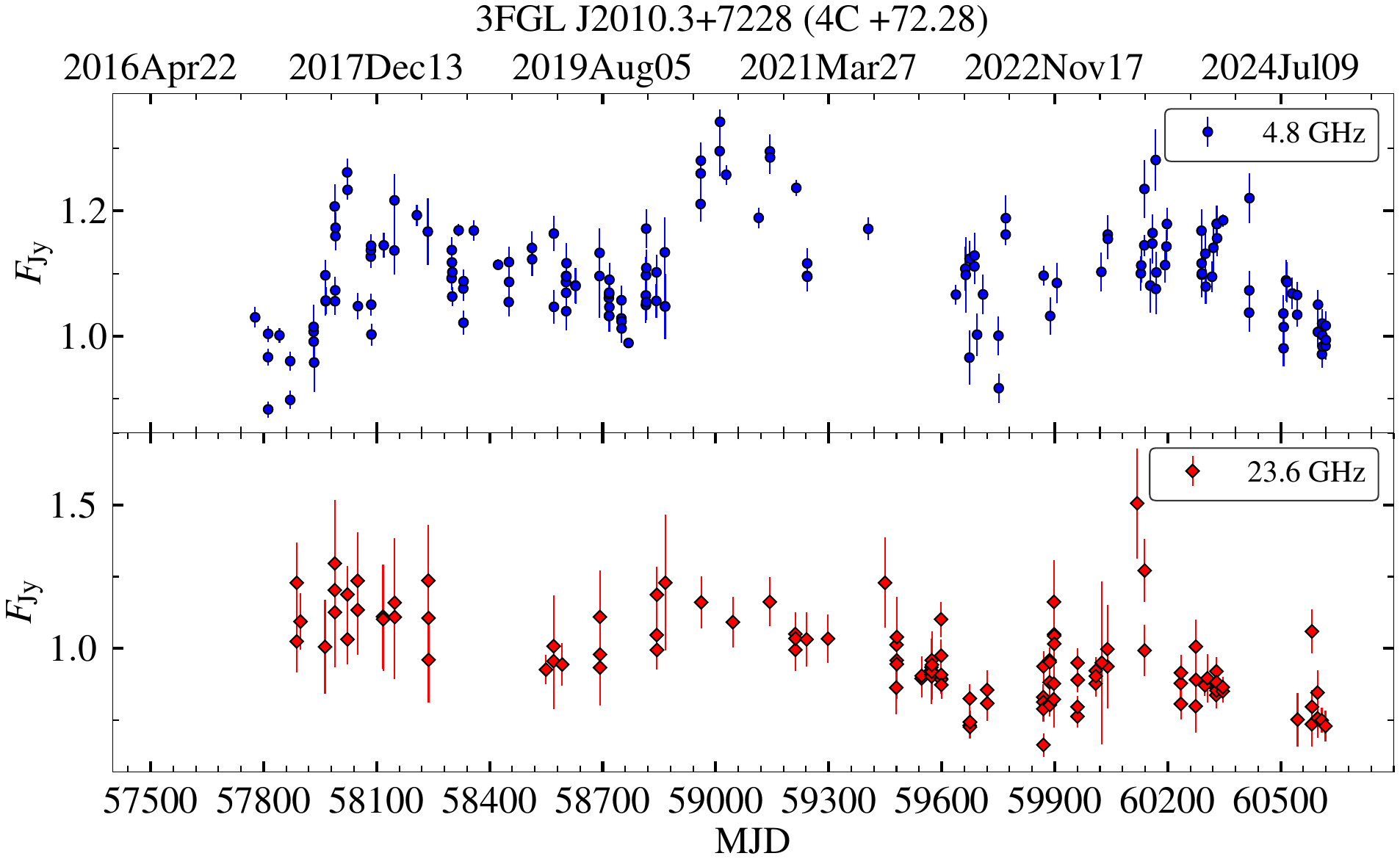}\\
\includegraphics[width=0.49\textwidth]{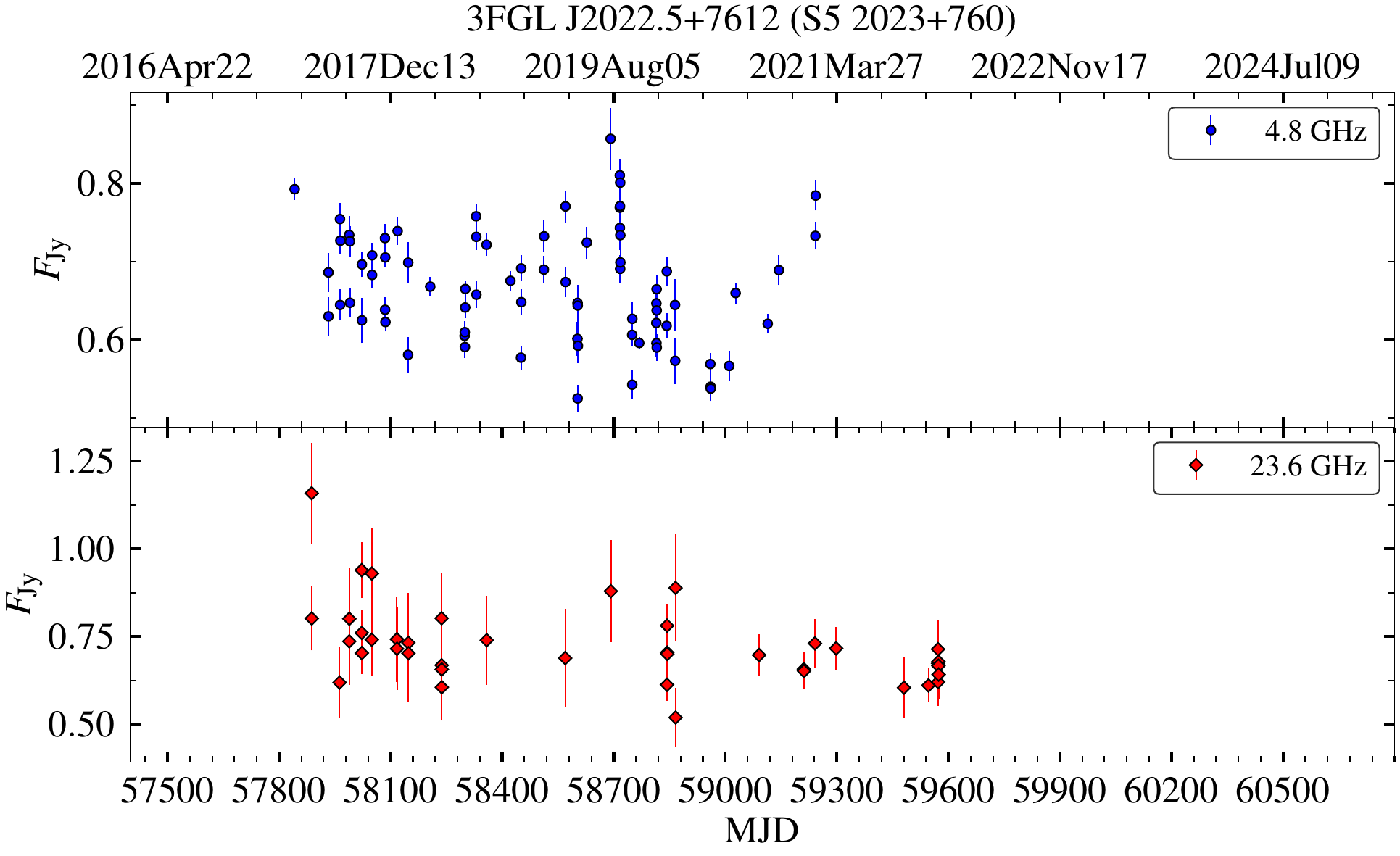}
\includegraphics[width=0.49\textwidth]{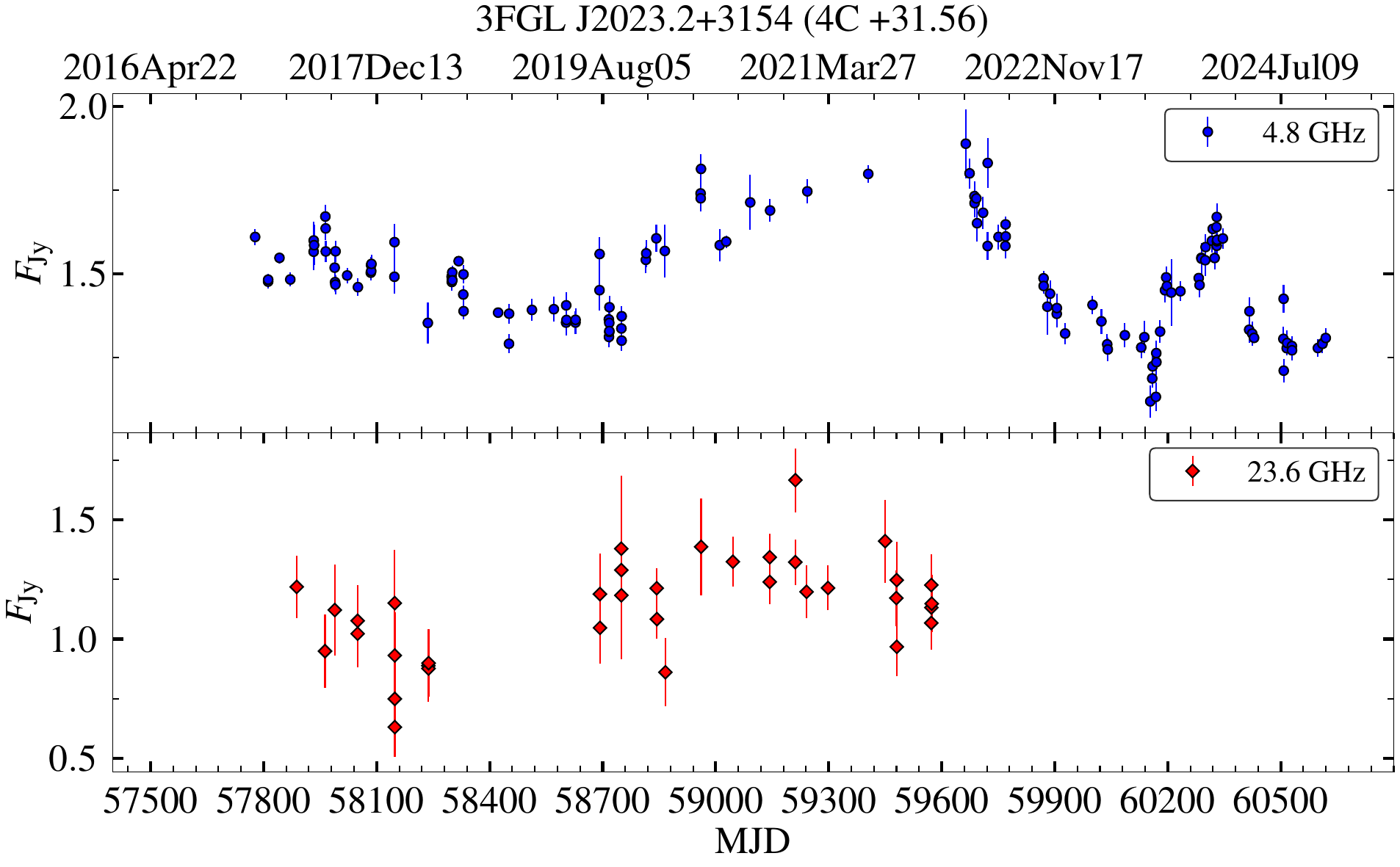}\\
\includegraphics[width=0.49\textwidth]{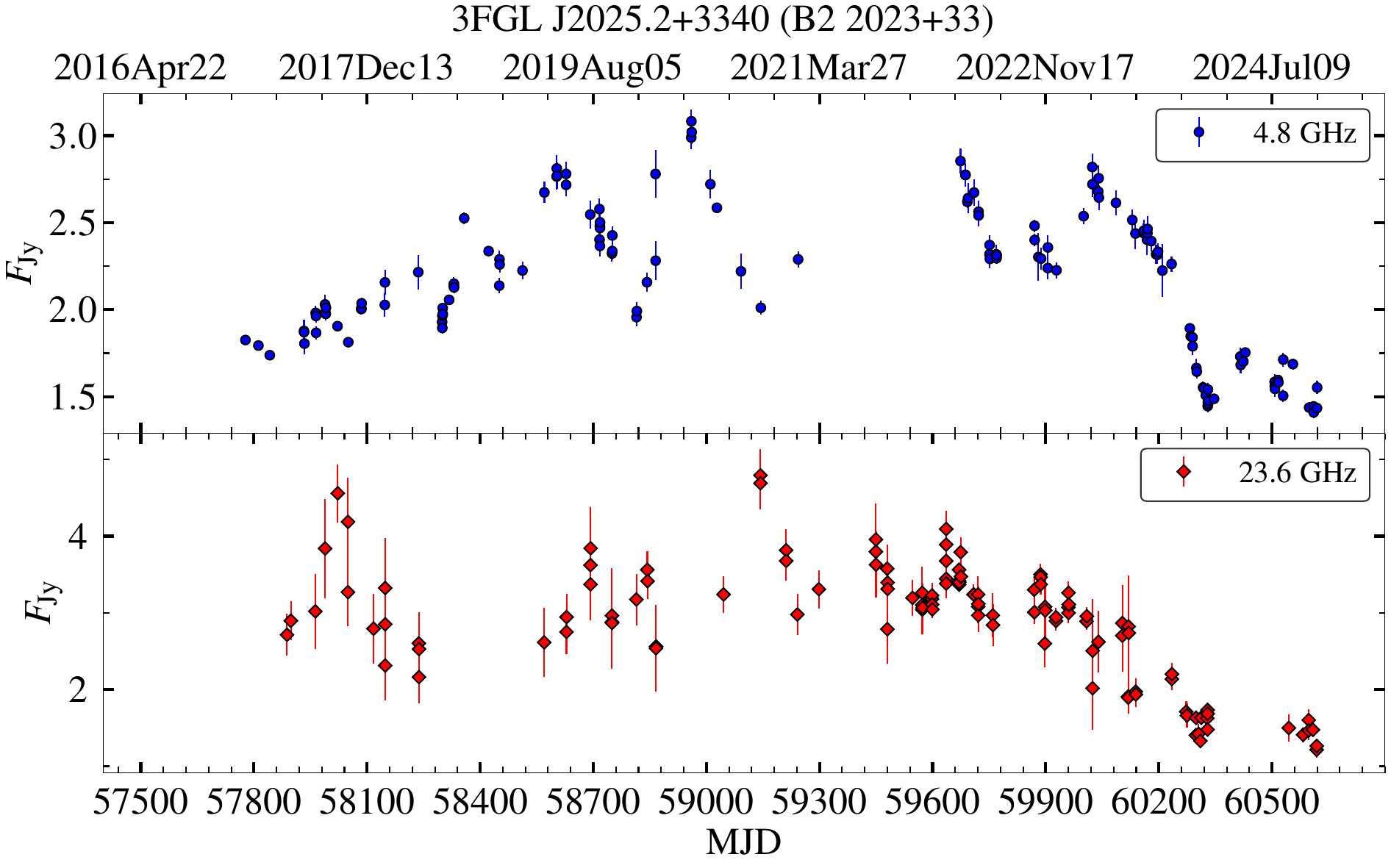}
\includegraphics[width=0.49\textwidth]{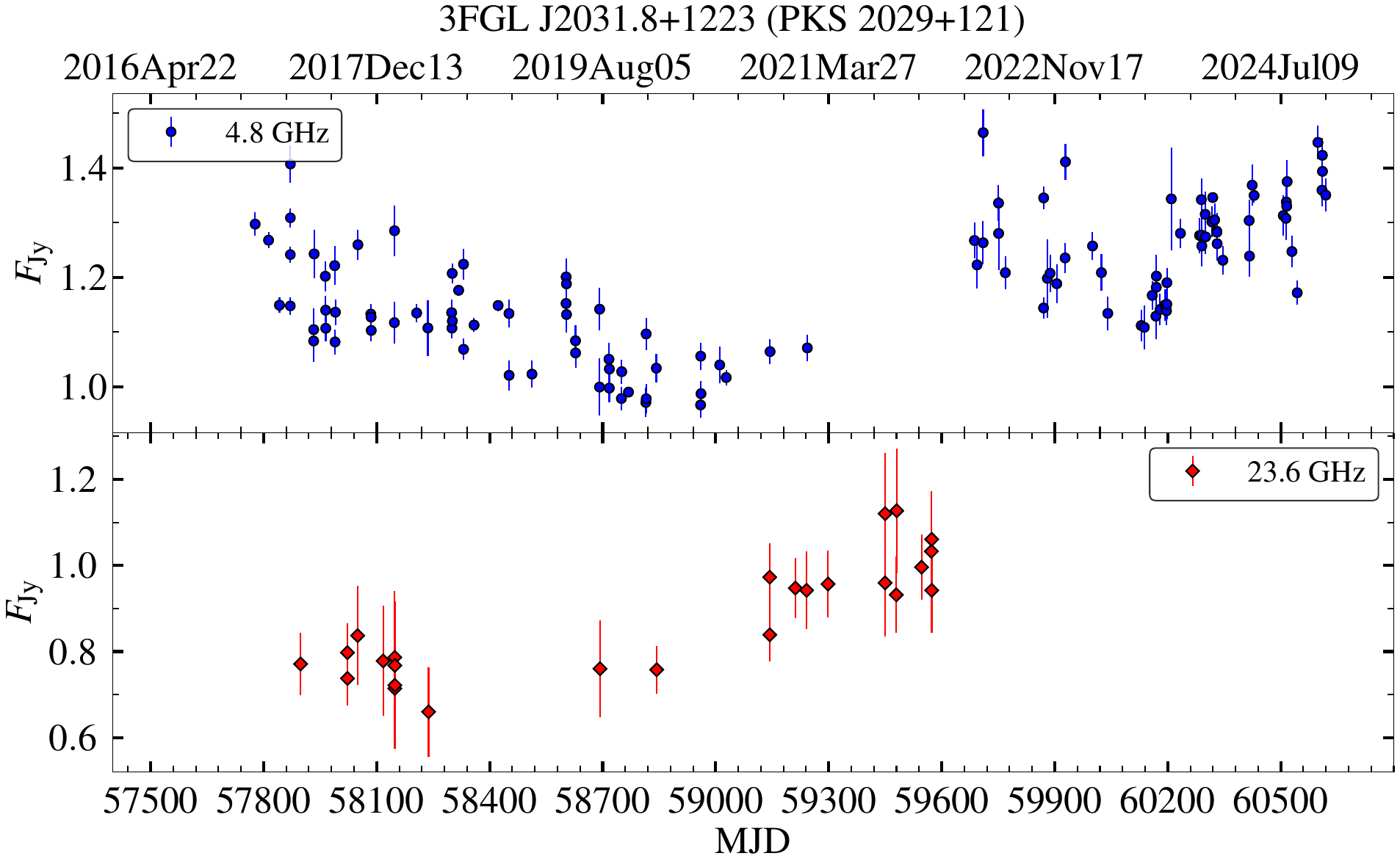}\\
\includegraphics[width=0.49\textwidth]{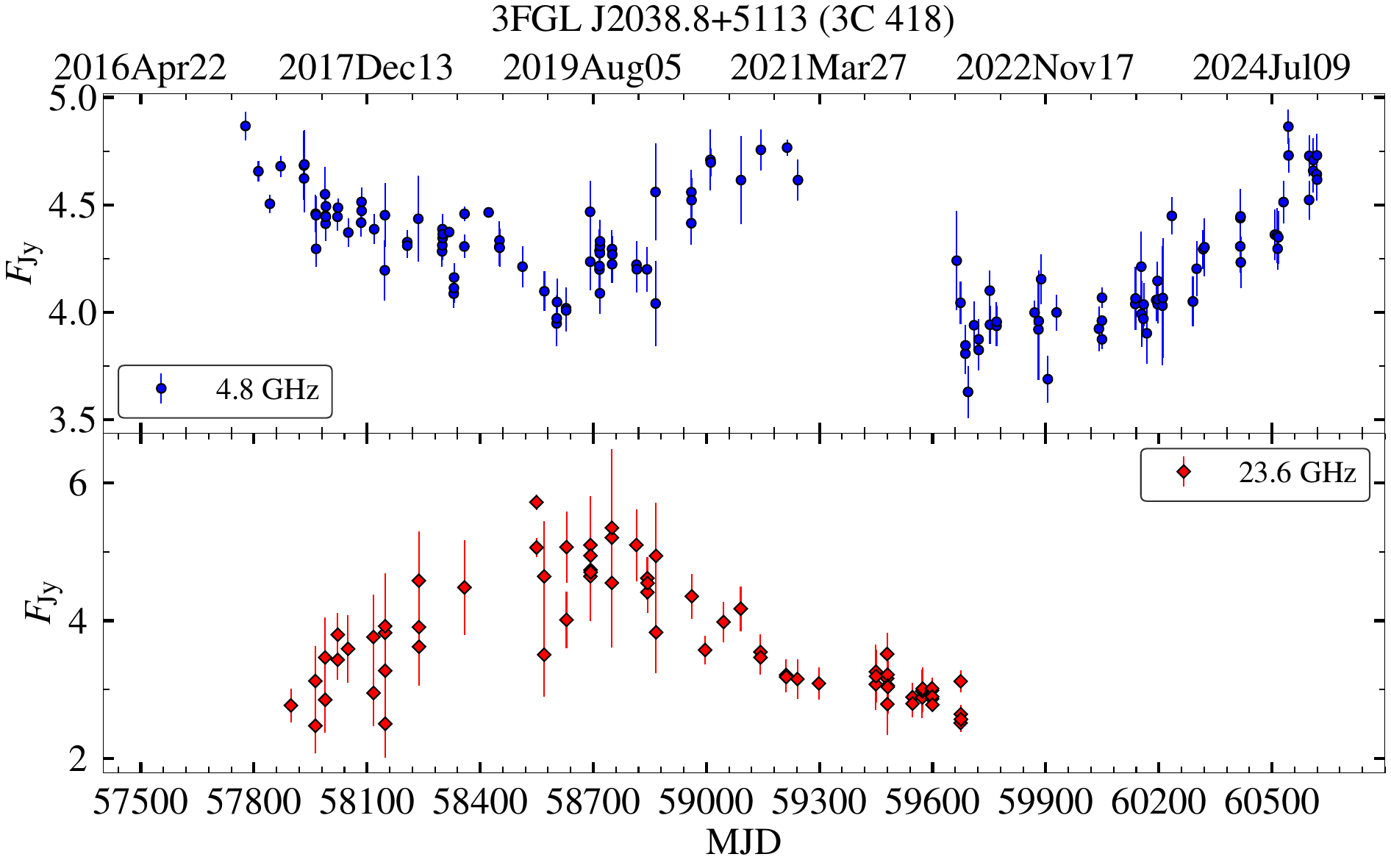}
\includegraphics[width=0.49\textwidth]{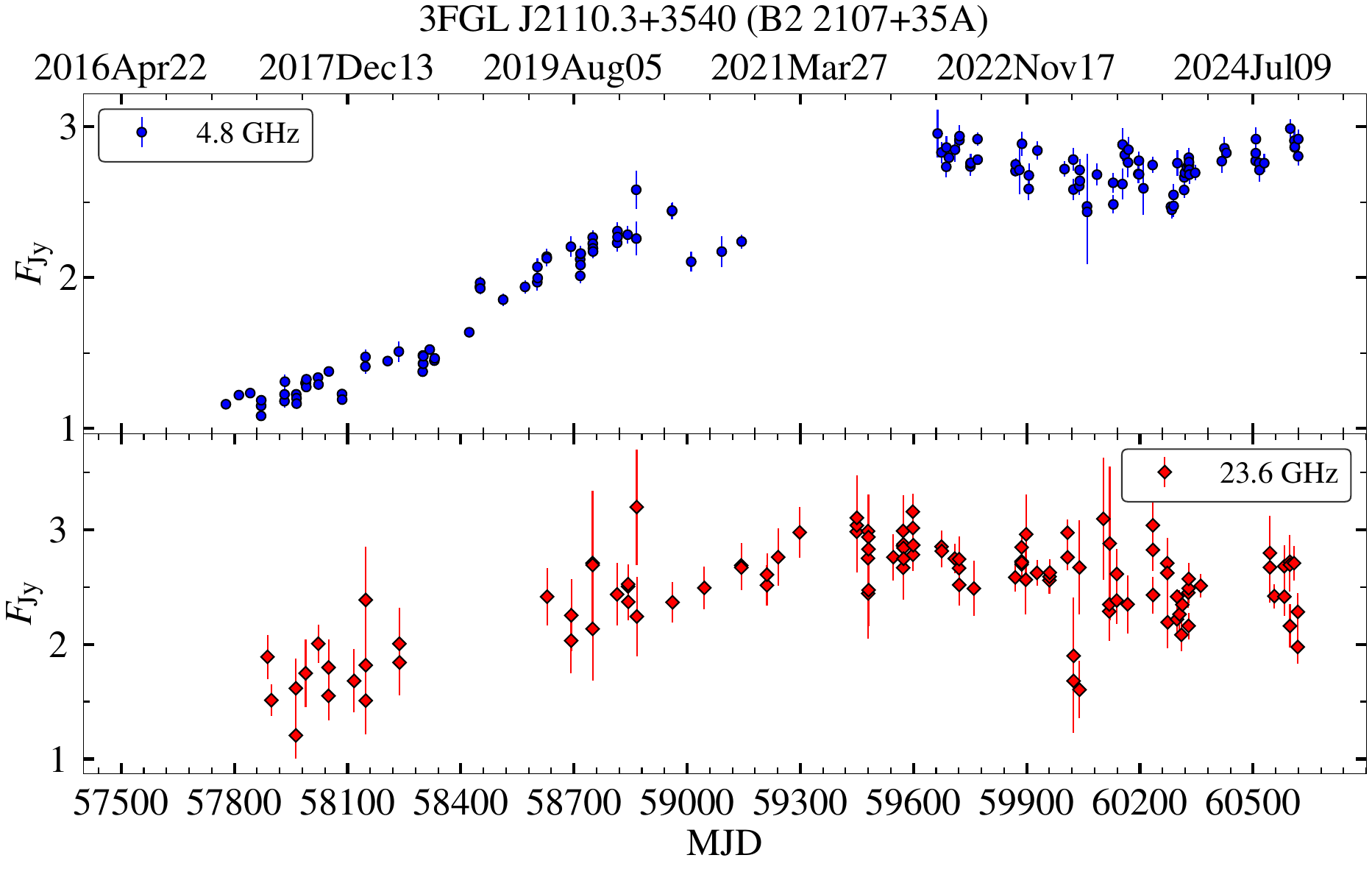}\\
\end{tabular}
\end{figure*}

\begin{figure*}[p]
\centering
\addtocounter{figure}{-1}
\caption{Continued.}
\begin{tabular}{cc}
\includegraphics[width=0.49\textwidth]{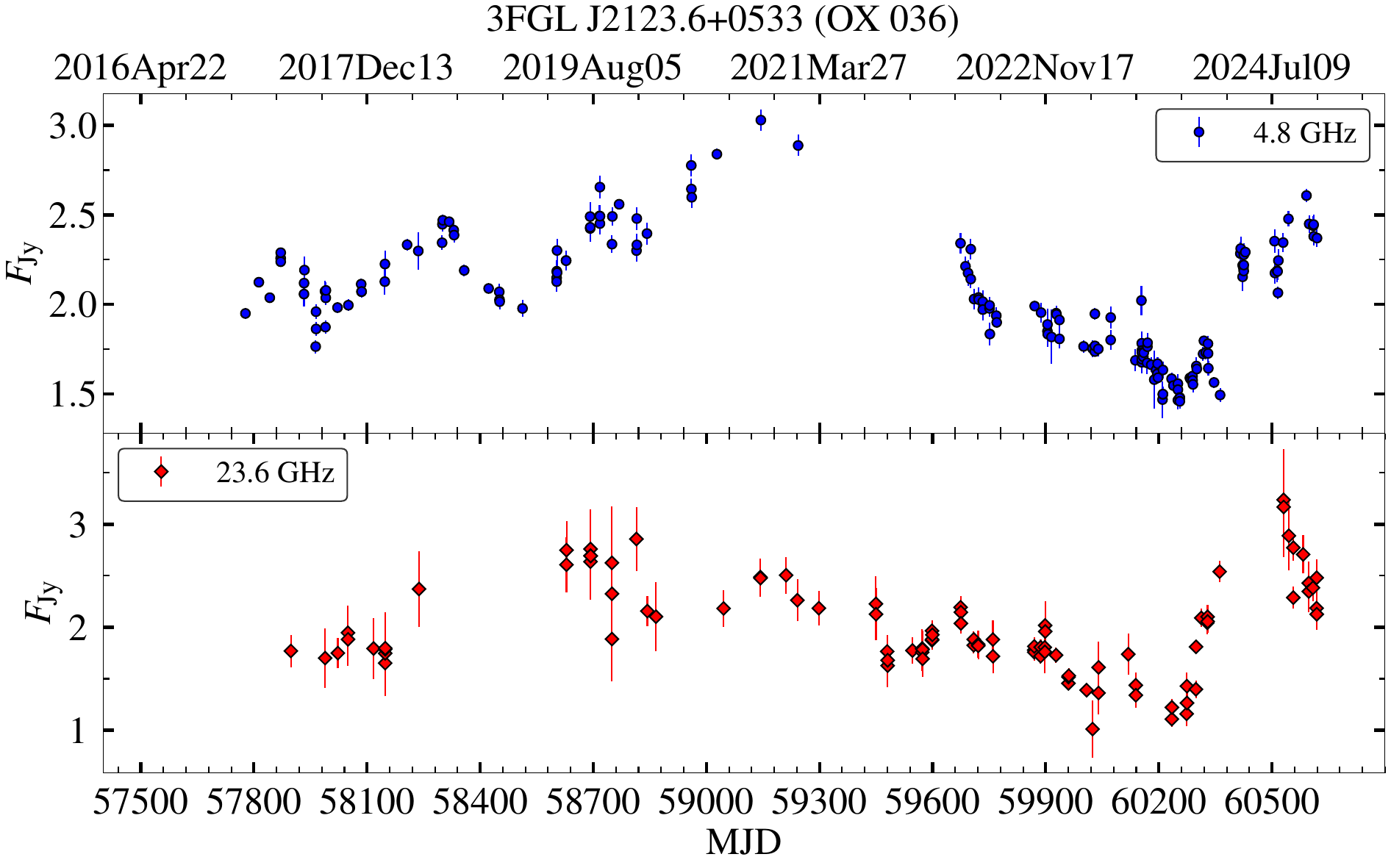}
\includegraphics[width=0.49\textwidth]{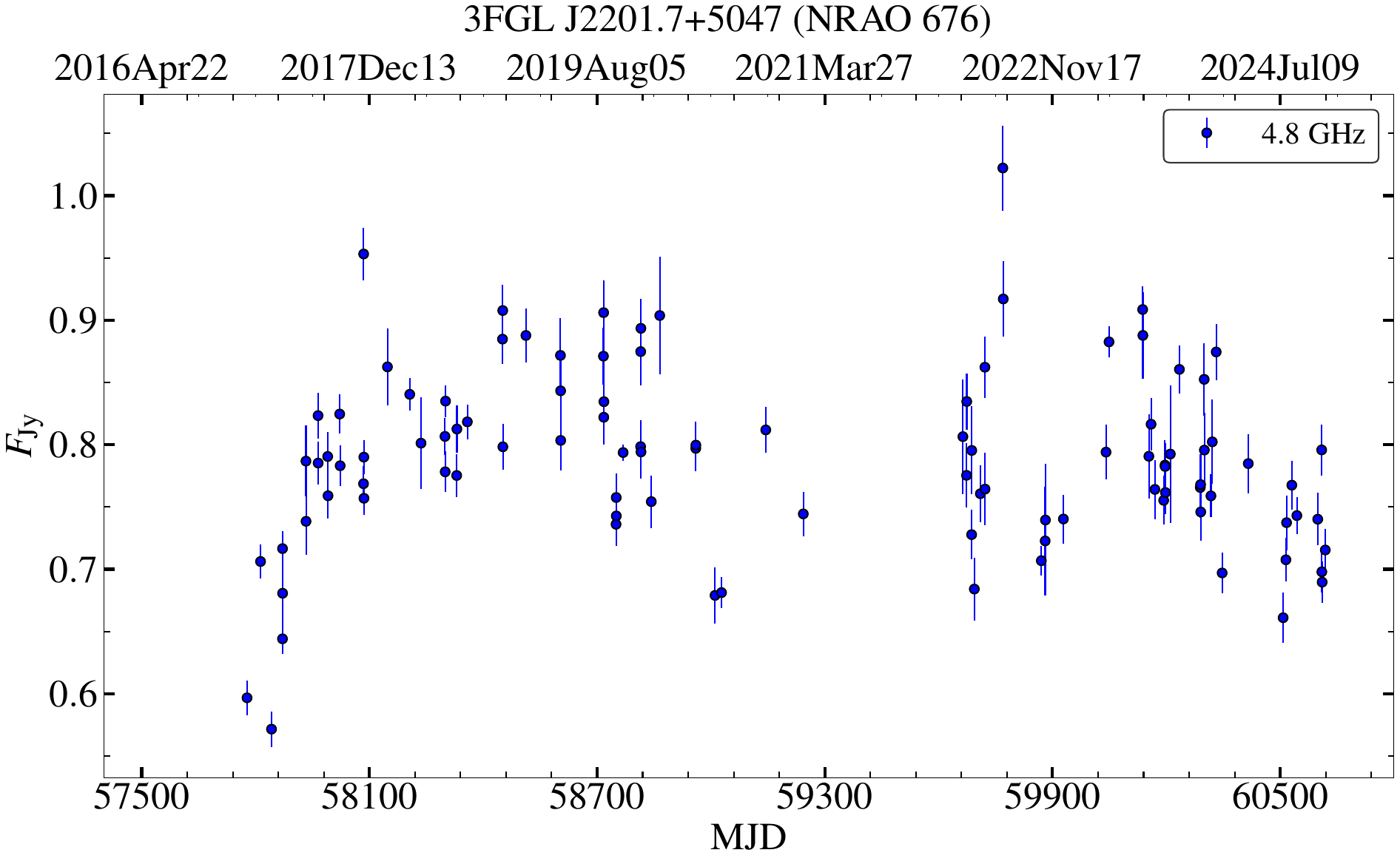}\\
\includegraphics[width=0.49\textwidth]{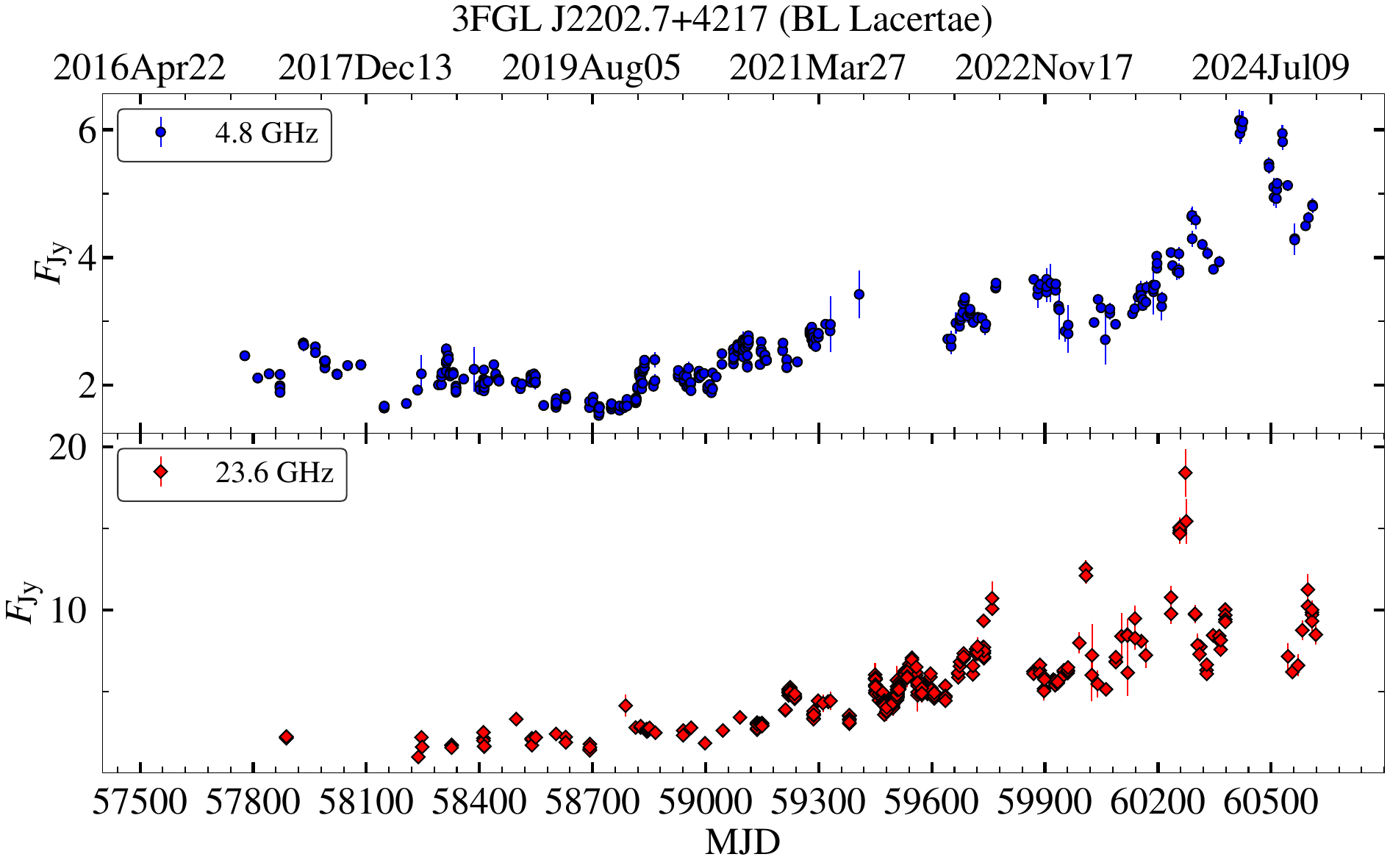}
\includegraphics[width=0.49\textwidth]{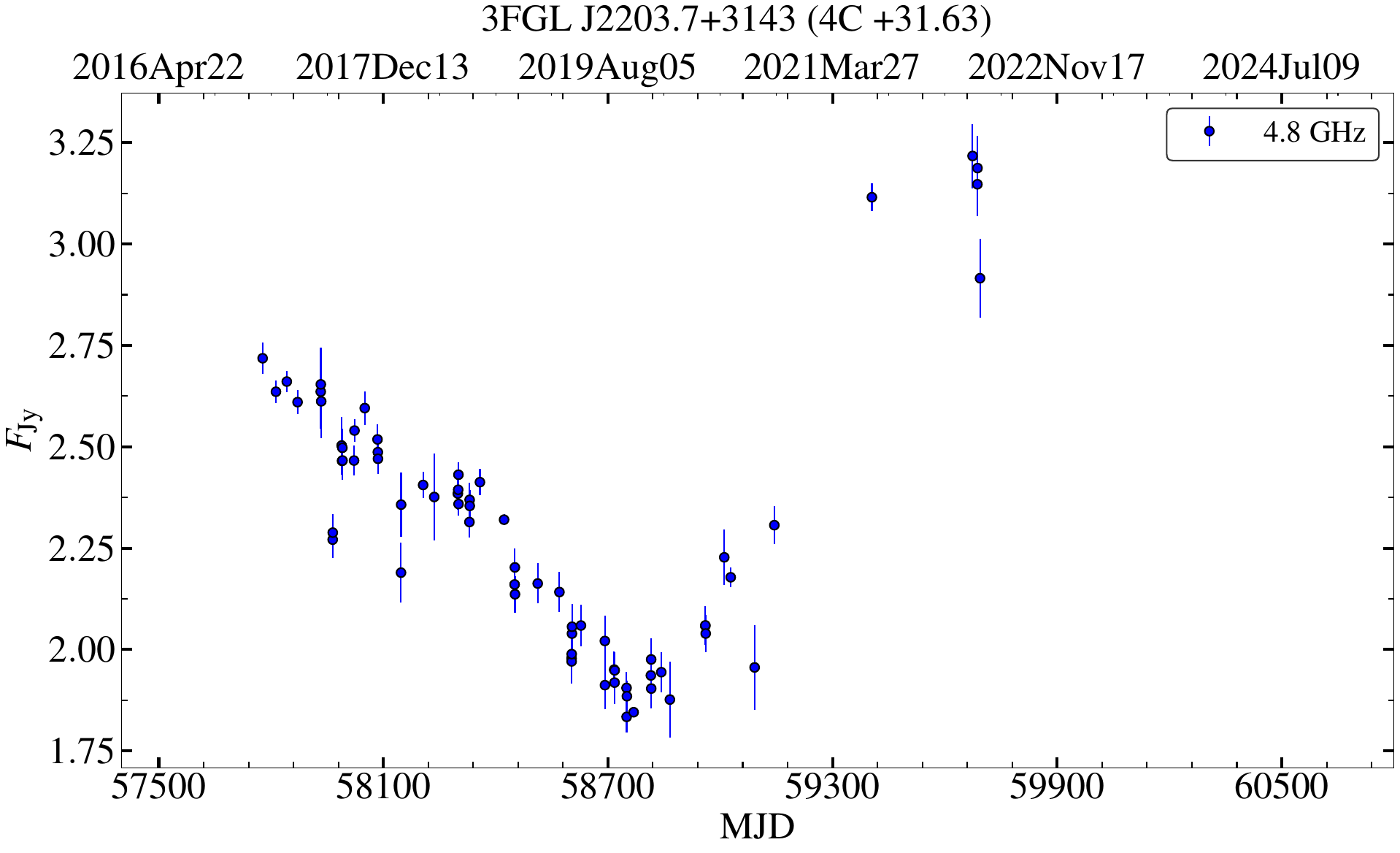}\\
\includegraphics[width=0.49\textwidth]{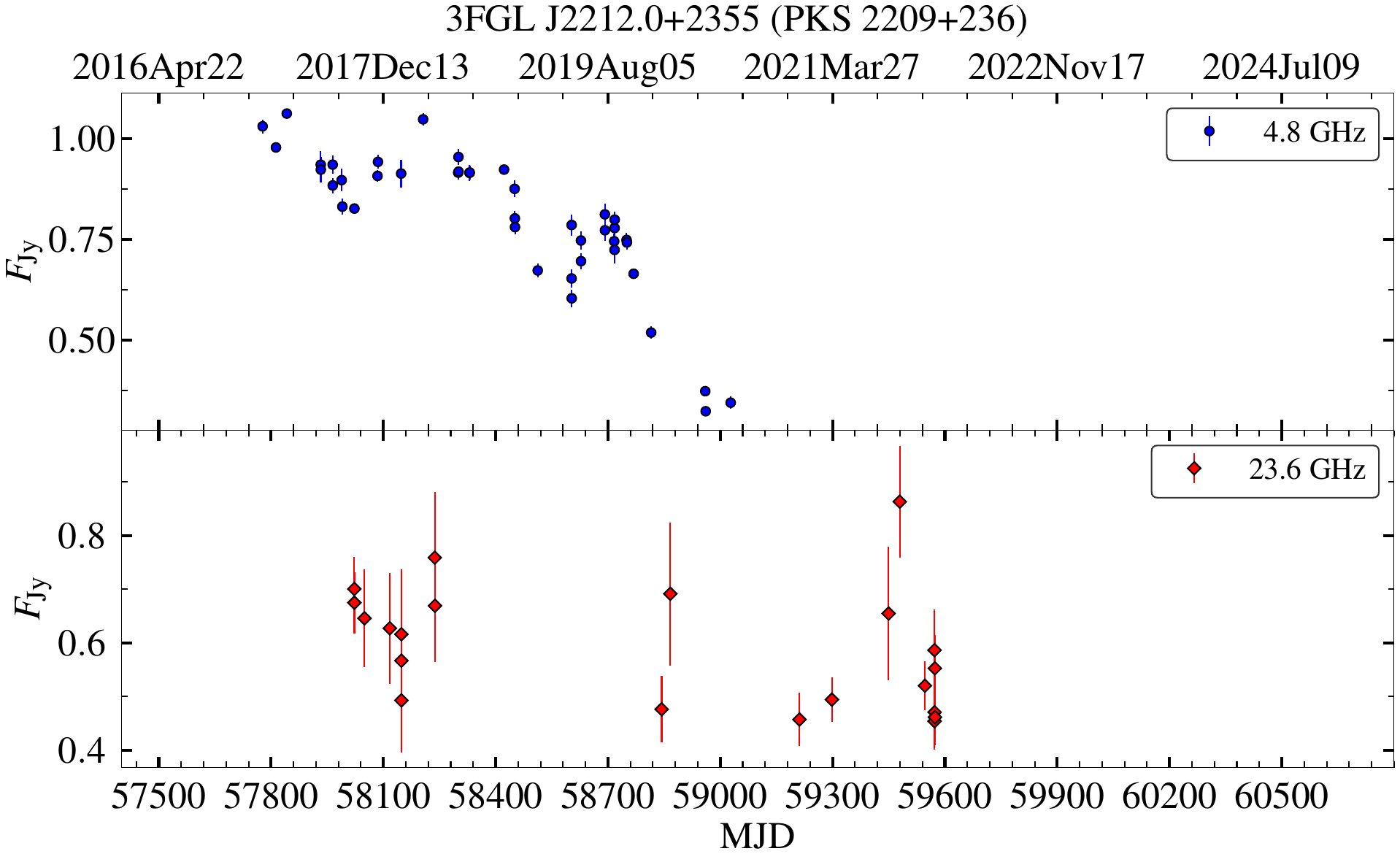}
\includegraphics[width=0.49\textwidth]{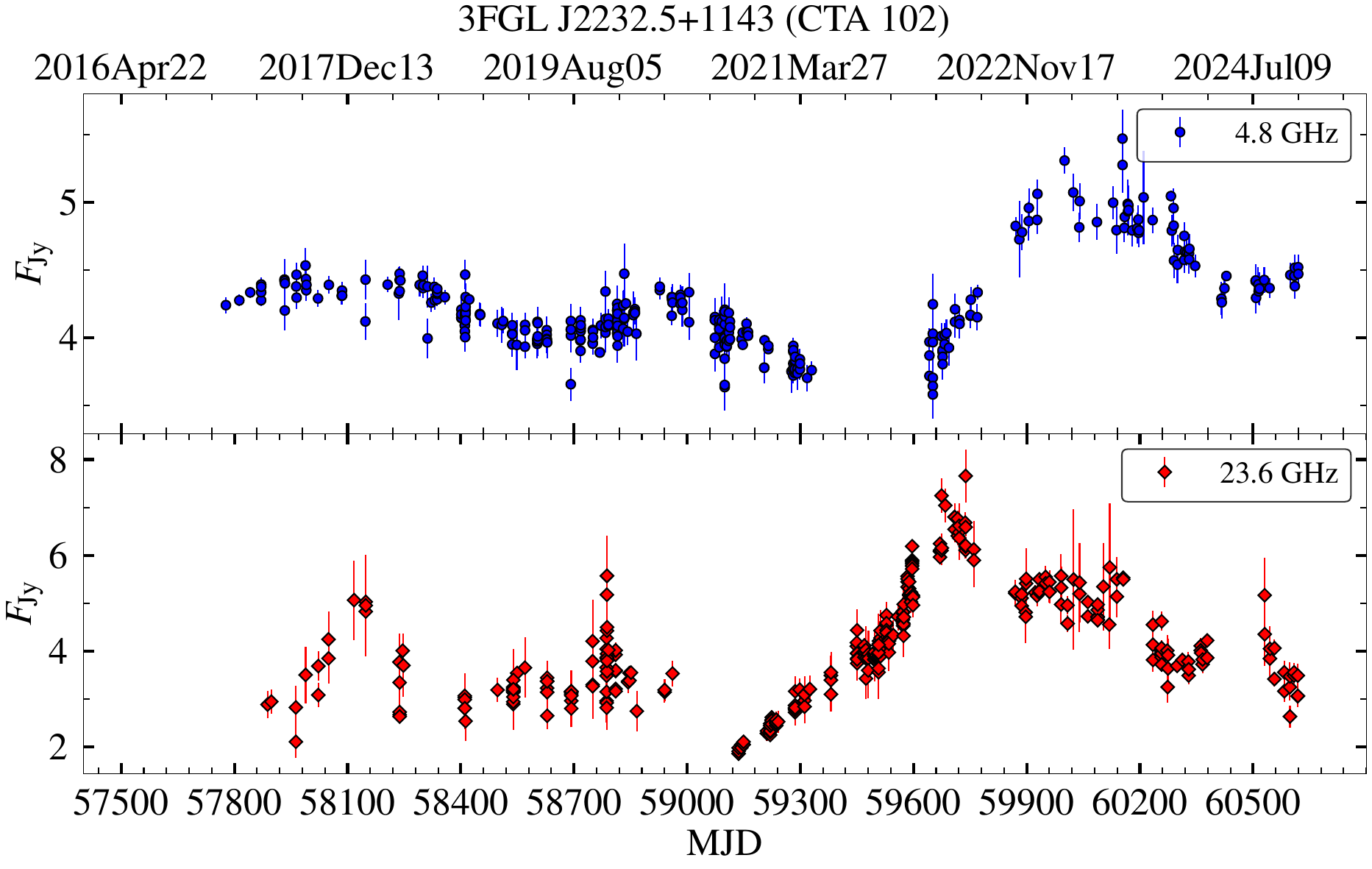}\\
\includegraphics[width=0.49\textwidth]{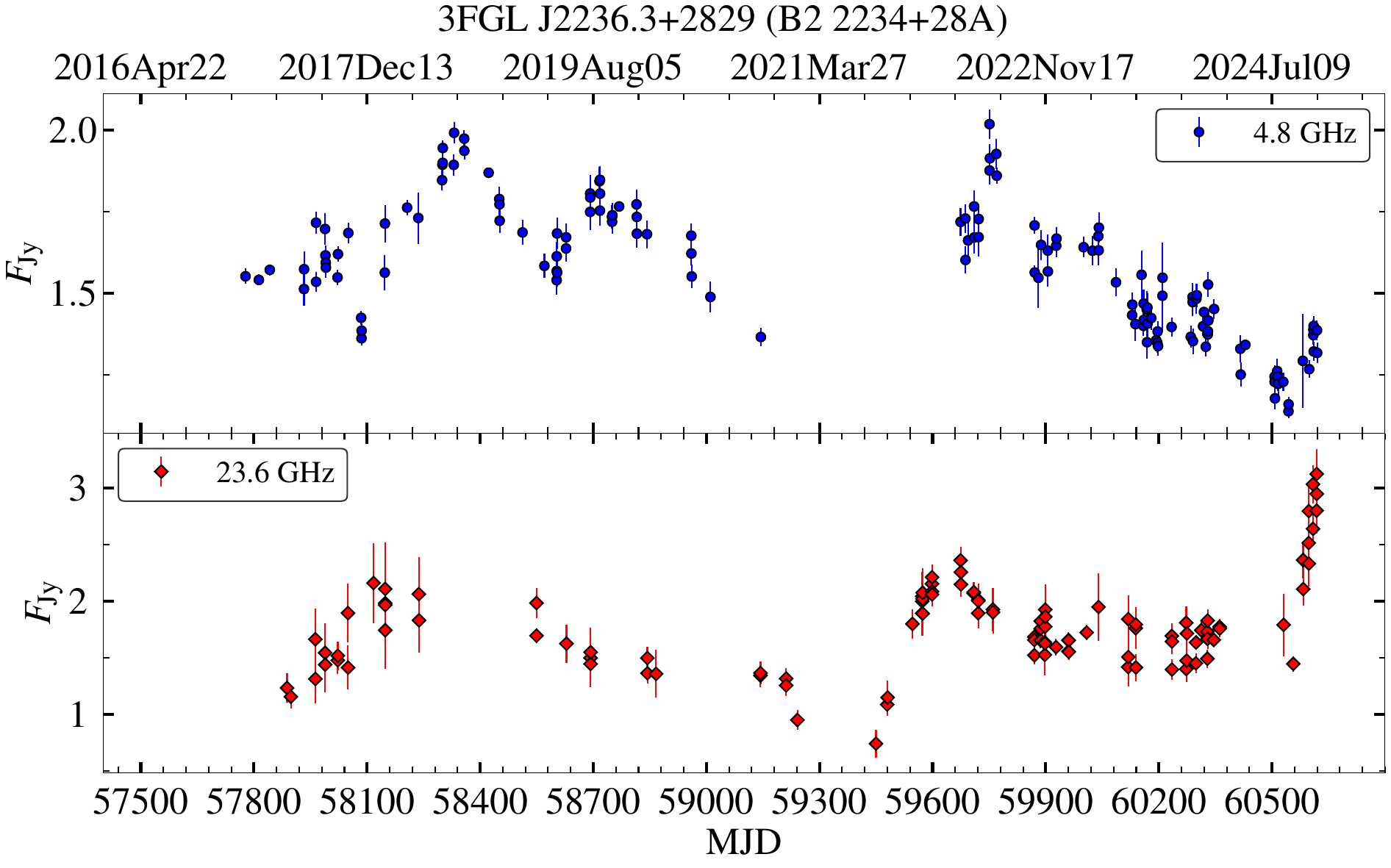}
\includegraphics[width=0.49\textwidth]{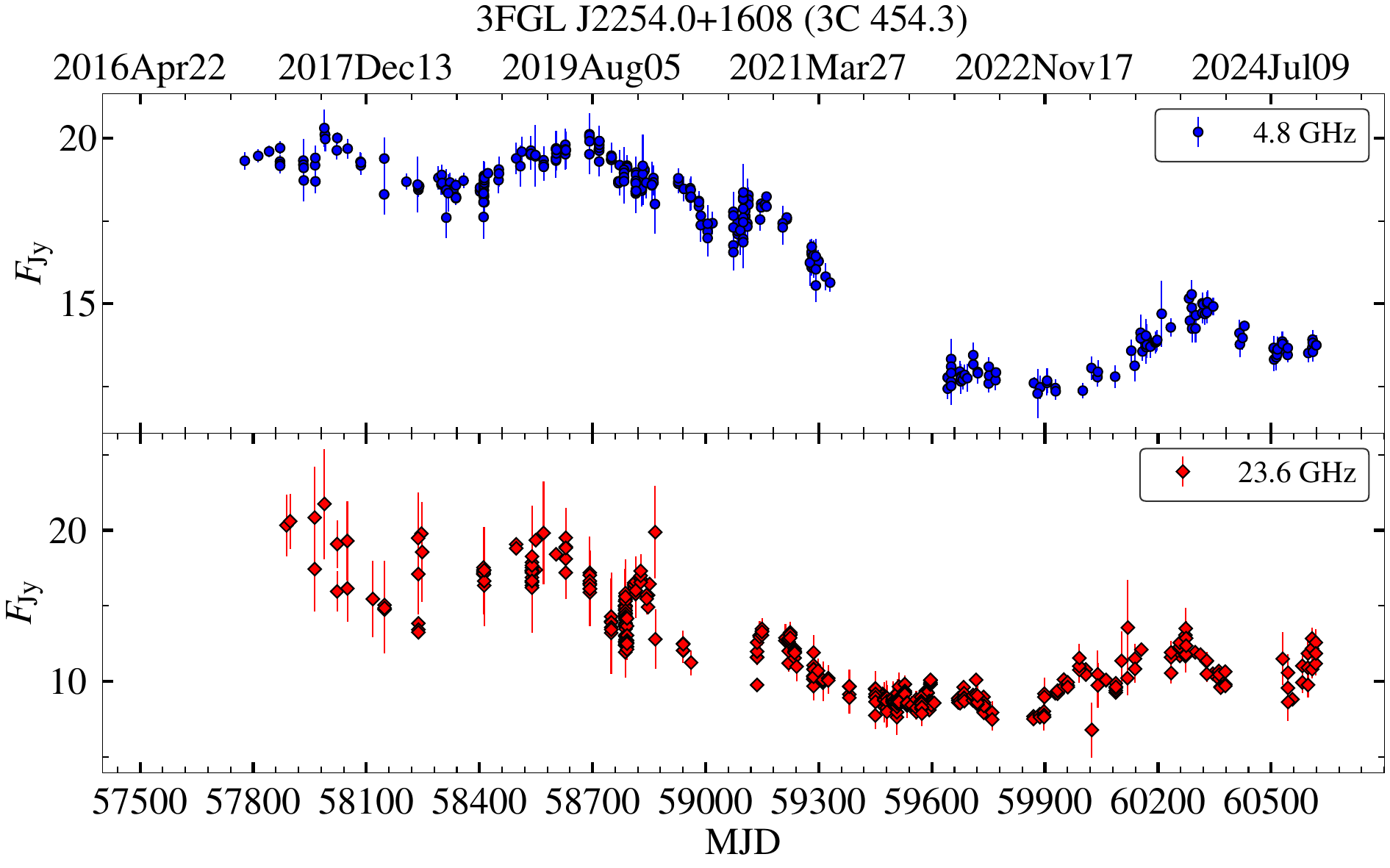}\\
\end{tabular}
\end{figure*}

\begin{figure*}[p]
\centering
\addtocounter{figure}{-1}
\caption{Continued.}
\begin{tabular}{cc}
\includegraphics[width=0.49\textwidth]{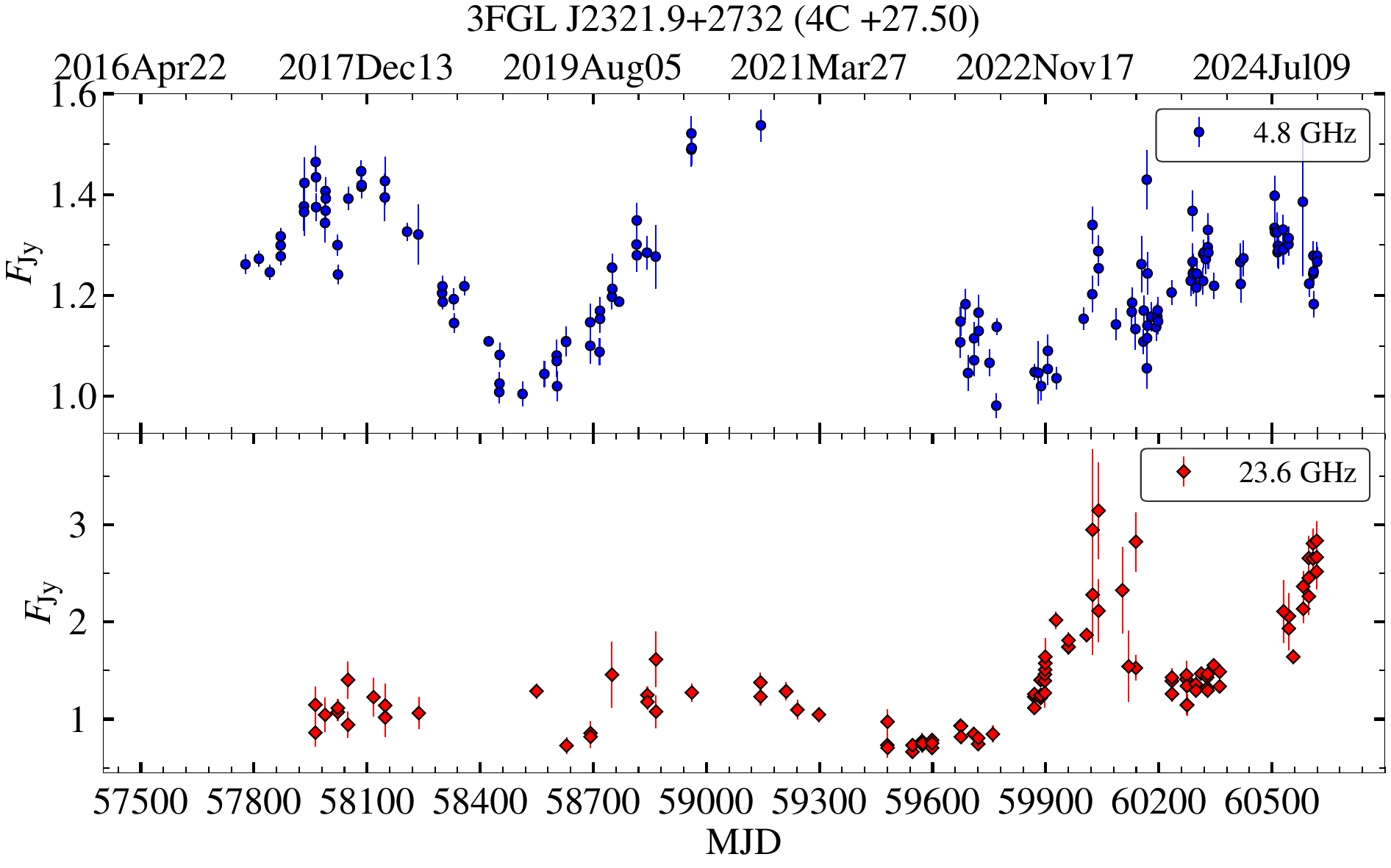}
\includegraphics[width=0.49\textwidth]{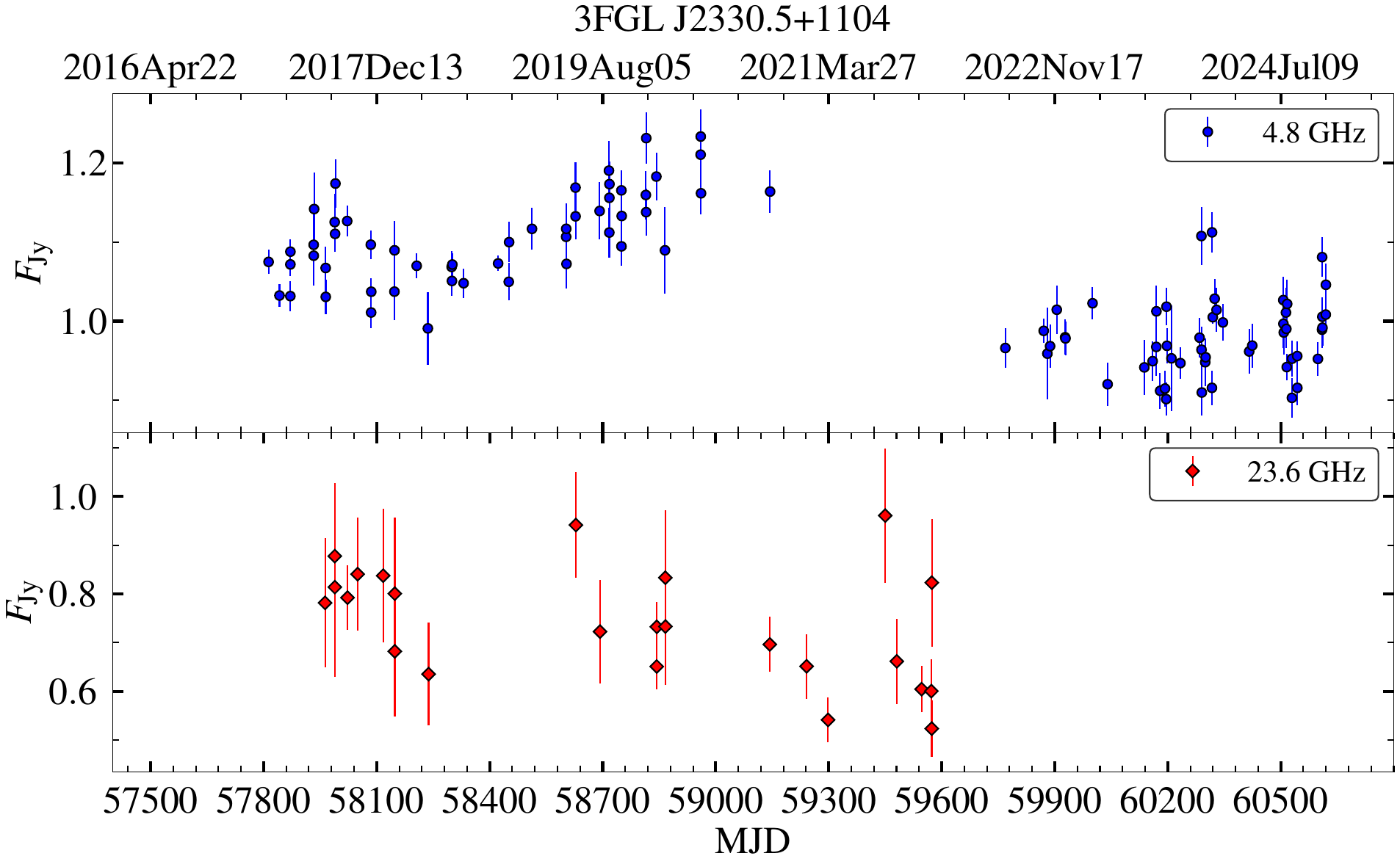}\\
\includegraphics[width=0.49\textwidth]{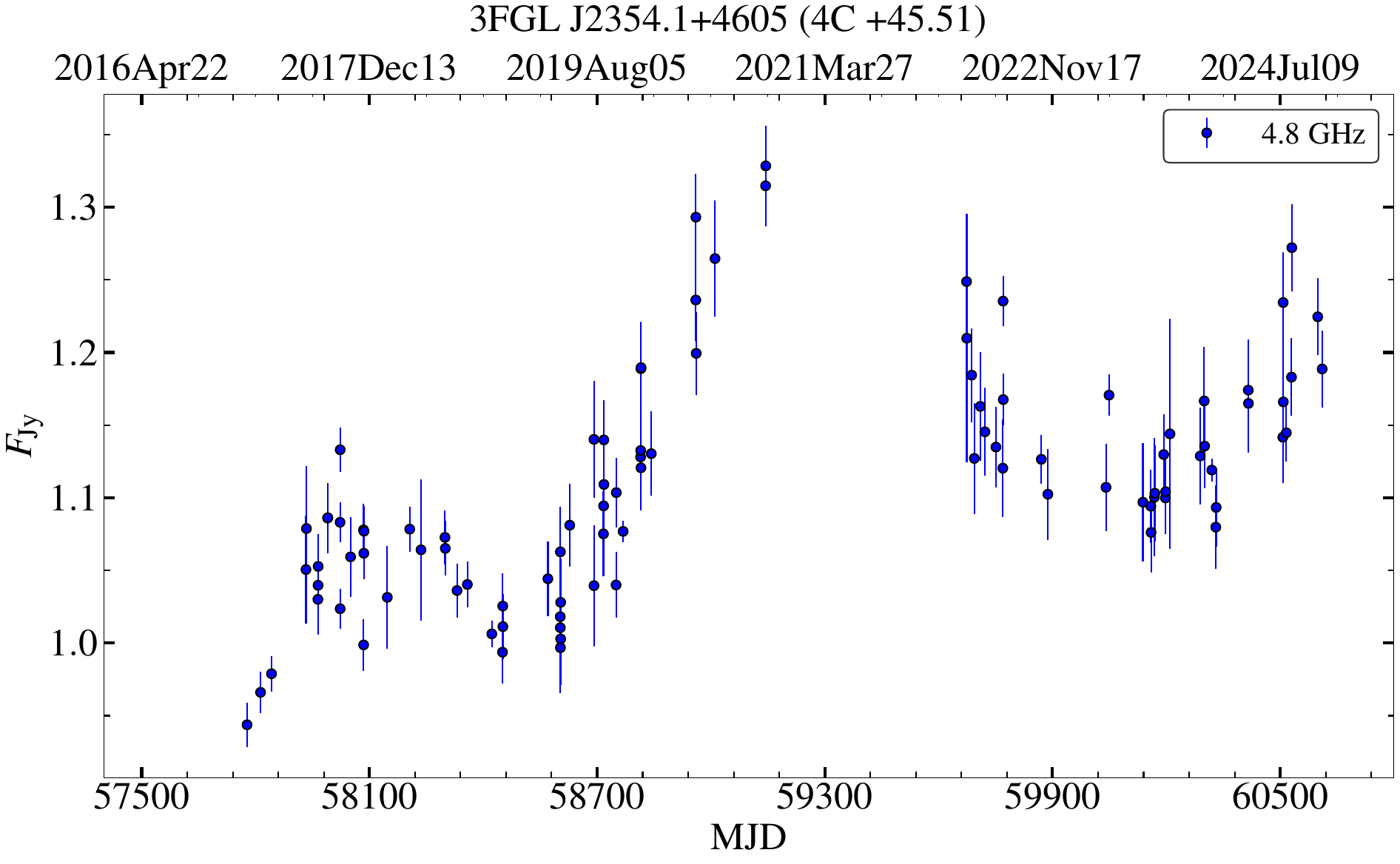}
\includegraphics[width=0.49\textwidth]{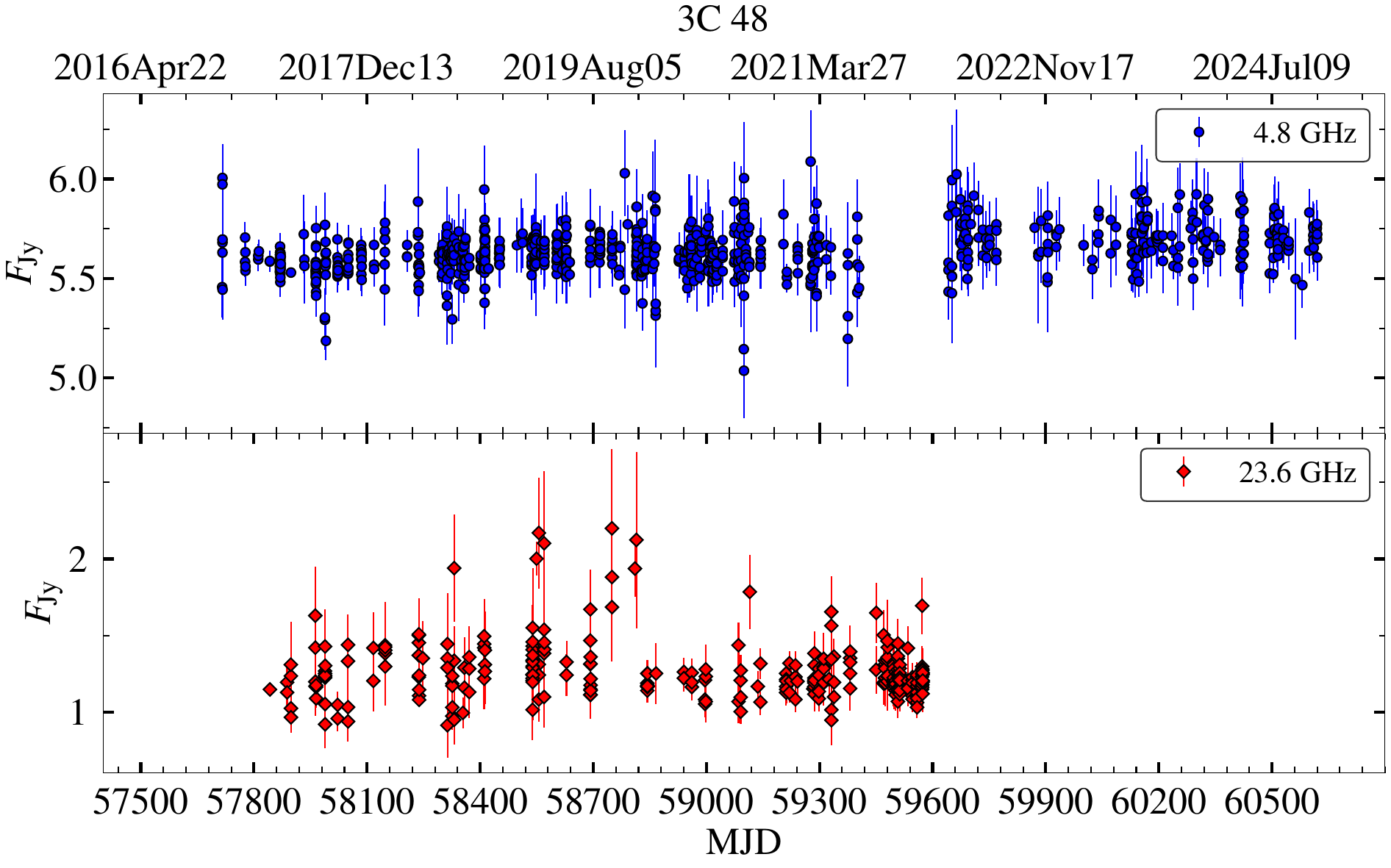}\\
\includegraphics[width=0.49\textwidth]{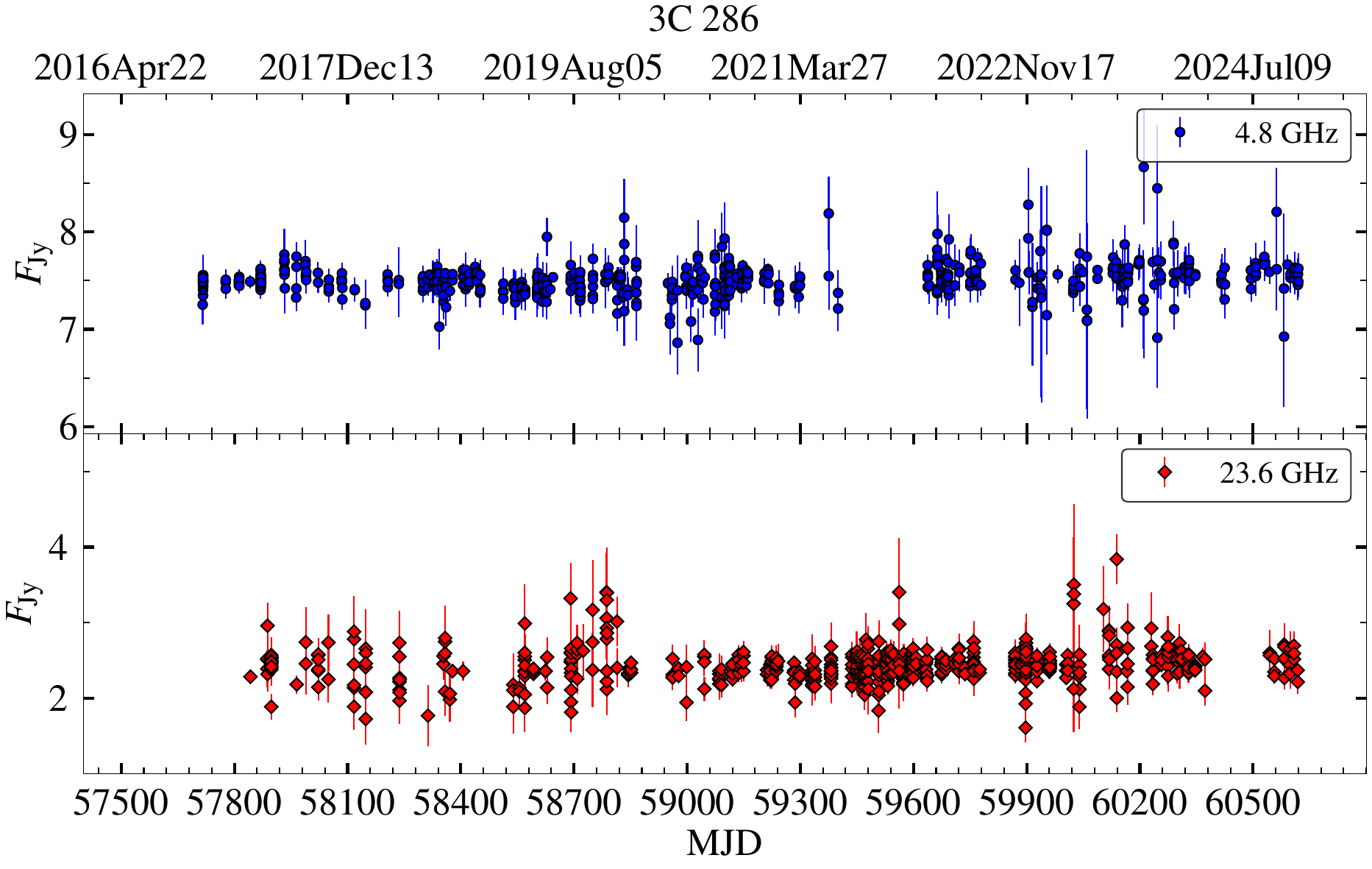}
\includegraphics[width=0.49\textwidth]{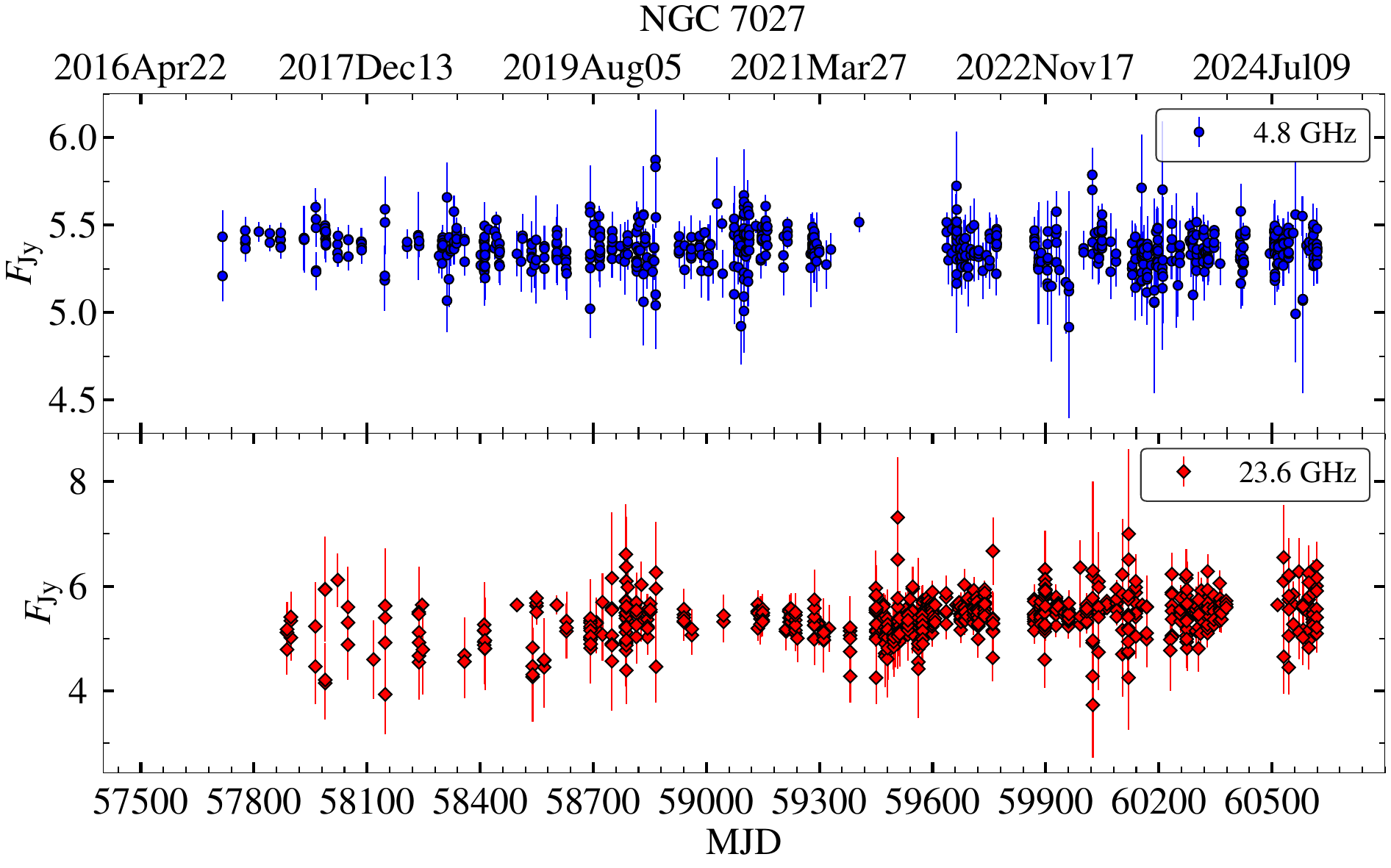}\\
\end{tabular}
\end{figure*}

%%%%%%%%%%%%%%%%%%%%%%%%%%%%%%%%%%%%%%%%%%%%%%%%%%%%%%%%%%%%%%%%%%%%%%

\bibliography{references}{}
\bibliographystyle{aasjournal}

%% This command is needed to show the entire author+affiliation list when
%% the collaboration and author truncation commands are used.  It has to
%% go at the end of the manuscript.
%\allauthors

%% Include this line if you are using the \added, \replaced, \deleted
%% commands to see a summary list of all changes at the end of the article.
%\listofchanges

\end{document}